\documentclass[11pt,a4paper]{article}

\usepackage[utf8]{inputenc}
\usepackage{fullpage}
\usepackage{graphicx}
\usepackage{slashed}
\usepackage{amsmath}
\usepackage{hyperref}
\usepackage[T1]{fontenc}
\usepackage{subcaption}
\usepackage{amssymb}
\usepackage{cite}

\graphicspath{{figures/}}
 
\def\beq{\begin{eqnarray}}
\def\eeq{\end{eqnarray}}
\def\eq#1{Eq.~(\ref{#1})}
\def\bra#1{%
  \left\langle\smash{#1}{\vphantom1}\right|}
\def\ket#1{%
  \left|\smash{#1}{\vphantom1}\right\rangle}
  \def\slash#1{#1 \hskip-0.45em /}

\newcommand{\secn}[1]{Section~\ref{#1}}
\newcommand{\as}{\ensuremath{\alpha_s}}
\newcommand{\eps}{\ensuremath{\epsilon}}
\newcommand{\e}{\ensuremath{\epsilon}}

\newcommand{\zbar}{\bar{z}}
\newcommand{\barpartial}{\partial_{\zbar}}

\newcommand{\ck}{\ensuremath{\mathcal K}}
\newcommand{\del}{\ensuremath{\partial}}
\newcommand{\ii}{1}
\newcommand{\jj}{2}
\newcommand{\kk}{3}
\newcommand{\Li}{\ensuremath {\textrm{Li}}}

\newcommand{\co}{\ensuremath{\mathcal O}}
\newcommand{\LO}{{\mbox{\tiny{LO}}}}
\newcommand{\NLO}{{\mbox{\tiny{NLO}}}}
\newcommand{\NNLO}{{\mbox{\tiny{NNLO}}}}
\newcommand{\npo}{{n+1}}
\newcommand{\npt}{{n+2}}
\newcommand{\pn}{\Phi_n}
\newcommand{\one}{\, (\mathbf{1})}
\newcommand{\two}{\, (\mathbf{2})}
\newcommand{\otwo}{\, (\mathbf{12})}
\newcommand{\E}{{\mbox{\tiny{E}}}}
\newcommand{\RV}{\, (\mathbf{RV})}

\newcommand{\ord}{{\cal O}}

\numberwithin{equation}{section}

\begin{document}


\begin{flushright}
  CERN-TH-2021-145 \\ \vspace{3mm}
  \today
\end{flushright}

\vspace{2.2cm}

\centerline{\Large \bf The Infrared Structure of Perturbative Gauge 
Theories\footnote{Partly based on graduate lectures given by L.M. at 
the Indian Institute of Technology in Hyderabad, under the GIAN Initiative 
of the Ministry of Human Resource Development, Government of India.}}

\vspace{12mm}

\centerline{\large Neelima Agarwal$^a$, Lorenzo Magnea$^{b,c}$,}
\vspace{1mm}
\centerline{\large Chiara Signorile-Signorile$^d$ and Anurag Tripathi$^e$}

\vspace{12mm}

\centerline{\textit{$^a$Department of Physics, Chaitanya Bharathi 
  Institute of Technology}}
\centerline{\textit{Gandipet, Hyderabad, 500075, Telangana, India}}

\vspace{2mm}

\centerline{\textit{$^b$Theoretical Physics Department, CERN, CH-1211 
  Geneva 23, Switzerland}}

\vspace{2mm}

\centerline{\textit{$^c$Dipartimento di Fisica and Arnold-Regge Center, 
  Universit\`a di Torino}}
\centerline{\textit{and INFN, Sezione di Torino}}
\centerline{\textit{Via P. Giuria 1, I-10125 Torino, Italy}}

\vspace{2mm}

\centerline{\textit{$^d$Institut f\"ur Astroteilchenphysik,
  Karlsruher Institut f\"ur Technologie (KIT),}}
\centerline{\textit{D-76021 Karlsruhe, Germany}}

\vspace{2mm}

\centerline{\textit{$^e$Department of Physics, IIT Hyderabad}}
\centerline{\textit{Kandi, Sangareddy, 502284, Telangana, India}}

\vspace{1.5cm}

\begin{abstract}

\noindent
Infrared divergences in the perturbative expansion of gauge theory amplitudes
and cross sections have been a focus of theoretical investigations for almost
a century. New insights still continue to emerge, as higher perturbative orders
are explored, and high-precision phenomenological applications demand an
ever more refined understanding. This review aims to provide a pedagogical
overview of the subject. We briefly cover some of the early historical results,
we provide some simple examples of low-order applications in the context of 
perturbative QCD, and discuss the necessary tools to extend these results to 
all perturbative orders. Finally, we describe recent developments concerning 
the calculation of soft anomalous dimensions in multi-particle scattering 
amplitudes at high orders, and we provide a brief introduction to the very 
active field of infrared subtraction for the calculation of differential distributions 
at colliders.
\end{abstract}


\newpage


\tableofcontents


\section{Introduction}
\label{Intro}

Students taking introductory classes in quantum field theory may be forgiven 
for believing, after perhaps the first half of their course, that they hold the keys
for a full understanding of particle interactions and countless phenomenological 
applications. Granted that coupling constants are not too big, armed with the
Feynman rules they have just derived, they may imagine that what lies ahead
is largely to learn a set of technical tools, to speed up calculations that they 
already have a fair idea of how to perform. This delusion is of course soon to 
be shattered when they are faced with their first radiative corrections, and 
they realise that `applying the rules' leads to apparently non-sensical results, 
uncovering serious conceptual problems, which need to be patiently and creatively 
addressed. This revelation strikes students, for example, when they reach Chapter 
6 of Ref.~\cite{Peskin:1995ev}, or Chapter 9 of Ref.~\cite{Sterman:1994ce}. 
Historically, the ultraviolet problem was indeed baffling enough that Nobel-Prize 
winning theoreticians went to their grave convinced that the entire edifice of 
quantum field theory should be discarded because of it~\cite{Farmelo}.

A more constructive point of view could be described as follows. If we have 
reason to believe that a certain quantum field theory is relevant to physics,
either because of its symmetry and aesthetic appeal, or because it is confirmed
by experiment, and yet we find that our calculations yield infinite answers,
the most likely explanation is surely that we {\it made a mistake}, and one of our
approximations is failing. It pays to look for the mistake, and try to patch the 
approximation, lest we `throw out the baby with the bath water'. As a matter 
of fact, finding, understanding and fixing such mistakes has invariably brought 
great progress, and a wealth of deeper physical understanding.

To illustrate this fact, consider a quantity $R$, that one assumes can be 
computed in perturbation theory, depending on some physical energy scale 
$Q$ and on a small dimensionless coupling $\alpha$. Our student would expect
that increasingly precise approximations to this quantity would take the form
\beq
  R (Q, \alpha) \, = \, R_0 \left[ 1 + \frac{\alpha}{\pi} \, c_1 (Q) +
  \left( \frac{\alpha}{\pi} \right)^2 c_2 (Q) + \ldots \right] \, ,
\label{naivepert}
\eeq
upon computing successive coefficients $c_k (Q)$. If the quantity $R$ really
depends only on a single scale $Q$, one further expects that $c_k$ must be 
independent of $Q$, on dimensional grounds. As we know, these expectations fail
drastically for almost all interesting theories, where one finds that the
expressions for $c_k$ are given by ultraviolet divergent integrals and are thus 
meaningless. Artificially introducing an ultraviolet cutoff $\Lambda$ typically
reveals that $c_k \sim \log^k (\Lambda/Q)$. The question then is: {\it what was 
our mistake}? In this case, it was {\it hybris}, an excess of confidence: we assumed 
that our theory would remain applicable to arbitrarily large energy scales, and to
arbitrarily small length scales; or, more precisely, we assumed that the effect
of large-energy, short-distance degrees of freedom would be negligible for 
our observable. This assumption is defeated by the rules of quantum mechanics,
which require us to sum over all unobserved field fluctuations, including those
that are far from classical -- far off-shell. Such fluctuations are of course 
suppressed in the path integral, but it turns out that, in $d=4$ space-time 
dimensions, there are just too many of them. This fact looks dangerous indeed: 
if unknown Planck-scale physics significantly affects our low-energy laboratory 
experiments, we could be facing a complete loss of predictivity. Fortunately, 
we have a well-established solution to this problem, provided by renormalisation, 
and more generally by the idea of effective field theory (see, for example,
\cite{Lepage:1989hf,Neubert:2019mrz}). We exploit the fact that ultraviolet 
contributions to our integrals correspond to highly localised fluctuations in 
space-time, which allows us to {\it parametrise our ignorance} of the UV completion 
of our theory by absorbing the unknown UV effects in the lagrangian couplings. 
Even when this is not sufficient (the theory is not renormalisable), in most cases 
we can still understand our theory in terms of a low-energy expansion, and 
increase the accuracy of our predictions by adding higher-dimensional operators 
to the Lagrangian. Operationally, we introduce a new scale $\mu$, which we 
may take to represent the boundaries of our current knowledge, and we promote 
\eq{naivepert} to
\beq
  R \left( \frac{Q}{\mu}, \alpha(\mu) \right) \, = \, R_0 \left[ 1 + 
  \frac{\alpha(\mu)}{\pi} \,\, c_1 \! \bigg( \frac{Q}{\mu} \bigg) +
  \bigg( \frac{\alpha (\mu)}{\pi} \bigg)^{\! 2} c_2 \bigg( \frac{Q}{\mu} \bigg) + 
  \ldots \right] \, .
\label{renpert}
\eeq
The perturbative coefficients now have a new, non-trivial scale dependence,
but this is tamed by renormalisation group equations, enforcing the independence
of physical observables on the artificial boundary set by $\mu$. Understanding 
our {\it UV mistake} has paid off handsomely: we now see that unknown high-energy 
physics does not necessarily spoil predictivity, we have at our disposal the tools
of the renormalisation group to study asymptotic behaviours, and the powerful
arsenal of effective field theories allows us to tackle systematically many problems
that defied earlier techniques.

\begin{figure}
    \centering
    \begin{subfigure}{0.3\textwidth}
    \centering
        \includegraphics[width=0.7\textwidth]{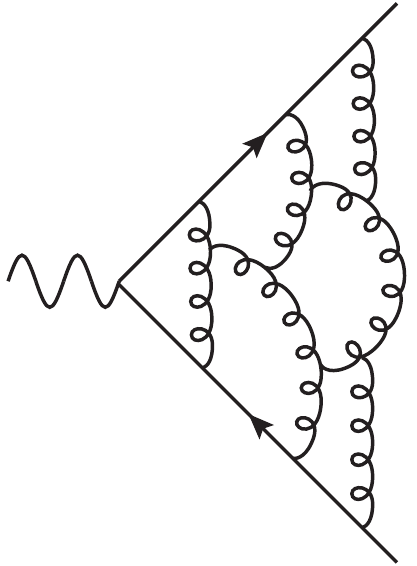}
        \caption{}
        \label{fig:LargeOrder_a}
    \end{subfigure}
    \quad 
      \begin{subfigure}{0.3\textwidth}
      \centering
        \includegraphics[width=0.7\textwidth]{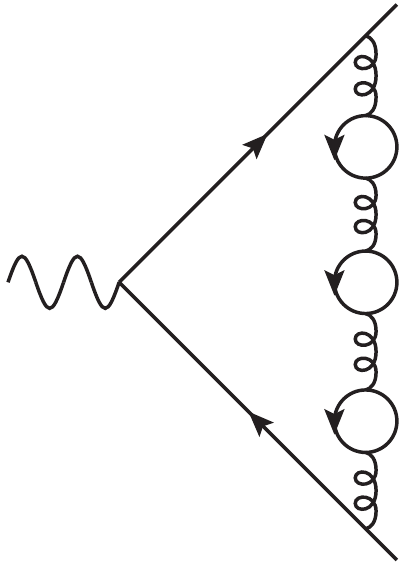}
        \caption{}
        \label{fig:LargeOrder_b}
    \end{subfigure}
      \caption{Examples of Feynman graphs representative of the factorial growth
  of perturbative coefficients at large orders.}
  \label{LargeOrder}
\end{figure}

A similar story can be told about another problem afflicting the naive expectations 
expressed by \eq{naivepert}, a problem which many students may not meet 
in their early courses at all. For almost all interesting quantities, it is a well-known 
fact~\cite{Dyson:1952tj,LeGuillou:1990nq} that the asymptotic behavior of the 
perturbative coefficients $c_k$ in \eq{naivepert} and in \eq{renpert} for large 
$k$ is $c_k \sim k!$, implying that the perturbative series is divergent -- in fact, 
it has vanishing radius of convergence, and it is often not amenable to be 
handled with standard summation methods, such as Borel summation. It is
not difficult to find evidence for this factorial growth in large-order Feynman 
graphs: first of all, the number of graphs grows factorially with the order, as 
illustrated in Fig.~\ref{fig:LargeOrder_a}; next, for example, evaluating graphs 
involving chains of fermion loops, such as the one shown in Fig.~\ref{fig:LargeOrder_b}, 
also results in factorial growth~\cite{Lautrup:1977hs}.
Once again, one may legitimately fear a loss of predictivity, and one may ask --
assuming one believes in the consistency of the underlying theory -- {\it what was
the mistake} that lead to the problem. Once again the answers are rich with 
insights and discoveries: in this case, the underlying assumption was that 
non-perturbative effects would be negligible, or in some sense decoupled from
the perturbative expansion. This, quite interestingly, turns out not to be the case.
The factorial growth in the number of diagrams can be precisely related to the 
existence of non-trivial classical solutions -- instantons -- which are not accessible
by means of a weak-coupling expansion~\cite{Bender:1973rz,Bender:1978vg};
renormalon singularities, which can be detected by means of fermion-loop 
chains, reveal contributions to observables originating from vacuum 
condensates, in many cases matching the results of the operator product 
expansion~\cite{Beneke:1998ui}. Remarkably, perturbation theory seems to
know about its own limitations, and can be used to infer the existence of 
non-perturbative effects and study their impact. While these insights are 
several decades old, new powerful techniques are currently being developed 
to refine existing tools and consolidate their mathematical foundations (see
for example~\cite{Marino:2012zq,Aniceto:2018bis,Costin:2020hwg}).

Finally, we come to the subject of our review. Even after renormalisation, and 
even if we confine ourselves to perturbative effects, \eq{renpert} is still `wrong'
for most interesting theories with massless particles (and in particular for 
unbroken gauge theories in $d = 4$): expressions for the coefficients 
$c_k(Q/\mu)$ involve integrals that diverge at low energies (in momentum 
space) or at large distances (in coordinate space). Furthermore, if we attempt 
to compute the probability for the emission of one or more massless particles 
in connection with a hard scattering process, this will typically also diverge.
This {\it infrared catastrophe} was actually the first problem to be clearly identified, 
among those we have discussed. Long before quantum field theory was fully 
developed, some of the earliest studies of the interactions of electrons with 
electromagnetic radiation showed that the frequency spectra of emitted photons
behave like $d \nu/\nu$, and thus are not integrable at low frequencies. This
emerged from analyses of electron scattering in a Coulomb field, when allowing
for extra photon radiation~\cite{Mott:1931,Sommerfeld:1931}, with results
subsequently refined in~\cite{Bethe:1934za}, where pair production in the 
Coulomb field was also considered.

Following the logic we proposed, we may ask one last time {\it what was the 
mistake} that led to this problem. The answer to this question has several layers 
of depth, that we will explore in the rest of this review, but it is interesting,
and truly remarkable, that a precise understanding of the underlying physical 
problem was developed as early as 1937, with the seminal paper by 
Bloch and Nordsieck~\cite{Bloch:1937pw}. It is worthwhile to reproduce here
the Abstract of that paper, which reads as follows.
\bigskip
\begin{center}
\parbox{0.9\textwidth}{\small 
Previous methods of treating radiative corrections in non-stationary processes 
such as the scattering of an electron in an atomic field or the emission of a 
$\beta$-ray, by an expansion in powers of $e^2/\hbar c$, are defective in that 
they predict infinite low frequency corrections to the transition probabilities. 
This difficulty can be avoided by a method developed here which is based 
on the alternative assumption that $e^2 \omega/m c^3$, $\hbar \omega/m c^2$ 
and $\hbar \omega/ c \Delta p$ ($\omega=$angular frequency of radiation, 
$\Delta p=$change in momentum of electron) are small compared to unity. 
In contrast to the expansion in powers of $e^2/\hbar c$, this permits the 
transition to the classical limit $\hbar = 0$.  External perturbations on the 
electron are treated in the Born approximation. It is shown that for frequencies 
such that the above three parameters are negligible the quantum mechanical 
calculation yields just the directly reinterpreted results of the classical formulae, 
namely that the total probability of a given change in the motion of the electron 
is unaffected by the interaction with radiation, and that the mean number of 
emitted quanta is infinite in such a way that the mean radiated energy is equal 
to the energy radiated classically in the corresponding trajectory.}
\end{center}
\bigskip
In essence, Bloch and Nordsieck argue that, in the soft photon limit, perturbation
theory (in powers of $\alpha= e^2/(\hbar c)$) must be abandoned, and they 
advocate a different approximation scheme (what we now call {\it eikonal 
approximation}), valid when the photon energy is much smaller than the
other energy scales in the problem (the electron mass and the momentum 
trasfer), and the photon wavelength is much larger than the classical electron 
radius $r_e = e^2/(m c^2)$. They then show that this approximation is 
semiclassical, in that the classical result for the mean radiated energy is
recovered, but this entails the radiation of an infinite number of low-energy
photons.

The Bloch and Nordsieck paper is truly remarkable because it engineers the
cancellation of divergences between virtual corrections and real-radiation
contributions, long before the treatment of virtual corrections could be formalised,
and furthermore it provides the first example of an all-order summation
of perturbation theory. In the following decades, it received several refinements,
which proved the complete generality of the results, placed it squarely in 
the context of (renormalised) QED~\cite{Jauch:1954,Yennie:1961ad}, and 
significantly streamlined the proof~\cite{Grammer:1973db}. We will
briefly summarise the argument, in modern language, in \secn{cataQED}.
We still need, however, to properly {\it diagnose the mistake} that lies at the origin 
of the problem: in this regard, there are two complementary viewpoints.

\begin{figure}
\centering
  \includegraphics[scale=0.9]{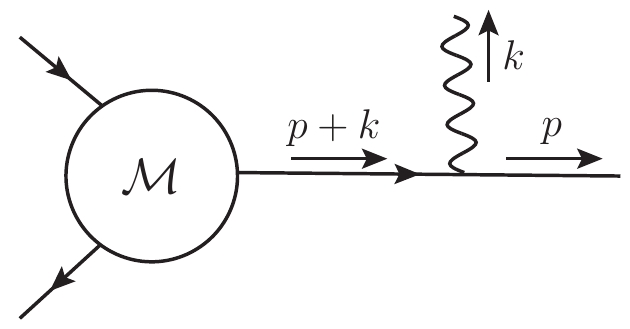}
  \caption{A photon emission diagram in massless QED, possibly responsible for 
  infrared divergences in soft and collinear configurations.}
\label{EasyIR}
\end{figure}

First, one can argue that the problem is {\it the proper definition of an observable}. 
In a theory with massless particles, in any scattering process one can produce 
particles with infinitesimal energy, as well as particles with infinitesimal angular 
separation; on the other hand, every physical detector has finite energy and 
angular resolutions. Quantum mechanics prescribes that we sum over all
configurations we do not observe, so, in principle, an arbitrary number of low-energy
or collinear particles must be included in a proper definition of an observable 
cross section. When this is done, the result is expected to be finite, as it 
corresponds to a truly measurable quantity. This line of reasoning led to the
most general theorem concerning the cancellation of infrared singularities,
the KLN theorem~\cite{Kinoshita:1962ur,Lee:1964is}. In particular, 
Ref.~\cite{Lee:1964is} shows that the cancellation is a completely general 
property of any quantum-mechanical system whose Hilbert space contains
sets of energy-degenerate states. We will sketch a proof of the KLN theorem
in \secn{KLN}. Before we continue, however, we need to make the argument 
a little sharper: in principle, the probability for emission of soft or collinear 
particles could be small, and the effect on the cross section negligible. 
In order to understand the physics of infrared enhancements, consider
a diagram for the emission of a photon from an external fermion leg in
massless QED, shown in Fig.~\ref{EasyIR}.
The QED Feynman rules give an expression of the form
\beq
  e \,  \overline{u} (p) \, \gamma^\mu \frac{\slashed{p} + \slashed{k}}{(p + k)^2 + 
  {\rm i} \eta} \;  {\cal M} \, ,
\label{EasyIReq}
\eeq
where ${\cal M}$ represent the rest of the scattering process, which may involve
many external legs and virtual corrections as well. If the photon is emitted in the
final state, so that $k^2 = p^2 = 0$, the denominator of the fermion propagator 
reads $2 p \cdot k + {\rm i} \eta = E_p \omega_k (1 - \cos \theta_{pk}) + 
{\rm i} \eta$, where $E_p = |{\bf p}|$ and $\omega_k = | {\bf k}|$. While 
the $\eta$ prescription protects from an outright singularity, one must expect 
enhancements from three sources: the soft photon limit $\omega_k \to 0$, the
soft fermion limit $E_p \to 0$, and the collinear limit $\theta_{pk} \to 0$. Whether
these enhancements translate into actual divergences will depend on the specific
observable one is computing, and more generally on the theory one is considering. 
To this end, one will need to develop appropriate power-counting techniques
(discussed here in \secn{AllOrd}), similarly to what is done for ultraviolet 
enhancements; for example, we can anticipate that the soft fermion limit never
leads to divergences in renormalisable theories, thanks to an extra power
of the energy arising from the wave functions of massless spinors. Note also
that, in case the fermion were massive, the collinear singularity would be regulated
by the fermion mass, since the angular factor in the denominator would
read $(1 - |{\bf v}| \cos \theta_{pk})$, with $|{\bf v}|<1$ the velocity of the fermion.

That being said, the origin of the enhancement is apparent: in the limits 
considered, the fermion propagator reaches the mass shell, $(p + k)^2 = 0$;
furthermore, since we are working in covariant perturbation theory, all four 
momentum components are conserved at each vertex; thus, the intermediate
fermion with momentum $p+k$ is a physical state in our theory, and can propagate 
freely for any length of space and interval of time. When deriving the momentum-space
Feynman rules, we have formally integrated over the position of the photon emission
vertex in spacetime, under the assumption that emission at long distances should
be sufficiently suppressed. Unfortunately, it is not, a consequence of the fact that
QED (like all unbroken gauge theories) has long-range interactions. The same
conclusion is reached, perhaps in an even more transparent way, if one uses
Time-Ordered Perturbation Theory (TOPT) (see, for example, Ref.~\cite{Sterman:1994ce}
for a detailed discussion): within that framework, all particles in intermediate 
states are on the mass shell; energy, on the other hand, is not in general conserved 
at the vertices, which forces the interactions to take place in a finite time. For soft 
or collinear emissions, the energy deficit at the emission vertex vanishes, so that
once again late-time emissions are unsuppressed. In this sense, both soft and 
collinear divergences are properly described as {\it long-distance effects}.

Importantly, these conclusions are essentially unaffected if the photon line
in Fig.~\ref{EasyIR} folds back and attaches to some other fermion line in
the amplitude, forming a loop, instead of being radiated into the final state.
In that case, the denominator of the fermion propagator has an additional 
$k^2$, which however is negligible with respect to $p \cdot k$ if all components 
of the photon momentum become small at the same rate.
Clearly, some power counting tools will have to be developed to weigh the
presence of the loop integral (as opposed to the phase-space integration over
the real photon momentum) and of the photon propagator, but the basic fact 
remains that, even for virtual corrections, soft and collinear emissions are
dangerously enhanced at large distances and times. This provides the seeds 
for the eventual cancellation of divergences between virtual and real corrections: 
both are enhanced by the same mechanism, and both must be included to allow 
for the finite energy and angle resolutions of detectors; once a properly defined 
observable is constructed, they must, and do, cancel each other.

Let us summarise this first viewpoint on the {\it mistake} that led to the rise of 
infrared divergences, as it emerges from the Bloch-Nordsieck analysis. Theories 
with massless particles have long-range interactions; as a consequence, late-
(and early-) time emissions are not sufficiently suppressed; in the circumstances,
our organisation of perturbation theory in powers of the coupling, which
distinguishes between virtual corrections and (undetected) real emissions
is inadequate, hence individual matrix elements are ill-defined. The proposed
solution is to introduce a temporary fix for the matrix elements (an infrared regulator 
such a particle mass), in the knowledge of the fact that the singular dependence 
on the regulator will cancel in properly defined observable cross sections.

The second viewpoint on the origin of the infrared problem is the following: 
when constructing our quantum theory, {\it we have mis-identified the asymptotic 
states}. In most textbook constructions of the $S$-matrix, one finds, somewhere 
along the way, a statement on the need to assume that interactions can be 
adiabatically {\it switched off} at large times. In theories with massless particles 
and long-distance interactions, this is simply unrealistic: after a hard interaction, 
electrons will continue to emit and absorb photons all the way into the asymptotic 
regime. Choosing as asymptotic states the eigenstates of the free hamiltonian 
(in QED, Fock states with a fixed number of electrons and photons) is inadequate, 
as such states are not a good approximation of the actual physical asymptotic 
states. Indeed, splitting the hamiltonian into `free' and `interaction' terms, as 
usually done, is inadequate, since interactions persist at late and early times, 
and such asymptotic interactions should be moved to the part of the hamiltonian 
that one will attempt to diagonalise exactly. With a better choice of asymptotic 
states, one may hope not only to improve the algorithm to compute physical 
observables, but also to rescue the $S$-matrix program, by defining scattering 
amplitudes that are automatically well-defined, even in the presence of 
massless particles. This idea was successfully pursued in QED, starting 
with preliminary studies in Refs.~\cite{Dollard:1964,Chung:1965zza,
Fradkin:1960lgt,Kibble:1969ep,Kibble:1969ip,Kibble:1969kd}, and culminating 
with the seminal paper by Kulish and Faddeev~\cite{Kulish:1970ut}, 
where a separable, Lorentz- and gauge-invariant Hilbert space of 
`coherent states' is defined, and the finiteness of the QED $S$-matrix in the
coherent state basis is proved to all orders. The coherent state approach 
will be introduced here in \secn{Coherent}. 

Even in the relatively simple case of the abelian theory, understanding our
{\it IR mistake} has been quite fruitful: the role of long-distance dynamics 
has been exposed, a sharper definition of a physical observable has emerged,
and a more accurate characterisation of asymptotic states has rescued the notion of
scattering amplitudes in the presence of massless particles. The generalisation
of these ideas to the much more intricate case of non-abelian theories
will occupy most of our review. It is to be expected  that this generalisation 
will be far from trivial, given what we know about the long-distance behaviour 
of unbroken non-abelian gauge theories: perturbative asymptotic states have 
very little to do with their non-perturbative, confined counterparts, and we 
can expect and hope that perturbation theory will carry at least some partial 
information about this breakdown.

Studies of the infrared problem in non-abelian theories started in the 
mid-seventies~\cite{Poggio:1976qr,Sterman:1976jh,Kinoshita:1977dd}.
It soon became evident that the simple cancellation mechanism for
soft singularities uncovered by Bloch and Nordsieck in QED, which
involves only a sum over degenerate final states, does not work for non-abelian 
theories~\cite{Doria:1980ak,DiLieto:1980nkq,Andrasi:1980qw,Carneiro:1980au,
Frenkel:1983di,Ciafaloni:2001vt}, and a full cancellation can happen only considering
degenerate initial states as well\footnote{The result was later extended 
to general non-abelian theories in~\cite{Catani:1987xy}, and was recently 
revisited in modern language in~\cite{Caola:2020xup}. Early papers discussing
infrared cancellations in the non-abelian theory for special cross sections
include~\cite{Libby:1978nr,Akhoury:1978vq,Ganapathi:1981fs,Collins:1981ta}.}. 
A more general problem arises because collinear divergences, which are regulated 
by fermion masses in QED, are intrinsic and unavoidable for non-abelian theories,
even when matter fields are massive: the three-gluon vertex involves 
three strictly massless particles and inevitably leads to collinear problems.
In the presence of collinear divergences, in particular those associated 
with radiation from initial-state coloured particles, it is clear from the outset
that  a simple pattern of cancellation \`a la Block-Nordsieck cannot work,
since hard collinear emissions from the initial state drastically alter the 
kinematic configuration of the hard scattering, while virtual corrections
do not affect it: the cancellation of singularities must therefore be spoiled.
At the level of scattering amplitudes, this means that the coherent state 
program for non-abelian theories is bound to be much more intricate 
than was the case for QED. Important results were however obtained
through the eighties, shedding light on many aspects of the fixed-order and 
all-order structure of infrared effects, in particular in QCD~\cite{Greco:1978te,
Curci:1978kj,Butler:1978rd,Curci:1979bg,Nelson:1980qs,Nelson:1980yt,
Muta:1981pe,Ciafaloni:1984zr,Catani:1985ta,Catani:1985xt,Catani:1987sp,
DelDuca:1989jt,Ciafaloni:1989vs}; these analyses culminated in a formal proof 
of the finiteness of the non-abelian $S$-matrix in the coeherent-state basis, 
including the case of collinear divergences, in Ref.~\cite{Giavarini:1987ts}. 
At the level of cross-sections, the problem of initial-state collinear divergences
was of course the starting point for the QCD factorisation program, 
reviewed for example in~\cite{Collins:1989gx,Sterman:1995fz}.

More recent studies have focused on two (overlapping) directions.
On the one hand, the demands of precision phenomenology have
required developing efficient tools for high-order calculations of observable
cross sections. In this regard, the coherent state approach, notwithstanding
its physical appeal, has not been the main way forward: rather, a `KLN'
approach has prevailed, in which virtual corrections and phase-space
integrals of unresolved real radiation are both computed with an infrared
regulator, and then combined to construct finite distributions. Not 
surprisingly, dimensional continuation, setting $d = 4 - 2 \epsilon$ with
$\epsilon < 0$, has been the regulator of choice, since it preserves
gauge and Lorentz invariance without adding to the complexity of the 
calculation by introducing non-trivial dependence on unphysical mass 
scales. Finite-order calculation will be discussed in \secn{FinOrd}, and 
detailed algorithms for the cancellation of IR poles will be introduced 
in \secn{Subtra}.

On the other hand, perhaps more interestingly from a theoretical viewpoint,
a great deal of activity has been devoted to elucidate the all-order structure
of infrared singularities, on the basis of the ideas of {\it factorisation} and 
{\it universality}. In principle, these ideas are natural, and they apply equally 
well to amplitudes and to cross sections: as we argued, IR singularities 
are associated with phenomena that take place at large times and distances
from the hard scattering center; this suggests that the singularity structure
should be largely independent on the details of the short-distance process
being considered; one may then hope to identify universal factors responsible
for the singular behaviour of the amplitude (or cross section). In practice,
proving these properties in a non-abelian theory by standard perturbative 
methods is very challenging: collinear configurations carry spin correlations
between different particles taking part in the scattering, while soft configurations
are responsible for very interesting but very intricate colour correlations at
large distances. Factorisation emerges only upon summing over Feynman 
diagrams, and only after the constraints of gauge invariance have been 
properly taken into account by means of Ward identities. At the level of
scattering amplitudes (which will be the focus of most of our discussion) 
the factorisation program was started with pioneering work on form factors,
first in QED~\cite{Sudakov:1954sw,Mueller:1979ih} and subsequently in 
QCD~\cite{Collins:1980ih,Sen:1981sd}, with the first extensions to fixed-angle 
scattering amplitudes coming shortly thereafter~\cite{Sen:1982bt}.
The generalisation of these results to multi-particle scattering amplitudes
in the modern language of dimensional regularisation will be the central
focus of our review.

In the remainder of this introductory Section, we will present short accounts 
of the Bloch-Nordsieck cancellation mechanism in QED, of the KLN theorem,
and of the coherent state method, with an eye to the long history of the subject,
but also to display in some more detail the different underlying viewpoints,
that we have just briefly introduced. In \secn{FinOrd}, we will present some
well-known one-loop results in QCD, in the modern language of dimensional 
regularisation: this is textbook material, but it will give us the opportunity to 
introduce some tools that will be useful in what follows, including in particular
the $d$-dimensional running coupling and the calculation of eikonal integrals 
in dimensional regularisation. In \secn{AllOrd}, we will introduce the methods
required to perform an all-order diagrammatic analysis of the IR problem,
which are a pre-requisite for the proof of any factorisation theorem. These
methods were reviewed in~\cite{Sterman:1995fz}, and are presented in
much greater detail in~\cite{Sterman:1994ce,Collins:2011zzd}; we decided 
however to include a pedagogical introduction and work out some key examples,
since this toolbox lies at the foundation of all subsequent developments.
Sections~\ref{FactEvo} and \ref{MultiPart} form the core of our review.
First, in \secn{FactEvo}, we consider the case of form factors, and we 
show how the tools developed in \secn{AllOrd} lead to the formulation of
an all-order factorisation theorem, isolating soft and collinear divergences
of form factors in universal functions, which can be defined in terms of
gauge-invariant matrix elements of fields and Wilson lines. Once factorisation
has been achieved, evolution equations are bound to follow, and solving
them leads to an all-order summation of perturbative contributions, in this
case of IR poles~\cite{Magnea:1990zb,Magnea:2000ss,Dixon:2008gr}. Form 
factors have the advantage of having a trivial colour structure: extending 
the analysis to general non-abelian scattering amplitudes is therefore less 
than trivial. This is pursued in \secn{MultiPart}, where the central concept 
of IR anomalous dimension matrix is introduced, and the most general form 
of the exponentiation of IR singularities is derived; known results up to three 
loops for the anomalous dimension matrix are reviewed~\cite{Aybat:2006mz,
Aybat:2006wq,Becher:2009cu,Gardi:2009qi,Becher:2009qa,Gardi:2009zv,
Almelid:2015jia}, and the most recent diagrammatic techniques for its
calculation are introduced. It is perhaps worthwhile to emphasise that
all the results of Sections~\ref{FactEvo} and \ref{MultiPart} are valid
not only for QCD, but in fact for any massless gauge theory (with computable
corrections in case the gauge fields are coupled to massive matter fields):
for example, they have found important application in the case of
conformal gauge theories, in particular ${\cal N} = 4$ Super-Yang-Mills
theory~\cite{Anastasiou:2003kj,Bern:2005iz}. On the other hand, Sections~\ref{FactEvo} 
and \ref{MultiPart} focus on virtual corrections to fixed-angle scattering amplitudes,
while the construction of measurable cross-sections must also include unresolved 
soft and collinear real radiation. Factorisation for real radiation is by itself a vast 
subject, and we don't do it justice by providing just a brief summary in \secn{Subtra}; 
there, we also introduce modern {\it subtraction} algorithms~\cite{Frixione:1995ms,
Catani:1996vz}, which are currently being developed for high-order perturbative 
calculations~\cite{TorresBobadilla:2020ekr}, in order to perform the cancellation 
of IR singularities in a general and efficient manner, even for the intricate
observables currently being measured at high-energy colliders.

As must be the case when reviewing a broad subject with a long history, many 
important developments and lines of research that are closely related to our 
topic were left out: we provide here a partial list, with some of the references that 
we believe can be useful for further exploration by the reader. First of all, we say 
very little in this review about the subject of {\it resummation} of large logarithms 
that arise in observable QCD cross sections and that are closely related to 
infrared singularities (for example threshold logarithms and transverse momentum 
logarithms): we only note in passing that the techniques used for resummations 
are tightly connected to the ones reviewed here, which indeed in some cases 
were developed directly for cross sections before being applied to amplitudes. 
QCD tools for resummation are reviewed, for example, in Refs.~\cite{Sterman:1995fz,
Laenen:2004pm,Luisoni:2015xha}, while automated methods are discussed
in~\cite{Banfi:2004yd,Banfi:2014sua}. Next, we recall that techniques to construct 
factorisation theorems in QCD, under certain general assumptions, and to derive the 
corresponding resummations, were made systematic in Refs.~\cite{Bauer:2000ew,
Bauer:2000yr,Bauer:2001ct,Bauer:2001yt,Beneke:2002ph,Beneke:2002ni,
Hill:2002vw}, by the construction of Soft-Collinear Effective Theory (SCET), a 
non-local effective field theory\footnote{The idea that infrared divergences could be 
organised at Lagrangian level in terms of a non-local effective theory was floated
for the first time (to the best of our knowledge) in Ref.~\cite{Parisi:1978az}.} for soft 
and collinear modes of quark and gluon fields. While the bulk of SCET applications 
concerns cross sections of phenomenological interest, SCET methods are of course 
fully applicable to the study of scattering amplitudes, and indeed some of the general 
results discussed here in \secn{MultiPart} were independently derived within the SCET 
framework~\cite{Becher:2009qa,Feige:2014wja}. SCET has generated countless important 
phenomenological applications, and is introduced and reviewed in Ref.~\cite{Becher:2014oda}; 
comparisons between the SCET approach and QCD factorisation were carried 
out, for example, in Refs.~\cite{Lee:2006nr,Bonvini:2012az,Sterman:2013nya,
Bonvini:2014qga,Almeida:2014uva,Bauer:2018svx,Bauer:2019bsp,vanBeekveld:2021hhv}. 
A third direction that we do not explore is the extension of factorisation theorems, 
and eventually resummation techniques, beyond leading power in the singular 
variables (for example the soft gluon energy). In QED, a classic result in this regard 
is the Low-Burnett-Kroll theorem~\cite{Low:1958sn,Burnett:1967km}, showing that 
next-to-leading power (NLP) contributions of soft photons to QED cross section are 
still universal, as an effect of gauge invariance; the theorem was later extended to 
collinear-enhanced configurations in Ref.~\cite{DelDuca:1990gz}. NLP contributions 
to amplitudes and cross sections do not induce divergences in renormalisable theories, 
but they are nevertheless very interesting for both theoretical and phenomenological 
reasons. In recent years, their factorisation and resummation properties have 
been studied intensively, both with direct QCD methods (see, for example, 
Refs.~\cite{Laenen:2008ux,Laenen:2008gt,Laenen:2010uz,Bonocore:2014wua,
Bonocore:2015esa,Bonocore:2016awd,Gervais:2017yxv,Bahjat-Abbas:2019fqa,
Laenen:2020nrt}) and in the context of SCET (see, for example, Refs.~\cite{Beneke:2002ph,
Larkoski:2014bxa,Feige:2017zci,Beneke:2017ztn,Beneke:2018gvs,Moult:2019mog,
Beneke:2019mua,Beneke:2019oqx,Liu:2019oav,Liu:2020wbn, Broggio:2021fnr})\footnote{For 
different approaches to the resummation of next-to-leading-power threshold logarithms, 
see, for example,~\cite{Akhoury:1998gs,Moch:2009my,Moch:2009hr,Almasy:2010wn,
LoPresti:2014ihe,Almasy:2015dyv,Ajjath:2020sjk,Ajjath:2020lwb,Ajjath:2021lvg}.}.  
A thorough review of these developments together with a survey of recent literature 
can be found in Ref.~\cite{Vita:2020ckn}. A final important connection that we do not 
develop here is that between gauge theories and gravity, first discussed in the pioneering 
work of Weinberg~\cite{Weinberg:1965nx}. Similarities and differences between 
massless gauge theories and gravity theories are intriguing, and have been 
addressed in the context of factorisation in a number of recent papers~\cite{Akhoury:2011kq,
White:2011yy,Beneke:2012xa,Oxburgh:2012zr,Saotome:2012vy,White:2014qia,
Stieberger:2015kia,Gervais:2017zky,Beneke:2021umj}. The connection between 
gauge theories and gravity theories is also at the root of radical re-interpretation of 
infrared phenomena, which originated in the work of Strominger~\cite{Strominger:2013lka,
Strominger:2013jfa}. Within this framework, infrared divergences are related to 
infinite-dimensional asymptotic symmetries of the (gauge or gravity) theory under 
consideration, acting on the {\it celestial sphere} intersecting the future (or past) light 
cone at asymptotic distances. For gravity, these symmetries were uncovered 
in~\cite{Bondi:1962px,Sachs:1962wk}, while for gauge theories they take 
the form of `large' gauge transformations with non-trivial action at future (or past) null 
infinity~\cite{Casali:2014xpa,He:2014cra,Lysov:2014csa,He:2015zea,Adamo:2015fwa,
Campiglia:2016hvg,Conde:2016csj,Gabai:2016kuf,Luna:2016idw,Miller:2021hty}. 
The universal form of soft and next-to-soft factors for tree-level radiative amplitudes 
emerges from the Ward identities of these symmetries. Early developments in this 
fast-developing field are reviewed in~\cite{Strominger:2017zoo}. While most of the 
work done so far within this framework concerns tree-level gravity or gauge-theory 
amplitudes (see, however, Refs.~\cite{Albayrak:2020saa,Gonzalez:2020tpi}), the 
connection between infrared properties of $d=4$ gauge theories and $d = 2$ conformal 
invariance on the celestial sphere is very intriguing, and represents a remarkable new 
point of view on an old problem. In \secn{MultiPart}, we will see that scale invariance 
plays an important role for infrared factorisation at any perturbative order: conformal 
invariance, however, is broken by quantum effects. Furthermore, we will see that 
the celestial framework allows for  a remarkable rephrasing of the all-order
expression for infrared-divergent colour-dipole correlations, and indeed an 
interpretation emerges in terms of a specific two-dimensional conformal theory on 
the celestial sphere~\cite{Magnea:2021fvy}. Further exploration of these ideas 
in the context of all-order perturbative calculations is undoubtedly of great
interest.

As is perhaps evident from this long introduction, the present review has a strong
pedagogical emphasis, and we hope that it may help students and junior researchers
to gain an orientation in a field with a long history, which however is still quickly 
progressing in both theoretical and phenomenological directions. At the same time,
we have included in \secn{MultiPart} and in \secn{Subtra} some very recent 
developments, which may be interesting to QCD practitioners, and give a 
flavour of current research. Possible future directions of research are briefly 
discussed in \secn{Persp}.


\subsection{Catastrophe and recovery in QED}
\label{cataQED}

In this Section, and in the following Sections \ref{KLN} and \ref{Coherent}, we take a 
largely historical perspective, and we present classic results on the cancellation of
infrared singularities, which also serve to introduce some of the key concepts to be 
developed later in the context of non-abelian theories. We begin by looking at the
relatively simple case of QED with massive fermions, and we derive the cancellation
of soft divergences between virtual correction and final-state real radiation, first at 
the lowest non-trivial order, and then to all-orders, where one can rather easily 
demonstrate the exponentiation of the lowest-order result. Our emphasis is on 
the concepts that will later be developed in greater detail, so the proof that we 
discuss is heuristic: for a detailed analysis one should refer, for example, 
to~\cite{Grammer:1973db}. Rather than using directly the arguments of 
Ref.~\cite{Bloch:1937pw}, we will take advantage of the more modern language 
of Ref.~\cite{Weinberg:1965nx}.

Choosing a simple example, which however displays all the general features of the
proof, we consider the process where an electron scatters from another heavy 
particle by means of $t$-channel photon exchange. The heavy particle simply plays 
the role of a source for an external photon field. We will first show the cancellation
at one loop, and then discuss how the mechanism generalises to all orders.

\begin{figure}[t]
\centering
     \begin{subfigure}{0.24\textwidth}
    \centering
        \includegraphics[scale=0.74]{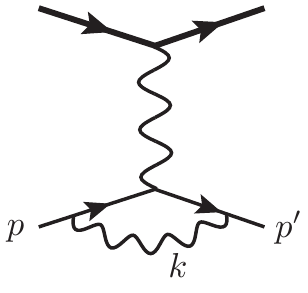}
        \caption{}
        \label{fig:vert1}
    \end{subfigure}
        \hspace{-15pt}
    \begin{subfigure}{0.24\textwidth}
    \centering
        \includegraphics[scale=0.74]{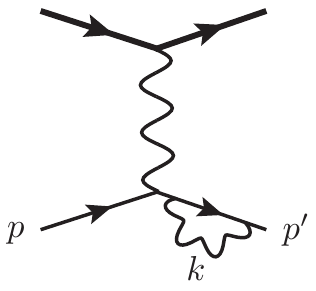}
        \caption{}
        \label{fig:vert2}
    \end{subfigure}
    \hspace{-15pt}
    \begin{subfigure}{0.24\textwidth}
    \centering
        \includegraphics[scale=0.76]{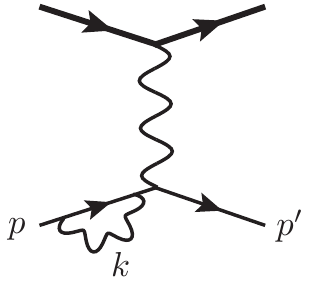}
        \caption{}
        \label{fig:vert3}
    \end{subfigure}
        \hspace{-15pt}
    \begin{subfigure}{0.24\textwidth}
        \centering
        \includegraphics[scale=0.75]{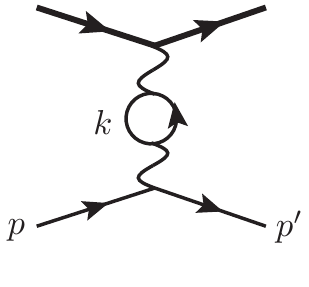}
        \caption{}
        \label{fig:vert4}
    \end{subfigure}
      \caption{Virtual corrections to electron scattering from a heavy particle, 
      at order $\alpha$.}
  \label{vertfig}
\end{figure}

\begin{figure}[t]
\centering
     \begin{subfigure}{0.24\textwidth}
    \centering
        \includegraphics[scale=0.74]{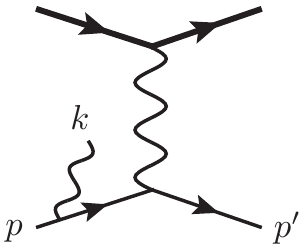}
        \caption{}
        \label{fig:brems_1}
    \end{subfigure}
        \hspace{-20pt}
    \begin{subfigure}{0.24\textwidth}
    \centering
        \includegraphics[scale=0.74]{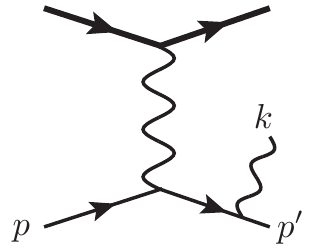}
        \caption{}
        \label{fig:brems_2}
    \end{subfigure}
      \caption{Bremsstrahlung diagrams for electron scattering from a heavy particle, 
      contributing to the cross section at order $\alpha$.}
  \label{bremsfig}
\end{figure}

At order $\alpha$, the diagrams contributing to virtual corrections are shown in 
Fig.~\ref{vertfig}, while those contributing to real photon radiation (bremsstralung) 
are shown in Figure~\ref{bremsfig}. 
We begin by considering the differential cross-section for radiating a photon with 
momentum $k$, when the electron scatters from a state of momentum $p$ to a state 
of momentum $p^{\prime}$, assuming $k^\mu \ll q^\mu \equiv (p^{\prime} - p)^\mu$. 
The matrix element for this process is the sum of two terms of the form depicted in 
Fig.~\ref{EasyIR}, and greatly simplifies when one expands in powers of the soft 
momentum $k$ and considers only the leading power (this {\it soft approximation} 
will be considered in greater detail in \secn{IRSafeObs} and in \secn{universal_fun}). 
At this order, referring to \eq{EasyIReq} but appropriately introducing the electron mass, 
one can perform the following manipulations:
\begin{itemize}
\item neglect $k$ in the numerator;
\item neglect $k^2$ in the denominator;
\item commute the hard momentum factor $\slash{p}$ (or $\slash{p}'$) in the numerator
of the electron propagator across the emission vertex, and use the Dirac equation to
simplify the expression.
\end{itemize}
Contracting the result with the photon polarisation vector $\varepsilon (k)$,
and denoting the lowest-order matrix element by ${\cal M}_0$, the leading power result 
for the radiative matrix element ${\cal M}_{\rm rad}$ takes the form
\beq
  {\cal M}_{\rm rad} \, = \, e \,\bigg( 
  \frac{p^{\prime} \cdot \varepsilon (k)}{p^{\prime} \cdot k} -
  \frac{p \cdot \varepsilon (k)}{p \cdot k} \bigg) \, {\cal M}_0 \, ,
\label{Mrad}
\eeq
where $e$ is the electron charge.
The soft approximation in \eq{Mrad} has several remarkable properties that we will 
explore later. Here we simply note that it is gauge invariant (it vanishes for longitudinally
polarised photons), independent of the spin of the emitter\footnote{It is instructive to 
derive the same result for scalar electrodynamics, and for gluon scattering in the
non-abelian theory.} and singular -- as expected -- for small $k$. It is formally easy 
to construct a soft approximation for the total radiative cross section: at leading 
power in $k$ the phase space factorises, and the flux factor reduces to the flux 
factor for the non-radiative process. Summing over polarisations, we find that the 
cross section for the radiation of a single real soft photon is given by
\beq
  \sigma_r \big( e(p) \rightarrow e(p^\prime) + \gamma \big) \, = \, 
  \sigma_0 \big( e(p) \rightarrow e(p^\prime) \big) \, \, 
  {\cal I}_r \bigg( \frac{m^2}{q^2}, \frac{\mu^2}{E^2}, \epsilon \bigg) \, ,
\label{realsoftxs}
\eeq
where we defined the soft factor for real radiation, ${\cal I}_r$, by
\beq
  {\cal I}_r \bigg( \frac{m^2}{q^2}, \frac{\mu^2}{E^2}, \epsilon \bigg) \, = \, 
 - \frac{e^2 (2 \pi \mu)^{2 \epsilon}}{16 \pi^3} \! \int_0^E d | {\bf k} | | {\bf k} |^{1 - 2 \epsilon} \!
  \int d \Omega_{2 - 2 \epsilon}
  \bigg(\frac{p^{\prime\mu}}{p^{\prime} \cdot k}-\frac{p^{\mu}}{p \cdot k} \bigg)
    \bigg(\frac{p^{\prime}_\mu}{p^{\prime} \cdot k}-\frac{p_{\mu}}{p \cdot k} \bigg) \, .
\label{softI}
\eeq
In \eq{softI} we have introduced dimensional regularisation, setting $d = 4 - 2 
\epsilon$, $\epsilon < 0$, since the integral is ill-defined in $d = 4$, diverging 
logarithmically as $|{\bf k}| \rightarrow 0$: this is the original {\it infrared catastrophe} 
of Ref.~\cite{Bloch:1937pw}. The fact that the divergence is logarithmic justifies 
{\it a posteriori} our choice to retain only the leading power in the soft expansion: 
sub-leading terms will contribute finite corrections. Furthermore, we integrate 
over photon energies up to an arbitrary cutoff $E \ll \sqrt{- q^2}$, which represents 
the minimum energy resolution of our detector: photons with $|{\bf k}| < E$ are 
unresolved, and their contribution must be included. The resulting integral is well 
known and not difficult to compute. One finds
\beq 
  {\cal I}_r \bigg( \frac{m^2}{q^2}, \frac{\mu^2}{E^2}, \epsilon \bigg) \, = \, 
  - \, \frac{\alpha}{\pi} \, \frac{1}{\epsilon} \bigg( \frac{4 \pi {\mu}^2}{E^2} 
  \bigg)^{\epsilon}   
  \left[ \left( \frac{1- 2 m^2/q^2}{\beta} \right) \log \frac{\beta+1}{\beta-1} - 1 \right] 
  + {\cal O} \big( \epsilon^0 \big) \, ,
\eeq 
where
\beq
  \beta \, = \, \sqrt{1 - 4 m^2/q^2} \, > \, 1 \, .
\label{fourvel}
\eeq
It is interesting to consider explicitly the limit of large momentum transfer, $ - q^2 
\rightarrow \infty$, focusing on terms enhanced by logarithms of $q^2$. We obtain
\beq
  {\cal I}_r \bigg( \frac{m^2}{q^2}, \frac{\mu^2}{E^2}, \epsilon \bigg) \, \simeq \,
  \frac{\alpha}{\pi} \left[ - \frac{1}{\epsilon} \log\left(\frac{- q^2}{m^2}\right) + 
  \log\left(\frac{E^2}
  {4 \pi \mu^2}\right) \log \left(\frac{- q^2}{m^2}\right) \right] \, .
\label{radiad}
\eeq
As we will see, the soft pole in $\epsilon$ will cancel against the virtual correction; the
second term displays the characteristic `Sudakov'  double-logarithmic enhancement, 
with one logarithm of soft origin (here parametrised by the resolution scale $E$) and 
one logarithm of collinear origin (which becomes a divergence in the massless limit, 
$m \to 0$).

Let us now turn to the evaluation of virtual corrections. The problem is significantly 
more intricate because of the concurrent presence of ultraviolet divergences, which 
must be dealt with by means of renormalisation: in particular, one must make sure 
that no infrared divergences survive in the UV counterterms appropriate to the chosen 
renormalisation scheme. This can be achieved for example by using a minimal scheme.
On the other hand, in the on-shell scheme typically used in QED, both the electron field 
renormalisation constant $Z_\psi$ and the vertex renormalisation constant $Z_1$
contain IR divergences, which however cancel in the scattering amplitude thanks to 
the QED Ward identity~\cite{Sterman:1994ce}. Armed with this preliminary result, 
we can concentrate on the vertex correction diagram of Fig.~\ref{vertfig}(a), which,
as we will see, gives the only surviving infrared-singular contribution. A straightforward 
power-counting argument shows that only terms with no powers of the photon 
momentum in the numerator can give infrared singularities, and those will be 
logarithmic. In this approximation, diagram (a) in Fig.~\ref{vertfig} gives
\beq
  (a)_{\rm soft} \, = \, - \, e^3 \mu^{3 \epsilon} \int \frac{d^d k}{(2 \pi)^d} 
  \frac{\overline{u}(p^\prime) \gamma^\alpha (\slashed{p}^\prime + m) \gamma^\mu 
  (\slashed{p} + m) \gamma_\alpha u(p)}{(k^2 + {\rm i} \eta) 
  (k^2 - 2 p^\prime \cdot k + {\rm i} \eta)(k^2 - 2 p \cdot k + {\rm i} \eta)} \, .
\label{vertir}
\eeq
The integral in \eq{vertir} can be performed with standard methods, yielding a 
single infrared pole in $\epsilon$: potentially, a {\it second} infrared catastrophe.
Instead of directly computing the integral, in view of the generalisation to higher 
orders, it is more instructive to take the soft approximation further, and perform 
on the integrand the same steps that we applied to the real emission diagrams,
{\it i.e.} neglecting $k^2$ in the denominator, and using the Dirac equation to 
simplify the numerator. Recognising that the lowest-order matrix element 
factorises, we write then
\beq
  (a)_{\rm soft} \, = \, \big( {\rm i} e^2 \mu^{2 \epsilon} \big) \, {\cal M}_0
  \int \frac{d^d k}{(2 \pi)^d} 
  \frac{p \cdot p^\prime}{(k^2 + {\rm i} \eta) 
  (- p^\prime \cdot k + {\rm i} \eta)(- p \cdot k + {\rm i} \eta)} \, .
\label{simpvertir}
\eeq
We note that the step between \eq{vertir} and \eq{simpvertir} has apparently generated
a new problem: the integral in \eq{simpvertir} is now also divergent in the UV, a
region where our approximation breaks down. This UV singularity is not present in
the original QED calculation: it is a singularity of a low-energy effective theory for
the infrared sector of QED, and it will be of great interest for the developments
discussed in \secn{FactEvo}. For the moment, we will simply introduce a UV regulator
and focus on the infrared pole. 

Examining \eq{simpvertir}, one begins to see some similarities with the integral in 
\eq{softI}: these can be sharpened by taking two further steps. First of all, we include 
the UV counterterms appropriate to the original QED calculation, picking the on-shell 
renormalisation scheme, which makes the calculation particularly transparent. In the 
on-shell scheme, the self-energy counterterm for the graphs in Fig.~\ref{vertfig}(b) 
and \ref{vertfig}(c) is such that the sum of the graph plus the counterterm vanishes 
on shell: these graphs therefore do not contribute to our calculation. The vertex 
counterterm, on the other hand, is fixed by the requirement that the renormalised 
vertex correction vanish for $q^2 = 0$, {\it i.e.} for $p^\prime = p$. This can be 
enforced directly in our soft approximation by writing~\cite{Sterman:1994ce}
\beq
  (a + {\rm counterterm})_{\rm soft} & = & {\cal M}_0 \,\, 
  \bigg( \!\! - \frac{{\rm i} e^2 \mu^{2 \epsilon}}{2} \bigg)
  \int \frac{d^d k}{(2 \pi)^d} 
  \frac{1}{(k^2 + {\rm i} \eta)} \left( \frac{p^{\prime \mu}}{- p^\prime \cdot k + 
  {\rm i} \eta} - 
  \frac{p^\mu}{- p \cdot k + {\rm i} \eta} \right)^2 \nonumber \\
  & \equiv & {\cal M}_0 \,\, {\cal I}_v \bigg( \frac{m^2}{q^2}, \frac{\mu^2}{q^2}, 
  \epsilon \bigg) \, ,
\label{simpvertirct}
\eeq
which manifestly vanishes for $p^\prime = p$, and where the cross term in 
the square of the parenthesis gives \eq{simpvertir}, while the squares of the 
two terms (which depend only on $m^2$) give the counterterm. The final step 
to bring \eq{simpvertirct} to a form directly comparable to \eq{softI} is to note 
(following Ref.~\cite{Weinberg:1965nx}) that the real part of the integral 
${\cal I}_v$ has support only on the photon mass shell\footnote{In fact, in 
space-like kinematics $(q^2 <0)$ the integral ${\cal I}_v$ is real, whereas 
in time-like kinematics $(q^2 >0)$ it has an imaginary part, responsible for 
a divergent phase in the scattering amplitude~\cite{Weinberg:1965nx}, and 
arising from configurations with an off-shell photon, but on-shell fermion lines.}, 
$k^2 = 0$. This can be seen by focusing on the $k_0$ integration, and observing 
the locations of the poles in the $k_0$ complex plane in \eq{simpvertir}: the poles 
associated with the fermion lines are in the upper-half plane, as is one of the 
two poles associated with the photon propagator. One can therefore compute 
the integral by closing the contour in the lower-half plane, and picking the residue 
of the photon pole, which effectively places the photon on the mass shell. The 
result is now precisely of the form of \eq{softI}, except for an overall factor of 
$-1/2$, and the fact that the resolution parameter $E$ must be replaced with 
an ultraviolet cutoff, which can naturally be chosen as $\sqrt{- q^2}$. We 
have then
\beq
  {\cal I}_v \bigg( \frac{m^2}{q^2}, \frac{\mu^2}{q^2}, \epsilon \bigg) \, = \, 
  - \frac{1}{2} \, {\cal I}_r \bigg( \frac{m^2}{q^2}, \frac{\mu^2}{- q^2}, \epsilon \bigg) \, .
\label{IrIv}
\eeq
The observed physical cross section is obtained by summing the real-emission
cross section in \eq{realsoftxs}, which gives the leading-order probability for the 
emission of an undetected soft photon, with the virtual correction to the elastic
scattering process, which is proportional to the tree-level cross section, with a 
factor given by twice the real part of ${\cal I}_v$. One finds
\beq
  \sigma_{\rm obs} \big( e(p) \rightarrow e(p^\prime) \big) \, = \, 
  \sigma_0 \big( e(p) \rightarrow e(p^\prime) \big) \bigg[ 1 + 
  {\cal I}_r \bigg( \frac{m^2}{q^2}, \frac{\mu^2}{E^2}, \epsilon \bigg) 
  + 2 \, {\cal I}_v \bigg( \frac{m^2}{q^2}, \frac{\mu^2}{q^2}, \epsilon \bigg) \bigg] \, .
\label{sigmaobsnlo}
\eeq
Using \eq{IrIv}, we see that all singular terms as $\epsilon \to 0$ cancel, as 
announced, leaving behind finite logarithms. In the limit of large negative $q^2$,
for example, the result is
\beq
  \sigma_{\rm obs} \big( e(p) \rightarrow e(p^\prime) \big) \, \simeq \, 
  \sigma_0 \big( e(p) \rightarrow e( p^{\prime} ) \big)
  \left[1- \frac{\alpha}{\pi} \log\left(\frac{-q^2}{E^2}\right) \log\left(\frac{-q^2}{m^2}\right) 
  + \mathcal{O}({\alpha}^{2})\right] \, ,
\label{ver}
\eeq
displaying the product of a soft logarithm and a collinear logarithm. It is appealing that
the cancellation we have observed is made possible by the fact that the divergent part 
of the loop integral originates from configurations in which the virtual photon is on-shell:
this resonates with the qualitative arguments given in the Introduction, identifying 
infrared divergences as long-distance effects. We note however that the correction 
we find in \eq{ver} is finite but negative, and it can lead to a negative cross section 
at very large momentum transfer. It is clear that the recovery from the IR catastrophe 
at NLO is only a partial solution, and we need to explore higher-order corrections.

A general and detailed proof of the cancellation of infrared divergences in QED 
would require developing power-counting tools (discussed here in \secn{AllOrd}), 
and analysing the impact of renormalisation to all orders~\cite{Yennie:1961ad,
Grammer:1973db}. Fortunately, it is possible to understand the general pattern
of the cancellation with simple diagrammatic and combinatorial arguments, discussed
for example in~\cite{Weinberg:1965nx}, which are sufficient to deal with terms 
with the largest logarithmic enhancements at each order of perturbation theory.
These tools will provide us with a first glimpse of two essential features of 
infrared enhancements, {\it factorisation} and {\it exponentiation}, which will be 
discussed at length in later Sections.
\begin{figure}
\centering
  \includegraphics[scale=0.7]{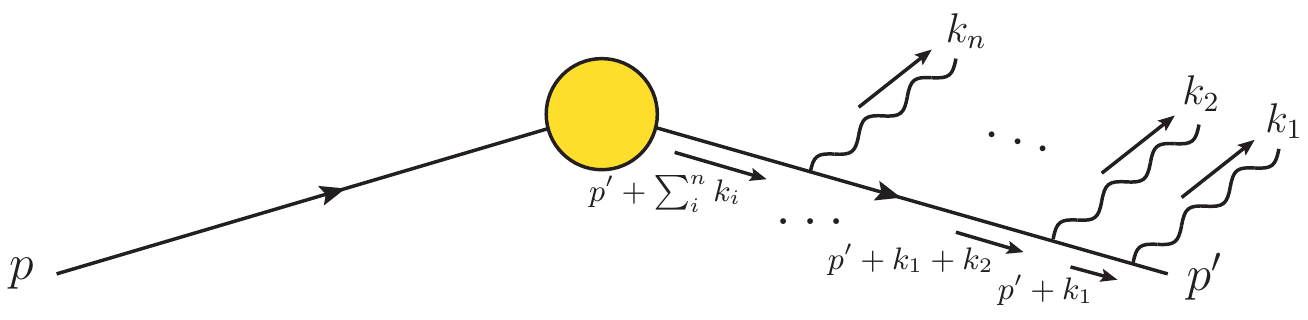}
  \caption{Emission of $n$ photons attached to outgoing electron line.}
\centering
\label{outrealfig}
\end{figure}   
Consider first the emission of $n$ photons with momenta $k_1, \ldots, k_n$ 
and polarisation vectors $\varepsilon_1, \ldots, \varepsilon_n$, attached to 
the outgoing electron line carrying momentum $p^{\prime}$, as depicted in 
Fig.~\ref{outrealfig}. The contribution to the transition amplitude from this 
graph is given by
\beq
  G \big(1, \ldots, n; p^\prime \big) \, = \, e^n \, 
  \overline{u} (p^\prime) \, \slash{\varepsilon}_1 \,
  \frac{\slash{p}^\prime + \slash{k}_1 + m}{2 p^{\prime} \cdot k_{1}} \, \ldots \,
  \slash{\varepsilon}_n \, \frac{\slash{p}^\prime + \slash{k}_1 \ldots \slash{k}_n + 
  m}{2 p^\prime \cdot \big( k_1 + \ldots + k_n \big)} \, {\cal H} \, ,
\label{ubarin}
\eeq
where ${\cal H}$ is the hard scattering subgraph represented by the circle in 
Fig.~\ref{outrealfig}. At leading power in each gluon momentum, we can iterate 
the procedure applied to the single-radiative amplitude, dropping all dependence 
on $k_i$ in the numerator, and repeatedly using the Dirac equation after commuting 
the electron momentum $p^\prime$ to the left. This leads to
\beq
  G \big(1, \ldots, n; p^\prime \big) \, \simeq \, e^n \,\, 
  \frac{p^\prime \cdot \varepsilon_1}{p^\prime \cdot k_1} \,
  \frac{p^\prime \cdot \varepsilon_2}{p^\prime \cdot \big( k_1 + k_2 \big)} \,
  \ldots \,  \frac{p^\prime \cdot \varepsilon_n}{p^{\prime} \cdot 
  \big( k_1 + k_2 + \ldots + k_n \big)} \, {\cal M} \, ,
\label{ubar}
\eeq
where the outgoing electron spinor has been reabsorbed in ${\cal M}$, the 
fixed-order matrix element without photon radiation. Crucially, the contribution 
to the physical amplitude where $n$ photons are radiated from the outgoing 
electron line is obtained by summing over $n!$ permutations of the photon 
lines. Bose symmetry then requires that the result be symmetric under 
permutations of the photon labels: this happens as a consequence of 
the {\it eikonal identity}
\beq
 \sum_{\sigma} \Bigg[ \prod_{i=1}^n \frac{1}{p \cdot 
 \sum_{j=1}^i k_{\sigma(j)} } \Bigg] \, = \,
 \prod_{i=1}^n \, \frac{1}{p \cdot k_i} \,\, ,
\label{eikonal_identity}
\eeq
where $\sigma$ enumerates photon permutations. The identity is easily verified for 
$n=2$, and can subsequently be proved by induction. It leads to 
\beq
  \sum_\sigma G \big( \sigma(1), \ldots, \sigma(n); p^\prime \big) 
  \, \simeq \, e^n \,\, \prod_{i = 1}^{n} \,
  \frac{p^\prime \cdot \varepsilon_i}{p^\prime \cdot k_i} \, {\cal M} \, ,
\label{eq:eikonalisation_qed}
\eeq
representing the uncorrelated emissions of $n$ soft photons without recoil. Notice
that soft divergences will arise from the phase-space integration of the emitted photon
momenta in \eq{ubarin} only if no hard photons are emitted further out along 
the electron line, closer to the outgoing electron spinor: any such emission 
would move all propagators closer to the hard scattering off their mass shell, so that
the approximation leading to \eq{ubarin} would not be applicable\footnote{For example,
for $n=2$, the last term in $(p^\prime + k_1 + k_2)^2 - m^2 = 2 p^\prime \cdot (k_1 + k_2) 
+ 2 k_1 \cdot k_2$ would not be negligible.}. We conclude that soft divergences arise
from real photon emission only when photons attach to {\it external lines}, defined as
lines connected to the external states from which every emission is soft: that is why we 
have factorised the non-radiative matrix element ${\cal M}$ in \eq{eq:eikonalisation_qed}, 
rather than its tree-level approximation ${\cal M}_0$. In a space-time picture, this 
factorisation of `external line' contributions nicely matches our understanding of soft 
divergences as long-distance phenomena, and it provides a first example of more 
intricate factorisation theorems to be discussed in later Sections.
\begin{figure}
\centering
  \includegraphics[scale=0.72]{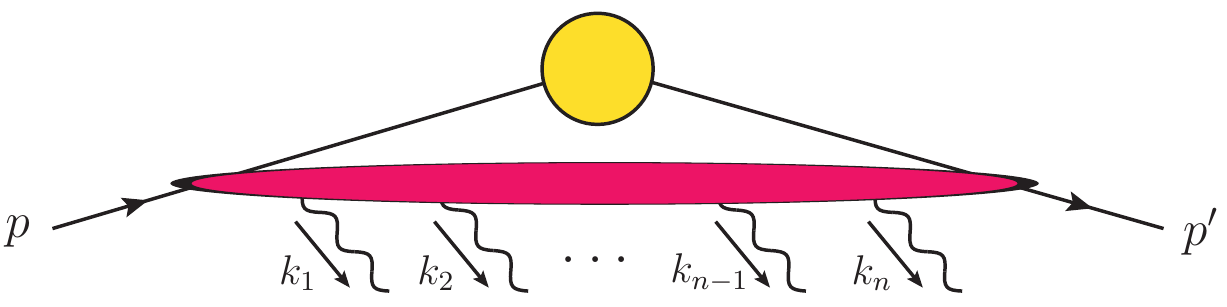}
  \caption{Emission of $n$ photons connected in any order to the initial and 
  final electron lines.}
\centering
\label{inoutrealfig}
\end{figure}
Naturally, an identical argument applies to the incoming electron line, where however 
each photon contributes with a relative minus sign, since $(p - \Sigma_i {k_i})^2 - m^2 
= - 2 p \cdot \Sigma_i {k_i}$. Considering together all the diagrams containing $n$ soft 
photons, connected in any possible order to the incoming and outgoing electron lines, 
as shown in Fig.~\ref{inoutrealfig}, we find that each photon contributes precisely an 
eikonal factor, as was the case in \eq{Mrad}. The result is then
\beq
  {\cal M}_{\rm rad}^{(n)} \, = \, e^n \, \prod_{i=1}^{n} \bigg( 
  \frac{p^{\prime} \cdot \varepsilon_i}{p^{\prime} \cdot k_i} -
  \frac{p \cdot \varepsilon_i}{p \cdot k_i} \bigg) \, {\cal M} \, .
\label{Mnrad}
\eeq
At leading power in each photon momentum $k_i$, the phase space for real radiation
factorises into a product of phase spaces for individual photons\footnote{Notice that 
some care is needed when treating the resolution scale $E$: here we are allowing 
each photon to have an energy up to $E$, whereas the limit should apply to the 
sum of the energies of individual photons; a more refined treatment can be found,
for example, in Ref.~\cite{Grammer:1973db}.}; furthermore, the phase-space integral 
must include a factor of $1/n!$ for $n$ identical bosons. Summing over polarisations 
one finds
\beq
  \sigma_r^{(n)} \big( e(p) \rightarrow e(p^\prime) + n \gamma \big) \, = \, 
  \sigma_0 \big( e(p) \rightarrow e(p^\prime) \big) \, \, \frac{1}{n!} \left[
  {\cal I}_r \bigg( \frac{m^2}{q^2}, \frac{\mu^2}{E^2}, \epsilon \bigg) \right]^n \, ,
\label{realsoftxsn}
\eeq
It is now straightforward to sum the series in \eq{realsoftxsn} over the number of
soft photons $n$, which results in {\it exponentiation} of the single-photon result,
\beq
  \sigma_r \big( e(p) \rightarrow e(p^\prime) \big) \, = \, 
  \sigma_0 \big( e(p) \rightarrow e(p^\prime) \big) \, \exp \Bigg[ \,
  {\cal I}_r \bigg( \frac{m^2}{q^2}, \frac{\mu^2}{E^2}, \epsilon \bigg) \Bigg] \, .
\label{realsoftxsexp}
\eeq
We can now observe that most of the steps leading to \eq{realsoftxsexp}  also apply
to virtual corrections: only combinatorial factors must be carefully considered. First of 
all, only virtual photons attaching to {\it external lines} will generate soft divergences. Next, 
focusing on vertex corrections, {\it i.e.} on photons connecting the two electron lines,
one may apply the eikonal identity in \eq{eikonal_identity}, by simply considering $n!$
copies of all diagrams contributing to the $n$-photon vertex correction, and relabelling
the photon momenta in each copy: one must then divide the result by $n!$. Notice 
that one needs to perform this sum and normalisation on only one of the two electron 
lines, since repeating the operation on the second electron line would reproduce the 
same diagrams\footnote{In case more than two charged lines are present, the 
combinatorial factors become somewhat more intricate, but the final exponentiation
of the one-loop result remains true.}. This step leads to a product of factors of 
the form of \eq{simpvertir}. Self-energy corrections on each electron line can then 
be renormalised to vanish on-shell, and the condition that  the vertex correction 
should vanish at $q^2 = 0$ turns each eikonal factor in the vertex into the 
form of \eq{simpvertirct}, introducing the appropriate factor of $1/2$ for each 
photon. We conclude that virtual corrections also {\it exponentiate}, and
\beq
  \sigma_v \big( e(p) \rightarrow e(p^\prime) \big) \, = \, 
  \sigma_0 \big( e(p) \rightarrow e(p^\prime) \big) \, \exp \Bigg[ \, 2 \,
  {\cal I}_v \bigg( \frac{m^2}{q^2}, \frac{\mu^2}{q^2}, \epsilon \bigg) \Bigg] \, .
\label{virtsoftxsexp}
\eeq
Combining real and virtual corrections, we see that, thanks to \eq{IrIv}, the cancellation 
of soft singularities is replicated to all orders in perturbation theory, and one finds
\beq
  \sigma_{\rm obs} \big( e(p) \rightarrow e(p^\prime) \big) & = & 
  \sigma_0 \big( e(p) \rightarrow e(p^\prime) \big) \, \exp \bigg[
  {\cal I}_r \bigg( \frac{m^2}{q^2}, \frac{\mu^2}{E^2}, \epsilon \bigg) 
  + 2 \, {\cal I}_v \bigg( \frac{m^2}{q^2}, \frac{\mu^2}{q^2}, \epsilon \bigg) \bigg] 
  \nonumber \\
  &\simeq &
  \sigma_0 \big( e(p) \rightarrow e(p^\prime) \big) \, \exp \bigg[ \!
  - \frac{\alpha}{\pi} \log\left(\frac{-q^2}{E^2}\right) \log\left(\frac{-q^2}{m^2}\right) 
  \bigg] \, ,
\label{sigmaobsall}
\eeq
where in the second line we have reported the leading double-logarithmic behaviour
in the limit of large momentum transfer. Upon combining real and virtual corrrections, 
{\it and} resumming the perturbative expansion, we achieved a finite and 
well-behaved result: the cross section is positive definite, and it exhibits the classic
`Sudakov' behaviour~\cite{Sudakov:1954sw}, vanishing exponentially at large 
momentum transfer, or equivalently for small values of the resolution parameter 
$E$ and of the fermion mass $m$.


\subsection{Observable cross sections are finite: the KLN theorem}
\label{KLN}

We have seen that in quantum electrodynamics soft divergences cancel out 
order by order in perturbation theory, when the transition rates are summed over 
final states that are physically indistinguishable. From the derivation in \secn{cataQED},
it is not however clear how general the mechanism is, and indeed the cancellation
may appear fortuitous. As a matter of fact, as briefly discussed in the Introduction,
the Bloch-Nordsieck theorem is specific to QED, and breaks down for non-abelian 
gauge theories~\cite{Doria:1980ak,DiLieto:1980nkq,Andrasi:1980qw,Carneiro:1980au,
Frenkel:1983di,Catani:1987xy,Caola:2020xup}. It is important, therefore, to 
develop a more general understanding of the cancellation, as well as a more 
intuitive picture of the underlying physical mechanism. The framework
is provided by two general results, known respectively as the Kinoshita-Poggio-Quinn 
(KPQ) theorem~\cite{Poggio:1976qr,Sterman:1976jh,Kinoshita:1977dd} and the
Kinoshita-Lee-Nauenberg (KLN) theorem~\cite{Kinoshita:1962ur,Lee:1964is,
Sterman:1978bi,Sterman:1978bj}, which apply directly to any quantum theory
with massless particles. The KPQ theorem establishes that momentum-space
Green functions with external momenta which are off the mass shell are infrared 
safe. This is a natural consequence of the physical picture that we have been 
developing: the fluctuations of off-shell fields are confined to limited volumes
of space-time, and cannot be affected by long-distance singularities. The KLN
theorem, on the other hand, establishes the general framework for the cancellation
of singularities when on-shell quantities are evaluated: one finds that, in any theory
involving massless particles, infrared divergences can be traced to the presence
of sets of quantum states that are degenerate in energy, and the divergences
cancel when the transition rates are summed over the sets of degenerate 
initial {\it and} final states. In what follows, we will sketch a proof of the KLN
theorem, following the line of argument presented in \cite{Lee:1964is}. We find 
this approach particularly suited to show the complete generality of the cancellation,
which indeed applies to any quantum system with sets of states which are
degenerate in energy, including non-relativistic theories and effective field theories.

Consider then a quantum-mechanical system, characterised by a Hilbert space 
of states $\mathcal H$, and by a quantum hamiltonian $H$ which we split into a 
solvable, quadratic part, and an interaction term, as
\beq
  H (t) \, = \, H_0 + H_I (t) \, ,
\label{myham}
\eeq
Crucially, we work in the {\it interaction picture}, where operators evolve in time
by the action of the free hamiltonian $H_0$, while the time evolution of quantum
states is dictated by the interaction hamiltonian $H_I (t)$. The time evolution operator
is then given by
\beq
  U \big( t_2, t_1 \big) \, = \,  {\rm T} \exp \left[ - {\rm i} 
  \int_{t_1}^{t_2} d t \, H_I (t) \right] \, ,
\label{eq:U(t,t')}
\eeq
where $H_I (t)$ is the interaction hamiltonian in the interaction picture,
\beq
  H_I (t) \, = \, {\rm e}^{ {\rm i} H_0 t} \, H_I (0) \, {\rm e}^{- {\rm i}  H_0 t} \, .
\label{intint}
\eeq
The $S$-matrix of the theory can then be formally defined by the limit
\beq
  S \, \equiv \, \lim_{t, t' \to \infty} U(t', 0) \, U(0, - t) \, \equiv \, \Omega_-^\dagger \, 
  \Omega_+ \, ,
\label{Sdef}
\eeq
where the $\Omega_\pm$ operators, sometimes called {\it M\"oller operators},
are in turn defined by
\beq
  \Omega_\pm \, \equiv \, \lim_{t \to \infty} \, T \exp \left[ - {\rm i} \int_{\mp t}^{0} d t \, 
  H_I (t) \right]  \, .
\label{Molldefi}
\eeq
Armed with this definition, we can compute transition amplitudes between incoming
and outgoing states in terms of the M\"oller operators, as
\beq
  \big\langle b, {\rm out}  \big| a, {\rm in} \big\rangle \, \equiv \,
  \big\langle b, {\rm in} \big| \, S \, \big| a, {\rm in} \big \rangle \, = \, 
  \sum_c \big \langle b \big| \, \Omega_-^\dagger \, \big| c \big\rangle 
  \big \langle c \big| \, \Omega_+ \, \big| a \big\rangle \, ,
\label{smatrel} 
\eeq   
where we used the completeness of the incoming states, and, in the last expression, 
we omitted for simplicity the initial-state label, as we will do in the following. We are 
tacitly assuming that the asymptotic states are eigenstates of the free hamiltonian 
$H_0$: if there are sets of energy-degenerate states, this is in general not a good 
approximation (as we will see in detail in the next section), since energy degeneracy 
signals long-range interactions. In the present setting, the problem emerges in the 
form of singularities in perturbation theory, which we will have to try to cancel. In order 
to proceed, we define the transition probabilities (per unit volume and per unit time)
\beq
  P \big( a \to b \big) \, = \, 
  \big| \bra{b} S \ket{a} \big|^2 & = & \sum \limits_{c, d} 
  \left( \bra{b} \Omega_-^\dagger \ket{c} \bra{c}
  \Omega_+ \ket{a} \right)
  \left( \bra {b} \Omega_-^\dagger \ket{d} \bra{d}
  \Omega_+ \ket{a} \right)^*  \nonumber \\
  & = &  \sum \limits_{c, d} \bra{d} \Omega_+ \ket{a}^* 
  \bra{c} \Omega_+ \ket{a}  
  \bra{b} \Omega_-^\dagger \ket{c} 
  \bra{b} \Omega_-^\dagger \ket{d}^* \nonumber \\
  & \equiv & \sum \limits_{c, d} W_{a, c d}^+ 
  \left( W_{b, c d}^- \right)^* \, ,
\label{Pab} 
\eeq
where in the last equality we defined the matrices
\beq
  W_{l, c d}^\pm \, \equiv \, \bra{d} \Omega_\pm \ket{l}^* 
  \bra{c} \Omega_\pm \ket{l} \, ,
\label{Wabc}
\eeq
providing a factorisation of  the transition rate into two parts, carrying the dependence 
on the initial state $\ket{a}$, and on the final state $\ket{b}$, respectively. 

The KLN theorem can be stated in terms of the transition probabilities $P(a \to 
b)$ as follows. Consider the eigenstates $\ket{\alpha}$ of the hamiltonian $H$, and 
let $E_\alpha$ be their energies. Focus on the set of eigenstates with energies falling 
in a fixed interval $E_0 - \epsilon < E_\alpha < E_0 + \epsilon$, and denote that set by 
$D_\epsilon (E_0)$. Transition probabilities $P(\alpha \to \beta)$ are in general 
singular in perturbation theory if $E_\alpha$ or $E_\beta$ are degenerate. In such 
cases, one must introduce a regulator for the singularity, for example a mass $m$
for all particles involved in the scattering. In the presence of the regulator $m$, one 
can then construct the {\it inclusive} transition probability
\beq
  P_m \Big[ D_\epsilon(E_a) \to D_\epsilon(E_b) \Big] \, \equiv \,
  \sum \limits_{a \in D_\epsilon(E_a)} \sum \limits_{b \in D_\epsilon (E_b)} 
  P_m (a \to b) \, ,
\eeq    
and the theorem states that this quantity remains finite order by order in perturbation 
theory when the regulator is removed by taking $m \to 0$. Given \eq{Pab}, this can 
be proved by considering the matrices $W_{l, c d}^\pm$, and showing that the 
combination
\beq
  \sum \limits_{a \in D_\epsilon(E_a)} W^{\pm}_{a, c d} \, \equiv \, W^{\pm}_{c d}(a)
\label{Wsum} 
\eeq
is finite. In order to see how the cancellation comes about, let us begin by considering
the first non-trivial perturbative order. Expanding \eq{Molldefi} to first order in $H_I$ we
find, for example
\beq
\label{eq:U}
  \bra{i} \Omega_- \ket{j} & = & \bra{i} \left[ 1 - {\rm i} \int_\infty^0 d t  \, H_I (t) 
  \right] \ket{j} \nonumber \\
  & = &  \delta_{ij} + {\rm i} \! \int_{0}^{\infty} d t \bra{i} {\rm e}^{{\rm i} H_0 t}
  \, H_I {\rm e}^{- {\rm i} H_0 t} \ket{j} \nonumber \\
  & = &  \delta_{ij} + {\rm i}  \bra{i} H_I \ket{j} 
  \int_0^\infty \! d t \, {\rm e}^{{\rm i} \big[ \left( E_i -E_j \right) + {\rm i} \epsilon \big] t }
  \nonumber \\
  & = &  \delta_{ij} - \frac{\bra{i} H_I \ket{j}}{E_i - E_j + {\rm i} \epsilon} \, ,
\eeq
where the infinitesimal real constant $\epsilon$ ensures the convergence of the integral 
at infinity, and $E_{i}$ and $E_{j}$ are the eigenvalues of $H_{0}$ for the states $\ket{i}$ 
and $\ket{j}$, respectively. \eq{eq:U}, and the analogous result for the matrix elements of
$\Omega_+$, can be substituted in \eq{Wabc}, yielding
\beq
  W^{\pm}_{a, ij} \, = \, \delta_{ia} \delta_{ja} - \frac{\bra{i} H_I \ket{a}}{E_i - E_a 
  \mp {\rm i} \epsilon} \, \delta_{ja} - \frac{\bra{j} H_I \ket{a}^*}{E_j - E_a \pm {\rm i} 
  \epsilon} \, \delta_{ia} \, ,
\label{outW}
\eeq
up to corrections of second order in $H_I$. If the energy levels $E_i$ and $E_j$
become degenerate when the regulator $m$ is removed, \eq{outW} exhibits 
singularities whenever $\ket{a}$ coincides with either $\ket{i}$ or $\ket{j}$.
It is easy to verify that the singularities cancel at this order in the combination
$W^{\pm}_{c d}(a)$ defined in \eq{Wsum}. Four different cases arise naturally
\begin{itemize}
\item Both states $\ket{i}$ and $\ket{j}$ lie in $D_\epsilon(E_a)$. This is the 
potentially singular case, but one easily verifies that the contribution with
$\ket{a} = \ket{i}$ cancels the one with $\ket{a} = \ket{j}$. So long as
$\ket{i} \neq \ket{j}$, one then finds $W_{ij}^{\pm}(a) = 0$.
\item Only state $\ket{i}$ belongs to $D_\epsilon(E_a)$, while $j \notin 
D_\epsilon(E_a)$. In this case only one term appears in the sum, but it 
is not singular
\beq
  W_{ij}^{\pm}(a) \, = \, - \frac{\bra{i} H_I \ket{j}}{E_j - E_i \pm {\rm i} \epsilon} \, , 
  \quad E_j \neq E_i \, ,
\label{seccase}
\eeq
\item If $i \notin D_\epsilon(E_a)$, $j \in D_\epsilon(E_a)$, by the same token 
one finds
\beq
  W_{ij}^{\pm}(a) \, = \, - \frac{\bra{i} H_I \ket{j}}{E_{i} - E_j \mp {\rm i}\epsilon} \, , 
  \quad E_i \neq E_j \, ,
\label{thicase}
\eeq
\item Finally, if $i \notin D_\epsilon(E_a),j \notin D_\epsilon(E_a)$, the sum receives 
no contributions, and one simply finds $W_{ij}^{\pm} (a) = 0$.
\end{itemize}
This proves that $W_{ij}^{\pm}(a)$ is infrared-finite to first order in the perturbation
$H_I$: the generalisation of the proof to arbitrary perturbative order can be achieved
by induction. To begin with, in order to keep track of the order of the perturbation, we
extract a factor of a (small) coupling $g$ from the interaction hamiltonian, changing
$H_I \to g H_I$. Next, we expand both the M\"oller operators $\Omega_\pm$
and  the matrices $W_{a, ij}^\pm$ and $W_{ij}^\pm (a)$ in powers of $g$. In 
particular, using the definitions in \eq{Wabc} and \eq{Wsum}, we can write
\beq
  W^\pm_{ij} (a) & \equiv &  \sum_{n = 0}^\infty \, g^n \, W^\pm_{n, \, ij} (a) \, , 
  \qquad
  \Omega_\pm \, \equiv \,  \sum_{n = 0}^\infty \, g^n \, \Omega_{\pm, \, n} \, , 
  \nonumber \\
  W_{n, \, ij}^\pm (a) & = & \sum_{r=0}^n \, \sum\limits_{a \in D_\epsilon (E_a)} 
  \bra{j} \Omega_{\pm, \, r} \ket{a}^* \bra{i} \Omega_{\pm, \, n - r} \ket{a} \, .
\label{Wn}
\eeq
Finally, we need to keep in mind some fundamental properties of the M\"oller
operators: first of all, unitarity
\beq
 \Omega_\pm^\dagger \, \Omega_\pm \, = \, {\bf 1} \, ,
\label{unitarMol}
\eeq 
which implies and reflects the unitarity of the $S$ matrix. Next, as we 
assume the asymptotic states to be eigenstates of the unperturbed hamiltonian
$H_0$, we conclude that the M\"oller operators must diagonalise the complete 
hamiltonian
\beq
  \Omega_\pm^{\dagger} \, H \, \Omega_\pm \, = \, \hat{H}_0 \, ,
\label{Molldiag}
\eeq 
where $\hat{H}_0$ is diagonal in the same basis as $H_0$, but we allow
for a difference in eigenvalues due to interactions: the effects of renormalisation, 
and more generally of virtual corrections, shifting particle masses, contribute to 
this difference. Combining \eq{Molldiag} with \eq{myham} we can write
\beq
  \Big[ \Omega_\pm, \hat{H}_0 \Big] \, = \, \left( H - \hat{H_{0}} \right) \Omega_\pm 
  \, = \, \big( \Delta + g H_I \big) \Omega_\pm \, ,
\label{Uia}
\eeq 
where $\Delta \equiv H_{0} - \hat{H}_0$. The diagonal operator $\Delta$ can also
be expanded in powers of the coupling as $\Delta = \sum_n g^{n} \Delta_{n}$.

With these tools in hand, we can now proceed to build the inductive argument.
We have already shown that $W_{n, \, ij}^{\pm}(a)$ is finite for $n=1$, and we 
know that $\Delta$ vanishes at lowest order; furthermore, we can safely assume
that $\Delta_n$ is infrared finite, which can be achieved order by order by a 
suitable choice of renormalisation scheme. With these results, we can now 
prove that, if $W_{n, \, ij}^{\pm}(a)$ is infrared finite for $n \leq N$, then 
$W_{N+1, \, ij}^{\pm}(a)$ will also be infrared-finite. As before there are  four 
cases to be considered, which in fact can be reduced to three.
\begin{itemize}
\item Consider first the situation in which $\ket{i} \notin D_\epsilon(E_a)$, with 
no assumption on $\ket{j}$. Then, taking the matrix element of \eq{Uia} between
the states $\bra{i}$ and $\ket{a}$, and using the fact that in this case $E_a 
\neq E_i$, we find
\beq
  \Big[ \,\Omega_{\pm, \, n} \Big]_{ia} \, = \, \frac{1}{E_a - E_i} 
  \left\{ \sum\limits_{k} \big( H_I \big)_{ik} \Big[ \Omega_{\pm, \, n-1} \Big]_{ka} 
  + \sum_{s=1}^n \big(\Delta_s \big)_{ii} \Big[ \Omega_{\pm, \, n - s} \Big]_{i a} 
  \right\} \, ,
\label{Uria}
\eeq
where we made use of the fact that $\Delta$ is diagonal and vanishes at lowest 
order. Crucially, the $n$-th order contribution to the M\"oller operator matrix 
element is expressed in terms of lower-order terms, multiplied times finite
factors. We can now substitute \eq{Uria} for one of the two matrix elements
in the definition of $W_{ij}^{\pm}(a)$, \eq{Wn}. Taking the contribution at
order $N+1$, and suitably shifting the summation indices, we find
\beq
  W_{N+1, \, ij}^\pm (a) \, = \, \frac{1}{E_a - E_i} 
  \left[ \sum_k \big( H_I \big)_{ik} W_{N, \, kj}^{\pm}(a) + 
  \sum_{s=1}^{N} \big(\Delta_s \big)_{ii} W_{N - s, \, ij}^{\pm}(a) \right] \, .
\label{finind}
\eeq
which is manifestly finite under the induction hypothesis.  
\item The symmetric case, in which $\ket{j} \notin D_\epsilon(E_a)$, with 
no assumption on $\ket{i}$, yields a finite result thanks to the hermiticity
properties of the matrices $W_{ij}^{\pm}(a)$. Indeed,
\beq
  W_{N, \, ij}^\pm (a) \, = \, \left[ W_{N, \, ji}^\pm (a) \right]^*\, ,
\label{hermW} 
\eeq
as easily seen from \eq{Wn}.
\item The last case to be considered, when both states $\ket{i}$ and $\ket{j}$ 
belong to the set $D_\epsilon(E_a)$, is potentially the most intricate. It is 
remarkable and suggestive that finiteness in this case can be established
using the unitarity of the M\"oller operators, and thus of the $S$ matrix.
Taking the $n$-th perturbative order in the expansion of the matrix elements of
\eq{unitarMol}, one finds
\beq
  \sum_{r = 0}^n \, \sum_{a \in D_\epsilon(E_a)} 
  \Big[ \,\Omega_{\pm, \, r} \Big]_{ja}^*  \, \Big[ \,\Omega_{\pm, \, n - r} \Big]_{ia} 
  + \sum_{r = 0}^n \, \sum_{a \notin D_\epsilon(E_a)}
  \Big[ \,\Omega_{\pm, \, r} \Big]_{ja}^*  \, \Big[ \,\Omega_{\pm, \, n - r} \Big]_{ia}
  \, = \, 0 \, . 
\label{uninoutD}
\eeq
This allows us to express the matrix elements of $W^\pm (a)$ between states
in the degenerate set in terms of those between states that lie outside the set,
which have already been shown to be finite. At order $N+1$,
\beq
  W_{N+1, \, ij}^{\pm}(a) \, = \, - \sum_{r=0}^{N+1} \, \sum\limits_{a \notin D_\eps(E_a)} 
  \Big[ \,\Omega_{\pm, \, r} \Big]_{ja}^*  \, \Big[ \,\Omega_{\pm, \, M + 1 - r} 
  \Big]_{ia} \, ,
\eeq
which completes the proof of the KLN theorem.
\end{itemize}
The simple quantum-mechanical setup of the proof that we have outlined
highlights the complete generality of the cancellation mechanism. On the other
hand, in a quantum field theory context, a detailed implementation of the proof
requires further work: one must verify for example that the renormalisation
procedure does not interfere with the cancellation, and identify the relevant
contributions in terms of Feynman graphs, where the splitting of the $S$-matrix 
in terms of M\"oller operators is related to unitarity cuts~\cite{Sterman:1978bi,
Sterman:1978bj,Akhoury:1997pb}. Finally, one needs to face the difficulties 
associated with summing over initial state degenerate configurations in the 
context of scattering experiments. In the next Section, we will explore an 
alternative approach, which attempts to construct directly a finite $S$-matrix, 
instead of relying upon a cancellation at the level of observed cross sections.


\subsection{Saving the $S$-matrix: coherent states}
\label{Coherent}
 
The KLN theorem provides us, in principle, with a practical way out of the infrared
problem: while it remains true that the fundamental theoretical objects for scattering
predictions, the $S$-matrix elements, are ill-defined in theories with massless 
particles, a careful construction of observable cross sections leads to finite
predictions, order by order in perturbation theory. As we will see, however, there 
is quite some distance between this solution `in principle' and an effective method 
to make reliable predictions, especially for non abelian theories. Furthermore, one
feels that it should be possible to be more ambitious: indeed, the physics of the
problem is quite clear, and suggests directions for refining the quantum field theory
framework in the massless limit, in order to deal with finite quantities at all stages
of the calculations.

To state again the basic fact, infrared singularities arise in massless theories 
because emission and absorption probabilities for massless particles do not
decrease fast enough at large distances and long times. This means that a
basic assumption in the construction of the $S$-matrix fails: interactions never 
become negligible in the distant past or future, and therefore the description
of the asymptotic states as Fock states built out of isolated non-interacting
particles is (to say the least) not sufficiently precise. This sounds, hopefully,
like a problem that can be fixed: after all, when given a quantum Hamiltonian
$H$, the distinction between what we call the `free', or `integrable' part, $H_0$,
and the `interaction' part $H_I$ has a degree of arbitrariness. This could 
be exploited, in principle, by identifying interactions that are non-negligible at 
long distances, and re-assigning them to $H_0$, which is supposed to be 
treated exactly at all distance scales. This would translate into a more accurate
characterisation of the asymptotic states, which would no longer be defined as 
eigenstates of the free Hamiltonian $H_0$, but rather as eigenstates of the proper 
asymptotic Hamiltonian. Such states have indeed been introduced and extensively 
studied, under the label of {\it coherent states}: they have been shown to provide a 
consistent definition of the $S$-matrix for theories with massless particles, and 
their construction has been exploited to uncover deep and interesting properties
of the infrared dynamics of perturbative gauge theories.

After a series of early suggestions and preliminary studies~\cite{Dollard:1964,
Fradkin:1960lgt,Chung:1965zza,Kibble:1969ip,Kibble:1969ep,Kibble:1969kd}, 
the breakthrough came with the landmark paper by Faddeev and 
Kulish~\cite{Kulish:1970ut} in 1970, where the problem of the construction 
of an infrared-finite $S$-matrix for QED was formally solved in a definitive manner.
In the rest of this Section, we will provide an introduction to the formalism of
coherent states for massless quantum field theories, following the general
logic of Ref.~\cite{Kulish:1970ut}, but using the concrete implementation and examples
from Ref.~\cite{DelDuca:1989jt}, and we will give a few pointers to later developments
along these lines. As will become clear, coherent states do not (as yet) provide
a fully practical tool for the evaluation of gauge theory observables for collider
applications, but they do give a precise theoretical framework, with a clear and
transparent physical interpretation, which must to some extent underpin and help
organise all subsequent developments.

As was the case for the KLN theorem, the starting point is the Hamiltonian of
our quantum field theory, which we split into a solvable, quadratic part, and 
an interaction, according to \eq{myham}. Time evolution is described in the 
interaction picture, so that Eqs.~(\ref{intint}) and (\ref{Molldefi}) hold. The crucial 
step is now to use the explicit knowledge of the time dependence of the evolution 
operator in the interaction picture to identify and isolate the leading behaviour 
of the interaction Hamiltonian at large times (we will see an explicit example of 
this step in \secn{phi3coh}). We need, of course, to introduce a scale $E$, in 
order to define what we mean by long and short times: roughly speaking, 
we consider times $t \geq \hbar/ E$ as asymptotic. We then write
\beq
  H_I (t) \, = \, H_R^E (t) + H_A^E (t) \, ,
\label{asimhal}
\eeq
where $H_A^E(t)$ denotes the {\it asymptotic Hamiltonian}, responsible for the 
leading large-time behaviour of the evolution operator. In turn, using $H_A^E(t)$, 
we can define asymptotic M\"oller operators
\beq
  \Omega_{A, \, \pm} (E) \, \equiv \, \lim_{t \to \infty} \, 
  T \exp \left[ - {\rm i} \int_{\mp t}^{0} d t \, H_A^E (t) \right]  \, ,
\label{asimMoll}
\eeq
which allows us to isolate the large-time contributions to the $S$-matrix by writing
\beq
  S  & = & \Omega_{A, \, -}^\dagger (E) \, \Omega_{R, \, -}^\dagger (E) \, 
  \Omega_{R, \, +} (E) \, \Omega_{A, \, +} (E) \nonumber \\
  & \equiv & \Omega_{A, \, -}^\dagger (E) \, S_R (E) \, \Omega_{A, \, +} (E) \, ,
\label{Sfact} 
\eeq
where we have introduced regular M\"oller operators $\Omega_{R, \, \pm}(E)$, 
and we have (somewhat optimistically) defined a {\it regular} $S$-matrix
\beq
  S_R (E)  \, = \, \Omega_{A, \, -} (E) \, S \, \Omega_{A, \, +}^\dagger (E) \, .
\label{Sreg} 
\eeq
Note that, in general, the regular Hamiltonian $H_R^E (t)$ does not commute 
with the asymptotic one, so the regular M\"oller operator is not simply computed
by exponentiating $H_R^E(t)$, but involves commutator terms: the fact that
the regular $S$-matrix is indeed free of infrared singularities must therefore 
be checked with explicit definitions in hand for the operators involved. Specifically,
with the definitions above, we expect that the regular $S$-matrix elements
in the usual Fock space will be free of infrared singularities. Alternatively, and 
perhaps more appropriately, one may define a basis of {\it coherent states}
\beq
  \ket{h, \, \pm}_E \, = \, \Omega_{A, \, \pm}^\dagger (E) \, \ket{h} \, ,
\label{cohsta}
\eeq
by acting with the asymptotic M\"oller operators on a generic Fock state 
$\ket{h}$. The expectation is then that the {\it usual} $S$-matrix will be
free of infrared singularities in the coherent state basis.

In QED, with massive fermions, this expectation was turned into a theorem 
by Faddeev and Kulish in Ref.~\cite{Kulish:1970ut}. As we will note below, the form
of the coherent state operator is particularly simple in the soft limit for abelian
gauge theories: this allows for a full treatment, and one can prove not only 
the perturbative finiteness of the coherent state $S$-matrix, but also that 
it is possible to build a Hilbert space of coherent states which is separable, 
gauge invariant, and containing a gauge-invariant subspace of positive norm 
states. In some sense, for the soft problem in the massive abelian theory, the book is 
closed. In the non-abelian case, the situation is considerably more complicated,
because there are unavoidable collinear singularities associated with the 
self-interactions of massless gluons: as we will see below, the coherent 
state operator is qualitatively much more intricate in the collinear limit, and
this makes it much more difficult to prove all-order statements. A number 
of papers between the late seventies and the early eighties explored the
construction of non-abelian coherent states, their gauge properties, and
the mechanism for the cancellation of divergences~\cite{Greco:1978te,
Curci:1978kj,Butler:1978rd,Curci:1979bg,Nelson:1980yt,Nelson:1980qs,
Muta:1981pe,Ciafaloni:1984zr,Catani:1985xt,Catani:1985ta,Catani:1987sp}, 
and finally an elegant formal proof of the finiteness of the non-abelian 
coherent-state $S$-matrix, including collinear singularities, was given 
by Giavarini and Marchesini in Ref.~\cite{Giavarini:1987ts}.

In the remainder of this Section, rather than focusing on all-order proofs, 
we would like to flesh out the rather formal arguments given above, showing
in concrete examples how the asymptotic Hamiltonian is constructed, and
how the coherent state operator engineers the cancellation of infrared
singularities. To this end, following Ref.~\cite{DelDuca:1989jt}, we begin by 
considering the simple case of a scalar theory, affected by collinear
divergences only, and then we briefly highlight similarities and differences
with the physically more interesting case of four-dimensional gauge theories.


\subsubsection{Collinear divergences in a scalar theory}
\label{phi3coh}

Our toy model for the construction of coherent states is a massless scalar 
theory with cubic interaction, which we take to live in dimension $d = 6$, where 
the theory is renormalisable. The Lagrangian is then simply
\beq
  {\cal L} \, = \, \frac{1}{2} \, \partial_\mu \phi \, \partial^\mu \phi - 
  \frac{\lambda}{6} \, \phi^3 \, .
\label{Lphi3} 
\eeq
A detailed power-counting argument at the diagrammatic level shows that 
scattering amplitudes in this theory can be affected by collinear singularities,
but not by soft ones. Here we will simply assume that this is the case, and 
we will study the asymptotic behaviour of scattering amplitudes focusing
on the collinear limit.

As explained in \secn{Coherent}, the construction of coherent states is 
particularly simple and transparent in the interaction picture, where
quantum operators evolve with the free Hamiltonian. The first step is then
to expand the scalar field in Fourier modes, each of which evolves
independently in time, with a fixed energy given by the classical mass-shell
condition. We write
\beq
  \phi \left( {\bf x}, t \right) \, = \, \int \widetilde{d k} \, \Big[ a({\bf k}) 
  {\rm e}^{\, {\rm i} \, {\bf k} \cdot {\bf x} \, - \, {\rm i} \, w({\bf k}) t} \, + \, 
  a^\dagger ({\bf k}) {\rm e}^{- {\rm i} \, {\bf k} \cdot {\bf x} \, + \, {\rm i} \, w({\bf k}) t}
  \Big] \, , 
\label{Fourphi} 
\eeq
where we defined
\beq
  \widetilde{d k} \, = \, \frac{d^5 k}{(2 \pi)^5 \, 2 w ({\bf k})} \, , \qquad 
  w ({\bf k}) \, = \, \left| {\bf k} \right| \, ,
\label{6dconv}
\eeq
and the creation and annihilation operators satisfy the normalisation condition
\beq
  \left[ a({\bf k}), a^\dagger ({\bf k}') \right] \, = \, (2 \pi)^5 \, 2 w({\bf k}) \, 
  \delta^5 \left( {\bf k} - {\bf k}' \right) \, .
\label{norma}
\eeq
It is straightforward to compute the interaction Hamiltonian, which is given by
\beq
  H_I (t) & = & \frac{\lambda}{6} \int d^5 x \, \phi^3 ({\bf x}, t) \nonumber \\
  & = & \frac{\lambda}{2}  \int \frac{\widetilde{d k}_1 
  \widetilde{d k}_2}{2 w_3 ({\bf k}_1, {\bf k}_2)} \Big[ a^\dagger ({\bf k}_1) a({\bf k}_2) 
  a ({\bf k}_1 - {\bf k}_2) \, {\rm e}^{ {\rm i} \, \left( w_1 - w_2 - w_3 \right) \, t} \nonumber \\
  && \hspace{2.8cm}  +  \, a^\dagger ({\bf k}_1) a^\dagger({\bf k}_2) 
  a^\dagger (- {\bf k}_1 - {\bf k}_2) \, {\rm e}^{ {\rm i} \, \left( w_1 + w_2 + w_3 \right) \, t} 
  \, + \, {\it h. c.} \Big] \, ,
\label{Hamphi3}
\eeq
where we introduced the shorthand notation $w_i \equiv w ({\bf k}_i)$, and 
where the spatial integration led to enforcing momentum conservation. In 
the spirit of time-ordered perturbation theory (TOPT), the Hamiltonian in 
\eq{Hamphi3} mediates interactions between on-shell particles, and, as 
a consequence, energy is not conserved at the interaction vertices. The 
coefficients of $t$ in the exponential factors are given by the energy deficits 
in the corresponding interactions, and each term in \eq{Hamphi3} has a 
transparent physical interpretation: for example, the first term acts on
a two-particle state, replacing it with a one-particle state, whereas the
second term corresponds to a quantum fluctuation in which three particles
are created out of the vacuum.

One now comes to the crucial part of the argument: contributions to $H_I (t)$ at large 
times, $|t| \to \infty$, are suppressed by the rapid oscillations of the exponential factors, 
except for configurations where the coefficients of $t$ become vanishingly small.
For the second term in \eq{Hamphi3}, and its hermitian conjugate, the only candidate
configurations with this property are the rather uninteresting ones in which all particles
involved have vanishing energy. On the other hand, it is clear that the first term and
its hermitian conjugate involve physically relevant configuration with a vanishing 
energy deficit: for example, soft configurations, say with $w_2 = 0$, where an 
energetic on-shell scalar quantum emits or absorbs a soft scalar without changing 
its own energy, and remaining on-shell, or, as we detail below, collinear configurations.

A power-counting analysis would show that soft configurations in this theory would
not lead to divergences in $S$-matrix elements. Bypassing the detailed argument,
here we concentrate directly on the collinear limit. To study it, we introduce a simple
parametrisation of the momenta entering the interaction vertex, writing
\beq
  {\bf k}_1 & = & w_1 \, \widehat{{\bf k}}_1 \, , \nonumber \\
  {\bf k}_2 & = & w_1 \big( \alpha \widehat{{\bf k}}_1 + 
  \beta \, \widehat{{\bf k}}_\perp \big) \, , \nonumber \\
  {\bf k}_3 & = & w_1 \big( (1 - \alpha) \widehat{{\bf k}}_1 -
  \beta \, \widehat{{\bf k}}_\perp \big) \, ,  
\label{momparam}
\eeq
where we picked ${\bf k}_1$ as collinear direction, and identified the perpendicular
direction in the scattering plane by ${\bf k}_\perp$, with $|\widehat{\bf k}_1| =
|\widehat{\bf k}_\perp | = 1$, and $\widehat{\bf k}_1 \cdot \widehat{\bf k}_\perp = 0$.
The collinear limit is clearly identified by $\beta \to 0$, while the limits $\alpha \to 0$
and $\alpha \to 1$ are soft. For small $\beta$, and generic $\alpha$, we easily find
\beq
 \left| w_1 - w_2 - w_3 \right| \, = \, \frac{\beta^2}{2 \alpha (1 - \alpha)} \, w_1 + 
 {\cal O} \left( \beta^4 \right) \, .
\label{colllim}
\eeq
At this stage we are assuming that soft configurations are not problematic:
the presence of singularities in the soft limits in \eq{colllim} is a harbinger of many
future problems, in cases where soft and collinear configurations have a singular
overlap. 
In the present case, writing the collinear asymptotic Hamiltonian is immediate:
we simply introduce an appropriate angular cutoff, and take the leading power of
the interaction Hamiltonian as $\beta \to 0$. We find
\beq
  H_A^\Delta (t) \, = \, \frac{\lambda}{2} \int \frac{\widetilde{d k}_1 
  \widetilde{d k}_2}{2 (1 - \alpha) w_1} \, \theta_\Delta ({\bf k}_1, {\bf k}_2)
  \Big[ a^\dagger ({\bf k}_1) a({\bf k}_2) 
  a ({\bf k}_1 - {\bf k}_2) \, {\rm e}^{ {\rm i} \, \frac{\beta^2}{2 \alpha (1 - \alpha)} w_1 t} 
  \, + \, {\it h. c.} \Big] \, ,
\label{Hasphi3}
\eeq
where $\theta_\Delta ({\bf k}_1, {\bf k}_2)$ is any set of $\theta$ functions constraining
the three momenta meeting at the vertex to be in the collinear region, say lying within 
a cone of angular size $\Delta$. Armed with the asymptotic Hamiltonian in \eq{Hasphi3},
one can easily compute the coherent state operators $\Omega_{A, \, \pm} (\Delta)$
order by order in perturbation theory. At leading order, performing the time integration,
one finds
\beq
  \Omega_{A, \, \pm}^{(1)} (\Delta) \, = \, \frac{\lambda}{2} \int \widetilde{d k}_1 
  \widetilde{d k}_2 \, \frac{\alpha}{\beta^2 w_1^2} \, \theta_\Delta ({\bf k}_1, {\bf k}_2)
  \Big[ a^\dagger ({\bf k}_1) a({\bf k}_2) 
  a ({\bf k}_1 - {\bf k}_2) \, + \, {\it h. c.} \Big] \, ,
\label{Omegaphi3}
\eeq
where the collinear enhancement as $\beta \to 0$ is clearly displayed in the 
first factor of the integrand. By construction, when applied to a single-particle 
Fock state, the operator in \eq{Omegaphi3} simulates the collinearly-enhanced 
part of the interaction, and generates a quasi-collinear pair; similarly, when acting
on a Fock state describing two quasi-collinear incoming particles, the operator
merges them into a single one.

\begin{figure}
\centering
  \includegraphics[height=2.7cm,width=8cm]{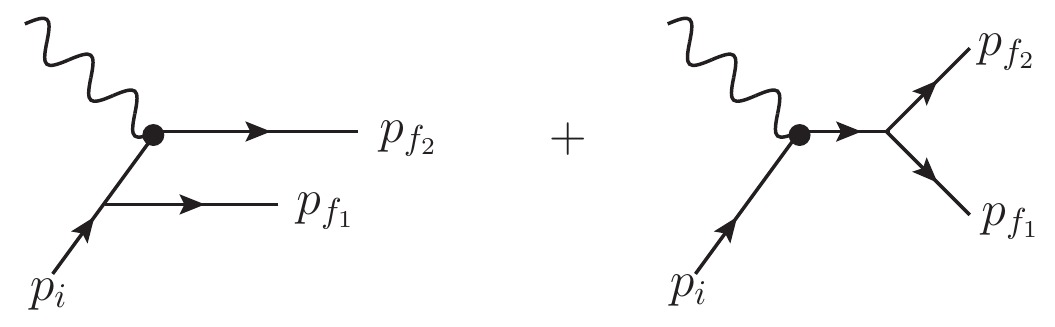}
  \caption{Single-radiation graphs contributing to scalar scattering from an external 
  current, at first order in the coupling $\lambda$, denoted by ${\cal A}_{\, S}$ in the 
  text.
  \label{disphifig}}
\end{figure}

In order to visualise more explicitly the mechanism of the cancellation, following
Ref.~\cite{DelDuca:1989jt}, we can work out the simple example of scattering from
a classical current, mimicking the well-known QCD calculation of collinear effects 
in DIS. Focusing on the real radiation correction, we compute the {\it regular}
$S$-matrix element 
\beq
  \bra{{\bf p}_{f_1}, {\bf p}_{f_2}} S_R (\Delta) \ket{{\bf p}_i} \, = \, 
  \bra{0} a ({\bf p}_{f_1}) a({\bf p}_{f_2}) \, \Omega_{A, \, -} (\Delta) \, S^{(0)} \,
  \Omega^\dagger_{A, \, +} (\Delta) \, a^\dagger ({\bf p}_i) \ket{0} \, ,
\label{Sdisphi3}
\eeq
where $S^{(0)}$ represent the scattering by the external current. We expect collinear
enhancements when ${\bf p}_{f_1}$ becomes collinear to ${\bf p}_{f_2}$, and when
either ${\bf p}_{f_1}$ or ${\bf p}_{f_2}$ become collinear to ${\bf p}_i$: for the sake
of this example, let us pick this second case, choosing ${\bf p}_{f_1}$ along a 
direction close to ${\bf p}_i$. In this case, ${\bf p}_{f_1}$ and ${\bf p}_{f_2}$ cannot
be collinear to each other, since the hard scattering by the external current imparts
a large momentum to the system. As a consequence, at ${\cal O} (\lambda)$ the
coherent state operator to the left of the classical scattering in \eq{Sdisphi3} cannot
contribute, and one is left with
\beq
  \bra{{\bf p}_{f_1}, {\bf p}_{f_2}} S_R (\Delta) \ket{{\bf p}_i}^{(1)} & = & 
  \bra{0} a ({\bf p}_{f_1}) a({\bf p}_{f_2}) \, S^{(0)} \,
  \Omega^{\dagger \, (1)}_{A, \, +} (\Delta) \, a^\dagger ({\bf p}_i) \ket{0} 
  \nonumber \\ && \hspace{5mm} + \, 
  \bra{0} a ({\bf p}_{f_1}) a({\bf p}_{f_2}) \, S^{(1)} \, a^\dagger ({\bf p}_i) \ket{0} 
  \nonumber \\ 
  & \equiv & {\cal A}^{(1)}_{ \, \Omega} + {\cal A}^{(1)}_{\, S} \, .
\label{Sdisphi31}
\eeq
In the first part of the amplitude, ${\cal A}^{(1)}_{ \, \Omega}$, the coherent state
operator acts on the initial particle, turning into a collinear pair, and one member
of the pair is then identified with the final-state particle carrying momentum 
${\bf p}_{f_1}$: indeed
\beq
  \Omega^{\dagger \, (1)}_{A, \, +} (\Delta) a^\dagger ({\bf p}_i) \ket{0}
  \, = \, \frac{\lambda}{2} \int \widetilde{d k} \, \frac{\alpha}{\beta^2 w_i^2} \, 
  \theta_\Delta ({\bf p}_i, {\bf k}) \, a^\dagger ({\bf k}) a^\dagger ({\bf p}_i - {\bf k}) 
  \ket{0} \, ,
\label{Omegaphi31}
\eeq
so that one finds
\beq
  {\cal A}^{(1)}_{ \, \Omega} \, = \, \lambda \, {\cal A}^{(0)} \left( {\bf p}_{f_2}, 
  {\bf p}_i - {\bf p}_{f_1} \right) \frac{\alpha_{f_1}}{\beta_{f_1}^2 w_i^2} \,
  \theta_\Delta ({\bf p}_i, {\bf p}_{f_1}) \, ,
\label{AOmega}
\eeq
where ${\cal A}^{(0)} \left( {\bf p}, {\bf q} \right)$ is the Born amplitude for scattering
by the external current. The second part of the amplitude in \eq{Sdisphi31}, 
${\cal A}^{(1)}_{ \, S}$, is just the ordinary $S$-matrix element for the scattering
accompanied by a single radiation: it comprises the two Feynman diagrams 
depicted in Fig.~\ref{disphifig} and yields
\beq
  {\cal A}^{(1)}_{ \, S} \, = \, \lambda \, {\cal A}^{(0)} \left( {\bf p}_{f_2}, 
  {\bf p}_i - {\bf p}_{f_1} \right) \, \frac{1}{(p_i - p_{f_1})^2} \, + \, \lambda \,
  {\cal A}^{(0)} \left( {\bf p}_{f_1} + {\bf p}_{f_2}, {\bf p}_i \right) 
  \, \frac{1}{(p_{f_1} + p_{f_2})^2} \, .
\label{AnonOmega}
\eeq
In our chosen configuration, the first term of \eq{AnonOmega} is collinearly 
enhanced, but one easily verifies that the collinear enhancement is precisely
cancelled at leading power in $\beta$ by the contribution of \eq{AOmega}: at this
order, we have indeed built a non-singular version of the scattering matrix in 
the collinear limit. Ref.~\cite{Kulish:1970ut} showed that this cancellation persist to 
all order in perturbation theory for soft enhancements in QED, while 
Ref.~\cite{Giavarini:1987ts} proved it in general for a non-abelian theory. A few 
observations are in order.
\begin{itemize}
\item Perhaps most remarkably, in the coherent state picture the cancellation
of singularities does not involve combining real and virtual corrections to scattering
amplitudes. Indeed, as illustrated in the simple example above, matrix elements
for real radiation and virtual corrections to the Born process are treated separately:
for the first, the coherent state contribution subtracts the leading-power enhancement,
that would lead to divergences upon phase-space integration; for the second, the
leading-power enhancement is subtracted at the level of the loop {\it integrand},
so that loop integrals becomes finite.
\item It is clear from the procedure leading to \eq{Hasphi3} that there is ample 
freedom in choosing what to include in the `asymptotic' behaviour of the Hamiltonian,
and therefore in the coherent state operator: depending on the application one has 
in mind, one might want to include in the asymptotic dynamics also terms that 
do not lead to divergences in the amplitudes, or sub-leading powers in the resolution 
parameter~\cite{Choi:2019rlz,Bonocore:2020xuj}. This could provide an interesting 
avenue to explore infrared enhancements, and eventually resummations, beyond 
leading power, a subject that has recently received considerable attention, and 
could lead to relevant phenomenological applications~(see, for example, 
\cite{Vita:2020ckn} and references therein).
\item It should also be clear from \eq{Hasphi3} and \eq{Omegaphi3} that the 
perturbative expansion of the collinear coherent state operator at higher orders 
will not display any immediate simplifications: it will involve increasingly longer 
strings of creation and annihilation operators, and, since all particles meeting
at collinear vertices carry non-vanishing momenta, the commutation properties 
do not lead to any obvious rearrangement at high orders. This is to be contrasted 
with the soft case in QED massive fermions, to be discussed briefly below, in 
\secn{QEDcoh}. The point is perhaps best made by noting that the states 
generated by acting with $\Omega_{A, \, \pm}^\dagger (\Delta)$ on the Fock 
states are not really `coherent' in the traditional sense: the asymptotic 
operator is not of the form $\exp (\alpha a^\dagger)$, with $\alpha$ a $c$-number.
Rather, it has a much more intricate structure, akin to $\exp (a^\dagger a a + 
{\it h.c.})$: in the presence of collinear dynamics, one should properly talk 
of {\it generalised} coherent states.
\end{itemize}


\subsubsection{Soft divergences in QED}
\label{QEDcoh}

\begin{figure}
\centering
  \includegraphics[height=5.5cm,width=11.5cm]{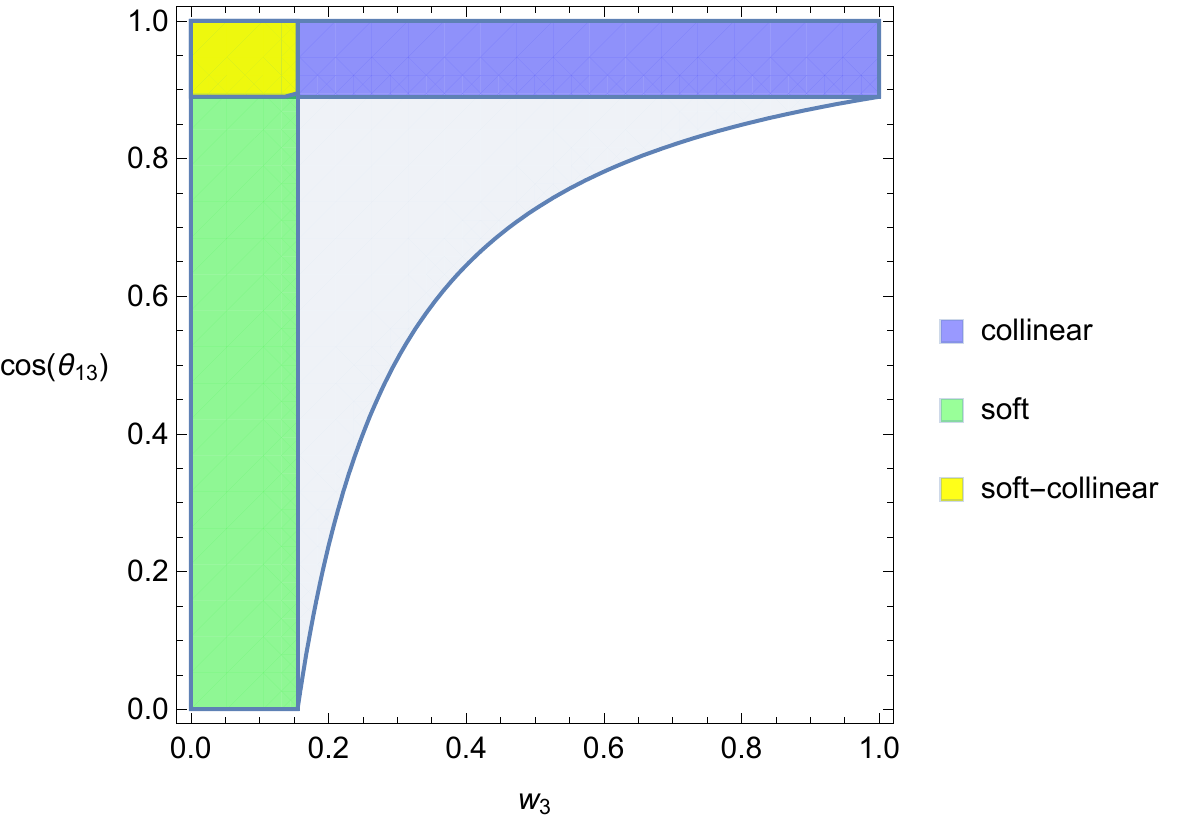}
  \caption{Asymptotic and hard regions for a massless gauge theory, with a 
  soft-collinear overlap.
  \label{hyperbfig}}
\end{figure}

When moving on to the much more interesting case of massless gauge theories,
it becomes unavoidable to grapple with issues related to the superposition of soft 
and collinear singularities. In QED, the latter are only relevant in the presence of
massless charged particles, while in the non-abelian case they are unavoidable,
as gluons must be massless and are charged. In such cases, the natural thing to 
do is to split the asymptotic hamiltonian into soft and collinear contributions, as
\beq
  H_A^E (t) \, = \,  H_S^E (t) + H_C^E (t) \, ,
\label{Hasqed}
\eeq
assigning the soft-collinear sector either to $H_S^E$  or to $H_C^E$, and avoiding
double counting. To get a feeling of how that might  work, notice that the boundary
of the asymptotic region is defined by imposing a `small' violation of energy 
conservation at the interaction vertex in \eq{Hamphi3},
\beq
  \left| w_1 - w_2 - w_3 \right| \, < \, E \, ,
\label{asvinc}
\eeq
where, for massless QED, we can for example take $w_3$ to be the photon energy.
Upon enforcing momentum conservation in the massless case, this becomes
\beq
  \left| w_1 - w_3 - \sqrt{ w_1^2 + w_3^2 - 2 w_1 w_3 \cos \theta_{13} } \, 
  \right| \, < \, E
\label{asvincexp}
\eeq
where $\theta_{13}$ is the angle between the momentum ${\bf k}_3$ of the photon 
and the momentum ${\bf k}_1$ of the charged particle. Considering for example
the case in which the photon is emitted by an incoming `electron', the physically
relevant branch of the inequality in \eq{asvincexp} establishes that the boundary
of the asymptotic region is a hyperbola in the $w_3$-$\cos \theta_{13}$ plane,
so that \eq{asvincexp} becomes
\beq
  \cos \theta_{13} \, > \, \left( 1 + \frac{E}{w_1} \right) - 
  \frac{E}{w_3} \left( 1 + \frac{E}{2 w_1} \right) \, ,
\label{hiperb}
\eeq
depicted in Fig.~\ref{hyperbfig}. Clearly, the precise shape of the boundary 
between the asymptotic region and the hard-scattering region is not
relevant, and one could just as well introduce separate cutoffs in energy and 
angle, as also suggested in Fig.~\ref{hyperbfig}: what matters is that the soft 
and collinear regions overlap. As a consequence, $H_S^E$ and $H_C^E$
in \eq{Hasqed} do not commute, and the coherent state operator can 
only be factorised into soft and collinear contributions up to commutator 
terms~\cite{DelDuca:1989jt}.
In the case of QED with massive fermions, one need not consider the collinear 
region at all; furthermore, one may exploit the fact that the soft part of the asymptotic 
hamiltonian is considerably simpler than its collinear counterpart, since, at
leading power in the soft momentum, one may neglect the recoil of the hard 
emitting fermion. A straightforward analysis for massive QED~\cite{Kulish:1970ut} 
leads to the expression
\beq
  H_S^E (t) \, = \, \int \widetilde{d k} \, \widetilde {d p} \, \frac{e}{w_p} \, 
  \theta \left(E - {\bf p} \cdot {\bf k} \right) {\cal H}_{\rm eik} ({\bf p}, {\bf k}; t) \, ,
\label{Hsqed}
\eeq
where ${\bf p}$ is the momentum of the emitting fermion, ${\bf k}$ is the photon
momentum, and we have introduced the (eikonal) momentum-space Hamiltonian 
density
\beq
  {\cal H}_{\rm eik} ({\bf p}, {\bf k}; t) \, = \, \rho ( {\bf p} ) \, \sum_\lambda
  {\bf p} \cdot {\bf \epsilon}_\lambda ({\bf k}) \, \Big[ a_\lambda ( {\bf k} )
  {\rm e}^{- {\rm i} \, \widehat{{\bf p}} \cdot {\bf k} \, t} + {\it h.c.} \Big] \, ,
\label{heik}
\eeq
where $\lambda$ is the photon polarisation and $\epsilon_\lambda$ the corresponding
polarisation vector. Crucially, we have been able to neglect ${\bf k}$ in all factors 
that are regular as $w_k \to 0$: creation and annihilation operators for the fermions 
are then evaluated at the same momentum ${\bf p}$, and they organise themselves 
to construct the fermion number operator
\beq
  \rho ( {\bf p} ) \, = \, \sum_s \Big[ b_s^\dagger ( {\bf p} ) b_s  ( {\bf p} ) -
  d_s^\dagger ( {\bf p} ) d_s  ( {\bf p} ) \Big] \, .
\label{fermnum}
\eeq
This simplification is at the heart of all subsequent developments in QED: it means, 
in particular, that for ordinary Fock states involving a fixed finite number of charged 
particles, which are eigenstates of $\rho({\bf p})$, the coherent state operator
takes the standard form $\exp(\alpha a^\dagger)$, with $\alpha$ a $c$-number.
The well-known properties of ordinary coherent states can then be exploited to
perform an all-order analysis. It is important to notice that the crucial simplification
leading to \eq{heik} is spin-independent: minimal coupling of the photon to a conserved 
matter current always leads, in the soft limit, to a Hamiltonian expressed in the terms 
of the appropriate number operator. The spin-independence of the soft approximation
will play an important role in many of the developments discussed in the coming 
sections.

It is clear that the coherent state approach provides in principle a conceptually 
satisfactory solution to the infrared problem, directly emerging from correcting
the inadequate approximation of asymptotic dynamics that is at the root of
soft and collinear divergences of ordinary $S$ matrix elements. Over the 
years, a number of authors have contributed to elucidating the interpretation 
of the method, providing precise definitions of the finite matrix elements that 
emerge, and working towards practical applications~\cite{Contopanagos:1991yb,
Contopanagos:1992fm,Contopanagos:1992fr,Misra:1993ps,Forde:2003jt,
More:2012ey,More:2013lua,More:2014rna,Hannesdottir:2019opa,Frye:2018xjj}. 
As we will see, however, phenomenological work in QCD has mostly followed 
a different approach, which we will sketch in \secn{QCD} and develop in the rest 
of our review. On the other hand, quite interestingly, the recent symmetry-based 
interpretation of infrared limits on the `celestial sphere'~\cite{Strominger:2017zoo} 
can be linked to coherent states~\cite{Kapec:2017tkm}. This has lead to a 
number of new studies and applications~\cite{Gabai:2016kuf,Carney:2017jut,
Carney:2017oxp,He:2020ifr,Anupam:2020fjp,Hirai:2020kzx}, including the case 
of gravity~\cite{Ware:2013zja,Choi:2017bna,Choi:2017ylo,Himwich:2020rro}, and 
it is to be expected that further developments will be forthcoming.


\subsection{What to do in QCD}
\label{QCD}

In the previous sections we have sketched three partial solutions to the infrared 
problem: an explicit all-order calculation in the abelian case, displaying exponentiation 
and cancellation of singular contributions, in \secn{cataQED}; a general proof of the 
cancellation of singularities in any theory with massless particles, in \secn{KLN}; 
and the outline of the construction of an improved Hilbert space, where the 
$S$-matrix becomes well defined, in \secn{Coherent}. Unfortunately, none of 
these arguments is fully adequate to tackle the practical problems that arise 
in the most relevant physical application, QCD at colliders; furthermore, many 
beautiful and interesting structural and  theoretical aspects of the problem 
remain hidden. 

To be more specific, the QED `solution' does not generalise to the non-abelian 
theory because the presence of massless interacting gluons inevitably brings
collinear divergences into the game, even if one couples only massive matter 
fields to the Yang-Mills theory. This, in turn, makes it necessary to consider 
initial-state degeneracies in the KLN sum. Next, of course, come the problems 
related to confinement, and the non-perturbative aspects of the non-abelian 
theory. In QED, the true asymptotic states are not Fock states, but they are 
not drastically different either, since at least they can be expressed in terms
of the same degrees of freedom, and we expect that a perturbative construction
of asymptotic coherent states, starting from Fock states, will remain close to the
true result. In QCD, the true asymptotic states are hadrons, and they are not
perturbatively related to Fock states built out of quarks and gluons. This makes
it essentially unfeasible to resort to the KLN theorem to compute physical QCD
cross sections: we would need to include in our calculations multi-parton
states, but in practice we do not know how such multi-parton states contribute
to the wave functions of the hadrons we are colliding\footnote{That being 
said, the contribution of two-parton states to QCD collider processes is 
phenomenologically relevant, and has been under study for several years, 
see for example~\cite{Gaunt:2009re,Diehl:2011yj,Manohar:2012jr,Diehl:2017kgu}.}. 
Similarly, a perturbative construction of QCD coherent states will mix Fock 
states with different parton content, with specific weights and phases dictated 
by the perturbative asymptotic hamiltonian: these weights and phases are 
bound to be drastically altered by non-perturbative effects, in ways that are 
not under control.

With limited help from the general cancellation theorems that we have outlined,
perturbative QCD relies upon the pillar of asymptotic freedom, and a mix of 
rigourous all-order perturbative analyses with physical understanding strongly 
supported by experiment. There are basically two categories of hadronic 
observables that are accessible by means of perturbative tools, and we briefly
outline them below.


\subsubsection{Infrared-safe cross sections}
\label{IRsafe}

In a situation without strongly-interacting particles in the initial state (in other 
words, at lepton colliders) one can hope to mimic the `QED solution', looking
for observables that are sufficiently inclusive in the soft and collinear radiation,
in order for the associated divergences to cancel. The first step is then to 
introduce soft and collinear regulators, in order to construct a perturbative
estimate of the chosen observable. In the renormalised theory, quark masses
$m_i (\mu)$, running with the renormalisation scale $\mu$, would act as 
a partial collinear regulator, and a gluon mass $m_g$ would regulate all infrared
divergences, were it not for the technical difficulties associated with the 
breaking of gauge invariance. In practice, the only workable and universally
applied infrared regularisation scheme for non-abelian theories is dimensional 
regularisation\footnote{In special circumstances, other possibilities have been 
explored, see for example~\cite{Barbieri:1973ni,Barbieri:1973mt,Alday:2009zm,
Henn:2010ir,Henn:2011by,Gardi:2011yz}.}, where, after renormalisation, one takes 
$d = 4 - 2 \epsilon$, with  $\epsilon < 0$. One then proceeds to compute a 
parton-level prediction for the observable, which we take to depend upon a 
set of momenta $\{p_i\}$. We write it as
\beq
  \sigma_{\rm part} \, = \, \sigma_0 \, {\cal F}_{\rm part} 
  \left( \frac{p_i \cdot p_j}{\mu^2}, 
  \alpha_s(\mu); \frac{m}{\mu}, 
  \epsilon \right)\, ,
\label{sigmapart}
\eeq
where $m$ is a parton-level mass scale, used as an infrared regulator. If the observable 
is sufficiently inclusive, in other words sufficiently insensitive to soft and collinear 
radiation, the perturbative parton-level prediction will  have a finite limit when 
the infrared regulator(s) are removed. One can then 
write~\cite{Sterman:1995fz}
\beq
  \sigma_{\rm part} \, = \, \sigma_0 \, {\cal F}_{\rm part} 
  \left( \frac{p_i \cdot p_j}{\mu^2}, \alpha_s(\mu); 0, 0 \right) \, + \, 
  {\cal O} \left[ \left( \frac{m}{\mu} \right)^{\! p}, \epsilon \right] \, ,
\label{sigmapart0}
\eeq
with $p$ a positive integer. Observables with this property are called {\it infrared 
safe}, and general criteria are available to ascertain the existence of the finite limit 
in \eq{sigmapart0} (some examples will be given in \secn{IRSafeObs}), which
can be established to all orders in perturbation theory. At this point, two further 
steps are needed in order to treat the leading term of \eq{sigmapart0} as a 
reliable estimate of the corresponding hadronic observable. First, we need 
to be able to trust the perturbative expansion, at  least as an asymptotic 
expansion. This happens in QCD if it is appropriate to choose the renormalisation 
scale $\mu$ as a hard scale, $\mu \gg \Lambda_{\rm QCD}$, which is the 
case if all Mandelstam invariants $p_i \cdot  p_j$ in \eq{sigmapart0} are large
in the same sense. The second step is more subtle, and ultimately involves a
degree of assumption: we {\it interpret} the weak dependence of $\sigma_{\rm 
part}$ on the infrared regulators as indicating a weak dependence on
long-time, long-distance physical processes, happening at length scales
$\Delta x \gg \hbar/\Lambda_{\rm QCD}$. These are the length scales at 
which partons hadronise and color is neutralised. We conclude that our
parton-level perturbative estimate of the observable will only be weakly 
affected by hadronisation corrections, which will be expected to be of
parametric size $\left( \Lambda_{\rm QCD}/\mu \right)^q$, with $q$ a 
positive integer. In many well-motivated models of power corrections,
where $m$ in \eq{sigmapart} is related to a gluon mass (see for 
example~\cite{Dokshitzer:1995qm,Gardi:2001di}), one finds that $q = p$,
which is generally supported by arguments based on the OPE (when 
available), and consistent with experimental data.

It is worth pointing out that this last step, translating from parton to hadron 
language, relies on a picture of confinement which is very strongly supported 
by an immense body of experimental data, but cannot be proved within 
QCD until better non-pertubative techniques become available. One could, 
in principle, imagine a world in which, as partons `cool', flowing away from 
the hard scattering, they meet a sharp confining transition which drastically
redistributes their momenta. The  data show instead  that confinement and
hadronisation after a hard collision happen {\it locally} in momentum space,
so that the parton configuration emerging from the hard scattering is closely
mirrored by the hadron distribution detected at large distances, with corrections
that are parametrically negligible at high energies -- a picture that was described 
as `Local Parton-Hadron Duality' in the  early decades of QCD~\cite{Azimov:1984np,
Dokshitzer:1991eq}.


\subsubsection{Factorisable cross sections}
\label{Factorsafe}

Whenever strongly interacting particles are present in the initial state\footnote{Or,
for that matter, whenever one considers processes with {\it identified} hadrons,
either in the initial or in the final state.}, the relatively simple picture of infrared 
safety discussed in \secn{IRsafe} is bound to break down. As discussed in a 
simple example in \secn{CollDiv}, processes initiated by massless partons are 
affected by collinear enhancements, which can be associated with the radiation 
of energetic gluons form the initiating parton. Such radiation changes the 
kinematics of the subsequent hard scattering, and therefore the singularity 
cannot possibly be cancelled by virtual corrections to the Born process with 
the given initial state: the KLN cancellation will require the inclusion of 
degenerate initial configurations.

Since, as discussed above, implementing the initial-state KLN sum is, to say
the least, impractical, we would reach an impasse. Fortunately, for high-energy
scattering, a new ingredient comes to the rescue: the separation of scales
between the hard scattering and hadronisation. Intuitively, we understand 
that the collinear divergence associated with radiation from the initial state
is due to the fact that emission at very early times is not sufficiently suppressed,
since the particles involved remain close to the mass shell after the splitting.
On the other hand, one would like to argue that such early emissions should
properly be associated with the wave function of the initial state, rather than 
with the hard scattering. In general, this intuitive, semi-classical picture breaks
down when quantum corrections are included, since all sub-processes contributing
to the cross section interfere. If, however, hadrons are formed on space-time 
scales that are widely separated from the short-distance hard-scattering 
process, one may expect the quantum-mechanical interference to be 
suppressed. Proving that this is indeed the case in a relativistic quantum 
field theory requires an intricate diagrammatic analysis, which in some cases
can be short-circuited with powerful effective field theory arguments. We 
will introduce, at amplitude level, some of the techniques entering 
factorisation proofs in \secn{AllOrd}. At the level of observable cross 
sections, the relevant arguments are summarised in excellent existing 
reviews, such as Refs.~\cite{Collins:1989gx,Sterman:1995fz,Becher:2014oda}.
In this Introduction, we simply present very briefly the result of these arguments.

Since the requirements of infrared safety discussed in \secn{IRsafe} are 
unchanged, we consider again a sufficiently inclusive cross section $\sigma_{\rm 
part}$ for a partonic process, this time involving identified partons in the
initial, and possibly the final state. We assume that the Mandelstam
invariants $p_i \cdot p_j$ of the partonic process are all much larger that
the hadronic scale $\Lambda_{\rm QCD}^2$. We then introduce at parton
level a mass parameter $m$, acting as an infrared cutoff, and a {\it factorisation 
scale} $\mu_f$, separating the energies characteristic of the hard scattering,
$\mu^2 \sim p_i \cdot p_j$ from the hadronic scale. One then finds that
the partonic cross section can be {\it factorised}, in general in the form of 
a convolution, which we write, somewhat formally, as
\beq
  \sigma_{\rm part} \, = \, {\cal S}_{\rm long} \left( \frac{m^2}{\mu_f^2} \right) \star
  {\cal H}_{\rm short} \left( \frac{p_i \cdot p_j}{\mu^2}, \frac{\mu_f^2}{\mu^2} \right)
  \, + \, {\cal O} \left[ \left( \frac{m^2}{\mu_f^2} \right)^{\! p} \, \right] \, ,
\label{sigmafact}
\eeq
where $\star$ denotes the appropriate convolution and, again, $p$ is a positive 
integer. Up to power-suppressed corrections, \eq{sigmafact} separates
the long-distance dynamics, governed by the factor ${\cal S}_{\rm long}$,
which contains all the singular dependence on the cutoff $m$, from the 
short-distance hard scattering factor ${\cal H}_{\rm short}$, which is free of 
infrared singularities. Intuitively, ${\cal S}_{\rm long}$ collects contributions
from (soft and) collinear radiation at early and late times, which build up probability
distributions for initial and final states  of the hard scattering: such contributions are 
expected to be universal, {\it i.e.} independent of the particular hard process
being considered; on the other hand, they are not calculable in perturbative 
QCD, and must be extracted from experiments. On the contrary, ${\cal H}_{\rm 
short}$ collects contributions from short-distance, off-shell exchanges: these
are infrared finite, and thus we expect them to be reliably calculable in perturbation 
theory; on the other hand, they are clearly process-dependent.

The factorisation of un-cancelled infrared singularities described by \eq{sigmafact}, 
and based on separation of scales, has evident analogies with the handling of
ultraviolet singularities by means of renormalisation: also in this case, singular
contributions have been assigned to universal factors, which must be determined
experimentally, and finite contributions from processes happening at laboratory 
scales have become calculable; we will pursue this analogy a little further 
in \secn{FactEvo}. As was the case for infrared safe observables, also for 
factorisable cross sections the application of \eq{sigmafact} to hadronic cross 
sections relies to some extent on the assumption, well supported by experiments,
that non-perturbative effects do not drastically disrupt the dynamical pattern 
emerging at parton level: we {\it interpret} \eq{sigmafact} as stating that long-distance
dynamics is factorisable for inclusive cross sections, up to power-suppressed 
contributions; we then substitute the divergent contributions to ${\cal S}_{\rm long}$
with the appropriate finite, experimentally measured, combinations of parton
distributions and fragmentation functions; we expect the result to be accurate
up to corrections suppressed by $(\Lambda_{\rm QCD}/\mu)^q$, with strong  
arguments suggesting that $q = p$ in many cases. We recall that \eq{sigmafact}
follows from the OPE for highly inclusive cross sections such as DIS structure
functions, and it was proved to all orders in perturbation theory for parton
annihilation into colour-singlet final states in Refs.~\cite{Collins:1985ue,
Collins:1988ig}. A crucial argument to extend this all-order proof to inclusive 
jet cross sections was given in Ref.~\cite{Aybat:2008ct}, while for general 
collider observables the debate on the boundaries of applicability of expression
of the form of \eq{sigmafact} continues (see, for  example, Ref.~\cite{Catani:2011st}
and the extensive discussions in~\cite{Collins:2011zzd}).


\section{Finite orders: tools and results}
\label{FinOrd}

After discussing the (pre-)history of the infrared problem in our Introduction, 
we now turn to the `modern' viewpoint, which is predominantly driven by the practical 
requirements of large-scale perturbative calculations for applications at high-energy
colliders. This means that the focus is the non-abelian theory, and it largely implies
that the infrared regulation scheme is dimensional regularisation. In the present  
section, we will briefly present the treatment of soft and collinear singularities in
simple one-loop QCD calculations. This is of course well-known textbook material,
and we will only sketch the relevant calculations: this simple context will, however,
allow us to introduce some of the tools of the trade, that we will employ in a more
abstract context in later Sections of this review. Following the arguments in 
\secn{QCD}, we will first present the case of infrared-safe cross sections at lepton
colliders, introducing concepts relevant to \secn{FactEvo}, and then we will sketch 
the derivation of collinear factorisation in one-loop Deep Inelastic Scattering, which 
will prepare some grounds for the discussion in \secn{Subtra}.


\subsection{Infrared safety: the total cross section}
\label{TotCross}

The prototype infrared-safe observable at lepton colliders is the total cross section
for lepton annihilation into hadrons. By definition, it is expected to be insensitive 
to the radiation of soft and collinear massless particles, and the parton-level result 
bears the simplest possible relation to the hadronic observable, since all radiated 
partons must turn into hadrons with unit probability. These expectations are borne 
out by the explicit calculation in dimensional regularisation, which we now outline: 
the Bloch-Nordsieck-like cancellation of soft and collinear poles between real
and virtual contributions will be on display.

Working in the perturbative regime, with a center-of-mass squared energy $s$ 
much larger than hadron masses, we consider the process in which a massless
lepton of momentum $k_1$ annihilates with an anti-lepton of momentum $k_2$,
with  $k_1^\mu  + k_2^\mu = q^\mu$, and $q^2 = s$. The annihilation of the two 
leptons produces a photon which later decays into massless partons. Since we 
will remain at leading order in QED, but will later consider all orders in the strong 
coupling, it is worthwhile to give general expressions for the ingredients entering 
the cross section. We start from the definition
\beq
  \sigma_{\rm tot} \left( q^2 \right) \, = \, \frac{1}{2 q^2} \, \sum_X
  \int d \Phi_X  \, \frac{1}{4} \sum_{\rm spin} \Big| {\cal A} 
  \big( k_1 + k_2 \rightarrow X \big) \Big|^2 \, ,
\label{sigmatot} 
\eeq
where $X$ is a generic strongly-interacting final state, $\Phi_X$ is the corresponding
phase space measure, and $2 q^2$ is the flux factor for massless particles. At 
leading order in QED, one can split the calculation into leptonic and hadronic 
tensors, according to
\beq
  \sigma_{\rm tot} \left( q^2 \right) \, \equiv \, \frac{1}{2 q^2} \,
  L_{\mu \nu} (k_1, k_2) \, H^{\mu \nu} (q) \, ,
\label{sigmatotLH} 
\eeq
where the leading-order leptonic tensor, including the photon propagator, is 
\beq
  L^{\mu \nu} (k_1, k_2) \, = \, \frac{e^2}{q^4} \, \Big( k_1^\mu k_2^\nu +
  k_1^\nu k_2^\mu - k_1 \cdot k_2 \, g^{\mu \nu} \Big) \, .
\label{Lmunu}
\eeq
The hadronic tensor, on the other hand, can be expressed (in principle 
non-perturbatively in the strong interactions) in terms of matrix elements of the 
electromagnetic current, as
\beq
  H_{\mu \nu} (q) & = & e^2 Q_f^2 \, \sum_X \bra{0} J_\mu (0) \ket{X} \, 
  \bra{X} J_\nu (0) \ket{0} \, (2 \pi)^4 \delta^4 \left(q - p_X \right) 
  \nonumber \\
  & \equiv & \big( q_\mu q_\nu - q^2 g_{\mu \nu} \big) \, H(q^2) \, .
\label{Hmunu}
\eeq
At parton level, the current $J_\mu (x)$ in (single-flavour) QCD is simply the 
conserved Dirac current $J_\mu (x) = \overline{\psi}(x) \gamma_\mu \psi(x)$,
and the simplest state contributing to the sum is a two-particle state containing
a quark-antiquark pair. In the second line of \eq{Hmunu} we have used current 
conservation to extract the transverse tensor structure, expressing the result 
in terms of the single scalar function $H(q^2) = H^\mu_{\,\,\, \mu}(q)/((d - 1) q^2)$.
Using now the fact that the leptonic tensor is also transverse, we express
the total cross section in terms of the trace of the hadronic tensor as
\beq
  \sigma_{\rm tot} \left( q^2 \right) \, \equiv \, \frac{e^2}{2 q^4} \,
  \frac{1}{3} \big(- g _{\mu \nu} \, H^{\mu \nu} (q) \big) \, .
\label{sigmatotH} 
\eeq
The calculation of the trace of the hadronic tensor up to ${\cal O} (\alpha_s)$ is 
a classic result (see for example~\cite{Sterman:1994ce}), and, in the meantime,  
the perturbative calculation has been pushed to an extraordinary ${\cal O} 
(\alpha_s^4)$ accuracy~\cite{Baikov:2008jh,Baikov:2010je,Baikov:2012er,
Herzog:2017dtz}. To summarise the ${\cal O} (\alpha_s)$ calculation, we 
introduce the first tool in our toolbox.


\subsection{A tool: cut diagrams}
\label{Cutdia}

\begin{figure}
\centering
  \includegraphics[height=2cm,width=10cm]{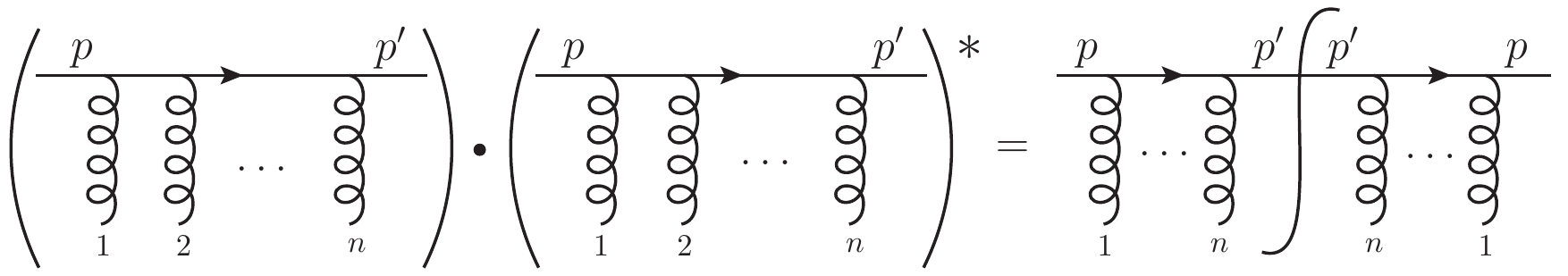}
  \caption{Expressing the squared modulus of fermion line contribution to a 
  Feynman diagram in terms of a cut diagram.
  \label{cutline}}
\end{figure}

Unitarity tells us that the imaginary part of a Feynman diagram $G$ for a 
scattering amplitude is given by the sum of all {\it cut} diagrams that can be 
constructed from it. A cut diagram is built by partitioning the original graph 
into two subgraphs, one containing initial state particles, while the second one 
contains the final state; propagators that are thus `cut' must be replaced with 
the mass-shell condition. This foundational result suggests that it should be
possible to construct a set of Feynman rules to compute directly squared
amplitudes, and thus cross sections: it is easy to convince oneself that this 
is indeed the case. The crucial identity is displayed in Fig.~\ref{cutline}, where
the continuous line can be taken to represent any charged particle - for
definiteness, we consider a quark radiating a set of gluons. In words, a charged 
line in a Feynman diagram can be complex-conjugated by reversing the
charge flow, with the understanding that the signs of explicit factors of $i$
must also be changed. This allows, for fixed final-state momenta, to `join'
the end of the line in the original diagram with the beginning of the 
complex-conjugate line. 
In the case of a quark radiating gluons, the identity in Fig.~\ref{cutline} follows
from the Clifford algebra identity
\beq
 \Big[ \overline{u}_{s'} (p') \, \gamma_{\mu_n} \, \ldots \, \gamma_{\mu_1} \,
 u_{s} (p) \Big]^* \, = \, \overline{u}_{s} (p) \, \gamma_{\mu_1} \, \ldots \, 
 \gamma_{\mu_n} \, u_{s'} (p') \, ,
\label{cliffid}
\eeq
with similar identities holding for other combinations of particle and anti-particle 
spinors. Colour indices also obey the line-reversal rule, thanks to the hermiticity
of the generators of the Lie algebra
\beq
  \Big[ T^a_{\phantom{a} ij} \Big]^* \, = \, T^a_{\phantom{a} ji} \, .
\label{hermi}
\eeq
Identities like \eq{cliffid} or \eq{hermi} are, of course, not accidents of the 
Clifford algebra, or of a specific symmetry group: they follow from the hermiticity
of the Hamiltonian, which in turn underpins the unitarity of the theory.

One can now go on and observe that the cut in Fig.~\ref{cutline} can be  
interpreted as imposing the mass shell condition (accompanied by the constraint
establishing the positivity of the energy of the outgoing particle). If one further 
performs the sum over final-state spins, the cut-line Feynman rule corresponds to 
replacing a would-be propagator with momentum $p$ with its imaginary 
part, accompanied by the propagator numerator arising from the spin sum:
\beq
 {\rm i} \frac{\slash{p} + m}{p^2 - m^2 + {\rm i} \eta} \, \to \,
 (2 \pi) \, \delta (p^2 - m^2) \, \theta(p_0) \sum_{s} u_s (p) \overline{u}_s (p) 
 \, \equiv \, (2 \pi) \delta_+ (p^2 - m^2)  \left( \slash{p} + m \right) \, .
\label{cutprop}
\eeq
Finally, one may wish to integrate final-state momenta over some range, 
in order to build an observable cross section. One notes then that, upon 
interfering a diagram with $k$ loops and $p$ final-state particles with
another diagram with $k'$ loops, and of course the same final particles,
the graph constructed according to the rules just described will have
$k + k' + p - 1$ loops, including $p-1$ cut loops. Integrating over the
$p - 1$ loop momenta, with $p$ mass-shell conditions from the cut propagators,
builds up the phase space integration appropriate for an inclusive cross 
section: one can then, eventually, insert in that phase space a weight function 
identifying the desired observable. In a sense, the procedure just outlined
reverses the standard proofs of unitarity for Feynman graphs: we start out
with an attempt to build Feynman rules for squared matrix elements, and
we end up with cut graphs. The underlying unitarity justifies the most delicate
aspects of the procedure: indeed, it is far from obvious that symmetry factors
and signs associated with fermion loops will match between cut and uncut 
graphs, but it can be shown in full generality that they do~\cite{Sterman:1994ce}.
\begin{figure}
\centering
  \includegraphics[height=2cm,width=5.3cm]{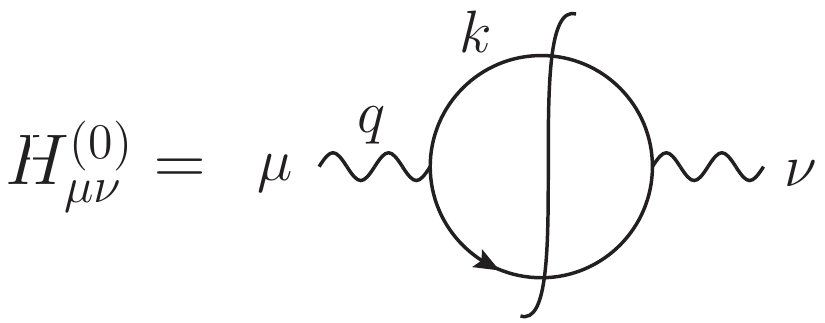}
  \caption{The tree-level hadronic tensor as a cut diagram.
  \label{Rtreefig}}
\end{figure}   
In order to illustrate how this works in practice, we sketch the computation 
of the trace of the hadronic tensor, \eq{Hmunu}, at lowest order in perturbation 
theory; in \secn{RealR} and in \secn{FormFac} we will outline the corresponding 
one-loop calculation. Looking at the cut Feynman diagram in Fig.~\ref{Rtreefig}, 
and noting that our observable is fully inclusive, we readily find
\beq
  H_{\mu \nu}^{(0)} (q) \, = \, - e^2 Q_f^2 \int \frac{d^4 k}{(2 \pi)^4} 
  ( 2 \pi ) \delta_+ (k^2) (2 \pi) \delta_+ \left( (k - q)^2 \right)
  {\rm Tr} \left[ \slash{k} \gamma_\mu \left( \slash{k} - \slash{q} \right) 
  \gamma_\nu \right] \, ,
\label{H0}
\eeq
where $Q_f$ is the electric charge of quarks of flavour $f$ in  units of the
electron charge. Crucially, we included in \eq{H0} a negative sign for the 
fermion cut loop: failure to do so would lead to a negative total cross section. 
The calculation of \eq{H0} is now immediate: the mass-shell $\delta$ function 
for momentum $k$ turns the loop integral into a Lorentz-invariant phase-space 
integral; the second mass-shell $\delta$ function then sets $k \cdot q  = q^2/2$; 
dimensional analysis dictates that the result must be proportional to $q^2$; the 
trivial angular integration determines the overall constant. Inserting the result
in \eq{sigmatotH}, and summing over colours and active flavours, one finds of 
course the classic value of the leading-order $R$ ratio
\beq
\label{Rratio}
  R \left( \frac{q^2}{\mu^2}, \as(\mu^2) \right) & \equiv & \frac{\sigma_{\rm tot}
  (e^+ e^- \to {\rm hadrons})}{\sigma_{\rm tot} (e^+ e^- \to \mu^+ \mu^-)} \, = \, R_0 \,
  \sum_{n = 0}^\infty \bigg(\frac{\as(\mu^2)}{\pi}\bigg)^{\! n} \Delta_n \left( \frac{q^2}{\mu^2} 
  \right) \, , \nonumber \\[3mm]
  R_0 & = & N_c \sum_f Q_f^2 \, , \qquad \Delta_0 \, = \, 1 \, .
\eeq  
As we will see in the next section, cut diagrams organise higher-orders calculations
in a natural way, and in particular they nicely highlight the mechanism for the 
cancellation of infrared singularity in the spirit of the KLN theorem.


\subsection{Real radiation at one loop}
\label{RealR}

\begin{figure}
\centering
  \includegraphics[scale=0.6]{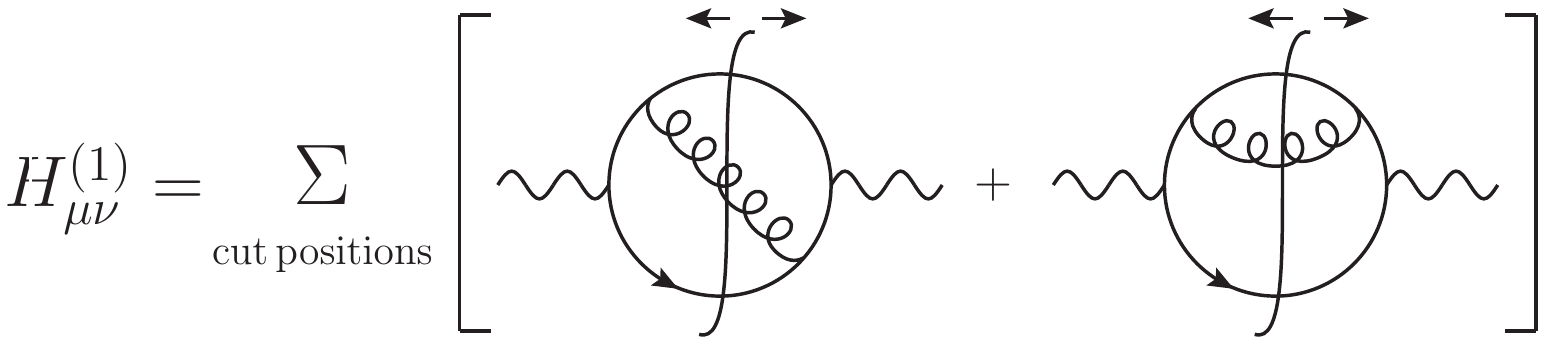}
  \caption{The one-loop hadronic tensor as a sum of cut diagrams.
  \label{Rolofig}}
\end{figure}

Following the reasoning presented in \secn{Cutdia}, the calculation of the hadronic 
tensor at higher orders is naturally organised by writing down the diagrams contributing
to the photon-induced vacuum polarisation, and then summing over the possible 
positions for the cut, as illustrated at one loop in Fig.~\ref{Rolofig}. 
It is important to note that the organisation of the calculation of a cross section 
in terms of cut diagrams is particularly well adapted to study the cancellation of 
infrared singularities. The three-particle cut shown in Fig.~\ref{Rolofig} corresponds
to a contribution from real gluon radiation, integrated over phase space; 
when the cut is moved across a gluon emission vertex, one finds a virtual
one-loop contribution, interfered with the Born process. The cancellation of
infrared singularities happens, for every uncut diagram, in the sum over cut 
positions. This bears an intuitive relation with the mechanism for cancellation
highlighted by the KLN theorem: when the radiated gluon becomes soft or collinear,
and thus unresolved, the real-radiation contribution becomes physically indistinguishable
from the virtual correction, and the rules of quantum mechanics require that the 
two configurations be summed. The KLN sum over degenerate states is thus 
precisely implemented by the sum over cut positions.

It is clearly desirable to have a systematic and practical method to perform 
the sum over cut positions at the integrand level, in order to cancel the contributions
of singular configurations without having to introduce external regulators for virtual 
poles and phase-space integrals. This would in principle allow for a fully numerical 
evaluation of infrared-safe distributions in the renormalised theory. Indeed, a method 
along these lines was proposed in Refs.~\cite{Catani:2008xa,Bierenbaum:2010cy}, 
starting from the so-called Feynman Tree Theorem~\cite{Feynman:1963ax} (see 
also~\cite{Brandhuber:2005kd}), and is steadily being developed~\cite{Sborlini:2016gbr,
Driencourt-Mangin:2019aix,Aguilera-Verdugo:2019kbz,Verdugo:2020kzh,Capatti:2019edf,
Capatti:2020ytd,Capatti:2020xjc,TorresBobadilla:2020ekr,Bobadilla:2021pvr}. For 
the sake of our current simple example, we will follow the standard approach of 
treating separately the real radiation contribution, integrated over phase space, 
and the virtual correction: singularities in both cases will be regulated by dimensional 
regularisation, taking $d = 4 - 2 \epsilon$, with $\epsilon < 0$, and they will, as 
expected, cancel in the sum.

Let us begin by writing down the general form of the three-particle cut contribution
to the hadronic tensor. It reads
\beq
  H_{\mu \nu}^{(1, R)} (q) \, \equiv \, \int \frac{d^d k \, d^d p}{(2 \pi)^{2 d - 3}} 
  \, \delta_+ (k^2) \, \delta_+ (p^2) \, \delta_+ \left( (q - p - k)^2 \right)
  {\cal H}^{(1)}_{\mu \nu} (p,k) \, ,
\label{H1R}
\eeq
where we take $k$ to be the gluon momentum, $p$ to be (say) the quark momentum,
while ${\cal H}^{(1)}_{\mu \nu}$ is the actual squared matrix element for $\gamma^*
\to q \bar{q} g$, computed from the sum of the three-particle cuts of the diagrams 
in Fig.~\ref{Rolofig}; furthermore, as announced, we have promoted the loop
integrations to $d$ dimensions. 

It is perhaps worthwhile discussing briefly here the application of dimensional 
regularisation to the infrared problem. In the case of \eq{H1R}, there are no UV 
divergences, so one can directly take $\epsilon < 0$. It is fairly obvious that
this choice will properly regulate soft divergences, which arise from the
uniform scaling limit $k^\mu \to \lambda k^\mu$, $\lambda \to 0$. Indeed,
the $d$-dimensional measure of integration in \eq{H1R} behaves as
\beq
  d^d k \, \delta (k^2) & \rightarrow & \frac{d^{d - 1} k}{k} \, \rightarrow \, d k \,
  k^{1 - 2 \eps} \, d \Omega_{d - 2} \, ,
\label{dregen}
\eeq
with $k = |{\bf k}|$, and $\Omega_{d - 2}$ the solid angle in $d-1$ space dimensions: 
a negative value of $\epsilon$ clearly improves the behavior of the integrand as 
$k^\mu \to 0$. It is less obvious that taking $d > 4$ will help with collinear divergences:
radiated collinear particles can indeed be very energetic, and only after developing 
appropriate power counting tools (as we do here in \secn{AllOrd}) one can be sure 
that the method will work\footnote{To further emphasise this point, we note 
that working in coordinate (as opposed to momentum) space, collinear divergences
are regulated by taking $\epsilon > 0$, and are thus assimilated to ultraviolet 
ones~\cite{Erdogan:2014gha}.}. In the present case, it is sufficient to write down 
the angular measure of integration in $d$ space-time dimensions, thus depending
on $d-3$ `co-latitudes' $0 \leq \theta_i \leq \pi$, $i = 1, \ldots, d-3$, and one azimuthal
angle $0 \leq \phi < 2 \pi$. The result is well known and, after simple manipulations, 
it can be written as
\beq
  d \Omega_{d - 2} & = & d \cos \theta_{d - 3} 
  \left(1 - \cos^2 \theta_{d - 3} \right)^{- \eps} \times \ldots \times d \theta_1
  \sin \theta_1 \, d \phi \, ,
\label{dregenan}
\eeq
which indeed displays a suppression of potential singularities as $\theta_{d-3} 
\to 0$ or as $\theta_{d-3} \to \pi$, for $\epsilon < 0$. Turning now again to \eq{H1R},
it is easy to convince oneself that the integrand in $d=4$ depends non-trivially
on just one angle, which can be taken to be the angle between the quark and 
gluon three-momenta. We can then set $\theta_{d-3} = \theta_{pk}$, and integrate
over the remaining angles. Choosing as integration variables
\beq
  z \, \equiv \, \frac{2 k}{\sqrt{q^2}} \, , \qquad \quad 
  y \, \equiv \frac{1 - \cos \theta_{pk}}{2} \, ,
\label{enang}
\eeq
so that $0 \leq \{z, y\} \leq 1$, and using again dimensional analysis, we can
rewrite \eq{H1R}, up to an overall constant, as
\beq
  H_{\mu \nu}^{(1, R)} (q) \, \propto \, \left( q^2 \right)^{1 - 2 \eps} 
  \int_0^1 dz  \, z^{1 - 2 \eps} \int_0^1 dy \big[ y (1 - y) \big]^{- \eps}
  {\cal H}^{(1)}_{\mu \nu} (p,k) \, .
\label{H1Rdreg}
\eeq
We need now to study the behaviour of the squared matrix element 
${\cal H}^{(1)}_{\mu \nu} (p,k)$ in the potentially singular limits, $z \to 0$ and
$y \to 0,1$. This will be done in a systematic way in \secn{AllOrd}, but, at 
this order, it is easy to extract the relevant information from the diagrams in 
Fig.~\ref{Rolofig}: when the radiated gluon becomes soft, both the matrix element
and its complex conjugate provide a factor of $1/z$; furthermore, when the gluon
becomes collinear to the (massless) quark, the quark propagator provides a factor
of $1/y$; when the emission is collinear to the antiquark, by momentum conservation
the antiquark propagator must provide a factor of $1/(1 - y)$. Altogether, the leading 
singularity of the squared matrix element is given by
\beq
  {\cal H}^{(1)}_{\mu \nu} (p,k) \sim \frac{1}{z^2 \, y \, (1 - y)} \, .
\label{leadsing}
\eeq
Noting that the $y$-dependent denominator can be partial-fractioned, we interpret
this as the superposition of a soft singularity with the {\it sum} of two non-overlapping
collinear singularities. Therefore both the $z$ and the $y$ integrations will yield
infrared poles. A complete calculation~\cite{Sterman:1994ce} gives the result
\beq
  \left( H^\mu_{\,\, \mu} \right)^{(1, R)} (q) \, \propto \, \frac{\alpha_s(\mu)}{\pi} \, 
  C_F \, q^2  \left( \frac{4 \pi \mu^2}{q^2} \right)^{\eps} \left[ \frac{2}{\eps^2} 
  + \frac{3}{\eps} - \pi^2 + \ldots \right] \, ,
\label{H1Rdregfin}
\eeq
where we performed the colour sums and introduced the renormalisation scale 
$\mu$. The result exhibits the typical double pole in $\epsilon$, representing
the superimposed soft and collinear singularities, and leading to double logarithms
of $q^2$ in the finite terms. The single pole in $\epsilon$ could in principle arise 
from soft gluon emission at wide angles with respect to the quark and the antiquark, 
or from hard collinear gluon emission: in the present case, as we will see, the 
single pole in \eq{H1Rdregfin} has the latter origin. Finally, we included for future 
reference the `large' transcendental constant $\pi^2$ (while not displaying the 
remaining rational numbers), arising from the expansion in powers of $\epsilon$ 
of the $\Gamma$ functions originating from the phase space integration. Such 
constants can play an important numerical role, amplifying the impact of radiative 
corrections, and in some cases they can be organised to all orders in perturbation 
theory, as first noted in Ref.~\cite{Parisi:1979xd}: we will return to their connection 
to infrared poles in later sections of our review.


\subsection{Virtual corrections: the quark form factor}
\label{FormFac}

According to the general theorems derived in \secn{Intro}, the infrared poles
in \eq{H1Rdregfin} must cancel against the contributions of two-particle cuts
to the diagrams in Fig.~\ref{Rolofig}. To any order in perturbation theory, such
cut diagrams build up the interference between the Born amplitude for $\gamma^*
\to q \bar{q}$ and the quark form factor, which is defined by
\beq
  \Gamma_\mu \! \left( p_1, p_2; \mu^2, \eps \right) \, \equiv \,  \bra{p_1, p_2}
  J_\mu (0) \ket{0} \, = \, \bar{u} (p_1) \gamma_\mu v(p_2) \,
  \Gamma \left( \frac{q^2}{\mu^2}, \alpha_s (\mu^2), \eps \right) \, ,
\label{defformfac}
\eeq
providing all purely virtual contributions to \eq{Hmunu}. The quark form factor 
is a fundamental object in perturbative QCD, and provides a crucial ingredient 
for the calculation of all cross sections involving only two hard quark jets, either 
in the initial or in the final state. As we will see in \secn{FactEvo}, it is also the 
simplest object for which the resummation of infrared poles to all orders can 
be completely carried out~\cite{Sudakov:1954sw,Mueller:1979ih,Collins:1980ih,
Sen:1981sd,Sterman:1986aj,Magnea:1990zb}. Its explicit form in dimensional 
regularisation is now known to three loops~\cite{Baikov:2009bg,Lee:2010cga,
Gehrmann:2010ue}, while all infrared poles are known to four loops~\cite{Henn:2019swt,
vonManteuffel:2020vjv}\footnote{The complete four-loop form factor has recently 
been computed in the case of ${\cal N} = 4$ Super-Yang-Mills theory in
Ref.~\cite{Lee:2021lkc}.}. Before making a few comments on the one-loop 
calculation, we will point out a few general properties of the form factor, which 
will be crucial in \secn{FactEvo}. We will then also take the opportunity to 
develop one more tool in our toolbox, the $d$-dimensional running coupling, 
discussed below in \secn{druncoupl}.

First of all, notice that, in the massless case, the vector form factor $\Gamma_\mu$
in \eq{defformfac} is characterised by a single scalar function $\Gamma$, with
the vector structure being fixed at tree-level. Next, observe that the quark form 
factor is defined as a matrix element of the conserved electromagnetic current. 
As such, it does not get renormalised: after the renormalisation of the QCD 
coupling there is no need for an overall multiplicative renormalisation associated
with the QED vertex. This is of course a consequence of the QED Ward identity,
but it has a clear physical motivation: if switching on the strong interactions were
to cause a change in electric charges, it would be hard(er) to explain the precise
experimental relations between the electric charges of quarks and leptons. The
non-renormalisation of the form factor implies that it obeys a renormalisation 
group equation with a vanishing anomalous dimension. In dimensional regularisation, 
we write
\beq
  \left( \mu \frac{\partial}{\partial \mu} + \beta \left( \eps, \alpha_s \right)
  \frac{\partial}{\partial \alpha_s} \right) 
  \Gamma \left( \frac{q^2}{\mu^2}, \alpha_s (\mu^2), \eps \right) \, = \, 0 \, .
\label{RGff}
\eeq
The crucial point in \eq{RGff} is that, since we are working  with a divergent 
quantity, which is mathematically well-defined only away from $d = 4$, we need
to use the $d$-dimensional version of the $\beta$ function. This will have pivotal 
consequences for the exponentiation of infrared poles, discussed in the following. 

For the moment, to complete our NLO example, we concentrate on summarising 
the evaluation of the form factor at one loop.
\begin{figure}
\centering
  \includegraphics[height=2.3cm,width=11cm]{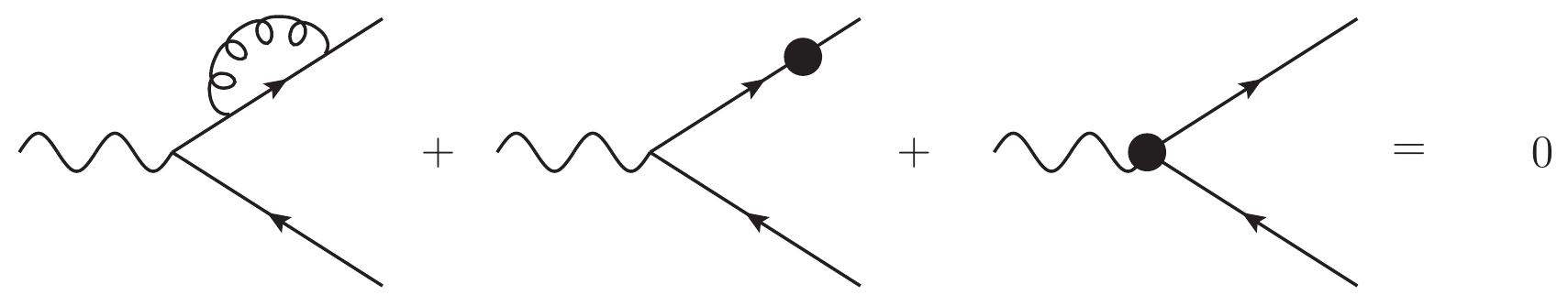}
  \caption{One-loop diagrams giving a vanishing contribution to the quark form 
  factor, including UV counterterms.
  \label{Feynff}}
\end{figure}   
The one-loop calculation in dimensional regularisation, in the massless case, is 
simplified by the fact that only the vertex-correction diagram contributes. The 
remaining diagrams, including ultraviolet counterterms, are displayed in 
Fig.~\ref{Feynff}: the fact that they give a vanishing contribution can be understood
in two slightly different ways. On the one hand, we can simply note that scale-less 
integrals are defined to vanish in dimensional regularisation, which applies to the 
self-energy correction to the massless quark propagator; the fact that the sum of 
the two UV counterterms vanishes is then a consequence of the QED Ward 
identity\footnote{Note that external leg corrections corresponding to 
one-particle-reducible diagrams must be counted only once, after subtracting 
the square root of the residue of the quark propagator on each external leg, 
according to the LSZ procedure.}. On the other hand, we can interpret the 
vanishing of the self-energy as the cancellation of a UV pole with an IR pole 
near $d=4$, an interpretation that can be checked by introducing auxiliary 
regulators; the quark field renormalisation counterterm in the MS scheme, 
which cancels the UV pole, must then be equal to the IR  pole of the 
self-energy diagram; the Ward identity thus forces the UV vertex counterterm 
to be equal  (with opposite sign) to the IR pole of the propagator correction.

The vertex correction diagram is of course a textbook exercise: here we just 
quickly pause to stress the power-counting properties of the terms it comprises. 
A straightforward application of the Feynman rules gives a decomposition
into tensor integrals in the form
\beq
  \Gamma_\mu^{(1)} \, = \, c_\mu I_0 + c_{\mu \alpha} I_1^\alpha +
  c_{\mu \alpha \beta} I_2^{\alpha \beta} \, ,
\label{tensdec}
\eeq
with the tensor integrals defined by
\beq
  \Big\{ I_0, I_1^{\alpha}, I_2^{\alpha \beta} \Big\} \, \equiv \,
  \int \frac{d^d k}{(2 \pi)^d} \, \frac{\big\{1, k^\alpha, k^\alpha k^\beta 
  \big\}}{\left( k^2 + {\rm i} \eta \right) \left[ \left( p_1 - k \right)^2 + 
  {\rm i} \eta \right] \left[ \left( p_2 + k \right)^2 + 
  {\rm i} \eta \right]} \, .
\label{tensint}
\eeq
Na\"ive power counting immediately shows that only the tensor integral
$I^{\alpha \beta}_2$ can contribute to the UV pole, while only the scalar
integral $I_0$ can contribute to the soft pole. The vector integral $I^\alpha_1$
appears free of both IR and UV singularities, but - since the suppression 
provided by the numerator is direction-dependent - it still can (and does) 
contribute a collinear pole. Clearly, one needs a more precise definition
of power counting in infrared regions, which we will address in \secn{AllOrd}.
When the dust settles, one finds the well known result
\beq
  \Gamma^{(1)} \, = \, - \, \frac{\alpha_s}{4 \pi} \, C_F \,  
  \left( \frac{4 \pi \mu^2}{- q^2 - {\rm i} \eta } \right)^{\!\! \eps} \, 
  \frac{\Gamma^2 (1 - \eps) \Gamma(1 + \eps)}{\Gamma(1 - 2 \eps)}
  \, \left[ \frac{2}{\eps^2} + \frac{3}{\eps} + \ldots \right] \, .
\label{olofofa}
\eeq
Once again, \eq{olofofa} exhibits the hallmarks of all calculations of IR singularities
of massless gauge-theory amplitudes in dimensional regularisation: every 
loop carries a double pole, corresponding to the superposition of a soft
singularity with the sum of (in this case) two non-overlapping collinear 
singularities. One then finds a single pole, which can be due to soft gluons
emitted at wide angles with respect to hard momenta, or (as is the case 
here) to the radiation of hard collinear gluons. We also note that, as we
are computing the time-like form factor ($q^2 > 0$), the amplitude has 
an imaginary part associated with the two-particle cut. In the presence
of a double infrared pole, this imaginary part provides a further source
of potentially large perturbative corrections to physical cross sections,
in the form of `large' transcendental constants such as $\pi^2$. Indeed,
expanding \eq{olofofa} in  powers of $\epsilon$, one must use
\beq
  \left( - q^2 - {\rm i} \eta \right)^{- \eps} \, = \, \big( q^2 \big)^{- \eps} 
  {\rm e}^{{\rm i} \pi \eps} \, = \, \big( q^2 \big)^{- \eps} 
  \left( 1 + {\rm i} \pi \eps - \frac{\pi^2}{2} \eps^2 + \ldots \right) \, .
\label{pisquare}
\eeq 
The ${\cal O} (\epsilon)$ imaginary part will cancel when computing
physical observables, but the ${\cal O} (\epsilon^2)$ contribution
will combine with the double infrared pole and can give a sizeable 
contribution. Such large constant contributions have been studied
for a long time~\cite{Parisi:1979xd}, and in many relevant cases they 
can be organised in exponential form to all orders in perturbation 
theory~\cite{Sterman:1986aj,Magnea:1990zb,Eynck:2003fn,Ahrens:2008qu}.
For the total cross section that we are computing, it turns out that at one loop
all transcendental constants cancel in the sum of real and virtual corrections: 
interfering \eq{olofofa} with the Born matrix element, and properly combining 
the result with \eq{H1Rdregfin}, one finally gets the well-known result
\beq
\label{Rratio1}
  R \left( \frac{q^2}{\mu^2}, \as(\mu^2) \right) \, = \,  R_0 \, 
  \bigg( 1 \, + \, \frac{\as(\mu^2)}{\pi} + {\cal O} \big( \alpha_s^2 \big) \bigg) \, .
\eeq  
The natural question at this point is to what extent one can maintain the
cancellation of infrared poles, while allowing for a more detailed (or less inclusive)
treatment of real radiation. Before tackling that question, we introduce
a second important tool in our toolbox.


\subsection{A tool: the $d$-dimensional running coupling}
\label{druncoupl}

\begin{figure}
\centering
  \includegraphics[height=4cm,width=7cm]{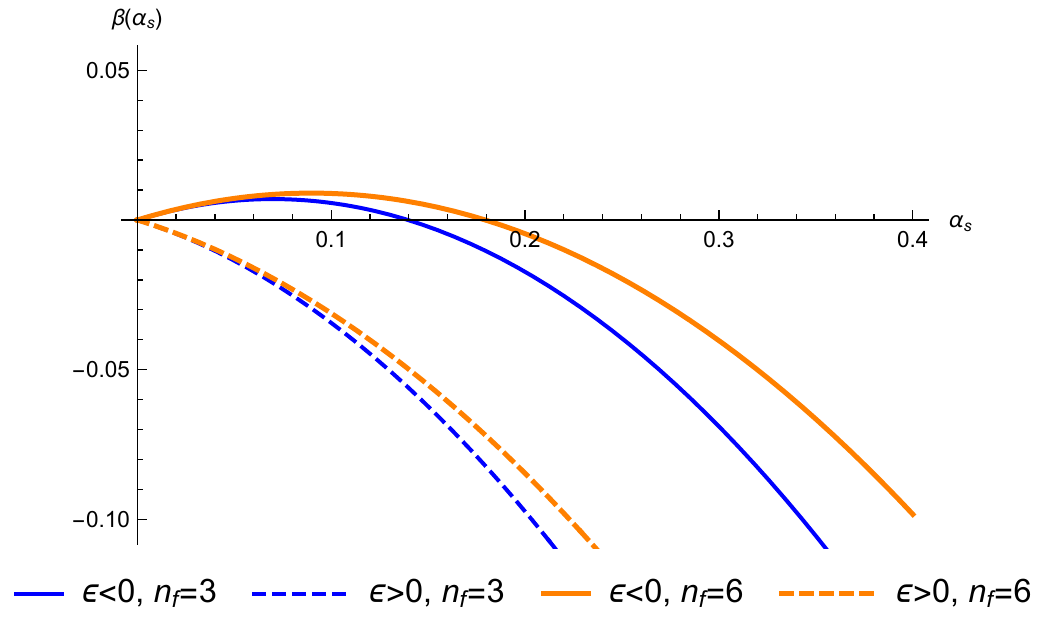}
  \caption{The $\beta$ function for different values of the regulator $\epsilon = \pm 0.1$ 
  and the number of flavour $n_f$.
  \label{betaeps}}
\end{figure}

Using the renormalisation group to explore quantum field theories away from 
dimension $d=4$ is a standard tool of statistical field theory~\cite{Wilson:1971dc,
Wilson:1972cf,Wilson:1973jj}, but not as widely used in the context of perturbative
QCD. It offers, however, a simple and powerful method to sum and exponentiate
infrared singularities, in a way which is directly comparable to diagrammatic 
calculations~\cite{Magnea:1990zb}. In particular, as we will see, it provides an
elegant way to bypass the Landau pole, thus leading naturally to a `purely
pertubative' resummation of infrared effects~\cite{Magnea:2000ss}. The starting 
point is the well-known relation between the bare and renormalised couplings 
in $d = 4 - 2 \epsilon$,
\beq
  \alpha_0 \, = \, \mu^{2 \epsilon} \alpha_s(\mu) Z_\alpha \, ,
\label{gren}
\eeq
which leads to the presence of an $\epsilon$-dependent term, linear in $\alpha_s$, 
in the $\beta$ function. We write
\beq
  \beta \left( \eps, \alpha_s \right) \, \equiv \, 
  \mu \frac{\partial}{\partial \mu} \,\alpha_s (\mu) \, = \,- 2 \eps \alpha_s
  - \frac{\alpha_s^2}{2 \pi} \, \sum_{n = 0}^\infty \, b_n 
  \left( \frac{\alpha_s}{\pi} \right)^n \, ,
\label{beta}
\eeq
with a normalisation chosen so that $b_0 = (11 C_A - 2 n_f)/3$ in QCD. Notice that,
in order to regularise infrared singularities, we must take $\epsilon < 0$: thus, differently
from what is commonly done in a statistical field theory context, we are working
at $d > 4$. The $\beta$ function in \eq{beta}, therefore, is positive for sufficiently
small values of $\alpha_s$. To get a feel for the consequences of \eq{beta}, we
may consider the one-loop approximation, setting $b_n = 0$ for $n \geq 1$,
and solve the equation for the running coupling, for fixed $\epsilon < 0$, evolving
the coupling from an initial scale $\mu_0$ to a final scale $\mu$. One easily finds
the solution
\beq
  \overline{\alpha} \left( \frac{\mu^2}{\mu_0^2}, \alpha_s (\mu_0^2), 
  \epsilon \right) \, = \, \frac{\alpha_s (\mu_0^2)}{\left(\frac{\mu^2}{\mu_0^2}
  \right)^\eps - \frac{1}{\eps} \left( \!1 - \left(\frac{\mu^2}{\mu_0^2}
  \right)^{\! \! \eps} \, \right) \frac{b_0}{4 \pi} \, \alpha_s (\mu_0^2)} \, ,
\label{oloruco}
\eeq
which is non-singular (as it must be), and gives the standard result, for $\epsilon \to 0$.
The most striking feature of the running coupling in \eq{oloruco} emerges already at
`tree level', setting also $b_0 = 0$. In order to keep the bare coupling in \eq{gren} 
independent of $\mu$, the renormalised coupling must scale as a {\it power}, so that
\beq
  \overline{\alpha} \left( \frac{\mu^2}{\mu_0^2}, \alpha_s (\mu_0^2), 
  \epsilon \right) \, = \,  \bigg( \frac{\mu^2}{\mu_0^2} \bigg)^{\! - \epsilon} 
  \alpha_s (\mu_0^2) \, ,
\label{treeruco}
\eeq
which, in particular, means that the running coupling  {\it vanishes} for $\mu \to  0$
in $d > 4$
\beq
  \overline{\alpha} \left( 0, \alpha_s (\mu_0^2), \epsilon < 0 \right) \, = \,  0 \, .
\label{treeruco0}
\eeq
This is not surprising, and can be understood by looking at Fig.~\ref{betaeps}: for 
$\epsilon < 0$, the ultraviolet fixed point responsible for asymptotic freedom has
moved away from the origin by a distance of order $\epsilon$; in the meantime,
at the origin one finds an infrared fixed point, which forces the coupling to vanish
for vanishing scale. 
One may wonder what happens to the Landau pole, at one 
loop and beyond: this can easily be verified, by setting the denominator of 
\eq{oloruco} to zero and solving for the scale $\mu$, which corresponds to the 
common definition of the Landau pole. One finds
\beq
  \mu^2 \, \equiv \, \Lambda^2 \, = \, \mu_0^2 
  \left( 1 + \frac{4 \pi \epsilon}{b_0 \, \alpha_s (\mu_0^2)} \right)^{- 1/ \epsilon} \, .
\label{Landaueps}
\eeq
Once again, this reduces to usual expression for $\epsilon \to 0$; on the other 
hand, for negative $\epsilon$, and in particular for generic $\epsilon < - 4 \pi/(b_0 
\alpha_s (\mu_0^2))$, the Landau pole moves away from the real axis and into 
the complex plane. This is particularly relevant when one solves renormalisation 
group equations in $d>4$, as we will do in \secn{FactEvo}: the solutions of these 
equations always involve integrals of anomalous dimensions, which are functions
of the running coupling, over a range of renormalisation scales. The discussion 
above allows to choose $\mu_0 = 0$ as the initial scale for evolution (where the
coupling vanishes and the strong interactions decouple), and integrate up to
our selected perturbative scale along the real axis, without encountering any 
singularities. It is a remarkable fact that {\it all} perturbative infrared singularities 
of gauge theory amplitudes, to all orders in perturbation theory, are generated by 
integrating finite anomalous dimensions in precisely this way, as we will see in 
detail in \secn{FactEvo} and in \secn{MultiPart}. Indeed, when one expresses the 
running coupling evaluated at the variable scale in terms of the coupling at the 
fixed hard scale $\mu$, one encounters integrals of the form
\beq
 \int_0^{\mu^2} \frac{d \lambda^2}{\lambda^2} \, \overline{\alpha} 
 \left( \frac{\lambda^2}{\mu^2}, \alpha_s (\mu^2), \eps \right) & = & 
 \alpha_s (\mu^2) \, \mu^{2 \eps} \int_0^{\mu^2} 
 \frac{d \lambda^2}{(\lambda^2)^{1 + \eps}} + {\cal O} (\alpha_s^2)
 \nonumber  \\ & = &   
 - \frac{1}{\eps} \, \alpha_s(\mu^2) + {\cal O} (\alpha_s^2) \, ,
\label{polefromruco}
\eeq
or, similarly,
\beq
 \int_0^{\mu^2} \frac{d \lambda^2}{\lambda^2} \,\, \overline{\alpha} 
 \left( \frac{\lambda^2}{\mu^2}, \alpha_s (\mu^2), \eps \right) \,
 \log \bigg( \frac{\lambda^2}{\mu^2 } \bigg) \, = \, 
 - \frac{1}{\eps^2} \, \alpha_s (\mu^2) + {\cal O} (\alpha_s^2) \, ,
\label{doupolefromruco}
\eeq
and their higher-order generalisations. In the special case of gauge theories
that are conformal in $d=4$, such as ${\cal N} = 4$ Super-Yang-Mills theory,
the $d$-dimensional running coupling obeys \eq{treeruco} exactly,
and anomalous dimension integrals can give at most double infrared 
poles, which greatly simplifies the infrared structure of scattering 
amplitudes~\cite{Bern:2005iz}.


\subsection{General infrared-safe observables}
\label{IRSafeObs}
 
\begin{figure}
\centering
  \includegraphics[height=3cm,width=8cm]{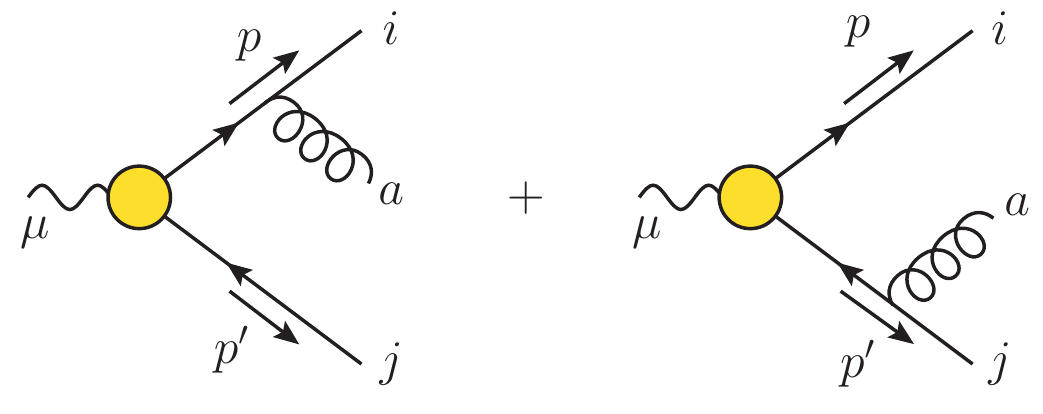}
  \caption{Diagrams for the radiation of a single gluon from a generic hard scattering 
  vertex producing a $q \bar{q}$ pair.
  \label{genvertfig}}
\end{figure}

The calculation leading to \eq{Rratio1} holds important lessons, and is amenable to
broad generalisations. In fact, it would not be very informative if the only observable
enjoying the cancellation of IR divergences was the total cross section, and this is
indeed not what the KLN theorem states: in order to cancel the divergences, one 
needs to sum (integrate) over states that are degenerate in energy. Operationally,
one may note that the virtual correction is proportional to the Born cross section. 
This is certainly not the case for the real radiation contribution, which is kinematically 
much more intricate, but it must become true in all singular limits, in order for the 
cancellation to take place. We have already seen, in \secn{cataQED}, how this 
happens in the soft approximation: slightly generalising that calculation, consider 
the radiation of a single gluon from an arbitrary hard-scattering vertex where a 
quark-antiquark pair is produced, as depicted in Fig.~\ref{genvertfig}. The 
amplitude is
\beq
  {\cal A}^{\mu \, a}_{i j} \, = \, g \, T^a_{i j} \,\, \overline{u} (p_1) 
  \left[ \frac{\slash{\varepsilon} (k) \left( \slash{p}_1 + \slash{k} \right) 
  \Gamma^\mu}{2 p_1 \cdot k} - \frac{\Gamma^\mu \left( \slash{p}_2 + 
  \slash{k} \right) \slash{\varepsilon} (k)}{ 2 p_2 \cdot k} \right] v (p_2) \, ,
\label{radmel}
\eeq
where $\Gamma_\mu$ is a generic Dirac structure at the hard vertex, $\varepsilon$  
is the gluon polarisation vector, and we have displayed all colour indices\footnote{Note 
that the colour structure is fixed, regardless of the nature of the hard interaction, 
since $T^a_{ij}$ is the only invariant tensor of the gauge group connecting the 
relevant representations.} .
We can now be more precise in defining the soft approximation: we rescale all 
components of the gluon momentum $k$ as
\beq
  k^\mu \, \to \, \lambda k^\mu \, , \qquad \quad \lambda \, \to \, 0 \, ,
\label{softscale}
\eeq 
and retain only the leading power in the Laurent expansion in powers of $\lambda$.
For the amplitude in \eq{radmel}, the only effect of this approximation is to drop
terms proportional to $k^\mu$ in the numerator. Then, as was done in \secn{cataQED},
we can use the Dirac equation, after commuting $\slash{p}_1$ and $\slash{p}_2$
across $\slash{\varepsilon} (k)$, to get
\beq
  {\cal A}^{\mu \, a}_{i j} \, = \, g \, T^a_{i j} \,
  \left( \frac{p_1 \cdot \varepsilon}{p_1 \cdot k} - 
  \frac{p_2 \cdot \varepsilon}{p_2 \cdot k} 
  \right) {\cal A}^{\mu}_{\rm Born} \, + \, {\cal O}(k^0) \, \equiv \, {\cal S}^a_{ij} \,
  {\cal A}^{\mu}_{\rm Born} \, + \, {\cal O}(k^0) \, ,
\label{radmelsoft}
\eeq
where ${\cal A}^{\mu}_{\rm Born} \equiv \overline{u} (p_1) \Gamma^\mu v (p_2)$.
It is worthwhile pausing briefly to note  the properties of the soft amplitude 
${\cal S}^a_{ij} \, {\cal A}^{\mu}_{\rm Born}$, which will generalise to higher orders.
\begin{itemize}
\item The soft amplitude is {\it gauge-invariant}, as can be seen by picking a longitudinal 
polarisation vector $\varepsilon^\mu \propto k^\mu$.
\item The soft amplitude is {\it universal}, in the sense that it does not depend upon 
the spin of the hard particles (which has been factored into the Born term), nor does it
depend on their energy: indeed, the soft factor ${\cal S}^a_{ij}$ is homogeneous in
the hard momenta $p_1$ and $p_2$, so that  in fact it is only sensitive to their direction
and not to their magnitude. One may rescale the momenta setting $p_i^\mu = Q \, 
\beta_i^\mu$, and the soft factor will depend only on the `four-velocities' $\beta_i$.
\item The soft amplitude can be interpreted as the action of a {\it soft colour
operator} ${\cal S}^a_{ij}$ on the Born amplitude, which is colour-diagonal in the 
fundamental representation, since the hard vertex is a colour-singlet.
\item Squaring the soft amplitude and summing over $N_c$ colours we get 
a differential cross section proportional to
\beq
  \sum_{\rm pol} \sum_{\rm col} \left| {\cal S}^a_{ij} \; {\cal A}^{\mu}_{\rm Born} \right|^2
  \, = \, g^2 \, N_c  \, C_F \, \left| {\cal A}^{\mu}_{\rm Born} \right|^2
  \frac{2 \beta_1 \cdot \beta_2}{\beta_1 \cdot k \, \beta_2 \cdot k} \, ,
\label{radmelsoft2}
\eeq
which, as expected, is proportional to the Born cross section, and correctly reproduces
the soft (though not the collinear) singularities of the full radiative cross sections computed 
from \eq{radmel}.
\end{itemize}
A similar, slightly more complicated analysis (see \secn{FactEvo}) shows that a 
factorisation of this kind happens also in the two collinear limits: in that case, the
collinear factorisation kernel is a colour singlet, but, in general, a spin matrix
acting on the Born amplitudes. Once again, in the singular limit, one finds a 
result proportional to the Born cross section, which enables the KLN cancellation
of singularities.

Since the cancellation of singularities happens locally in the radiative phase 
space, as stated by the KLN theorem, in order to construct a safe observable 
it must be sufficient to integrate the radiation only over a neighbourhood of
the singular regions (soft and collinear configuration). This has led to the
definiton and usage of {\it weighted cross sections}, or {\it event shapes}: one 
proceeds by defining a potential observable in the $m$-particle phase space
as a function $E_m(p_1, \ldots, p_m)$; one then computes the distribution
for that observable by weighing all possible final states with that function, as
\beq
  \frac{d \sigma}{d e} \, = \, \frac{1}{2 q^2} \sum_m \int d \Phi_m \, \overline{\left| 
  {\cal A}_m \right|^2} \, \delta \big(e - E_m(p_1, \ldots, p_m)  \big) \, ,
\label{evshape}
\eeq
where $\Phi_m$ is the Lorentz-invariant phase space for the $m$ particles carrying
momenta $p_i$, $i = 1, \ldots, m$, and ${\cal A}_m$ is the corresponding amplitude.
The cancellation of infrared singularities will happen, to all orders in perturbation
theory, if the chosen function $E_m$ is insensitive to soft and collinear radiation,
so that final states that differ by such radiation are weighted equally. The precise
requirement is
\beq
  \lim_{p_i \to 0} \, E_{m+1} \big( p_1, \ldots, p_i, \ldots, p_{m+1} \big) & = & 
  E_m \big( p_1, \ldots, p_{i-1}, p_{i+1}, \ldots, p_{m+1} \big) \, ,
  \nonumber \\
  \lim_{p_i \parallel \, p_j} \, E_{m+1} \big( p_1, \ldots, p_i, \ldots, p_j, \ldots 
  p_{m+1} \big) & = & 
  E_m \big( p_1, \ldots, p_i + p_j, \ldots, p_{m+1} \big) \, ,
\label{irsafe}
\eeq
defining an infrared-safe, or soft-collinear-safe observable~\cite{Ellis:1991qj,
Ellis:1980nc}. Jet cross sections in e$^+$ e$^-$ annihilation belong to this category, 
where the functions $E_m$ are, for example, sets of step functions that force soft 
and collinear radiation to be integrated over~\cite{Sterman:1977wj}, while 
hard partons are `measured' (so that the observable $e$ becomes a collection 
of energies and angles); alternatively $E_m$ can be defined iteratively, building 
up an (infrared-safe) jet algorithm (see, for example~\cite{Cacciari:2008gp}).


\subsection{A tool: eikonal integrals}
\label{eikint}

As discussed above, the soft approximation for real radiation is a fairly straightforward 
affair: it basically amounts to a Laurent expansion in the soft momentum, where, by
picking the leading power, one locates the soft singularities and uncovers interesting
universality properties. Not surprisingly, the situation is significantly more intricate
when it comes to taking the soft approximation for virtual corrections, where the
soft momentum is an integration variable. The soft limit and the loop integration
do not commute, and things must be handled with care. In order to illustrate the
problems that arise, consider taking the soft limit, at the level of the integrand,
on the scalar integral $I_0$ defined in \eq{tensint}, which is responsible for the 
soft poles of the form factor. Applying \eq{softscale}, and restoring the prefactors
arising from the Feynman rules and from the colour sum, we define
\beq
  I_{\rm eik} \, \equiv \,  {\rm i} g^2 \mu^{2 \eps} \, N_c \, C_F \, \beta_1 \cdot \beta_2
  \int \frac{d^d k}{(2 \pi)^d} \, \frac{1}{\left( k^2 + {\rm i} \eta \right) 
  \left( - \beta_1 \cdot k + {\rm i} \eta \right) \left( \beta_2 \cdot k
  +  {\rm i} \eta \right)} \, ,
\label{eq:eikint}
\eeq
By inspection, there are several things to note about \eq{eq:eikint}.
\begin{itemize}
\item As expected, also in the case of virtual corrections the soft approximation is 
independent of spin and energy: the integral is homogeneous in both external momenta, 
and thus depends only on the dimensionless four-velocities $\beta_i$.
\item The integral is highly singular, in a rather intricate way: it is simultaneously 
affected by ultraviolet, soft and collinear divergences. Before even attempting to 
evaluate it, it will be necessary to understand the origin of the different singularities,
and how to treat them.
\item On the other hand, while highly singular, $I_{\rm eik}$ is readily `evaluated' 
in dimensional regularisation: since it does not depend on any scale, it is defined 
to vanish.
\end{itemize}
The proper way to handle this vanishing  integral emerges upon studying the nature
of its singularities. First of all, as should be clear  from the analysis of \secn{FormFac},
we observe that the UV divergence of $I_{\rm eik}$ is {\it not} inherited from QCD: 
indeed, the UV divergence of the original QCD diagram is completely contained in 
the tensor integral $I^{\alpha \beta}$. The fact is, taking  the limit in~\eq{softscale}, 
we have constructed an integrand which is a good approximation of the form factor 
integrand in the region $k^\mu \to 0$, but a very poor approximation in the UV 
region $k^\mu \to \infty$. As a consequence, our `effective low-energy theory' 
for the form factor has developed a {\it new} UV divergence, and, if we want to 
recover the proper low-energy result, we need to subtract (renormalise) this
spurious UV pole. In other words, the physics we are seeking to control is not
to be found in the `bare' version of $I_{\rm eik}$, with which  we have been 
working so far, but in the renormalised quantity
\beq
  I_{\rm eik}^{({\rm R})} \, = \, I_{\rm eik} + I_{\rm eik}^{({\rm ct})} \, = \, 
  I_{\rm eik}^{({\rm ct})} \, ,
\label{reneik}
\eeq
where $I_{\rm eik}^{({\rm ct})}$ is the UV counterterm, and we have used the 
fact that $I_{\rm eik} = 0$. What needs to be done is now clear: we need to introduce
auxiliary regulators to make the original $I_{\rm eik}$ integral finite in the soft and 
collinear regions; this will hopefully allow to unambiguously compute the UV pole; 
we then take the (minimal subtraction) UV counterterm as the the definition of 
$I_{\rm eik}^{({\rm R})}$, as in \eq{reneik}. Using a minimal scheme is essential 
in order to avoid dependence on the auxiliary regulators.

A detailed and consistent technique to perform this sequence of operations at
high orders in the loop expansion will be presented in \secn{CompMatr} (see 
Ref.~\cite{Gardi:2013saa}).  For a simple integral such as $I_{\rm eik}$, where 
all singular regions are easily identified, this is not needed, and we can get the 
right result by a sleight of hand. Proceed by introducing Feynman parameters 
in two steps: first combining the two linear denominators in $I_{\rm eik}$ with 
parameter $x$, and then combining the resulting (squared) denominator with 
the gluon propagator, with parameter $y$. One finds
\beq
\label{eikint2}
  I_{\rm eik} & = &  {\rm i} g^2 \mu^{2 \eps} \, N_c \, C_F \, \beta_1 \cdot \beta_2
  \int_0^1 \! dx \int_0^1 \! dy \int \frac{d^d k}{(2 \pi)^d} \, \frac{2 (1 - y)}{\Big[ y k^2 + 
  (1 - y) \big( x \beta_2 - (1 - x) \beta_1 \big) \cdot k +  {\rm i} \eta 
  \Big]^3} \nonumber \\ 
  & = & - \frac{\alpha_s}{2 \pi} \left[ \frac{8 \pi \mu^2}{- \beta_1 \cdot \beta_2
  - {\rm i} \eta} \right]^\eps \, N_c \, C_F \, \Gamma(1 + \eps) \, 
  B \left(- \eps, - \eps \right) \, \int_0^1 d y \, y^{-1 + 2 \eps} \, 
  (1 - y)^{-1 - 2 \eps} \, .
\eeq
The pole in $B \left(- \eps, - \eps \right) = - 2/\eps + {\cal O} (\eps)$ is of collinear 
origin: one can see this by noting that it arises from the $x$ integration, and the 
corresponding singular regions in Feynman-parameter space are $x \to \{0, 1\}$; 
the limit $x \to 0$ in the first line of \eq{eikint2} exposes the collinear singularity 
as $k^\mu \propto \beta_1$, which implies also $k^2 \to 0$; similarly, the limit 
$x \to 1$ exposes the collinear singularity as $k^\mu \propto \beta_2$. Next, 
one may be tempted  to replace the last integral in \eq{eikint2} with 
$B(2 \eps, - 2 \eps) = 0$. This looks consistent with the definition of the 
integral in dimensional regularisation, but actually it merely reminds us that 
the original integral is not well-defined for {\it any} values of $\eps$: it is 
plagued by either UV or soft divergences in any dimension. Once again, 
it is simple in this case to identify and disentangle the UV pole, which we 
are interested in. Looking  at the integrand in the first line of \eq{eikint2}, we 
see that the limit $y \to 0$ exposes a UV singularity (the integral diverges 
as $k^\mu \to \infty$ when $y = 0$), and is correspondingly regulated by taking 
$\eps > 0$. Conversely, the limit $y \to 1$ exposes an IR singularity (the integral 
diverges as $k^\mu \to 0$ for $y \to 1$), and is correspondingly regulated by taking 
$\eps < 0$. The two singular regions can be separated by inserting a factor of 
$1 = y  + (1 - y)$, getting
\beq
  \int_0^1 d y \, y^{-1 + 2 \eps} \, (1 - y)^{-1 - 2 \eps} & = & 
  \int_0^1 d y \, y^{-1 + 2 \eps} \, (1 - y)^{-1 - 2 \eps} \, \left[ y + (1 - y) \right]
\nonumber  \\ & = & \Gamma(1 + 2 \eps) \Gamma(- 2 \eps) + \Gamma(2 \eps)
  \Gamma(1 - 2 \eps) \, . 
\label{trick}
\eeq
The second term in \eq{trick} gives the UV pole, so that the one-loop counterterm 
in the MS scheme is 
\beq
  I_{\rm eik}^{({\rm R})} \, = \, 
  I_{\rm eik}^{({\rm ct})} 
  \, = \, - \frac{\alpha_s}{2 \pi} \, N_c \, C_F \, \frac{1}{\eps}
  \left[ \frac{1}{\eps} - \gamma_E + \ln (4 \pi) + 
  \ln \left( \! - \frac{\beta_1 \cdot \beta_2}{2} \right) \right] \, .
\label{oneloopct}
\eeq
Three observations are in order.
\begin{itemize}
\item The overall $1/\eps$ pole in \eq{oneloopct} is of ultraviolet origin. The
pole in the square bracket is the collinear divergence: if the hard partons were 
massive, it would be replaced by a logarithm of the Minkowskian angle 
between the two parton velocities, as discussed in detail in \secn{CompMatr}.
\item The UV pole in \eq{oneloopct}, which is the UV pole of $I_{\rm eik}$ with the 
opposite sign, can be directly interpreted as a soft singularity. The coefficient of the
ensuing double (soft-collinear) pole is the one-loop contribution to anomalous 
dimension associated with this UV divergence, which we will identify, in \secn{FactEvo}, 
with the QCD light-like cusp anomalous dimension~\cite{Korchemsky:1985xj,
Korchemsky:1987wg}.
\item In the presence of the collinear singularity, the rescaling symmetry of 
$I_{\rm eik}$ under $\beta_i \to \kappa_i \beta_i$ is broken by the logarithmic
term in \eq{oneloopct}, which multiplies the ultraviolet pole and thus cannot be 
discarded. This `anomalous' breaking of rescaling invariance will have very 
significant consequences for the structure of IR divergences of multi-parton 
amplitudes, to be discussed in \secn{MultiPart}.
\end{itemize}

\noindent
Most of the discussion in this section so far has focussed on the soft limit. 
Before turning, in \secn{AllOrd}, to the all-order generalisation of the one-loop 
results presented here, we briefly recall what happens in the presence of 
collinear divergences, when the Bloch-Nordsieck cancellation does not 
suffice, and  one needs to resort to collinear factorisation.


\subsection{Hadron scattering and collinear divergences}
\label{CollDiv}

\begin{figure}
    \centering
    \begin{subfigure}{0.25\textwidth}
    \centering
        \includegraphics[width=\textwidth]{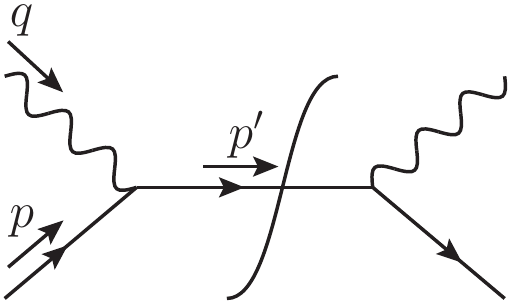}
        \caption{}
        \label{fig:DIS_NLO_a}
    \end{subfigure}
    \quad 
      \begin{subfigure}{0.25\textwidth}
      \centering
        \includegraphics[width=\textwidth]{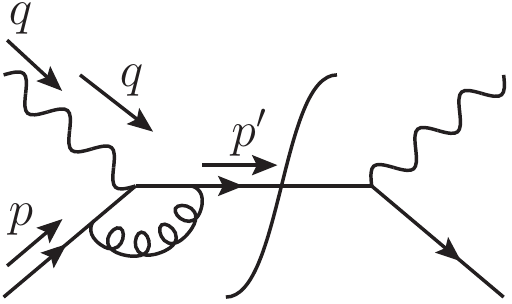}
        \caption{}
        \label{fig:DIS_NLO_b}
    \end{subfigure}
      \caption{Leading-order and virtual correction diagrams for Deep Inelastic Scattering 
      (DIS) up to order $\alpha_s$ (self-energy diagrams are omitted).}
  \label{fig:DIS_LO_NLO_virtual}
\end{figure}

\begin{figure}
    \centering
    \begin{subfigure}{0.25\textwidth}
    \centering
        \includegraphics[width=\textwidth]{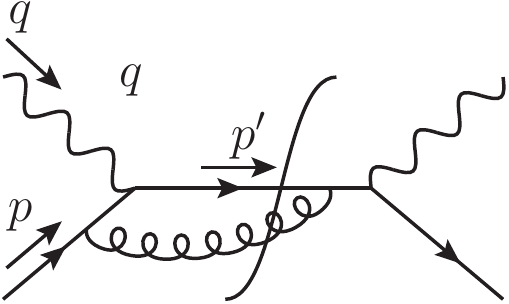}
        \caption{}
        \label{fig:DIS_NLO_c}
    \end{subfigure}
    \quad 
    \begin{subfigure}{0.25\textwidth}
    \centering
        \includegraphics[width=\textwidth]{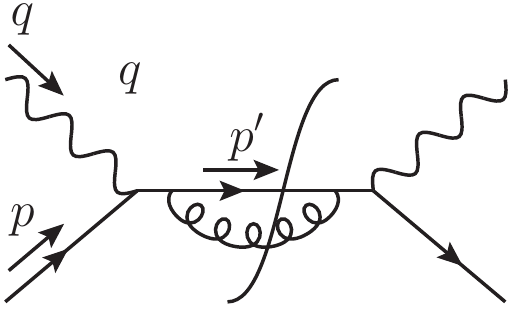}
        \caption{}
        \label{fig:DIS_NLO_d}
    \end{subfigure}
    \quad 
      \begin{subfigure}{0.25\textwidth}
      \centering
        \includegraphics[width=\textwidth]{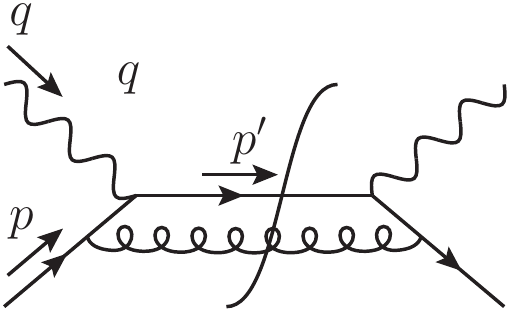}
        \caption{}
        \label{fig:DIS_NLO_e}
    \end{subfigure}
      \caption{Real-radiation diagrams for the DIS cross section at 
      order $\alpha_s$.}
  \label{fig:DIS_NLO_real}
\end{figure}

For the sake of completeness, in this Section we sketch the one-loop calculation of
Deep Inelastic Scattering (DIS) structure functions, the classic example of non-cancellation 
of infrared divergences in QCD, due to the presence of initial-state hadrons, and the 
starting point of the factorisation program. The calculation is of course well known, 
so we only briefly summarise it, and we focus on the origin and the treatment of 
uncanceled divergences, to be compared with \secn{RealR} and \secn{FormFac}. We 
consider lepton-proton scattering via photon exchange, neglecting masses, 
and specifically we focus on scattering off valence quarks, involving the diagrams 
displayed in Fig.~\ref{fig:DIS_LO_NLO_virtual}-\ref{fig:DIS_NLO_real} at LO and NLO. 
We assign momentum $l$ to the incoming lepton, momentum $l^\prime$ to the 
outgoing lepton, and momentum $p$ to the incoming hadron, and we employ the 
usual conventions for kinematic variables
\beq
  q^\mu \, \equiv \, \big( l^\prime - l \big)^\mu  \, , \qquad - q^2 \, \equiv \, Q^2 \, > \, 0 \, ,
  \qquad x  \, \equiv \,  \frac{Q^2}{2 p \cdot q} \, \rightarrow \, 
  s \, \equiv \, (p + q)^2 \, = \, Q^2 \, \frac{1 - x}{x} \, .
\label{Disvar}
\eeq
In analogy with \eq{sigmatotLH}, we can factor the cross section for this process
into the product of leptonic and hadronic tensors. For the spin-averaged differential
cross section we write then
\beq
  d^3 \sigma \, = \, \frac{1}{2s} \, \frac{d^3 l^\prime}{(2 \pi)^3 2 E^\prime} \, 
  L^{\mu \nu} (l,l^\prime) \, W_{\mu \nu} (p,q) \, 
\eeq
where $E^\prime = |{\bf l}^\prime|$, and, to leading order in the lepton charge, 
the lepton tensor (defined to include photon propagators) simply reads
\beq
  L^{\mu \nu} \, 
  \equiv \,
   \frac{e^2}{Q^4} \,\, {\rm Tr} 
   \Big( \slashed{l} \, \gamma^\mu \,
  \slashed{l}^\prime \, \gamma^\nu \Big) \, .
\eeq
In analogy to what we did for the annihilation cross section in \eq{Hmunu}, we 
can express the hadronic tensor $W_{\mu \nu}$ in terms of matrix elements of 
the electromagnetic current in the relevant hadron state. For a particle of electric 
charge $e Q_f$ we write
\beq
  W_{\mu \nu} (p, q) & = & 
  \frac{e^2 Q_f^2}{8 \pi} \sum_{{\rm spin}, \, X} \, \bra{p} J_\mu^\dagger (0) \ket{X} 
  \bra{X} J_\nu (0) \ket{p} \, (2 \pi)^4 \, \delta^4(p_X - q - p)  \nonumber \\
  & = &
  \frac{e^2 Q_f^2}{8 \pi} \, \sum_{\rm spin} \int d^4 x \, {\rm e}^{{\rm i} q \cdot x} \,   
  \bra{p} J_\mu^\dagger (x) \, J_\nu (0) \ket{p} \, .
\label{htdis}
\eeq
Lorentz covariance, gauge invariance and parity conservation constrain the tensor 
structure to be of the form
\beq 
  W_{\mu \nu} (p, q) \, = \, - \bigg( g_{\mu \nu} -\frac{q_\mu q_\nu}{q^2}\bigg) 
  F_1(x, Q^2) + \frac{1}{p\cdot q} \bigg( p_{\mu} - q_\mu \frac{p\cdot q}{q^2}\bigg)
  \bigg( p_{\nu} - q_\nu \frac{p\cdot q}{q^2}\bigg) F_2 (x, Q^2) \, ,
\label{F1F2}
\eeq
and one easily verifies that, if the state $\ket{p}$ in \eq{htdis} is a massless
spin $1/2$ fermion, at tree level one finds the {\it elastic scattering} result
\beq 
  F_2^{(0)} (x,Q^2) \, = \, 2 F_1^{(0)} (x,Q^2) \, = \, e^2 Q_f^2 \, \delta(1-x) \, ,
\label{F1F20}
\eeq
famously exhibiting {\it scaling} behavior~\cite{Bjorken:1968dy}, {\it i.e.} independence
on the scale $Q^2$, which is a hallmark of the parton model~\cite{Feynman:1969ej}, 
and was experimentally verified at SLAC half a century ago~\cite{Friedman:1972sy}. 

At NLO, the infrared content of the process becomes non-trivial. Since we know 
that $W_{\mu\nu}$ is built out of only two independent scalar functions, it is 
sufficient to compute the contractions of the hadronic tensor with $g_{\mu \nu}$ 
and with $p_\mu p_\nu$. For our purposes, the interesting quantity is the trace 
of the hadronic tensor, $W \equiv - g^{\mu \nu} W_{\mu \nu}$, which contains 
the non-trivial infrared information. Looking at the virtual diagrams in 
Fig~\ref{fig:DIS_LO_NLO_virtual}, panel {\it (b)}, one realises that they build 
up the quark form factor, \eq{olofofa}, this time for space-like kinematics, as 
can be expected. The precise result is
\beq
\label{Disvirt}
   W_{\rm virt}^{(1)} (p,q) & = & 2 \, {\bf Re} \left[ 
   \Gamma^{(1)} \bigg( \frac{Q^2}{\mu^2}, \epsilon \bigg) \right] W^{(0)} 
   \, = \, 2 \, {\bf Re} \left[ \Gamma^{(1)} \bigg( \frac{Q^2}{\mu^2}, \epsilon \bigg) \right]
   (1 - \eps) \, e^2 Q_f^2 \, \delta (1-x) \\ 
   & = & - \frac{\as}{2 \pi} \, C_F 
   \bigg( \frac{4 \pi \mu^2}{Q^2} \bigg)^{\! \eps} \,
   \frac{ \Gamma^2 (1 - \eps) \, \Gamma(1 + \eps)}{\Gamma(1 - 2 \eps)} 
   \bigg( \frac{2}{\eps^2} + \frac{3}{\eps} + 8 \bigg) \, 
   (1 - \eps) \, e^2 Q_f^2 \, \delta (1-x) \, , \qquad \nonumber 
\eeq
where in the first line we used the explicit expression for the tree-level trace of 
the hadronic tensor, $W^{(0)}$, easily derived from \eq{F1F2} and \eq{F1F20}.
Motivated by our success in cancelling infrared poles in \secn{cataQED} 
and in \secn{TotCross}, we can now compute the cross section for the emission of a 
real gluon with momentum $k$. The contribution to the trace of the hadronic tensor, 
in analogy to \eq{H1R}, is
\beq
  W_{\rm real}^{(1)} (p,q) & = &
  \frac{1}{8\pi} \int \frac{d^d p^\prime d^d k}{(2 \pi)^{d-2}} \, 
  \delta^d (p + q - p^\prime - k) \, \delta_+ (p^{\prime \,  2} ) \,
  \delta_+ (k^2) \, \Big| {\cal A} (p,q,k) \Big|^2 \, ,
\label{Wtoint} 
\eeq
where $| {\cal A} (p,q,k) |^2$ is a short-hand notation for the square of the real-radiation 
matrix element, summed over final state colours and polarisations, averaged over
initial-state colour, and traced over the open photon indices. ${\cal A}$ now characterises
a $2 \to 2$ process, with the Mandelstam invariant $s$ given by \eq{Disvar}, while
\beq
  t \, = \, (p - k)^2  \, = \, - 2 p \cdot k \, , \quad 
  u \, = \, ( p - p^\prime)^2 \, = \, - 2 p \cdot p^\prime \quad 
  \rightarrow \quad s +  t + u \, = \, - Q^2 \, . 
\label{moreDisvar}
\eeq
In $d = 4 - 2 \epsilon$, one finds
\beq
  \Big| {\cal A} (p,q,k) \Big|^2 \, = \, 32 \pi \, e^2 Q_f^2 \, \alpha_s^2 \mu^{2 \eps} 
  \, C_F \, (1 - \eps) \bigg[ (1 - \eps) \bigg( \frac{s}{-t} + \frac{-t}{s} \bigg)
  + \frac{2 Q^2 u}{s t} + 2 \eps \bigg] \, .
\label{radmatr}
\eeq
In order to perform the integrals in \eq{Wtoint}, it is useful to parametrise the 
invariants as
\beq
 t \, = \, - \frac{Q^2}{x} (1 - y) \, , \qquad u \, = \, - \frac{Q^2}{x} y \, ,
\label{inv}
\eeq
where $y$ plays the role of an angular variable, as in \eq{enang}: in the collinear limit
$y \to 1$ and $t \to 0$, and the squared matrix element in \eq{radmatr} displays the
expected collinear singularity. Performing the phase space integrals in \eq{Wtoint},
with the help of \eq{dregen} and \eq{dregenan}, one finally finds
\beq
\label{Disreal}
  W_{\rm real}^{(1)} (p,q) & = & e^2 Q_f^2 \,\,
  \frac{\as}{2 \pi}  C_F \,  \bigg( \frac{4 \pi \mu^2}{s} \bigg)^{\! \eps} \,
  \frac{\Gamma(2 - \eps)}{\Gamma(1 - 2 \eps)} 
  \bigg[ - \frac{1}{\eps} \, \frac{1+x^2}{1-x} - \frac{3 + \eps}{2 (1 - 2 \eps)} \, \frac{1}{1 - x} \\
  & & \hspace{6.5cm}
  + \, \frac{2 (1 - 2 \eps^2)}{(1 - \eps)(1 - 2 \eps)} + (1 - x) \bigg] \, . \nonumber 
\eeq
Had one expected a cancellation of the singularities of \eq{Disvirt}, \eq{Disreal} would
be a striking disappointment, displaying no obvious double pole, and a single pole with
non-trivial dependence on $x$. Let's look at these two aspects in turn.

Considering first the soft singularity, we note that the soft limit $k^\mu \to 0$
coincides with the elastic scattering limit $x \to 1$: the soft singularity is present, but
it is localised at $x = 1$. In order to emphasise this, we first express the prefactor in 
\eq{Disreal} in terms of the scale $Q^2$, using \eq{Disvar}: this turns the factors
of $(1 - x)^{-1}$ into $(1 - x)^{-1 - \eps}$, regularising the $x \to 1$ singularity
for $\eps < 0$, as expected. We then proceed to interpret this factor in terms of 
distributions, noting that integrals of the form 
\beq
  I(f) \, = \, \int_0^1 d x \, (1 - x)^{- 1 - \eps} f(x) \, ,
\label{intf}
\eeq
for any function $f(x)$ regular at $x \to 1$, can be computed as power series in $\eps$
using
\beq
  I(f) & = & \int_0^1 d x \, \frac{f(x) - f(1)}{(1 - x)^{1 + \eps}} + 
  f(1)  \int_0^1 d x \, \frac{1}{(1 - x)^{1 + \eps}} 
  \nonumber \\
  & = & - \frac{1}{\eps} f(1) \, + \, \sum_{n = 0}^\infty \, (-1)^n \, \frac{\eps^n}{n!} 
  \int_0^1 d x \, \frac{\ln^n (1 - x)}{1 - x} \, \big( f(x) - f(1) \big) \, .
\label{intf2}
\eeq
Introducing the distributions
\beq
  {\cal D}_n (x) \, \equiv \, \left[ \frac{\ln^n (1 - x)}{1-x} \right]_+ \,  \rightarrow \,\,
  \int_0^1 d x \, {\cal D}_n (x) f(x) \, \equiv \, \int_0^1 d x \, \frac{\ln^n (1 - x)}{1-x} \,
  \big( f(x) - f(1) \big) \, ,
\label{plusdef}
\eeq
one can formalise \eq{intf2} in terms of the distribution identity
\beq
  \frac{1}{(1 - x)^{1 + p \eps}} \, = \, - \frac{1}{p \eps} \, \delta(1 - x) \, + \,  
  \sum_{n = 0}^\infty \, (-1)^n \, \frac{(p \eps)^n}{n!} \, {\cal D}_n(x) \, .
\label{distrid}
\eeq
Substituting \eq{distrid} in the pole term of \eq{Disreal} we immediately see that a double
pole proportional to $\delta(1 - x)$ is generated, and indeed it cancels the double pole
of \eq{Disvirt}. Discarding contributions of order $\epsilon$, the final result for the
trace of the hadronic tensor can be written as
\beq
\label{eq: coll_pole}
  && W^{(1)} (p,q) \, = \, 
  \frac{\as}{2 \pi} \, \bigg( \frac{4\pi \mu^2}{Q^2} \bigg)^\eps \, e^2 Q_f^2 \,
  (1 - \eps) \bigg[ - \frac1\eps \, \frac{\Gamma(1 - \eps)}{\Gamma(1 - 2 \eps)} \, 
  P_{qq} (x) +
  \\ && \hspace{5mm} + \, C_F \bigg(
  \big( 1 + x^2 \big) {\cal D}_1 (x) - \frac32 \, {\cal D}_0 (x)
  - (1 + x^2) \, \frac{\log x}{1 - x} + 3 - x - 
  \Big( \frac92 + \frac{\pi^2}{3} \Big) \delta(1-x) \bigg) \bigg] , \nonumber
\eeq
where we introduced the leading-order DGLAP {\it splitting function} \cite{Altarelli:1977zs,
Gribov:1972ri,Dokshitzer:1977sg}
\beq
  P_{qq} (x) \, = \, C_F \bigg[ (1 + x^2) \, {\cal D}_0 (x) + \frac32 \, \delta (1 - x) \bigg] \, , 
\label{DGLAPpqq}
\eeq
with the virtual term normalised so that $\int_0^1 d x P_{qq}(x) = 0$. While the splitting
function was derived here in the context of DIS, it is a universal quantity, and we will
discuss it further, and give a process-independent derivation, in \secn{Real}.

After the cancellation of the soft singularity, \eq{eq: coll_pole} is left with a single 
pole, which is manifestly of collinear nature (since it has support for $x \neq 1$). It is 
important to note at the outset that this pole has nothing to do with the non-abelian 
nature of the interaction, and certainly nothing to do with confinement: indeed, precisely 
the same pole would arise in the abelian theory, for example in the scattering of a 
massless charged fermion from a classical source. In ordinary QED, the singularity 
is regularised by the electron mass, but this is of little help for perturbative calculations 
at large momentum transfer, since the pole is replaced at each order by large logarithms 
of the ratio $m_e/Q$, which spoil the behaviour of the perturbative series unless
they are resummed. Let us then discuss the origin and treatment of this collinear
singularity in light of the results of \secn{KLN} and \secn{Coherent}.

To understand the failure of the Bloch-Nordsieck mechanism, it helps to compare
the DIS process to the total annihilation cross section, considering the relevant
diagrams in the context of time-ordered perturbation theory. For the real emission 
diagrams in Fig.~\ref{Rolofig}, the dominant time ordering for the emission of soft
or collinear gluons is the one where the current creates the quark pair at time $t_c$, 
and the gluon is radiated at time $t_r \gg t_c$. Crucially, this late-time emission 
does not affect the Born process, which still consists of the creation of a pair with
total energy $Q$. As the virtual correction is always proportional to the Born process,
the cancellation of singularities is at least possible. In DIS kinematics, on the other 
hand, the dominant time ordering for collinear emission in the `handbag' diagram of 
Fig.~\ref{fig:DIS_NLO_e} is the one in which gluon radiation takes place at time $t_r 
\ll t_c$: the current therefore scatters a quark with an energy different from that of
the Born process, which is reflected by the fact that $x \neq 1$. Only when the radiated
gluon is soft one recovers the elastic configuration, and the cancellation of divergences
with the virtual correction becomes possible.

Any treatment of this problem must rely on the understanding that gluon (or photon)
emission at very early times is not to be associated with the hard scattering process,
but rather with the wave function of the initial state. Indeed, as we saw in \secn{phi3coh},
in the coherent-state picture the collinear singularity is cancelled by a contribution
in which the radiation originates from the initial-state coherent state operator; 
analogously, applying the KLN theorem requires in this case a sum over 
degenerate initial states. The inescapable conclusion is that a massless 
particle in the initial state must {\it always} be understood as a {\it beam}, 
containing different Fock state components. In general, different components 
should be evaluated at amplitude level, and will interfere at cross-section level; 
at large momentum transfer, the statement of collinear factorisation is that 
interference terms are suppressed by powers of the hard scale, and one is 
allowed to sum incoherently over different channels, weighted by the respective 
probabilities. 

We will not discuss techniques to prove factorisation here: they will be introduced, 
at the level of scattering amplitudes, in Sections~\ref{AllOrd}, \ref{FactEvo} and 
\ref{MultiPart}, and they have been thoroughly presented elsewhere (see, for 
example, Ref.~\cite{Collins:1989gx}). We take however the opportunity to make
the qualitative arguments of \secn{Factorsafe} slightly more precise. Consider 
a DIS-type process initiated by a massless particle of flavour $i$, which has 
lagrangian interactions with a set of massless particles with flavours $\{j\}$. 
Initial-state radiation can turn particle $i$ into any other particle $j$ (including 
of course the case $j = i$), with a degraded longitudinal momentum, and one 
can define a probability distribution $f_{j/i} (z, \epsilon)$ for detecting particle 
$j$ as a `constituent' of particle $i$, carrying a fraction $z$ of the longitudinal 
momentum of $i$. Such distributions can be constructed by means of (non-local)
matrix elements of field operators in the single-particle state containing particle 
$i$~\cite{Collins:1981uw,Collins:1989gx,Sterman:1994ce,Sterman:1995fz,
Collins:2011zzd}, in order to reproduce the collinear divergences of the hadronic 
tensor. Even without resorting to a formal definition, it is easy to illustrate
the mechanics of the factorisation procedure: any projection of the hadronic 
tensor $W^{\mu \nu}_i$ for the process initiated by particle $i$ can be written as
\beq
  W_i (p,q) \, = \, \sum_j \int_0^1 \frac{d z}{z} \, H_j (z p, q) \, f_{j/i} (z, \epsilon) \, ,
\label{partmod}
\eeq
with the goal of defining a set of collinear-finite hard functions $H_j$. At tree 
level, with no emissions, one can normalise the hadronic tensor (dropping 
coupling factors) so that
\beq
  W_i^{(0)} (p,q) \, = \, \delta(1 - x) \, , \! \quad \!
  f_{j/i}^{(0)} (z, \epsilon) \, = \, \delta_{ij} \delta(1 - z) \! \quad \! \rightarrow
  \quad H_j^{(0)} (z p, q) \, = \, \delta \left( 1 - \frac{x}{z} \right) \, .
\label{treeDIS}
\eeq
It is then straightforward to expand \eq{partmod} to NLO, with the result
\beq
  W_i^{(1)} (p,q) \, = \, \sum_j f_{j/i}^{(1)} (x, \epsilon) + 
  H_i^{(1)} (p, q) \, .
\label{nlodis}
\eeq
One sees that divergent contributions such as those emerging in \eq{eq: coll_pole}
can be collected as radiative corrections to the {\it parton distributions} $f_{j/i}(x, \eps)$; 
furthermore, one is free to choose a {\it factorisation scheme}, by assigning sets
of nonsingular contributions to either $f_{j/i}$ of $H_i$. The subsequent interpretation 
and usage of \eq{nlodis} and its higher-order generalisations will depend on 
the context. In a confining theory such as QCD, the perturbative parton distribution 
$f_{j/i}$ must be convoluted with its hadronic counterpart, which is non-perturbative 
and needs to be determined experimentally. In the case of QED, on the other 
hand, it is natural to replace the dimensional regulator with the physical one -- 
the mass of the charged particle: the distributions $f_{j/i}$ are then computable 
in perturbation theory~\cite{Kuraev:1985hb,Ellis:1986jba}. In either case, the
most striking consequence of \eq{eq: coll_pole} is the breaking of the scale 
invariance that was observed at tree level in \eq{F1F20}. Upon expanding 
\eq{eq: coll_pole} in powers of $\e$, and after removing the collinear pole to
the parton distribution, one observes a logarithmic dependence on the hard
scale $Q$, with a coefficient given by the splitting kernel $P_{qq}$. We note
now a pattern which will be recurring in the following Sections (see, in particular,
the discussion at the beginning of \secn{FactEvo}): once a factorisation, such 
\eq{partmod}, is achieved, extracting contributions that are singular as the 
regulator is removed entails the choice of a {\it factorisation scale}, $\mu_f$,
separating long-distance and short-distance contributions. In the present case, 
the scale $\mu_f$ can be introduced simply by splitting the logarithm emerging 
from \eq{eq: coll_pole} as 
\beq
  \ln \left( \frac{\mu^2}{Q^2} \right) \, = \,  \ln \left( \frac{\mu^2}{\mu_f^2} \right) 
  \, + \, \ln \left( \frac{\mu_f^2}{Q^2} \right) \, ;
\label{splitlog}
\eeq
one then assigns the first term to the parton distribution, which is taken to 
depend on energy scales below $\mu_f$, and the second term to the hard
function, which retains the dependence on the hard scale $Q$. 

As we will discuss in greater detail in \secn{FactEvo}, this artificial scale 
separation inevitably leads to the existence of an evolution equation, following 
from the requirement that physical quantities be independent of the arbitrary 
scale choice. The solution to this equation will resum scale logarithms, in this
case of collinear origin. At our present accuracy, using the tree-level result in 
\eq{treeDIS}, and applying \eq{splitlog}, one immediately verifies that the 
one-loop distribution satisfies the DGLAP equation
\beq
  \mu_f^2 \frac{\partial}{\partial \mu_f^2} \, f_{q/q} \big( x, \mu_f^2 \big) \, = \, 
  \frac{\alpha_s}{2 \pi} \int_x^1 \frac{d y}{y} \,
  P_{qq} \left( \frac{x}{y}, \alpha_s \right) f_{q/q} \big( y, \mu_f^2 \big) \, ,
\label{DGLAP}
\eeq
where \eq{DGLAPpqq} gives the leading order approximation to the kernel 
$P_{qq}$, and the lower limit of integration reflects the fact that collinear 
radiation can only degrade the longitudinal momentum fraction\footnote{When
applying the formal definitions of parton distributions in terms of non-local matrix 
elements of field operators, as in Refs.~\cite{Collins:1981uw}, the DGLAP 
equation~\ref{DGLAP} emerges as a {\it renormalisation group equation}: 
the {\it factorisation} scale for the complete theory (QCD in this case) coincides 
with the {\it renormalisation} scale for the low-energy collinear effective theory.}. 
Both in QED~\cite{Skrzypek:1990qs,Skrzypek:1992vk,Cacciari:1992pz,Frixione:2019lga,
Bertone:2019hks,Frixione:2021wzh} and in QCD~\cite{Altarelli:1977zs,Gribov:1972ri,
Dokshitzer:1977sg}, the resummation of collinear logarithms that follows 
from solving DGLAP equations like \eq{DGLAP} is a crucial step in order 
to obtain accurate predictions for high-energy processes~\cite{Ellis:1991qj}.

For inclusive DIS, the validity of the factorisation in \eq{partmod} to all 
orders in perturbation theory can be established both with diagrammatic 
methods~\cite{Sterman:1994ce,Collins:2011zzd} and using the Operator 
Product Expansion (OPE)~\cite{Peskin:1995ev}. The generalisation to collider 
processes, involving two hadrons in the initial state, is both conceptually and 
technically non trivial. Physically, one could readily imagine that, in a collider 
process, as the two colliding hadrons approach, colour forces from one hadron
might alter the parton configuration of the second one, spoiling the universality
of the distributions measured in single-hadron process. Soft gluons, with their
long wavelength, are prime candidates to be responsible for this failure of 
factorisation. In a general collider process, the OPE is not available, and 
establishing the cancellation of such soft-gluon effects to all orders in 
perturbation theory is a delicate exercise~\cite{Ellis:1978ty,Altarelli:1979ub,
Bodwin:1984hc,Collins:1985ue,Collins:1988ig}, reviewed in Ref.~\cite{Collins:1989gx}.
Such factorisation proofs require an all-order analysis of Feynman diagrams:
we now turn to describing some of the tools required to perform these 
studies.


\section{Infrared singularities to all orders}
\label{AllOrd}

The discussion in \secn{FinOrd} was mostly limited to the lowest non-trivial 
perturbative order -- one loop. Here we prepare the ground for studying the 
problem to all orders. Superficially, the complexity of the problem appears 
daunting -- how can one disentangle the structure of singularities of arbitrarily
intricate Feynman diagrams, and organise them in an intelligible pattern?
We can find some hope in the successful treatment of QED in \secn{cataQED},
which suggests that infrared singularities are perhaps relatively simple, as they are 
constrained by their semi-classical nature and by their origin in long-distance dynamics.
This hope is indeed well founded: research developed over several decades
has in fact, to a large extent, succeeded in uncovering the all-order infrared 
structure of perturbative non-abelian gauge theories, as we will see in some 
detail in Sections~\ref{FactEvo} and~\ref{MultiPart}.

Factorisation and exponentiation of soft and collinear singularities can be 
achieved in four steps. First, one needs to {\it locate potential singularities} in
generic Feynman integrals, establishing {\it necessary} conditions for singularities
to arise. These conditions are summarised by the {\it Landau equations}, discussed
below in \secn{Landau}. Next, one needs to ascertain in what cases the solutions
of Landau equations yield {\it actual singularities}, establishing {\it sufficient} conditions
for Feynman integrals to diverge. While the necessary conditions are largely 
universal, sufficient conditions are theory-specific, and require the development
of {\it power-counting tools}, similar to the ones used for the renormalisation of 
ultraviolet divergences. These tools are discussed in \secn{IRPowCo}. The third 
step leverages the universality of infrared singularities and their long-distance character
by {\it constructing operator matrix elements}, independent of the hard scattering 
process being considered, and yet capable of reproducing the leading-power
soft and collinear behaviour of the corresponding amplitudes. This step is the 
most technically intricate: while it is easy to derive the form of the desired operators
at low orders, proving their properties to all orders requires the machinery of
Ward identities and a careful handling of integration contours in the relevant 
limits for generic diagrams. We will present simple arguments for the identification
of the appropriate operators in \secn{universal_fun}. Once the divergences have 
been organised into operator matrix elements, the last step is to {\it determine their
all-order behaviour}, resulting in the {\it exponentiation} of soft and collinear poles.
This can be done by means of {\it evolution equations}, which are direct consequences
of factorisation, as discussed in \secn{FactEvo}; alternatively, and complementarily,
exponentiation can be achieved by means of combinatorial tools, developed
in the second part of \secn{MultiPart}.


\subsection{Locating potential singularities: the Landau equations}
\label{Landau}

Feynman integrals are analytic functions of external momenta, and their very
existence and meaning are tied to the ${\rm i} \eta$ prescription defining
particle propagators. The proper tools to study their properties are therefore
those of multi-variate complex analysis. A vast and growing body of research
has been devoted to this subject, but, for our purposes, we will barely need
to scratch the surface of the knowledge that has been accumulated. A wealth 
of information can be found in the classic reference~\cite{Eden:1966dnq}, and
more detailed discussions of the aspects that we cover below can be found in 
many textbooks, such as Ref.~\cite{Sterman:1994ce} and Ref.~\cite{Collins:2011zzd}.
It is worthwhile noticing that the singularity structure of amplitudes, and in 
particular the Landau equations, have been the focus of much recent research,
see for example Refs.~\cite{Dennen:2016mdk,Abreu:2017ptx,Mizera:2021fap,
Hannesdottir:2021kpd}; in particular, Ref.~\cite{Collins:2020euz} provided an 
important refinement of existing proofs.


\subsubsection{Singularities of complex functions defined by an integral}
\label{integrals}

We begin by illustrating in some simple examples how multivariate integrals in
 complex variables can become singular. The crucial point is that, thanks to
Cauchy's theorem, it is not sufficient for a singularity of the integrand to move
to the integration contour, since the contour can in general be deformed away 
from the singularity, thus providing an analytic continuation of the integral function.
Integrals can become singular only when the contours cannot be deformed away 
from the poles.

To be specific, consider a function $f(z)$ that is defined by an integral 
representation in the form
\beq
  f(z) \, = \, \int_\gamma d \omega \, \varphi (\omega, z) \, .
\eeq
The integration contour in the complex $\omega$-plane is denoted by $\gamma$, 
and we assume that $\varphi(\omega, z)$ is a meromorphic function in the complex 
$\omega$-plane, with a set of isolated poles located at points $ \omega_i = 
\omega_i (z)$. These singular points  move in the $\omega$-plane as $z$ is 
varied. Suppose that for some point $z_0$ in the $z$-plane, none of the singular 
points $\omega_i$ lies on the contour $\gamma$. This then implies the existence
of open neighbourhoods $Z$ of the point $z_0$, whose points share the same property.
For each such $Z$, there exists an open neighbourhood $\Omega$ in the $\omega$-plane 
that contains the contour $\gamma$, while none of the singular points $\omega_i$ lies 
in $\Omega$. The function $f(z)$ is then analytic in $Z$. Now, as $z$ moves out of 
$Z$, it may happen that some of the singular points $\omega_i (z)$ move towards 
$\gamma$ and ultimately land on it. An integral with a pole on its integration contour 
is ill-defined: in general, however, we can exploit Cauchy's theorem and deform the 
contour to avoid the approaching singular points. The integral over the deformed 
contour provides an analytic continuation of the original integral. Therefore, if such 
a deformation is possible, the integral $f(z)$ remains analytic for such values of $z$.

There are three distinct configurations in which the singularities $\omega_i$ cannot 
be avoided by contour deformations and, thus, leads to actual singularities in $f(z)$.
We examine them in turn.
\begin{itemize}
\item {\it End-point singularity}. A function $f(z)$ is singular at a point $z_0$, if, as 
$z$ approaches $z_0$, one of the singular points $\omega_i(z)$ approaches either of 
the end-points of the integration contour.
\begin{figure}
    \centering
    \begin{subfigure}{0.3\textwidth}
    \centering
        \includegraphics[width=0.8\textwidth]{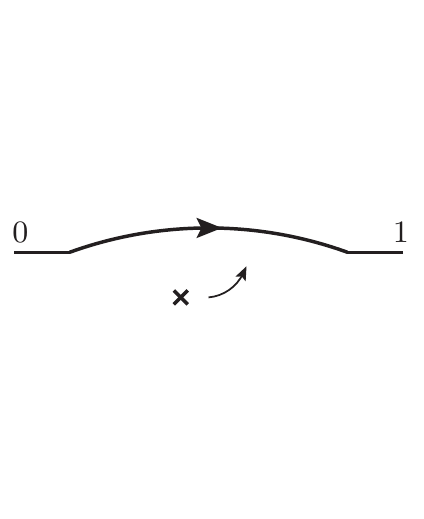}
        \caption{}
        \label{fig:path_a}
    \end{subfigure}
    \quad 
    \begin{subfigure}{0.3\textwidth}
    \centering
        \includegraphics[width=0.8\textwidth]{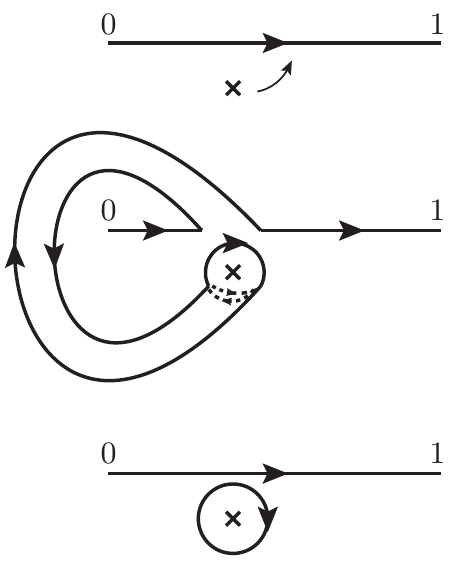}
        \caption{}
        \label{fig:path_b}
    \end{subfigure}
    \quad 
      \begin{subfigure}{0.3\textwidth}
      \centering
        \includegraphics[width=0.8\textwidth]{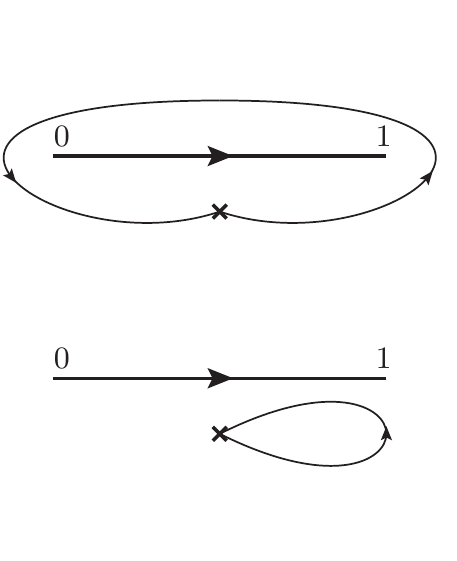}
        \caption{}
        \label{fig:path_c}
    \end{subfigure}
      \caption{Contour deformations: a) Contour deformation to avoid an approaching pole.
        b) Deformation when the pole makes a complete circle around the end point.
        c) No contour deformation is required if the pole encircles both the end points.}
  \label{fig:deform}
\end{figure}
A simple example is the integral
\beq
  f_1(z) \, = \, \int_0^1 \frac{d \omega}{\omega - z} \, .
\label{ex1}
\eeq
Here the integrand has a single pole in the $\omega$-plane at $\omega_1 (z) = z$. We 
expect therefore singularities at $z = 0$ and at $z = 1$, while, if $z$ approaches the real 
axis in the region $0 < z < 1$, we can deform the contour to avoid the approaching pole, 
as in Fig.~{\ref{fig:deform}(a)}. It is also easy to predict that the singularities will be branch 
points: indeed, if $z$ encircles the point $z=0$ and comes back to its original location, 
then the singular point $\omega_1(z)$ also makes a circle around the end point, but, in 
doing so, it drags the contour along with it, as shown in Fig.~{\ref{fig:deform}(b)}. This new 
contour can be written as a sum of two contours, one of which encloses the pole. Thus 
we see that a residue is picked up, and there is a discontinuity equal to $2 \pi \textrm{i} $ 
times that residue. The same residue is picked up each time the endpoint is circled, giving 
rise to the same amount of discontinuity: this shows that $z=0$ is a logarithmic branch 
point of $f(z)$, and the same of course holds true for the other end point, $z = 1$. If, on
the other hand, $z$  moves in such a manner that the pole never crosses the contour, then 
no residue is picked up: two such trajectories are shown in the Fig.~{\ref{fig:deform}(c)}; in 
both these cases we remain on the same branch of $f(z)$. All the above observations are 
obviously verified by the explicit result of integration which gives
\beq
  f_1(z) \, = \, \log \, \frac{z - 1}{z} \, .
\label{ans1}
\eeq
\item {\it Pinch singularity}. A function $f(z)$ is singular at a point $z_0$ if two (or more) 
singularities of the integrand, say $\omega_1(z)$ and  $\omega_2(z)$, approach the 
contour from opposite directions and eventually meet at the point $z_0$ preventing 
contour deformation. An example with a slight twist is given by
\beq
  f_2 (z) \, = \,  \int_0^1 \frac{d \omega}{(\omega - z) (\omega - a)} \, ,
\label{ex2}
\eeq
where we can take $1 < a < 2$ for the sake of the present argument.
Here the integrand has two poles, $\omega_1(z) = z$ and $\omega_2 (z) = a$. We
already know that the presence of $w_1$ will lead to branch points in $f_2 (z)$, at 
$z = 0$ and at $z = 1$, as in the previous example. The role of $\omega_2$, which 
is independent of $z$, is less obvious. To clarify it, take a point $z$, away from the 
contour, and lying on the principal sheet defined by the two branch points. The function
$f(z)$ is analytic at this point, since both singular points $\omega_1$ and $\omega_2$ 
are away from the contour. Now, recall that if we make a closed circuit around an end 
point, say $z = 1$, this will bring us to the second Riemann sheet. There is a new 
singularity present on this sheet, which is absent on the principal sheet: if for example 
one takes the point $z = a$ on the principal sheet and circles it around $z = 1$, as 
shown in Fig.~{\ref{newpinch}}, in the process the contour gets dragged along, and 
it finally gets pinched with the pole at $w = a$ on the second sheet. By the same 
argument, it is clear that all other sheets (except the principal sheet) will have a 
singularity at $z = 1 + (a-1) {\rm e}^{2 \pi {\rm i} k}$, where $k$ is a non-vanishing 
integer.  
\begin{figure}
        \centering
        {\includegraphics[height=2.5cm,width=3.5cm]{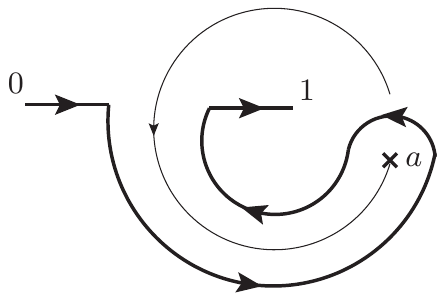} }
        \caption{The countour gets pinched on the second sheet producing a singularity 
        of the function at $z=2$. }
        \label{newpinch}
\end{figure}
As expected, these observations are verified by the explicit result of integration, 
which gives 
\beq
  f_2 (z) \, = \, \frac{1}{z - a} \, \ln \frac {a (z - 1)} {z (a - 1)} \, ,
\label{ans2}
\eeq
where the pole at $z = a$ on the principal sheet is regulated by the vanishing 
logarithmic factor, while in all other sheets the pole survives, as the logarithm 
provides a non-vanishing $2 \pi \textrm{i} k$ factor. Note that the pinch singularity 
examined here happens to involve a continuation to the second Riemann sheet, 
but of course pinches can arise on the principal sheet as well. It is easy to write 
the general form of a necessary condition for a pinch singularity to occur: if the 
integrand is a rational function (as is the case for Feynman integrals), the potential 
singularities will be located at the solutions of a polynomial equation of the form
\beq
  D \left(w, \left\{ z_i \right\} \right) \, = \, 0 \, ,
\label{genden}
\eeq
where $D(w, \{z_i\})$ is the denominator of the integrand function, which can 
depend on several `external' complex variables $z_i$. The condition for two 
roots of \eq{genden} to collide, on the surface defined by \eq{genden}, is then 
simply 
\beq
  \frac{\partial D}{\partial \omega} \, = \, 0 \, ,
\label{genpinch}
\eeq
as one readily verifies in the example in \eq{ex2}. Clearly, \eq{genpinch}, together 
with \eq{genden}, are only necessary conditions, since the two roots can for 
example collide away from the contour, or approach the contour from the same 
side, and furthermore a possible singularity may be cancelled by numerator factors.

\item {\it Pinch at infinity}. Consider now the integral
\beq
  f_3 (z) \, = \, \int_1^{a} \frac{d \omega}{1+ z \omega} \, .
\label{ex3}
\eeq
The integrand has a pole at $\omega_1 (z) = - 1/z$, which gives end-point singularities at 
$z = - 1/a$ and $z = -1$. The fact that the integrand decreases only as $w^{-1}$ for large 
$w$ does not seem to be relevant at first sight, since the integration contour is finite. In 
order to see the issue it is best to change variable to $y = 1/\omega$, which gives
\beq
  f_3 (z) \, = \, \int_{1/a}^1 \, \frac{d y}{y \, (y + z)} \, .
\label{alex3}
\eeq
This displays the same branch points as \eq{ex3}, but it is now also clear that we are in 
a situation similar to our previous example: if one takes a point $z$ away from the 
contour, and circles one of the branch points, one can then approach the pole at 
$y = 0$ and generate a pinch singularity, which will again be present on every Riemann 
sheet except the principal one. Indeed the integral gives
\beq
  f_3 (z) \, = \,  \frac{1}{z} \ln \frac{1 + a z}{1 + z} \, .
\label{ans3}
\eeq
\end{itemize}
These simple considerations can be extended to the case of several complex 
variables, though in full generality the mathematics becomes intricate (for a brief 
introductory discussion, and relevant references, see~\cite{Abreu:2017ptx}). 
Fortunately, the search for infrared divergences  in Feynman diagrams only 
requires simple tools, as we will see below.


\subsubsection{The case of Feynman diagrams: the Landau equations}
\label{landau}

We now use the simple tools described in \secn{integrals} to explore the singularities 
of a generic $L$-loop Feynman diagram  with $M$ external lines carrying momenta 
$p_i$, $i = 1, \ldots, M$. Let $G_M(p_i)$ denote the corresponding Feynman integral, 
which is an integral over loop momenta that we denote by $k_\ell$, $\ell = 1, \ldots, L$. 
We take the diagram to have $N$ internal lines, whose momenta we label by $q_j$, 
$j = 1, \ldots, N$, and we represent the integral introducing Feynman parameters 
$\alpha_j$ for each internal line. We can then write
\beq
  G_M (p_i) \, = \, \int_0^1 \prod_{j = 1}^N  d \alpha_j \; 
  \delta \bigg( \sum_{i = 1}^N \alpha_i -1 \bigg) 
  \int \prod_{\ell =1}^L \frac{d^d k_\ell}{(2 \pi)^d}
  \, \frac{ {\cal N} \left( \alpha_j, k_\ell, 
  p_i\right)}{\Big[ {\cal D} \left( \alpha_j, k_\ell, p_i\right) \Big]^N} \; ,
\label{gengraph}
\eeq
where the numerator ${\cal N}$ contains coupling and spin dependence, and 
does not affect our search for potential singularities\footnote{We use the mixed
representation in \eq{gengraph}, and do not perform the integration
over loop momenta, because we believe it makes the analysis of Landau equations
particularly transparent. On the other hand, the integration over loop momenta 
can be performed in complete generality, yielding a rational Feynman-parameter 
integrand expressed in terms of graph polynomials: see, for example,
\cite{Weinzierl:2013yn} for a recent review.}. A first necessary condition for 
the existence of a singularity is that the denominator should vanish, 
\beq
  {\cal D} \big( \alpha_j, k_\ell, p_i \big) \, \equiv \,  \sum_{j = 1}^N \alpha_j \big( 
  q_j^2  - m_j^2 \big) + {\rm i} \eta \, = \, 0 \; ,
\label{eq:denom}
\eeq
where the internal momenta $q_j$ are linear functions of external and 
loop momenta,
\beq
  q_j \, = \, \sum_{r = 1}^L \eta_{j r} \,  k_r + \sum_{m = 1}^M \beta_{j m} \, p_m \, .
\label{eq:l_decomp}
\eeq
For any consistent assignment of loop momenta to the graph edges, the incidence 
matrices $\eta_{j r}$ and $\beta_{j m}$ have entries in the set $\{-1,0,1\}$: for
example, $\eta_{j r} = 0$ if loop momentum $k_r$ does not flow through line $j$, 
while $\eta_{j r} = 1$ if the loop momentum $k_r$ flows through the $j$th line in 
the same direction as $q_j$, and $\eta_{j r} = - 1$ in the opposite case.

There are a total of $d L + N - 1$ independent integration variables in \eq{gengraph}, 
each one integrated along a path that can be modified, if required, to provide analytic 
continuation. A necessary condition for the existence of a singularity is therefore
that {\it all} integration variables be localised either at a pinch surface or at
an endpoint: if any variable is not constrained, the direction of that variable in 
the multi-dimensional complex integration space can be used to deform the contours. 
At least for non-exceptional configurations of the external momenta, since the 
denominator is a sum of propagator denominators with positive coefficients, the 
condition ${\cal D}(\alpha_j, k_\ell, p_i) = 0$ is realised when, {\it for each} internal 
line $j$, either the corresponding momentum $q_j$  is on the mass-shell, or the 
corresponding Feynman parameter $\alpha_j$ vanishes. That is, 
\begin{subequations}
\beq
{\rm either} \qquad q_j^2 - m_j^2 & = & 0 \, , \\
{\rm or} \hspace{2cm} \alpha_j & = & 0 \, .
\eeq
\label{allandau}
\end{subequations}
Next, we need to make sure that {\it all} variables are localised either at end-point
or at pinch singularities. Here the important point about the graph representation 
in \eq{gengraph} is that the denominator ${\cal D}$ is quadratic in the loop momenta 
$k_\ell$, and linear in the Feynman parameters $\alpha_j$. This  leads  immediately
to the following conclusions.
\begin{itemize}
\item Since ${\cal D}(\alpha_j, k_\ell, p_i)$  is linear in $\alpha_j$, there can only 
be a single pole in the $\alpha_j$ plane, so the only possible singularities are end-point 
singularities at $\alpha_j = 0$ or at $\alpha_j = 1$. On the other hand, if any of the 
$\alpha_j =1$, then all the other $\alpha_k$ must vanish simultaneously because 
of the $\delta$ function in \eq{gengraph}, a configuration which is not relevant for 
infrared divergences.
\item Pinch singularities, on the other hand, do arise for loop momentum components, 
since \eq{eq:denom} is quadratic in $k_j^\mu$. The condition for the two poles in the 
$k_j^\mu$ complex plane to coalesce is given by \eq{genpinch}, which we rewrite 
here as
\beq
  \frac{\partial}{\partial k_j^\mu} \, {\cal D} (\alpha_j, k_\ell, p_i) \, = \, 0 \; .
\label{eq:landau2}
\eeq
Upon substituting \eq{eq:denom} and \eq{eq:l_decomp}, \eq{eq:landau2} gives
\beq
  \displaystyle {\sum\limits_{i \, \in \, \textrm{loop} \, j} } \eta_{ij} \,
  \alpha_i q_i  \, = \,  0 \, ,\quad \textrm{for each} \,\, j  \, ,
\label{Landau circuit}
\eeq
where the sum includes all the lines $i$ through which loop momentum $k_j$ runs.
\item Finally, we note that the integration domain of loop momentum components 
extends to infinity, and one should also consider end-point singularities for these integrals.
The corresponding singularities, however, are ultraviolet, and we are assuming that they 
have already been taken care of by the inclusion of appropriate renormalisation 
counterterms. 
\end{itemize}

\noindent
Eqs.~\ref{eq:denom} and \ref{eq:landau2} or, equivalently, \eq{allandau} and
\eq{Landau circuit}, are known as Landau equations~\cite{Landau:1959fi}. The first 
set of Landau equations, setting ${\cal D} (\alpha_j, k_\ell, p_i) = 0$, defines a set of 
surfaces in the $(d L + N - 1)$-dimensional space of integration variables. The surfaces 
on which the second set of equations are also satisfied are called {\it pinch surfaces}. 
On such surfaces, all internal lines may be on the mass shell ($\alpha_i \neq 0, \forall 
\, i$), or some of the lines may be off the mass shell, provided the corresponding 
$\alpha_i = 0$. Lines with a vanishing Feynman parameter do not appear in the 
second set of Landau equations: as a consequence, since these lines are irrelevant 
for the conditions determining a pinch, when studying the corresponding pinch surfaces 
we can graphically reduce the lines to points. This procedure gives us a new set of 
diagrams, specific to the pinch surface under consideration, which are called {\it reduced 
diagrams}.

In this context, let's make here a simple observation that will be useful 
later~\cite{Sterman:1978bi}. If we single out a massless line $l$ in loop $j$, which 
carries momentum $q_l$, and the corresponding Feynman parameter is $\alpha_l$, 
then the second set of Landau equations for loop $j$ becomes
\beq
  \alpha_l q_l \, + \, \displaystyle {\sum\limits_{i \, \in \, \textrm{loop} \, j , \, i \neq l }} 
  \alpha_i q_i \, = \, 0 \, ,
\label{Landau Soft}
\eeq
where we chose the line momenta $q_i$ to flow in the same direction as the loop 
momentum $k_j$. If line $l$ is massless, and if $q_l^\mu = 0$ (so that $l$ is on 
shell), we see that this line drops out of the Landau equation, leaving no trace, 
as it is not even required for momentum conservation. This means that, once the 
integral is trapped on a pinch surface, adding an arbitrary number of soft massless 
lines does not alter the requirements on the other lines of the graph, in order for the 
integral to remain localised on the pinch surface.


\subsubsection{The Coleman-Norton physical picture}
\label{ColeNorto}

The Landau equations provide, in principle, a solution to the problem of locating all
potential singularities of Feynman integrals. The actual task, however, remains daunting,
especially if one is attempting to work to all orders in perturbation theory: without any
further information, it still appears that Feynman diagrams need to be treated one by 
one, and each one will present a plethora of pinch surfaces. As it turns out, the search 
for solutions of the Landau equations for external momenta {\it in the physical region}
is greatly simplified by a remarkably simple and intuitive criterion introduced by
Coleman and Norton (CN) in Ref.~\cite{Coleman:1965xm}. As we will see below, 
the CN criterion allows to quickly grasp the structure of solutions of the Landau 
equation for infinite classes of Feynman diagrams, paving the way for studies to
all orders in perturbation theory.

The CN criterion for the existence of a solution of the Landau equations for physical 
values of the external momenta can be stated as follows: a candidate pinch surface 
of a given Feynman integral provides a solution to the Landau equations if and only 
if the corresponding reduced Feynman graph can be interpreted as describing a 
classical scattering processes involving on-shell particles carrying the momenta 
assigned to each line of the reduced graph. 

To be more precise, let us consider a generic pinch surface and the corresponding
reduced diagram. Note that we have already localised the Feynman parameters
of off-shell lines by setting them to zero. It is now necessary to pinch the contour
of every loop momentum component on the configuration where all surviving lines
are on-shell: in this way, non-vanishing Feynman parameters cannot be used to
escape the pinch surface, since they multiply vanishing factors given by on-shell
propagator denominators. Consider now a line carrying momentum $q$, and let 
$\alpha$ be the Feynman parameter for that line. As the contour for every 
loop-momentum component $k_\ell^\mu$ is pinched and forced to be on the 
real axis, we see from \eq{eq:l_decomp} that $q$ will take real values for physical 
external momenta $p_j$. Since $q$ is on-shell, $q^2 = m^2$, we can write $q^\mu 
= m \, dx^\mu /d \tau$, where $\tau$ is the proper time and $x^\mu(\tau)$ parametrises
the classical trajectory of a free particle with momentum $q$, which of course 
moves on a straight line with constant four-velocity. We can rearrange this into
\beq
  \Delta x^\mu \, = \, \frac{\Delta \tau}{m} \, q^\mu \, .
\label{Deltax}
\eeq
If we now make the identification
\beq
  \frac{\Delta \tau}{m}  \, \equiv \, \alpha \, ,
\label{defalpha} 
\eeq
then the product $\alpha q^\mu$ can be interpreted as the space-time displacement 
of a classical particle of mass $m$ in the proper time $\Delta \tau = m \alpha$. 
Considering the Lorentz invariant ratio 
\beq
  \frac{\Delta t}{q^0} \, = \, \frac{ \Delta \tau}{ m} \, ,
\label{lorinvrat}
\eeq
we can also say that the time of propagation is equal to $\alpha q^0 $ in the reference 
frame in which the particle has energy $q^0$. This shows that the space-time 
picture we are constructing is not limited to massive particles: for a massless particle,
we can pick a frame in which the energy carried by the line has a given value
$q^0$, and we can replace \eq{Deltax} with $\Delta x^\mu \, = \, \Delta t \, \beta^\mu$,
where $\beta^\mu$ is the light-like four-velocity associated with the momentum 
$q^\mu$, $\beta^\mu = \{1, {\bf q}/q^0\}$. The Feynman parameter can now be 
identified with the {\it l.h.s} of \eq{lorinvrat}. Note that Feynman parameters are 
positive on the undeformed contours, which is consistent with particles moving
forward in time.

Within this framework, it is tempting to interpret the two endpoints of each line 
as space-time points separated by a distance $\alpha q^\mu$. For this interpretation 
to be consistent, however, the separation between any two vertices in the reduced 
diagram should be independent of the path taken to join them. Equivalently, if we start 
at a vertex, make a closed loop, and arrive back at the same vertex, the total 
displacement should be zero. It is easy to verify that the second Landau equation, 
\eq{Landau circuit}, precisely guarantees that this is the case. This gives us the
{\it Coleman-Norton physical picture}: the leading singularities of reduced diagrams 
in the physical region are supported on configurations where the internal lines can 
be regarded as the paths of classical particles moving freely between the vertices, 
while internal vertices can be regarded as points of interactions, where particle 
velocities change.
\begin{figure}
    \centering
    \begin{subfigure}{0.3\textwidth}
    \centering
        \includegraphics[width=\textwidth]{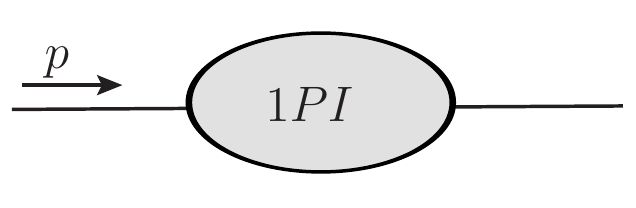}
        \caption{}
        \label{fig:OnePI}
    \end{subfigure}
    \begin{subfigure}{0.3\textwidth}
    \centering
        \includegraphics[width=\textwidth]{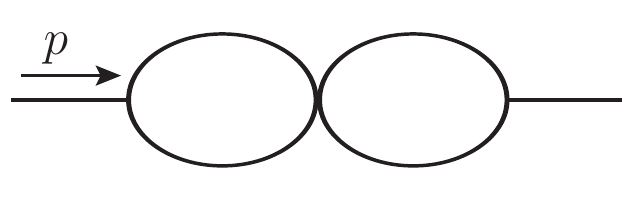}
        \caption{}
        \label{fig:PS2}
    \end{subfigure}
      \begin{subfigure}{0.3\textwidth}
      \centering
        \includegraphics[width=\textwidth]{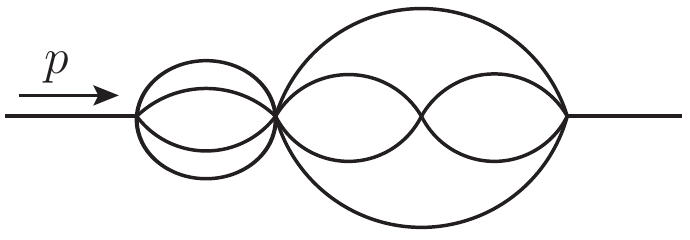}
        \caption{}
        \label{fig:PS3}
    \end{subfigure}
        \caption{(a)1PI contributions to a two-point function, (b) A two-loop 
        reduced diagram exhibiting a threshold at $p^2 = 4 m^2$, (c) A seven-loop 
        reduced diagram with a threshold at $p^2 = 16 m^2$. }
  \label{onePI}
\end{figure}

The intuitive power of the CN picture can hardly be overstated: it turns a rather 
abstract mathematical problem into a question about classical scattering, and 
very general results for entire classes of Feynman diagrams can readily be 
derived. To provide a simple, yet powerful example~\cite{Sterman:1994ce}, 
consider the two-point correlator for a theory with only one species of 
particles with mass $m$. In the leftmost panel of Fig.~\ref{onePI}, the 
shaded ellipse represents the one-particle-irreducible (1PI) contributions
to the correlator in this theory, assuming the external momentum to be $p$.
Following the CN picture, we must now seek classical processes in which
a source carrying (off-shell) momentum $p$ can split into a set of on-shell 
particles, which later reassemble into a single one. Clearly, any splitting
producing particles with non-vanishing momenta in directions transverse
to $p$ is not allowed, since, by momentum conservation, such particles
would propagate in opposite directions, and could never meet again. This
leaves us with the possibility of collinear splittings, however two on-shell
lines with three-momenta ${\bf k}_1$ and ${\bf k}_2$ cannot in general merge 
to produce an on-shell line with three-momentum ${\bf k} = {\bf k}_1 + 
{\bf k}_2$, as the invariant mass of that line would be greater than $m$.
Only one set of possibilities is left: for $p^2 = n^2 m^2 > 0$, with integer 
$n$, a simple CN process is easily identified in the rest frame of $p$: the
off-shell source particle can split into $n$ on-shell particles at rest, and these 
can further merge and split any number of times, provided particle number is
conserved; finally, they can merge again to reconstruct the momentum $p$. 
With remarkable generality, we can conclude that the only possible singularities
of two-point functions of massive particles are located at the thresholds 
for the creation of $n$ on-shell particles, which are called {\it normal 
thresholds}\footnote{Recall that the CN argument applies only to the 
physical Riemann sheet, so that further singularities are possible after 
analytic continuation to different sheets.}, and are clearly and directly related
to unitarity cuts. Thus, a single analysis has solved the problem for an infinite 
set of Feynman diagrams, regardless of the specific details of the interactions 
of the massive particles involved: Fig.~\ref{onePI}(b) and Fig.~\ref{onePI}(c) 
show two examples of reduced diagrams for $p^2$ equal to $4 m^2$ and 
$16 m^2$ respectively. 

Note that the argument can be rephrased in a boosted frame where $p$ is 
not at rest, shedding light on the values of Feynman parameters allowed 
by the CN constraints. Imagine one creates two particles of equal masses 
which then move together collinearly. If they have different energies, 
we can move to a frame where the one with the smaller momentum will be 
at the rest. Now the two particles can never meet again, unless the second 
particle scatters off a `wall' to reverse its direction; this circumstance can be 
realised only if we include an extra vertex, which is a situation we will encounter 
when we discuss three-point graphs. Thus, the two massive particles can 
meet at a later time without encountering a new vertex only if they have 
equal energies. In general, the CN time they take to merge is given by 
$\Delta t = \alpha_1 E_1 = \alpha_2 E_2$, therefore the Feynman parameters
are constrained to satisfy $\alpha_1 = \alpha_2$. The same physical picture 
for $n$ massive particles gives equal values for all $\alpha_i$, and finally
the presence of the factor $\delta(1 - \sum_i \alpha_i)$ enforces
\beq
  \alpha_1 \, = \, \alpha_2 \, \ldots \, = \, \alpha_n \, = \, 1/n \, .
\label{fixFpa}
\eeq 
Clearly, this simple analysis of two-point functions can be extended to 
higher-point functions as well, allowing the identification of at least {\it some}
of their potential singularities. As an example, one can consider a four-point
function in the $s$-channel physical region. When the two incoming momenta
have a total invariant mass $s = n^2 m^2$, with $n$ integer, in their
center-of-mass frame one finds that the same CN process that was 
identified for the two-point function is available also in this case: there 
is just enough energy to create $n$ on-shell particles at rest, and these 
particles can interact conserving particle number until they finally decay
into the final state pair. Note also that the discussion of two-point functions 
for massive particles has implications for the massless case: if the off-shell 
source carrying momentum $p$ couples only to massless particles, then all 
the normal thresholds collapse to the point $p^2 = 0$. One must however 
remember, as discussed at the end of \secn{landau}, that adding a soft particle 
with a vanishing momentum $q$ does not influence the Landau equations: in the 
CN picture, soft lines correspond to vanishing displacements $\Delta x^\mu$ 
and do not modify the classical scattering configuration\footnote{An alternative 
point of view on this fact is that particles with vanishing momenta have infinite 
De Broglie wavelengths, so, in a sense, they are fully delocalised and can 
mediate interactions at any distance.}. Two-point functions involving massless 
particles will therefore potentially be affected by soft singularities, and power 
counting techniques will need to be employed to check if these lead to 
actual divergences.

In order to prepare the ground for our analysis of massless form factors in 
\secn{FactEvo}, we need to consider in more detail the case of three-point 
functions. First of all, let us note that the general structure of singularities 
becomes significantly more complicated when $n>2$ external particles are 
involved, since the resulting Feynman integrals are functions of several 
complex variables: for example, a three-point function in a theory with only 
one species of particles with mass $m$ will depend on three dimensionless 
ratios, say $x = p_1^2/m^2$, $y = p_2^2/m^2$, and $z = p_3^2/m^2$, with 
$p_3^2 = (p_1 + p_2)^2$. 

\begin{figure}[h!]
\begin{center}
        {\includegraphics[scale=0.7]{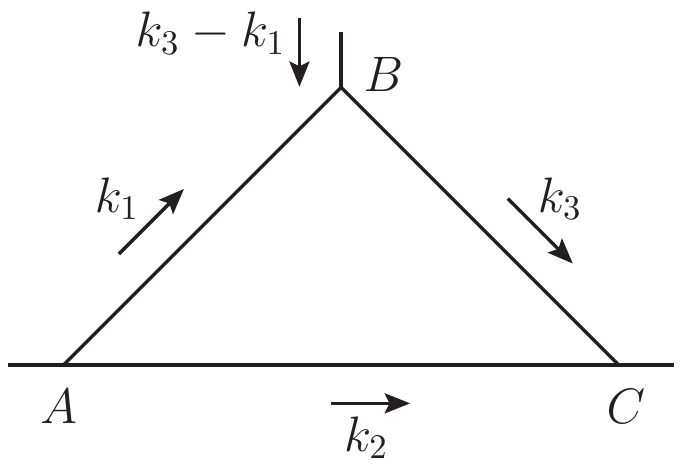} }
\caption{Momentum routing for a three-point graph displaying possible anomalous 
              thresholds, as discussed in the text.}
\label{anomThreshold}
\end{center}
\end{figure}

One can readily verify that the presence of a third particle (and therefore 
of at least one more interaction vertex, as compared to the two-point function) 
allows for singularities which are not related to normal thresholds, and are
usually referred to as {\it anomalous thresholds}. While unitarity cuts arise by
putting on the mass shell exactly the number of lines needed to split a given 
diagram into two disconnected subdiagrams, containing respectively initial-state 
and final-state particles, anomalous thresholds arise from configuration in which
a larger number of lines are on-shell, and the diagram may split into several
disconnected pieces (thus they are often referred to as {\it generalised unitarity}
cuts, while the configuration in which {\it all} internal lines are on-shell is often
described as {\it leading singularity}). To illustrate this, consider the simple case
of a one-loop triangle diagram, depicted in Fig.~\ref{anomThreshold}, 
which of course can be understood as a reduced diagram arising from a more 
complicated configuration. It is not difficult to find a CN process such that
the three (massive) lines be on shell: one may for example envisage two 
particles being created with momenta $k_1$ and $k_2$ at vertex $A$. If
the source momentum $p=k_1+k_2$ is timelike, $p^2 > 0$, one can consider the
frame in which $p$ is at rest, so that the two produced particles are emitted
back to back. Without further scattering, they could never meet again at any 
later time at vertex $C$. One can however envisage one of the two particles
bouncing off a `wall' at vertex $B$: if sufficient momentum is injected there, 
the particle can reverse its direction to annihilate the other one at vertex $C$. 
Clearly, the existence and location of this singularity will depend upon the 
values of the external invariants $x$, $y$ and $z$: in particular, the singularity
may or may not appear on the physical Riemann sheet.

As was the case for two-point functions, the massless limit introduces new
pinch surfaces, which are the ones of interest for the analysis of infrared 
divergences. In particular, given an existing configuration with on-shell lines,
one is always free to add soft massless lines, without affecting the Landau
equations. Furthermore, as discussed already in \secn{Intro}, a massless 
line can split into two or more collinear massless lines, while conserving
four-momentum and keeping all lines on the mass shell. We illustrate these
facts in some more detail in \secn{VerteGra} below.
%


\subsubsection{Pinch surfaces of massless vertex graphs}
\label{VerteGra}

\begin{figure}
        \centering
        {\includegraphics[scale=0.8]{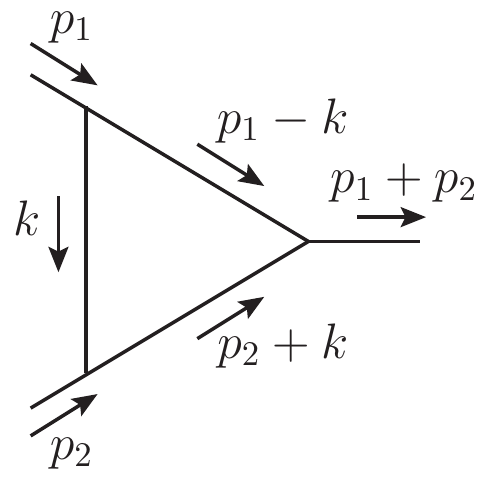}}
        \caption{One-loop vertex correction to the three-point function in a massive scalar theory.}
        \label{fig:diagram_for_anurag}
\end{figure}
To illustrate the situation for massless three-point functions, consider the simple 
one-loop triangle graph depicted in Fig.~\ref{fig:diagram_for_anurag}. We consider
a scalar graph, since for our present considerations the numerator structure is
not relevant. The expression for the graph after Feynman parametrisation is 
of the form of \eq{gengraph}, and can be written as
\beq
  G_3 (p_1, p_2) \, = \, \int_0^1 \prod_{j = 1}^3  d \alpha_j  
  \int \frac{d^d k}{(2 \pi)^d}
  \, \frac{\delta \big( \sum_{i = 1}^3 \alpha_i -1 \big)}{\Big[ 
  \alpha_1 k^2 + \alpha_2 (p_1 - k)^2 + \alpha_3 (p_2 + k)^2 + 
  {\rm i} \eta \Big]^3} \; ,
\label{gengraph3}
\eeq
where we take $p_1^2 = p_2^2 = 0$. The second Landau equation for 
this graph reads
\beq
  \alpha_1 k^\mu - \alpha_2 \big( p_1 - k \big)^\mu + 
  \alpha_3 \big( p_2 + k \big)^\mu \, = \, 0 \, ,
\label{trilaueq}
\eeq
and we need to pick solutions such that, if a given $\alpha_i$ is not vanishing,
then the corresponding four-momentum must be on shell. Setting for the moment
aside ordinary and anomalous thresholds, we see that two new sets of pinch 
surfaces arise in the massless case.

\begin{figure}
    \centering
    \begin{subfigure}{0.3\textwidth}
    \centering
        \includegraphics[width=0.5\textwidth]{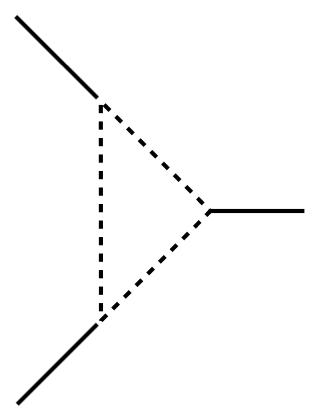}
        \caption{}
        \label{fig:VertexPinchA}
    \end{subfigure}
    \begin{subfigure}{0.3\textwidth}
    \centering
        \includegraphics[width=\textwidth]{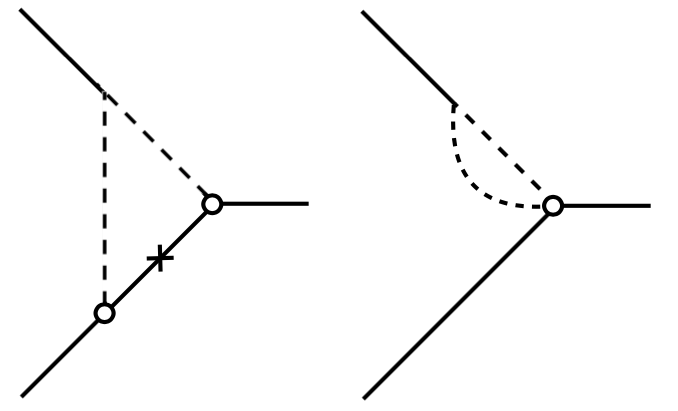}
        \caption{}
        \label{fig:VertexPinchB}
    \end{subfigure}
        \caption{(a) Reduced diagram for the soft pinch surface. (b) Building the reduced 
        diagram for a collinear pinch surface. In the left panel of (b) we highlighted with 
        a cross the off-shell propagator that shrinks to a point, yielding the right panel. 
        Dashed lines mark on-shell propagators.}
  \label{fig:VertexPinch}
\end{figure}

\begin{itemize}

\item {\bf Collinear pinch surfaces}. If, for example, $k^\mu$ is parallel to $p_1^\mu$,
so that $k^\mu = x \,p_1^\mu$, we see that the line carrying momentum $p_1 - k$ is 
on shell, while the line $p_2+ k$ is off shell for generic configurations of external 
momenta. We can thus find a solution of the Landau equations by setting
$\alpha_3 = 0$ and then requiring that
\beq
  x \alpha_1 \, = \, (1 - x) \alpha_2 \, ,
\label{collpconf}
\eeq
in order to solve \eq{trilaueq}. The reduced diagram for this pinch surface is 
shown in Fig.~\ref{fig:VertexPinchB}, left panel. The corresponding CN configuration 
has two light-like particles starting from the origin at $t = 0$ and moving in 
the direction of $p_1^\mu$ with different energies, $x \, p_1^0$ and $(1 - x) p_1^0$ 
respectively; they meet again when their CN displacements become equal, which 
happens when \eq{collpconf} is satisfied, see the second diagram in Fig.~\ref{fig:VertexPinchB}. 
Note that CN times are positive when $0 < x < 1$, so that the singular configuration 
can be on the integration contour and thus on the physical Riemann sheet. Needless
to say, an identical situation is reproduced when the loop momentum is 
collinear to $p_2$, say $k^\mu = - y \, p_2$. Then the line
carrying momentum $p_1 - k$ will be off-shell, and the Landau equations can 
be solved by setting $\alpha_2 = 0$ and then requiring that
\beq
  y \alpha_1 \, = \, (1 - y) \alpha_3 \, .
\label{collpconf2}
\eeq

\item {\bf Soft pinch surfaces}. As discussed in \secn{landau}, in a massless
theory one can always build solutions of the Landau equations by adding
particles with vanishing momenta, which do not affect the equations, nor the
relevant CN configurations. In the present case, one can for example solve
\eq{trilaueq} by picking any $\alpha_1 \neq 0 $ and setting
\beq
  k^\mu \, = \, 0 \, , \qquad \alpha_2 \, = \, \alpha_3 \, = \, 0 \, .  
\label{softkconf}
\eeq
In this limit, all particles circulating in the loop are on-shell, so that the reduced
diagram coincides with the original one, as depicted in Fig.~\ref{fig:VertexPinchA};
the soft particle carrying momentum $k$ does not contribute to the CN 
displacement sum, \eq{Landau circuit}, corresponding to the fact that a particle
with vanishing four-momentum is fully delocalised. Note that, while the 
diagram we are considering is not symmetric under the exchange of the 
external particles, since $(p_1 + p_2)^2 \neq 0$, there are nevertheless
soft pinch surfaces associated with the soft limit for any particle circulating in 
the loop: these are most easily detected by changing the parametrisation of 
the loop momenta, but even with our current choice we see that one can
solve \eq{trilaueq}, for example, by taking $k^\mu = p_1^\mu$ and $\alpha_1 
= \alpha_3 = 0$. Note also that all our discussion so far does not depend
on particle spins, nor on the specifics of their interactions: the one-loop
triangle diagram can arise as a reduced diagram in any theory, even if the 
interactions are not cubic in the fields. In order to ascertain if the pinch
surfaces we have identified give actual divergences, we will need the 
power-counting tools discussed in \secn{IRPowCo}.
\end{itemize}
It should be clear from the above discussion of the one-loop scalar diagram
that we have actually identified sets of pinch surfaces for an infinite set of Feynman
diagrams, to any perturbative order. For example, adding further loops involving
soft particles will automatically solve the Landau equations. Similarly, adding
more particles collinear to either $p_1$ or $p_2$ will provide new pinch 
surfaces, since each new particle will add a Feynman parameter, and a single
constraint of the form of \eq{collpconf}, so that the resulting system of equations
will remain solvable. In the CN picture, if two particles are produced back to back
at the hard interaction vertex, they can never meet again, however each 
particle can split into collinear multiplets which then recombine at later times, 
and these two {\it jets} of particles evolving in opposite directions can always
be connected by means of soft radiation. We are thus beginning to develop 
a general picture of the singularity structure of massless form factors to all 
orders: in a generic diagram, the hard interaction can be dressed with any 
number of off-shell virtual corrections, which shrink to a point in the corresponding
reduced diagram; infrared singularities may arise when, out of this hard effective 
vertex, two jets of collinear particles emerge, which in turn can be connected
to each other by a soft subdiagram, where all particles have vanishing 
four-momentum. This is a very significant simplification of the generic situation
we started with, but we will see in what follows that the final form of the factorisation
we are seeking is simpler still.


\subsection{Identifying actual singularities: infrared power counting}
\label{IRPowCo}

Up to this point, we have presented a set of conditions to identify all the 
potential sources of singularities in Feynman integrals, in particular 
infrared divergences. The derivation requires that the poles of the 
integrand must be forced to lie on the integration contour. Clearly,
this is only a necessary condition for actual divergences to occur: 
first of all, the detected singularity might be - for example - a branch cut 
and not a divergence; next, positive powers of the integration variables 
can appear in the numerator of the integrand and mitigate the potential 
divergence. Sufficient conditions to identify the kinematic configurations 
that result in IR singularities can only be derived from {\it power-counting 
techniques}, reminiscent of those employed in the UV to study renormalisation.

In the UV case, power counting is straightforward, since the singular
region is approached when all loop momentum components become very
large. One may thus detect the behaviour of the integral by rescaling
loop momenta as $k^\mu \to \Lambda \bar{k}^\mu$, and counting powers 
of $\Lambda$. Since a Feynman integrand is a product of rational factors,
one may expand each factor in powers of $\Lambda$, retaining only
the leading power in each factor: the resulting overall power of $\Lambda$
gives the superficial degree of UV divergence of the diagram under study.
The situation in the infrared is more intricate: as we have seen, potential 
singularities arise in the massless limit when loop momenta become 
soft or collinear to some of the external momenta. Different loop-momentum
components must then be treated differently, as only some of them are
related to singularities of the integral. It is not difficult to make this
analysis systematic, as we discuss below.


\subsubsection{Intrinsic and normal coordinates, and homogenous integrals}
\label{IntNorCoo}

To study infrared power counting, it is best to revert to momentum-space
integrals, before the introduction of Feynman parameters (see, however,
Ref.~\cite{Erdogan:2014gha} for a coordinate space representation).
In full generality, a $p$-dimensional pinch surface $\Sigma_p$ in the
$D$-dimensional loop momentum space ${\cal L}$ (where $D = d L$ for 
an $L$-loop diagram in $d$ space-time dimensions) will be characterised 
by $D - p$ conditions, which can be cast in the form
\beq
  F_i \left( k_\ell^\mu \right) \, = \, 0 \qquad i \, = 1, \ldots, D-p ; \quad 
  \ell \, = \, 1, \ldots, L \, .
\label{Sigmacond}
\eeq
For our purposes, the constraints $F_i$ will be linear, and the resulting 
surfaces will be hyperplanes; the discussion
can however be generalised to non-linear constraints and curved 
surfaces. Given any differentiable set of constraints $F_i$, one can 
pick a parametrisation of ${\cal L}$ such that $p$ coordinates $k_I$ 
parametrise the surface $\Sigma_p$, while the remaining $D-p$ 
coordinates $k_N$ serve to measure the distance of any point in 
${\cal L}$ from $\Sigma_p$. Changing the values of the coordinates
$k_I$ moves a point within the surface, so these coordinates are 
called {\it intrinsic}; changing the value of the coordinates $k_N$ 
moves points away from the pinch surface, and these coordinates are
correspondingly called {\it normal}. The strength of the singularity 
encountered on $\Sigma_p$ is controlled by the behavior of the
integrand as a function of {\it normal} coordinates, and one is free 
to define normal coordinates such that their zeros lie on the pinch 
surface $\Sigma_p$. 

Since we are interested in the behaviour of the reduced graph near 
the surface $\Sigma_p$, we can again take advantage of the structure
of the Feynman integrand, which is a product of rational factors, and
we can expand each factor in the denominator and in the numerator
in powers of the normal coordinates, retaining the 
lowest order approximation. Thus, given a reduced graph $G$, and
for every pinch surface $\Sigma$ of $G$, we can write 
\beq
  G_\Sigma (p_i) \, = \, \int d^p k_I \, d^{D-p} k_N \; {\cal I} 
  \big(p_i, k_I, k_N \big) \, ,
\label{GSigma}
\eeq
where $p_i$ denotes the external momenta. We then associate with $G_\Sigma$
a {\it homogenous integral}
\beq
  \overline{G}_\Sigma \, = \, \int d^p k_I \, d^{D-p} k_N \; 
  \overline{\cal I} \big( p_i, k_I, k_N \big) \, , 
\label{GSigmahom}
\eeq
where $\overline{\cal I}$ is the leading-power approximation of ${\cal I}$ with 
respect to the set of normal coordinates\footnote{In a more general setting, 
for example when considering effective theories, which may display power-like
divergences, one may need to retain sub-leading powers in the expansion of
the integrand ${\cal I}$, in order to capture all divergent contributions to the 
integral $G_\Sigma$.}. 

Moving internal lines 
away from the mass shell will take the graph off the pinch surface. Virtualities
of internal lines will therefore be proportional to normal coordinates, and the 
ratio of these virtualities to the hard scale $Q$ will determine how close or 
how far we are from the surface. In the next section, we will present the
analysis of two simple one-loop examples, using again a three-point graph
as a laboratory, and then we will proceed by illustrating how the one-loop
results generalise to all perturbative orders.


\subsubsection{One-loop examples: the QED vertex graph}
\label{oloexe}

Let us consider the one-loop QED vertex graph, adapting what we have
already presented in \secn{cataQED} and for the scalar theory in \secn{VerteGra}. 
In the massless limit, we can pick a frame in which the external momenta in are 
given by
\beq
\label{momCM}
  q  & = & (Q, 0, 0, 0) \, , \nonumber \\
  p_1 & = & \big( Q/2, 0, 0, Q/2 \big) \, , \\
  p_2 & = & \big( Q/2, 0, 0, - Q/2 \big) \; , \nonumber
\eeq 
where $Q$ is the photon virtuality, giving the hard scale of the problem. The 
expression of the graph (see also \eq{vertir}) is 
\beq
  G_3 (p_1, p_2) \, = \,  {\rm i} e^2 \mu^{2 \epsilon} \, \int \frac{d^dk}{(2 \pi)^d} \, 
  \frac{\mathcal{N}(p_1, p_2, k)}{ (k^2+ {\rm i} \eta) \big[ (p_1 - k)^2 + {\rm i} \eta \big]
  \big[ (p_2 + k)^2 + {\rm i} \eta \big] } \, ,
\label{QEDvertagain}
\eeq
where ${\cal N}(p_1, p_2, k)$ contains spinors and Lorentz structures.
We already know from 
the Landau equation analysis in \secn{VerteGra} that this graph has pinch
surfaces for configurations where the loop momentum $k$ is soft, and/or 
collinear to either fermion line. Let us now consider each case in turn.

Near the soft pinch surface, $k^\mu \to 0$,  {\it all} the components of the 
virtual photon momentum $k$ are normal coordinates, and there are {\it no} 
intrinsic coordinates. Upon approximating the factors in the denominator by 
their lowest order expressions in the expansion in powers of $k^\mu$, we
readily recognise, unsurprisingly, that the homogenous integral is given by 
the eikonal approximation, as in \eq{simpvertir} and \eq{eq:eikint},
\beq
  \overline{G}_{3, \, {\rm soft}} (p_1, p_2) \, = \, {\rm i} e^2 \mu^{2 \epsilon}  
  \int \frac{d^dk}{(2 \pi)^d} 
  \, \frac{ {\cal N} (p_1, p_2, 0)}{ (k^2 + {\rm i} \eta) 
  (2 p_2 \cdot k + {\rm i} \eta)
  ( - 2 p_1 \cdot k + {\rm i} \eta)} \, .
\label{eq:hom_soft}
\eeq
In the present case, it is easy to see directly that the soft homogeneus integral
is dimensionless and logarithmically divergent in $d = 4$. In a more general 
setting, in order to determine the distance from the soft pinch surface as a function 
of the normal coordinates, we can choose a scaling variable $\lambda_{\rm s}$,
and impose the same scaling for each momentum component, as was done in 
the UV region, but this time taking the limit $\lambda_{\rm s} \to 0$. Thus we 
can set
\beq
  k^\mu \, = \, \lambda_{\rm s} \bar{k}^\mu \, ,
\label{scalesoft}
\eeq
and count powers of $\lambda_{\rm s}$ in the homogeneous integral. The
scaling variable $\lambda_{\rm s}$ is often taken to be dimensionless, however,
in case one is interested in keeping track of powers of the hard scale $Q$, one
can also take $\lambda_{\rm s}$ to have dimensions of energy, and treat the
reduced components $\bar{k}^\mu$ as `angular' variables~\cite{Collins:2011zzd}.
In this case, one can choose, for example
\beq
  \lambda_{\rm s} \, = \, \sum_{\mu} \big| k^ \mu \big| \, ,
\label{normasoft1}
\eeq
so that the dimensionless variables $\bar{k}^\mu$ are normalised as
\beq
  \sum_{\mu} \big| \bar{k}^ \mu \big| \, = \, 1 \, .
\label{normasoft2}
\eeq
Treating $\lambda_{\rm s}$ as a dimensionless parameter, substituting 
\eq{scalesoft} in the soft homogeneous integral, \eq{eq:hom_soft}, and 
replacing every factor of $p_1$ and $p_2$ by the hard scale $Q$, we 
verify that
\beq
  \overline{G}_{3, \, {\rm soft}} (p_1, p_2) \, \sim \,
  \frac{1}{\lambda_{\rm s}^{4 - d}} \frac{1}{Q^2} \times 
  {\cal N} (p_1, p_2, 0) \, .
\eeq
In a gauge theory, as seen already in \eq{vertir}, the homogeneous numerator is 
proportional to $Q^2$, so, as announced, the integral is logarithmically divergent
in $d = 4$, and contributes at leading power in $Q$.

Turning now to the collinear limit, it is useful to introduce \emph{light-cone} 
coordinates, such that any four-vector is expressed as $x^\mu=(x^+, x^-, 
\bf{x}_\perp)$, with 
\beq
  x^+ \, = \, \frac{x_0 + x_3}{\sqrt{2}} \, , \qquad  
  x^- \, = \, \frac{x_0 - x_3}{\sqrt{2}} \, ,
\label{light_cone}
\eeq
so that
\beq 
  x^\mu y_\mu \, = \, x^+ y^- + x^- y^+ - \bf{x}_\perp \cdot \bf{y}_\perp \, ,
\label{scalar_prod_light_cone}
\eeq
while the measure of integration becomes
\beq
  d^d k \, = \,  d k^{+} d k^{-} d^{d-2} k_\perp \, .
\label{measurelc}
\eeq
When the photon is collinear to the electron carrying momentum $p_1$, in
the frame in which \eq{momCM} applies, $k_\perp$ and $k^-$ must vanish, 
while we can set $k^+ = z p_1^+$. Changing the value of $z$ leaves
the loop momentum on-shell, so that configurations with any $z$ in principle
belong to the collinear pinch surface: $z$ is an {\it intrinsic} coordinate; clearly,
the azimuthal angle $\phi$ for the transverse momentum ${\bf k}_\perp$ is
also an {\it intrinsic} coordinate. Note however that, using the CN physical picture, 
the graph can be interpreted as a classical collinear splitting of massless particles 
with on-shell momenta only if $0 \leq z \leq 1$. If $z$ is outside this range, the 
photon propagator is still on shell, but the CN interpretation is not valid, and we 
expect that the contour may be deformed to avoid the singularity. Considering 
the remaining variables, we see that generic non-vanishing values of $k^-$ and 
$k_\perp$ will move the photon away from the mass shell: $k^-$ and $k_\perp$ 
are {\it normal} coordinates.

Expressing $G_3 (p_1, p_2)$ in these variables, expanding each factor in powers 
of the normal variables, and retaining the leading power in each one of them, we
find the collinear homogeneous integral
\beq
\label{collhoi}
  \overline G_{3, \, {\rm coll}_1} (p_1, p_2) & = & \frac{{\rm i} e^2 
  \mu^{2 \epsilon}}{(2 \pi)^d} \int d z \,\, \frac{{\cal N} (p_1, p_2, z)}{2 p_2^- z 
  + {\rm i} \eta} \\
  && \times \, \int dk^- d^{d-2} k_\perp \, 
  \frac{1}{2 k^+ k^- - k_\perp^2 + {\rm i} \eta} \, 
  \frac{1}{- 2 (1 - z) p_1^+ k^- - k_\perp^2 + {\rm i} \eta} \, , \nonumber 
\eeq
where in the numerator we have set $k^- = k_\perp = 0$, but retained the
dependence on $k^+$. We see that the photon and electron denominators are 
linear in the normal coordinate $k^-$, but quadratic in $k_\perp$. The natural 
scaling choice, ensuring that these denominators are homogeneous in the 
scaling parameter, and linear in normal coordinates, is therefore to take 
\beq
  k^- \, = \lambda_{\rm c} \, \bar{k}^- \, ,  \qquad
  k_\perp^2 \, = \,  \lambda_{\rm c} \, {\bar{k}}_\perp^2 \, , 
\label{collscal}
\eeq
where we choose $\lambda_{\rm c}$ to be dimensionless\footnote{If one wishes 
to work with dimensionless rescaled momentum components $\bar{k}^\mu$ one 
may instead, for example, set $k^- \, = \lambda_{\rm c} \, \bar{k}^-/p^+$, with $\lambda_c$ 
having dimensions of mass squared.}.  We see that the collinear homogeneous 
integral is ${\cal O} (\lambda_c^{\epsilon})$, implying a logarithmic divergence in 
$d=4$. We also see that the off-shell positron line gives a $1/Q$ suppression, 
however the collinear numerator also grows linearly with $Q$, so that the 
collinear region contributes at leading power in $Q$, as was the case for the 
soft region. In the remainder of \secn{AllOrd}, we will omit the subscripts from the 
scaling parameters $\lambda_{\rm s}$ and $\lambda_{\rm c}$, with the understanding 
that soft and collinear scalings are defined by \eq{scalesoft} and in \eq{collscal}.

Let us conclude this section with two important notes. First, recall that the
soft and collinear regions overlap: this is reflected in the fact that the collinear
homogeneous integral in \eq{collhoi} still has a soft singularity, as $z \to 0$, 
and similarly the soft homogeneous integral in \eq{eq:hom_soft} has collinear
singularities when $k$ becomes parallel to either $p_1$ or $p_2$. At the 
moment, this is not a serious concern, as we are only interested in determining
the strength of the singularities. At a later stage, when building an explicit
factorised expression for amplitudes and cross sections, a careful treatment
of overlapping singularities will become necessary. A second important caveat
emerges from the observation that different components of loop momenta
may become small at different rates, as is the case in the collinear region,
\eq{collscal}. This raises the possibility that there could be different scalings
of momentum components (for example combining soft and collinear limits
with different strengths) that might need to be treated separately. We will 
briefly get back to this issue in \secn{MultiPart}: here we note that this is
in principle a process-dependent question, that can only be addressed via
a detailed diagrammatic analysis. The issue is particularly relevant when
factorisation theorems are constructed by means of effective field theories,
where the relevant momentum modes must be selected {\it a priori}. On
the other hand, once a particular scaling has been understood to be 
relevant, the effective theory can be extended to include it (see, for 
example, Refs.~\cite{Bauer:2002uv,Bauer:2010cc,Rothstein:2016bsq}).


\subsubsection{Soft and collinear power counting to all orders} 
\label{softcollord}

\begin{figure}
        \centering
        {\includegraphics[scale=0.7]{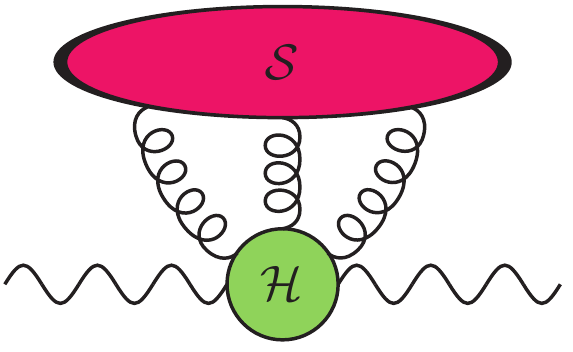} }
        \caption{Pinch surfaces for the  two-point function of an off-shell photon.}
\label{fintwop}   
\end{figure}
\begin{figure}
        \centering
        \begin{subfigure}{0.3\textwidth}
        \centering
        \includegraphics[scale=0.45]{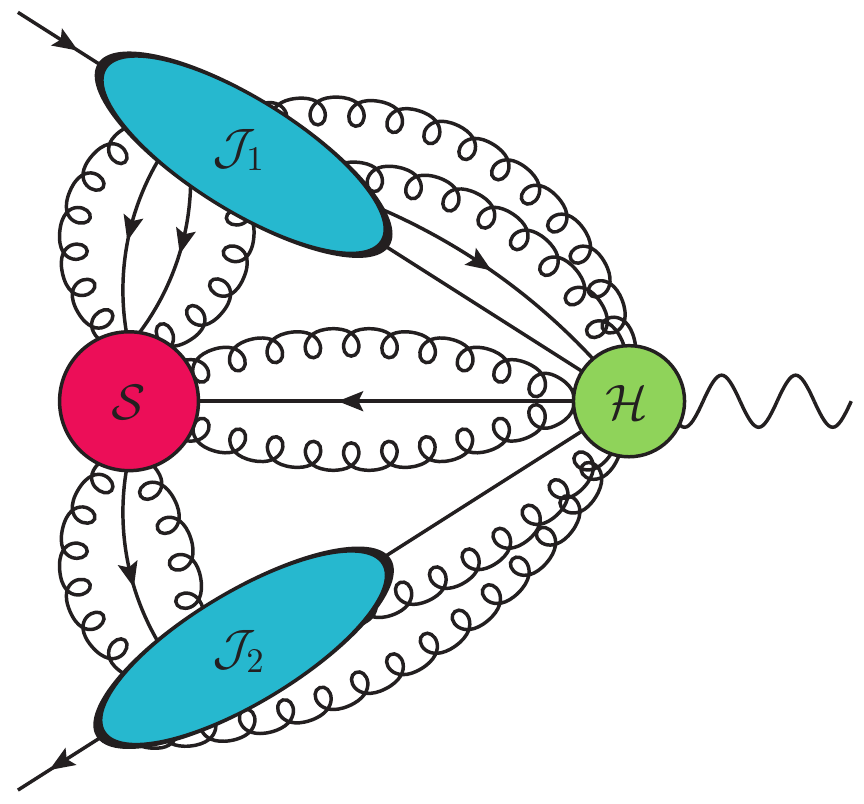} 
        \caption{}
        \label{fig:fact1}
        \end{subfigure}
        \begin{subfigure}{0.3\textwidth}
        \centering
        \includegraphics[scale=0.45]{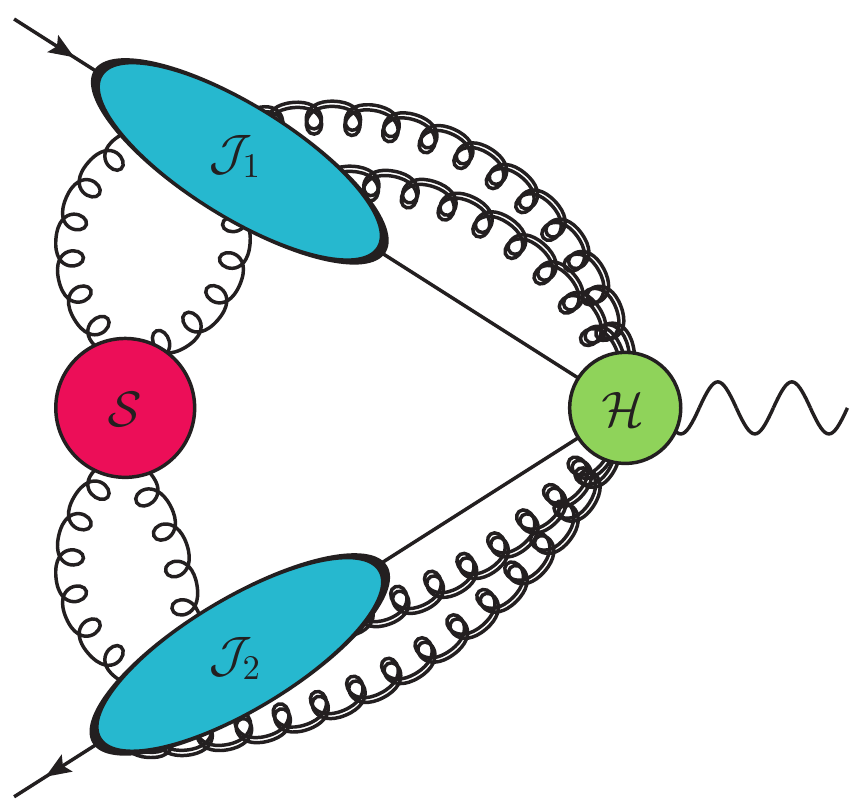} 
        \caption{}
        \label{fig:fact2}
        \end{subfigure}
        \begin{subfigure}{0.3\textwidth}
        \centering
        \includegraphics[scale=0.45]{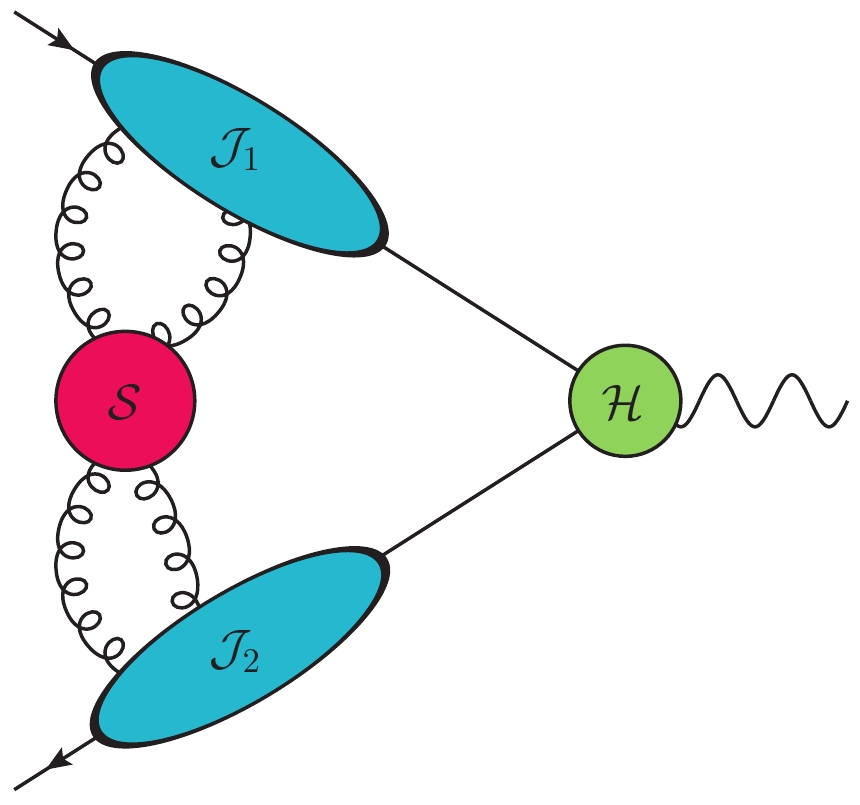} 
        \caption{}
        \label{fig:fact3}
        \end{subfigure}
        \caption{CN and power-counting analysis for the quark form factor. Panel $(a)$ 
        displays soft, jet and hard subgraphs, connected by an arbitrary number of lines. 
        Panel $(b)$ shows the subset of pinch surfaces contributing to IR divergences 
        in Feynman gauge: double curly lines linking $\mathcal{J}_i$ and $\mathcal{H}$ 
        denote scalar-polarised gluons. Panel $(c)$ displays singular surfaces in a physical 
        gauge: in contrast with panel $(b)$, only a single line can connect $\mathcal{J}_i$ 
        and $\mathcal{H}$.}
\label{redgendiag}   
\end{figure}

Extending soft and collinear power counting beyond one loop is not too cumbersome,
and can be done by using a suitable generalisation of the techniques familiar from the 
analysis of UV divergences. Graphically, it is useful to introduce a diagrammatic notation 
for subgraphs, which we employ in Fig.~\ref{fintwop} and in Fig.~\ref{redgendiag}, 
and illustrate in what follows. In general, a coloured `blob' denotes an arbitrary subgraph,
where all lines are supposed to share a common kinematic constraint: for example,
all lines in the {\it jet} subgraphs labelled $\mathcal{J}$ in Fig.~\ref{redgendiag}
are collinear to the initial-state line entering the subgraph; similarly, all lines in the
soft subgraphs labelled $\mathcal{S}$ in Figs.~\ref{fintwop} and~\ref{redgendiag} 
have vanishing four-momenta, and all lines in the hard subgraphs labelled $\mathcal{H}$ 
are off the mass shell, and thus are contracted to a point in the reduced graph. When 
two subgraphs are connected by more than one line, it is understood (unless otherwise 
stated) that the number of such lines is arbitrary, with only a few being drawn for illustrative 
purposes. 

To illustrate the rather formidable power of the apparently simple tools that we have 
just assembled, consider first the case of the two-point function for a massive particle,
or for an off-shell massless particle such as the photon in Fig.~\ref{fintwop}, allowing 
for couplings to other massless particles such as gluons~\cite{Sterman:1995fz}. 
In this case, the CN analysis allows for no collinear pinch surfaces, since there 
are no massless external particles, and, if all couplings are to massless particles, 
there are no normal thresholds. One is left with only potential soft singularities, 
whose pinch surfaces are portrayed in Fig.~\ref{fintwop}. The next crucial step 
consists in identifying which of the pinch surfaces described by Fig.~\ref{fintwop} 
give an actual infrared divergence. Thus, we need to find the superficial degree 
of infrared divergence $\omega_s$ for a generic soft pinch surface, adapting the 
power counting technique to the subgraph decomposition we have just discussed. 
In the case at hand, each soft-gluon loop in $d$ dimensions will provide $d$ powers 
of the soft momentum in the numerator, from the loop integration measure. Adding 
a gluon to the soft subgraph, however, is tantamount to adding a loop, and each 
gluon propagator will provide two powers of soft momentum in the denominator. 
Possible three-gluon couplings can only provide further powers of soft momentum 
in the numerator. For $N_s$ soft gluons we find then
\beq
  \omega_s \, \geq \, (d - 2) N_s \, .
\label{softpc}
\eeq
In the case of infrared power counting, a divergence will arise only if $\omega_s \leq 0$: 
thus we have just proved that the two-point function is infrared finite for any $d > 2$.
Since the total cross section for electron-positron annihilation is given by the imaginary
part of just this two-point function, thanks to unitarity, we have also proved that the
cancellation of infrared divergences in the total cross section, demonstrated at one
loop in \secn{FinOrd}, is actually verified to all perturbative orders. It is not difficult 
to extend this simple analysis to general massless correlators with euclidean 
momenta~\cite{Sterman:1994ce}, which results in the Poggio-Quinn finiteness 
theorem~\cite{Poggio:1976qr,Sterman:1976jh}.

In the same spirit, the CN analysis for the form factor brings us to Fig.~\ref{redgendiag}(a), 
which represents the generic diagram responsible for one of the pinch surfaces described 
in \secn{VerteGra}. Let us note at this point that the key aspects of the form factor analysis 
leading to Fig.~\ref{redgendiag}(a) are in fact considerably more general, and remain 
valid beyond our specific example. In particular, for massless fixed-angle multi-particle 
amplitudes,
\begin{itemize}
\item the number of {\it jet} subgraphs must be at most equal to the number of hard particles 
involved, since jets must connect to external on-shell legs, and both incoming and outgoing 
particles may be affected by collinear emissions; \\[-12pt]
\item with at most two incoming coloured particles, only one {\it hard} subgraph
appears in the reduced diagram, and gives the meeting point of incoming and
outgoing jets\footnote{In cases in which it is necessary to consider multiple incoming 
lines, for example for hard exclusive amplitudes~\cite{Lepage:1980fj,Botts:1989kf}, 
or when studying double parton scattering~\cite{Diehl:2017wew}, multiple hard 
subgraphs may occur. These cases typically involve {\it exceptional} momentum 
configurations, where some of the external momenta are parallel, or there are 
non-trivial vanishing partial sums of external momenta. For fixed-angle scattering 
amplitudes, the coordinate-space analysis in Ref.~\cite{Erdogan:2014gha} shows 
that graphs with multiple hard components are power-suppressed.}; \\[-12pt]
\item particles belonging to different jets may interact only through soft mediators, 
which can be merged in a single {\it soft} subgraph.
\end{itemize} 
In order to proceed to the power counting, it is important to devise a consistent
treatment of the lines connecting the various subgraphs, in order to avoid double 
counting. One possibility~\cite{Sterman:1978bi} is to assign to the soft subgraph
all loops containing {\it at least one} zero-momentum line, and to assign to jet 
subgraphs all loops containing {\it only} collinear lines with non-vanishing momentum. 
According to this criterion, for example, the loop in the reduced graph in 
Fig.~\ref{fig:VertexPinch}(a) belongs entirely to the soft subgraph, and there
are no collinear subgraphs, while the reduced diagram in the second panel of 
Fig~\ref{fig:VertexPinch}(b) has only a jet subgraph and no soft subgraph.
With these assignments, one can proceed to define soft and collinear superficial 
degrees of divergence, in terms of the numbers of loops and legs of each subgraph 
in Fig.~\ref{redgendiag}(a). Let $L_s$ be the number of soft loops, and $L_{c, \, i}$
with $i = 1,2$ the number of collinear loops in the $i$-th jet; similarly, let $N_r^b$ 
and $N_r^f$ be the numbers of bosonic and fermionic lines, respectively, in 
subgraph $r$, with $r = \{ s, j_1, j_2 \}$. According to the scaling rules introduced 
in \secn{oloexe}, in the soft limit bosonic propagators scale as $1/\lambda^2$,
while fermion propagators scale as $1/\lambda$. In the collinear limit, both boson 
and fermions provide a factor of $1/\lambda$ (which may be corrected by numerator 
factors, as we will see below). Considering also the integration volume, that involves 
two normal coordinates for each collinear loop and four normal coordinates for each 
soft loop, we can define
\beq
  \omega_s & = & 4 L_s - 2 N_s^b - N_s^f + n_s \, , \nonumber \\
  \omega_{c, \, i} & = & 2 L_{c, \, i} - N_{c, \, i}^b - N_{c, \, i} ^f + n_{c, \, i} \, ,
\label{supdegdiv}
\eeq
where $n_s$ and $n_{c, \, i}$ count positive powers of $\lambda$ arising from
numerator factors. With our assignments, one readily verifies that $\omega_s = 0$
for the reduced graph in Fig.~\ref{fig:VertexPinch}(a), and $\omega_c = 0$ for 
the reduced graph in the second panel of Fig.~\ref{fig:VertexPinch}(b). In general,
provided the lines connecting subgraphs have been consistently assigned, superficial
degrees of divergence for subgraphs are additive. For a generic pinch surface 
$\Sigma$ for form factor graphs one can then define
\beq
  \omega_\Sigma \, = \, \sum_{i = 1}^2 \omega_{c, \, i} + \omega_s \, .
\label{sumomega}
\eeq
The next step in the procedure is to rewrite \eq{supdegdiv} and \eq{sumomega} 
in terms of the numbers of vertices and external lines of the various subgraphs, 
using elementary graphical relations, such as the Euler identity connecting the 
numbers of loops, lines and vertices. There are a number of subtleties in the 
analysis: for example it is necessary to pay attention to the possible existence 
of lower-dimensional pinch surfaces in a given homogeneous integral, and to 
the fact that suppression factors for collinear numerators are polarisation-dependent, 
and thus gauge-dependent. The details of the procedure were worked out in 
Refs.~\cite{Sterman:1977cj,Libby:1978bx,Ellis:1978sf}, and are reviewed, for 
example, in Refs.~\cite{Sterman:1994ce,Sterman:1995fz,Collins:2011zzd,
Berger:2003zh, Bonocore:2016wur}. In what follows, we will just provide a set of simple and, 
hopefully, intuitive arguments for the structure of the constraints emerging 
from power counting, and we will summarise the final results.

Consider for example the lines directly connecting the soft and hard subgraphs
in Fig.~\ref{redgendiag}(a): each such line would provide a power suppression,
and they ultimately cannot contribute to divergent momentum configurations.
To see that, let a soft gluon with momentum $k$ attach directly to a hard line 
carrying momentum $q \sim Q$, and assume that the graph contributes at 
leading power before the soft gluon is inserted. In case the hard line is a gluon, 
the corresponding off-shell propagator contributes a factor of $1/Q^2$ before 
the soft gluon insertion. After the insertion, the hard propagator is doubled, 
contributing a factor of $1/Q^4$, while the new three-gluon vertex can contribute 
at most a factor of $Q$: one has effectively gained a power suppression, which 
is sufficient to make the new graph finite if the graph was logarithmically divergent 
before the insertion. One can similarly argue that a graph with a soft fermion 
line connecting soft and jet subgraphs is suppressed by one power of the 
hard scale with respect to the same graph with a soft gluon, since soft fermion 
denominators are linear in the soft scaling parameter.

Turning now to numerator suppressions, we first note that in the soft limit
the only source of suppression is the presence of three-gluon vertices, which
are linear in momenta and thus provide a single power of $\lambda$ in the
numerator: the suppression factor $n_s$ thus simply equals the number of 
soft three-gluon vertices. The situation concerning numerator suppressions 
in the collinear limit is much more subtle and interesting, since suppression 
factors are spin-dependent. In order to understand this, consider a fermion
line carrying momentum $p$ and emitting a collinear gluon with momentum 
$k$. Both $p$ and $k$ can depend on collinear loop momenta, and thus they 
can have transverse components, which will however vanish in the collinear 
limit, where both particles reach the mass shell. The numerator factor associated
with the emission vertex and the neighboring propagator reads
\beq
  \left( \slashed{p} + \slashed{k} \right) \gamma_\alpha \, \slashed{p} \, = \,
  - \gamma_\alpha \left( \slashed{p} + \slashed{k} \right) \, \slashed{p} \, + \,
  2 \left( p + k \right)_\alpha \, \slashed{p} \, .
\label{collpolvert}
\eeq
In the collinear limit, $p$ and $k$ become light-like and the first term vanishes;
as a consequence, away from the strict limit it will provide a power suppression
proportional to $k_\perp$, {\it i.e.} a factor of $\sqrt{\lambda}$ for every collinear 
vertex. The second term in \eq{collpolvert} does not provide any suppression,
but one should keep in mind that the index $\alpha$ will be contracted with the
neighbouring gluon propagator, whose numerator represents a sum over gluon
polarisations; in the collinear limit, assuming the jet to be in the $+$ direction, 
the only surviving polarisation coupling to the vertex will be a $-$ component,
which corresponds to an unphysical gluon. Such gluons would be absent in a
physical gauge, and must decouple, after summing over diagrams, in a covariant
gauge~\cite{Sterman:1995fz,Berger:2003zh}. To illustrate what happens in a 
physical gauge, note that a gluon connecting the jet subgraph to the hard (or 
soft) subgraph will carry a factor of the axial-gauge gluon propagator
\beq
  G^{\mu \nu} (k) \, = \, \frac{1}{k^2 + {\rm i} \eta} \left( - g^{\mu \nu} + 
  \frac{n^\mu k^\nu + n^\nu k^\mu}{n \cdot k} - n^2 \, 
  \frac{k^\mu k^\nu}{(n \cdot k)^2} \right) \, ,
\label{axprop}
\eeq
where $n^\mu$ is the gauge vector, and the gauge condition imposes $n \cdot A = 0$. 
As illustrated by \eq{collpolvert}, if the gluon connects the jet to the hard subgraph, 
at the vertex where $G^{\mu \nu}(k)$ attaches to the jet, it will be contracted with 
a collinear momentum, proportional to the gluon momentum $k$. This will yield
a factor\cite{Sterman:1995fz}
\beq
  k_\mu G^{\mu \nu}(k) \, = \, \frac{1}{n \cdot k} \left( n^\nu - 
  \frac{n^2k^\nu}{n \cdot k}\right) \, . 
\label{axcontr}
\eeq
This factor has no pole at $k^2 = 0$ (except in the soft limit $k \sim 0$), so that 
such a graph is power-suppressed. In covariant gauges, the argument breaks down, 
since the contraction $k_\mu G^{\mu \nu}(k)$ does not feature the same suppression. 
For instance, in Feynman gauge, we have 
\beq
  k_{\mu} G^{\mu \nu}(k) \, = \, - \frac{k^{\nu}}{k^2} \, .
\label{Feyncontr}
\eeq
We see that in Feynman gauge multiple gluons may connect the jets subgraphs 
to hard subgraph $\mathcal{H}$, but \eq{Feyncontr} suggests that such gluons 
must carry unphysical polarisations. This, in turn, means that, when computing 
a gauge-invariant quantity, these configurations will be suppressed by means of 
the Ward Identity, when all the relevant diagrams have been summed. On a 
diagram-by-diagram basis, this is not guaranteed, and one needs to further 
manipulate the longitudinally polarised gluons to factor $\mathcal{J}_i$ from 
$\mathcal{H}$. 

It is hopefully clear from these arguments that only a small subset of the pinch 
surfaces depicted in Fig.~\ref{redgendiag}(a) will actually yield divergent contributions. 
A detailed analysis~\cite{Sterman:1994ce,Sterman:1995fz,Collins:2011zzd,
Berger:2003zh} yields the following results.
\begin{enumerate}
\item The superficial degree of divergence $\omega_\Sigma$ is bounded by
$\omega_\Sigma \geq 0$, implying that all infrared divergences are logarithmic.
This has important consequences in what follows, since it means that even a single
extra power suppression, such as those discussed above, makes the diagram 
superficially finite.
\item As a first consequence, pinch surfaces with lines directly connecting the soft 
subgraph $\mathcal{S}$ and the hard subgraph $\mathcal{H}$ are finite. 
\item A second consequence is that only gluons can directly connect the soft 
subgraph $\mathcal{S}$ with the jet subgraphs $\mathcal{J}_i$.
\item Finally, in a physical gauge, in order to generate a divergence, every jet 
subgraph $\mathcal{J}_i$ must be connected to the hard subgraph $\mathcal{H}$ only by a single 
line (which must therefore carry the same quantum numbers as the external line 
entering or exiting the graph). In Feynman gauge, further additional lines linking 
$\mathcal{J}_i$ and $\mathcal{H}$ are allowed, but they must be scalar-polarised gluons. Furthermore 
one finds that, for each jet $\mathcal{J}_i$, the number of external soft gluons plus the number 
of jet gluons attached to the hard part cannot be greater than the total number of 
three-point vertices in that jet.
\end{enumerate}
Graphically, these rules mean that pinch surfaces giving rise to divergences are 
considerably simpler than the general configuration depicted in Fig.~\ref{redgendiag}(a), 
and they can be described by Fig.~\ref{redgendiag}(b) in a covariant gauge (with 
double curly lines denoting scalar-polarised gluons), and by Fig.~\ref{redgendiag}(c)
in a physical gauge.

To summarise, we have learnt from power counting that sufficient conditions for
the presence of infrared divergences are considerably more restrictive that the
corresponding necessary conditions arising from the Landau equations and the 
CN physical picture. In axial gauges, Fig.~\ref{redgendiag}(c) is already, effectively, 
a partially factorised expression for the form factor, separating the divergent
subgraphs $\mathcal{S}$ and $\mathcal{J}_i$ from the finite subgraph $\mathcal{H}$. 
In Feynman gauge, we intuitively expect that Ward identities will drive a cancellation 
among diagrams that will ultimately lead to to a similar result. We can also hope 
to apply similar arguments to tame soft gluons attaching the soft subgraph to 
jet subgraphs. The third part of the analysis consists in examining in detail
the actual Feynman rules in the relevant soft and collinear limits: this will
enable us to leverage the universality of long-distance contributions, identifying
operator matrix element that generate soft and collinear divergences for all 
fixed-angle scattering amplitudes, regardless of the particular hard process 
under consideration.


\subsection{Diagrammatic construction of soft and collinear functions}
\label{universal_fun}

Our next task is to turn the ``quasi-factorisation'' depicted in Fig.~\ref{redgendiag} (b)
or (c) into an exact statement, writing the form factor (and later more general scattering 
amplitudes) as a product of functions, each responsible for the enhancements associated
with specific kinematic configurations. This is still far from trivial, since what we have 
achieved so far is true for individual pinch surfaces, and the task ahead is to organise
the contributions of all these pinch surfaces into universal operator matrix elements,
that can be computed independently of the specific hard process under consideration.
In other words, we want to identify matrix elements which share with the scattering
amplitude specific sets of singularities, soft and/or collinear, so that the product of
these matrix elements will contain all the infrared poles of the amplitude, leaving
behind only a finite `matching coefficient'. If we succeed in identifying such matrix 
elements, we can then study infrared singularities to all orders, in a process-independent
way, by means of evolution equations, as we will discuss in \secn{FactEvo}.

One way to identify the matrix elements we need is to consider soft and collinear 
approximations at the diagrammatic level, observe the systematic simplifications that
arise in the relevant limits, and build the appropriate operators accordingly. A prominent
role in this game will be played by Wilson-line operators, as we will see in some detail 
in \secn{WiliEik}.


\subsubsection{Soft and collinear approximations at one loop} 
\label{factorisation_LP}

We begin by revisiting our one-loop results in \secn{oloexe}, in particular the 
approximations leading to the soft and collinear homogeneous integrals in 
\eq{eq:hom_soft} and \eq{collhoi}. At this level, we can easily generalise the 
results to the non-abelian theory, and, this time, we will pay closer attention to 
the numerator structures, which carry information about gluon polarisations, 
providing a more precise chracterisation of the result in \eq{collpolvert}. For
this purpose, consider the annihilation process $q(p_1) \, \bar{q}(p_2) \rightarrow 
\gamma^* $. The un-approximated integral corresponding to the one-loop QCD 
correction to the QED vertex reads
\beq
  \Gamma_\mu^{(1)} (p_1, p_2) \, = \, - e Q_q \int \frac{d^d k}{(2 \pi)^d} \;
  \frac{\bar{v}(p_2) \, \big( g \,  T_{ij}^a \, \gamma^\alpha \big) \,  
  \big( - \slash{p}_2 - \slash{k} \big) \, \gamma_\mu \, 
  \big( \slash{p}_1 - \slash{k} \big) \, 
  \big(g \, T_{jk}^a \, \gamma_\alpha \big) \, u(p_1)
  }{
  \big( k^2 + {\rm i}  \eta \big) \,
  \big[ (p_2 + k)^2 + {\rm i}  \eta \big] \,  
  \big[ (p_1 - k)^2 + {\rm i}  \eta \big] }  \, .
\eeq
\begin{figure}
        \centering
        \begin{subfigure}{0.45\textwidth}
        \centering
        {\includegraphics[scale=0.6]{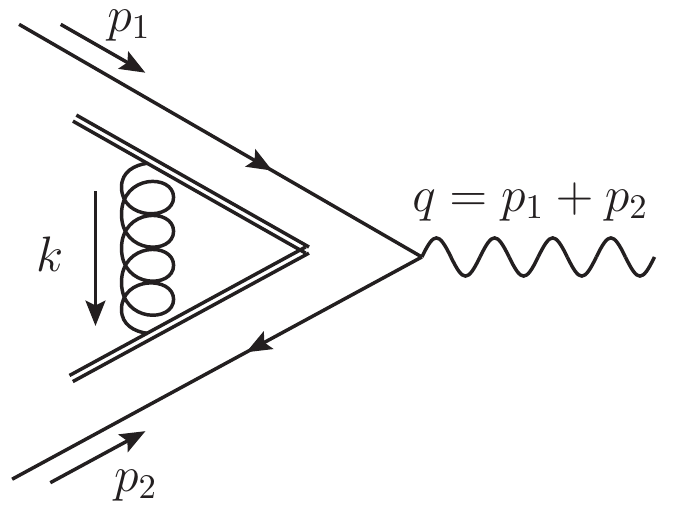} }
        \caption{}
        \label{fig:vertex_correction_eikonal}
         \end{subfigure}
\hspace{1pt}
        \begin{subfigure}{0.45\textwidth}
        \centering
        {\includegraphics[scale=0.6]{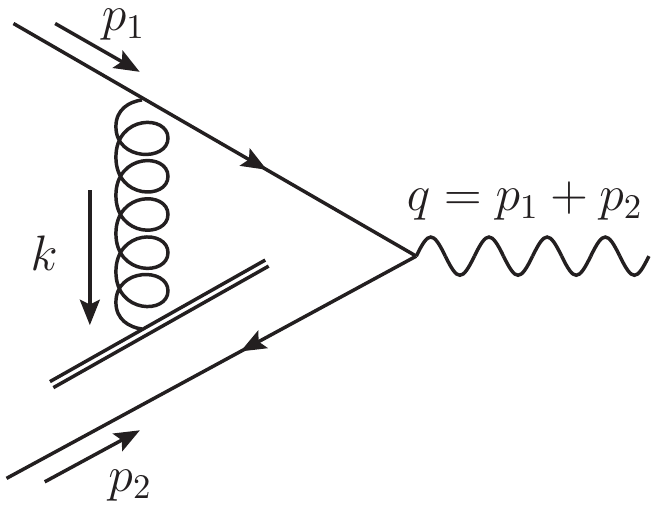} }
        \caption{}
        \label{fig:vertex_correction_coll}
         \end{subfigure}
        \caption{IR interactions modelled by eikonal Feynman rules. (a) Soft gluon 
        exchange: both interaction vertices become eikonal. (b) Collinear exchange, with 
        $k \parallel p_1$: only the vertex involving the antiquark line becomes eikonal.}
\label{fig:reduced_vertex}
\end{figure}
In the soft limit, as was discussed already in \secn{cataQED}, the homogeneous integral
arises upon taking the {\it eikonal approximation}, amounting to
\beq
  \frac{\big( \slashed{p}_1 - \slashed{k} \big) \big( g \, T^a  \,  \gamma_\alpha \big) 
  u(p_1)}{(p_1-k)^{2} + {\rm i} \eta} 
  &\overset{k^{\mu} \to \, 0}{\longrightarrow} &
  \frac{g \,  T^a \beta_{1, \alpha}}{- \beta_1 \cdot k + {\rm i} \eta} \, u(p_1) \, , 
  \nonumber \\
  \frac{\bar{v}(p_2) \big( g \, T_a  \,  \gamma^\alpha  \big)  \big( - \slashed{p}_2 - 
  \slashed{k} \big)}{(p_2 + k)^{2} + {\rm i} \eta} 
  &\overset{k^{\mu} \to \, 0}{\longrightarrow} &
  \bar{v}(p_2) \, \frac{- g \,  T_a  \, \beta_2^{\alpha}}{\beta_2 \cdot k 
  + {\rm i} \eta} \; , 
\label{eq:soft_eikonal}
\eeq
where the Dirac equation has been used, and we took advantage of the scaling 
invariance of the approximation to replace the momenta $p_1$ and $p_2$ 
with the corresponding four-velocities using $p_1 = \mu \beta_1$ and $p_2 = 
\mu \beta_2$. Analogous replacements can be made for final-state fermions 
and anti-fermions, with due attention to overall signs and to the relative sign 
of the ${\rm i} \eta$ prescription: the results can be simply summarised in the 
{\it eikonal Feynman rules}
\beq
  {\rm propagator:} \quad \frac{\rm i}{\beta \cdot k + {\rm i} \eta} \, , \qquad \qquad
  {\rm vertex:} \quad {\rm i} g \,  T_a \beta^\alpha  \, ,
\label{eikFeyn}
\eeq
with the convention that $\beta$ is the velocity flowing along the arrow of the fermion 
line, and $k$ is the soft momentum flowing in the same direction\footnote{Thus in the
first line of \eq{eq:soft_eikonal} one must take $\beta \to p_1/\mu$, and $k \to - k$, while 
in the second line $\beta \to - p_2/\mu$, and $k \to -k$.}. Applying these Feynman 
rules one gets directly the expression for the soft homogeneous integral, \eq{eq:hom_soft},
with the appropriate non-abelian prefactor.

As already noted in \secn{cataQED}, these Feynman rules are independent of the 
spin and energy of the hard emitter, and only sensitive to its direction and colour 
charge. They apply at leading power in all soft momenta, which, by the power-counting 
arguments of \secn{IRPowCo}, is all that is needed to capture soft divergences. 
We note now two further important  properties of the eikonal approximation. First, 
the coupling to the colour charge is universal, in the sense that it is sufficient to 
replace the generator $T_a$ in the fundamental representation in \eq{eikFeyn} 
with the corresponding matrix $T^{r}_a$ in the representation $r$ of the gauge 
group, in order to reproduce the leading-power result if the emitter belongs to 
$r$ (for example, $(T_a)_{bc} \to - {\rm i} f_{abc}$ in the adjoint representation): 
we will take advantage of this universality property in \secn{MultiPart}. Second, 
we note that, taking the hard momentum in a  fixed direction, say $\beta = \{ \beta^+, 
0^-, {\bf 0}_\perp \}$ in light-cone coordinates, the eikonal coupling of the soft 
gluon to the hard line involves only the $-$ component of its polarisation at leading 
power. Furthermore, if all components of the soft gluon momentum scale equally 
in the soft limit, according to \eq{scalesoft}, the only component of the gluon momentum 
which is relevant at leading power is $k^-$. Soft gluons thus couple longitudinally 
to hard lines, a fact which will be instrumental in what follows. The simplified coupling 
of soft lines to hard lines is shown pictorially in Fig~\ref{fig:reduced_vertex} (a), where 
eikonal vertices are represented as the merging of gluon propagator with a double line.

In the same spirit, let's revisit the collinear approximation, leading to the collinear
homogeneous integral in \eq{collhoi}, focusing on the numerator structure. In the 
frame in which $p_1^\mu$ in in the $+$ direction, while $p_2^\mu$ is in the 
$-$ direction, the integrand reads
\beq
T^a \, T_a  \, 
  \frac{\bar{v} (p_2) \gamma^- \big( \slash{p}_2 + \slashed{k} \big)  \,
  \gamma^\mu \big( \slash{p}_1 - \slash{k} \big) \, 
  \gamma^+ u(p_1)
  }{
  \big[k^2 + {\rm i} \eta \big] \;
  \big[(p_1-k)^2 + {\rm i} \eta \big] \;   
  \big[(p_2+k)^2 + {\rm i} \eta \big] } \, ,
\label{eq:I_coll}
\eeq
where we used the massless Dirac equations in this frame, which read $\bar{v} 
(p_2) \gamma^+ = \gamma^- u(p_1) = 0$. By the same token, and using the
fact that, in the region collinear to $p_1$, the $k^-$ component of the loop 
momentum is power-suppressed, we can approximate the numerator 
of \eq{eq:I_coll} using
\beq
  \bar{v} (p_2) \gamma^- \big( \slash{p}_2 + \slashed{k} \big) 
  \, \simeq \, (p_2 + k)^- \; \bar{v} (p_2) \gamma^-  \gamma^+
  \, = \, 2 (p_2+k)^- \, \bar{v} (p_2) \, \simeq \, 
  2 \, p_2^- \, \bar{v} (p_2) \, .
\label{numapprocoll}
\eeq
In the collinear approximation, \eq{eq:I_coll} can then be rewritten as
\beq
\label{finnumcoll}
T^a \, T_a  \, 
  \frac{\bar{v} (p_2) \,
   \gamma^\mu \big( \slashed{p}_1 - \slashed{k}) 
  \gamma_\alpha \, 
  u(p_1)}{k^2 (p_1-k)^2} \; \frac{\beta_2^\alpha}{\beta_2
  \cdot k} \; .
\eeq
Remarkably, the vertex attaching the collinear gluon to the anti-quark has become 
eikonal, and can be expressed in terms of the effective Feynman rules of \eq{eikFeyn}, 
as represented pictorially in Fig.~\ref{fig:reduced_vertex} (b). As before, the gluon 
coupling to the anti-quark has become independent of the anti-quark energy; 
furthermore, the gluon couples to the antiquark with a polarisation parallel to 
its momentum: it is longitudinal, and thus unphysical. On the other hand, the 
coupling of the collinear gluon to the quark has not simplified, and retains its spin 
and energy dependence.
\begin{figure}
        \centering
        {\includegraphics[scale=0.7]{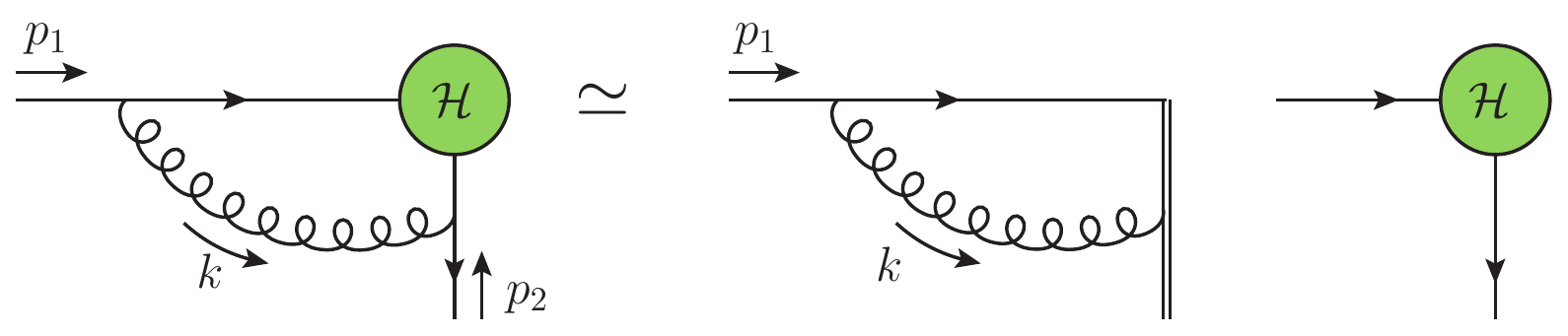} }
        \caption{Factorisation of the hard subgraph after a collinear emission. Notice that
        the collinear prefactor is a spin matrix, as discussed in the text.}
\label{collward}
\end{figure}
In view of generalising this result to higher perturbative orders, it is useful
to note that the sequence of approximations leading to \eq{finnumcoll}
can be reformulated as the application of a tree-level Ward identity.
Let us assume to have a longitudinally polarised gluon moving in the $+$
direction, and attaching to a generic hard interaction subgraph $H$. Writing 
the (Feynman-gauge) numerator of the gluon propagator as a sum over 
polarisations, including unphysical ones, the amplitude for this process is 
of the form
\beq
  \sum_\lambda \, \bar{v} (p_2) \gamma_\alpha
  \frac{\slashed{p}_2 + \slashed{k}}{(p_2 + k)^2 + {\rm i} \eta} \, 
  \varepsilon^\alpha_{(\lambda)} (k) \, H \, \varepsilon^{* \beta}_{(\lambda)} (k)
   \frac{\slashed{p}_1 - \slashed{k}}{(p_1 - k)^2 + {\rm i} \eta} \, \gamma_\beta \, 
  u(p_1) \, ,
\label{eq:H}
\eeq
where the hard subgraph $H$ is defined here to include the denominator of the 
gluon propagator and we can focus on the case of longitudinal polarisation. We 
can then substitute $\slashed{\varepsilon}_{(\lambda)}(k) = \gamma^- \varepsilon^+ 
(k)$, and insert a factor of $1 = k^+ \beta_2^-/k^+ \beta_2^-$, so that \eq{eq:H}
becomes~\cite{Collins:1989bt}
\beq
  && \hspace{-8mm}
  \frac{k^+ \beta_2^-}{k^+ \beta_2^-} \,  
  \bar{v} (p_2) \, \gamma^-
  \frac{\slashed{p}_2 + \slashed{k}}{(p_2 + k)^2 + {\rm i} \eta} \, 
  \varepsilon^+(k) \, H \, 
  \frac{\slashed{p}_1 - \slashed{k}}{(p_1 - k)^2 + {\rm i} \eta} \, \slashed{\varepsilon}(k) 
  \, u(p_1) \nonumber \\ & = & 
  \frac{\varepsilon(k) \cdot \beta_2}{k \cdot \beta_2} \;
  \bar{v} (p_2) \, \slashed{k} \, 
  \frac{\slashed{p}_2 + \slashed{k}}{(p_2 + k)^2 + {\rm i} \eta} \, H \,
  \frac{\slashed{p}_1 - \slashed{k}}{(p_1 - k)^2 + {\rm i} \eta} \, 
  \slashed{\varepsilon}(k) \, u(p_1) 
  \nonumber \\
  & = & 
  \bar{v} (p_2) \, H \, \frac{\varepsilon(k) \cdot 
  \beta_2}{k \cdot \beta_2} \frac{\slashed{p}_1 - 
  \slashed{k}}{(p_1 - k)^2 + {\rm i} \eta} \, 
  \slashed{\varepsilon}(k) \, u(p_1) \, ,
\label{eq:Ward}
\eeq
where in the last step we have used $\slashed{k} = - \slashed{p}_2 + 
(\slashed{p}_2 + \slashed{k})$: the first term then vanishes by the Dirac 
equation, and the second term cancels the denominator of the antiquark 
propagator, effectively replacing it with the eikonal denominator. One then 
readily recognises \eq{finnumcoll}. Thanks to this `Ward identity', we can 
effectively write the amplitude as the product of a spin factor with an eikonal 
coupling to the antiquark, multiplying the hard subgraph, as depicted in 
Fig.~\ref{collward}. 

To summarise the results of this one-loop analysis, we have seen that a soft
gluon couples eikonally to all hard lines, while a collinear gluon couples eikonally
to anti-collinear hard lines. Eikonal couplings are energy- and spin-independent,
and, more interestingly, involve unphysical polarisations at leading power. This
observation is the key to the generalisation of this result to higher orders in 
perturbation theory: the only effect of unphysical gluons is to dress the hard 
particles with a gauge rotation. The framing of the one-loop collinear limit in 
the language of diagrammatic Ward identities suggests how to proceed at 
higher orders: in case of multiple radiations, the number of possible collinear 
connections to the hard subgraph increases, however the Ward identity will 
guarantee that the eikonal approximation on the anticollinear line still holds, 
and all emissions can be expressed in terms of effective Feynman rules. 
This was already established for abelian soft radiation in \secn{cataQED}.
In the non-abelian theory, further care must be taken in order to insure that 
the colour ordering of the attachments is consistently reproduced: as we will 
see in the next section, there is a natural solution to this problem. We must 
however note an important caveat in the non-abelian case: at higher orders, 
soft gluons will couple to {\it a set} of collinear lines (rather than a single line), 
which in general will have momenta with small transverse components. 
Couplings of soft gluons to these components will be negligible only if the 
soft limit is taken at the same rate for all components of the soft momentum 
$k$, as in \eq{scalesoft}. In a momentum region where $k^-$ is much smaller that 
$k_\perp$, the approximation that we have made in order to effectively factorise 
soft lines from `jet' lines may fail. We will briefly return to this issue in \secn{MultiPart}. 
A more detailed description of the all-order proof of the factorisation of soft lines 
from hard collinear ones is given in Ref.~\cite{Sterman:1995fz}, while a full analysis 
can be found in Ref.~\cite{Collins:2011zzd}.


\subsubsection{Wilson lines and the eikonal approximation}
\label{WiliEik}

Power counting and diagrammatic analyses have given us a much-simplified picture
of soft and collinear radiation. Specifically, we have seen that only limited physical 
information links soft, hard and collinear subgraphs. Our next step is to identify
operator matrix elements that contain this same information, and therefore share
the same singular regions in loop-momentum space. Before we proceed, however,
it is important to pause and focus on the physical underpinnings of our conclusions
so far. A crucial ingredient of the factorisation is the fact that soft radiation is insensitive
to the nature of hard exchanges, as well as to the internal structure of collinear jets 
of fast particles: only the overall direction of the collinear beam and its overall colour
charge can be detected by soft gluons. This is a consequence of the fact that soft 
radiation has long wavelength, which cannot discriminate the short-distance 
features of the scattering process, or the fine structure of collimated jets. This 
simple physical fact is not directly apparent from individual Feynman diagrams, 
and only emerges, at high orders, after intricate cancellations. It has however 
been well-understood for a long time, under the heading of {\it colour transparency},
first discussed in the non-abelian theory in Refs.~\cite{Low:1975sv,Nussinov:1975mw},
and reviewed for example in Refs.~\cite{Nikolaev:1990ja,Dokshitzer:1991wu}. This
in turns forms the basis for the {\it colour coherence} approach, one of the first powerful
tools for all-order analyses in perturbative QCD (see, for example,~\cite{Dokshitzer:1978hw,
Bassetto:1984ik,Dokshitzer:1987nm,Forshaw:2021rlk,Pathak:2021wdr}). For collinear 
radiation, we have observed a similar phenomenon: collinear gluons are sensitive to 
the spin and energy of the emitter, but cannot resolve the quantum numbers of 
anti-collinear particles. The axial-gauge result, showing that only a single particle 
can connect jets to the hard scattering, which in turn is reflected by the eikonal 
couplings of scalar-polarized collinear gluons in covariant gauges, also implies that 
hard collinear radiation depends on the colour charge of the collinear beam, but
is not sensitive to colour exchanges with other hard particles: the only long-distance
colour exchanges between different hard particles are effected by soft gluons.

We now turn to the construction of matrix elements reproducing these physical
results, expressed by eikonal Feynman rules. In the soft case, a natural guess 
leads to the correct ansatz: at leading power in the soft radiation, hard particles
do not recoil, and follow straight-line trajectories from the hard scattering point
out to time-like infinity, in a specified direction. Along these trajectories, soft interactions
can only dress the hard particle with a  gauge phase. This phase in turn is naturally
expressed by integrating the gauge connection along the trajectory. The required
operator, for each hard particle, is therefore a straight Wilson line, defined in 
general by
\beq
\label{def_wilson}
  \Phi_n (\lambda_2, \lambda_1) \, = \, \mathcal{P} \exp \bigg[
  {\rm i} g \int_{\lambda_1}^{\lambda_2} d \lambda \; n 
  \cdot A^a(\lambda n) \, T_a  \bigg] \; ,
\eeq
where $n^\mu$ is the direction of the line, and, in the non-abelian case, one needs
to introduce the path ordering operator $\mathcal{P}$. \eq{def_wilson} is a colour 
matrix in the representation of the selected hard particle. In the present
context, we will only need semi-infinite lines, with $\lambda_1 = 0$ and $\lambda_2
= \infty$. It is not difficult to convince oneself that Wilson lines reproduce eikonal 
Feynman rules, and the path-ordering provides the correct non-abelian generalisation
of the eikonal approximation.

To this end, consider, for example, a graph $G$ featuring an outgoing on-shell hard 
quark line carrying momentum $p$, emitting an arbitrary number $n$ of low-energy 
gluons with colour indices $a_i$ and carrying momenta $k_i$. At leading power in 
all soft momenta, we can simply adapt \eq{ubar} to include color operators, 
and we obtain the expression
\beq
  G^{\, a_1 \dots a_n}_{\mu_1 \dots \mu_n} \big( \, p, \{k_{i}\} \big)
  \, \simeq \, \prod_{i=1}^n \;  \left[ \big( {\rm i} g \,  T^{a_i} \big) 
  \frac{i \, p^{\mu_i}}{p \cdot \left( \sum_{j=1}^i k_j \right)} \right] 
  \bar{u}(p) \, {\cal M} \big(p\big) \; .
\label{eq:M_eikonal}
\eeq
In the abelian case, it was possible to sum explicitly over the permutations of the
outgoing gluons, so that \eq{eq:M_eikonal} could be simplified by applying the 
eikonal identity, \eq{eikonal_identity}.  The identity, however, expresses  the fact 
that successive photon emissions are independent and uncorrelated in the abelian 
theory, and is no longer valid in the non-abelian case, since the colour matrices 
do not commute. This, in turn, requires the path-ordering prescription in 
\eq{eq:M_eikonal}.
\begin{figure}
        \centering
        \begin{subfigure}{0.45\textwidth}
        \centering
        {\includegraphics[scale=0.47]{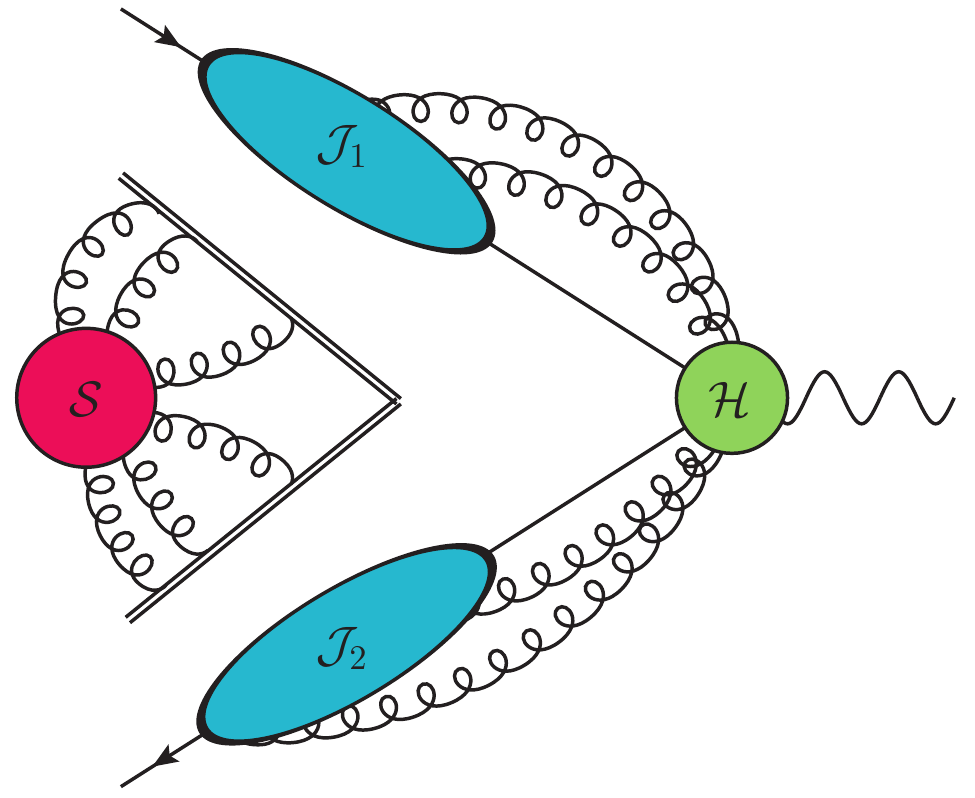} }
        \caption{}
        \label{fig:reduced_diagram_soft_hard+jet}
         \end{subfigure}
         \hspace{-15pt}
          \begin{subfigure}{0.45\textwidth}
        \centering
        {\includegraphics[scale=0.45]{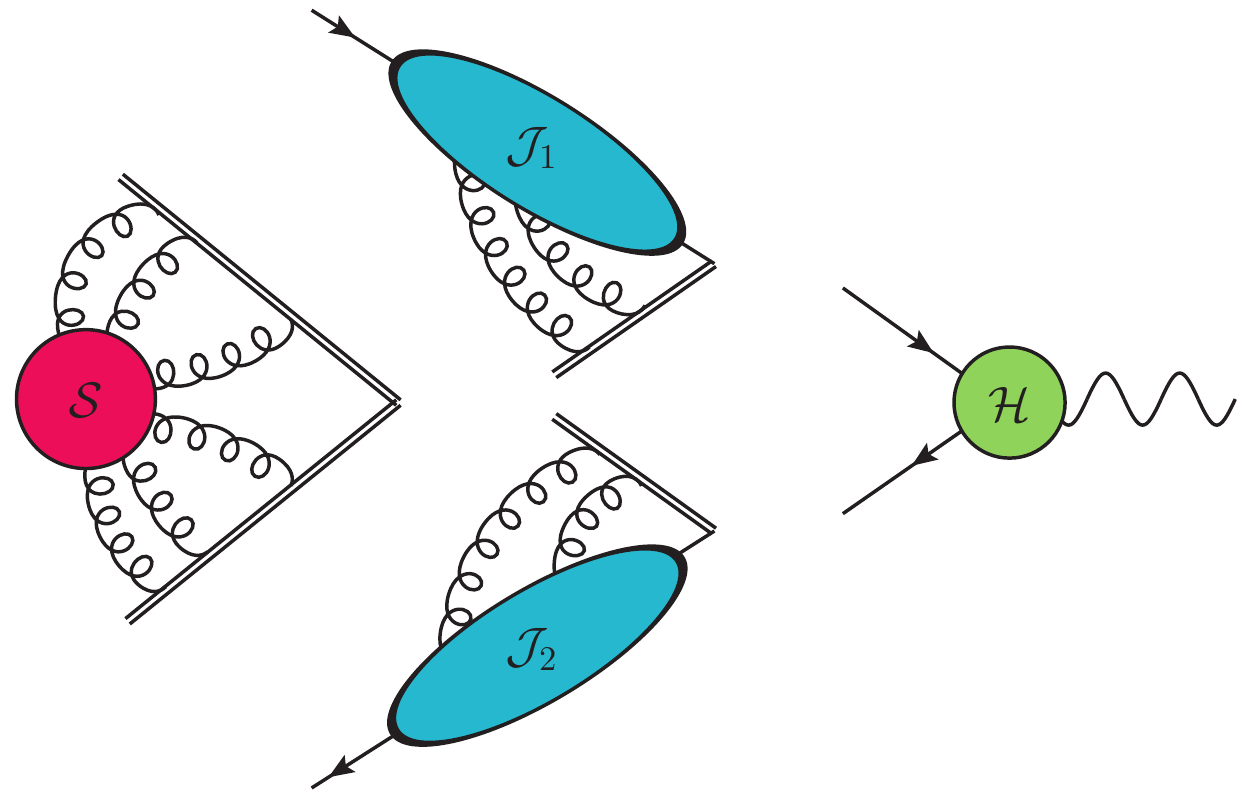} }
         \caption{}
        \label{fig:reduced_diagram5}
         \end{subfigure}
        \caption{Different steps of the factorisation procedure. a) Factorisation of the soft 
        subgraph: multiple soft gluon emissions are modelled {\it via} eikonal Feynman rules.
        b) Factorisation of the jet subgraph from the hard part: collinear gluons attach to
        eikonal vertices}
\label{fig:reduced_regions}
\end{figure}
To verify that the momentum-space eikonal Feynman rules are reproduced, it is sufficient 
to expand the definition in \eq{def_wilson}, and Fourier transform the gauge field, according 
to
\beq
  {A_\mu^a (\lambda n) \, = \, \int \frac{d^d k}{(2\pi)^d} \; e^{ {\rm i} k \cdot \lambda n} 
  \tilde{A}_\mu^a (k)} \, .
\label{Fourtraf}
\eeq 
At leading order the expansion gives
\beq
\label{expansion_wilson}
  {\cal P} \exp \bigg[ \, {\rm i} g \int_0^\infty d \lambda \, 
  n \cdot A^a (\lambda n) \, T_a \bigg] \, = \, 
  1 - g \int \frac{d^dk}{(2 \pi)^d} \, 
  \frac{n \cdot \tilde{A}^a (k)}{n \cdot k + {\rm i} \eta} \,\,  T_a  + {\cal O} (g^2) \, ,
\eeq
where we used Feynman's prescription to define the parameter integral at large 
distances, resulting in
\beq
  \int_0^\infty d \lambda \, {\rm e}^{{\rm i} ( k \cdot n ) \lambda} \, \to \,
  \int_0^\infty d \lambda \, {\rm e}^{{\rm i} ( k \cdot n + {\rm i} \eta) \lambda} \, = \, 
  \frac{{\rm i}}{k \cdot n + {\rm i} \eta} \, ,
\label{inttoinf}
\eeq
reflecting the Feynman rules in \eq{eikFeyn}. At the next order in the expansion the
path-ordering prescription becomes relevant, yielding the correct partial denominators
of \eq{eq:M_eikonal}. Indeed one finds
\beq
  && \hspace{-5mm} \big( {\rm i} g \big)^2 \int_0^\infty \! d \lambda_1 
  \int_0^{\lambda_1} \! d \lambda_2 \, n \cdot  A^a (\lambda_1 n) \,  
  n \cdot  A^b (\lambda_2 n) \, T_a \, T_b \nonumber \\
  &&
  = \, \big( {\rm i} g \big)^2 \, \int \frac{d^d k_1}{(2 \pi)^d} \frac{d^d k_2}{(2 \pi)^d}
  \int_0^\infty \! d \lambda_1 \int_0^{\lambda_1} \! d \lambda_2 \, 
  {\rm e}^{ {\rm i} ( \lambda_1 k_1 + \lambda_2 k_2 ) \cdot n} \, 
  n \cdot  \tilde{A}^a (k_1) \, n \cdot  \tilde{A}^b (k_2) \, T_a \, T_b
  \nonumber \\
  &&
  = \, g^2 \! \int \frac{d^d k_1}{(2 \pi)^d} \frac{d^d k_2}{(2 \pi)^d} \,
  \frac{n \cdot \tilde{A}^a (k_1)}{k_1 \cdot n + {\rm i} \eta} \, 
  \frac{n \cdot \tilde{A}^b (k_2)}{(k_1 + k_2) \cdot n + {\rm i} \eta} \, T_a \, T_b \, ,
\label{eq:WL_2order}
\eeq
which is fully consistent with the diagrammatic expression of a double emission. 
The pattern in \eq{eq:WL_2order} generalises to all orders, yielding 
\beq
  \mathcal{P} \exp \left[ \, {\rm i} g \, T_a \int_0^\infty d \lambda \, 
  n \cdot A^a (\lambda n) \right] \, = \, 1 + \sum_{n = 1}^\infty \left[ \,
  \prod_{i = 1}^n \left( \int \frac{d^d k_i}{(2 \pi)^d} \frac{g \, T_{a_i} \, 
  n \cdot \tilde{A}^{a_i} (k_i)}{\sum_{j = 1}^i n \cdot k_i + {\rm i} \eta} \right) \, \right] \, ,
\label{allordwil}
\eeq
which reproduces the leading-power result for soft gluon attachments to a hard
line, exemplified in \eq{eq:M_eikonal}.

These results confirm our intuition, that the interactions of a hard particle as it
propagates in a background of soft gluons without recoil are correctly reproduced by
replacing the particle with an appropriate Wilson-line operator. Interactions between 
different hard particles propagating in different directions and exchanging soft gluons
will similarly be reproduced by taking a vacuum expectation value of a set of Wilson
lines, each in the appropriate representation of the gauge group, and defined along 
the classical straight-line trajectory of the hard emitter. The path-integral evaluation
of the resulting correlator will automatically generate all the radiative corrections
building up the generic soft subgraphs discussed in the previous sections. To illustrate
these facts in the simplest case, we can easily reproduce the expression of the 
one-loop eikonal integral in \eq{eq:eikint} by considering the correlator of two Wilson
lines. Writing explicitly the open colour indices, we find
\beq
\label{twolines}
  && \hspace{-5mm}
  \bra{0} {\rm T} \, \Big[ \Phi^{f_1 g_1 }_{\beta_1} (\infty,0) \,
  \Phi^{f_2 g_2}_{\beta_2} (\infty,0) \Big] \ket{0}
  \nonumber \\
  && = \, 1 \, - \, g^2 \, \big( T_1^a \big)^{f_1 g_1} \, \big( T_2^b \big)^{f_2 g_2} \, 
  \beta_1^\mu \beta_2^\nu \int_0^\infty d \lambda_1 \, d \lambda_2 \,
  \bra{0} {\rm T} \, \Big[ A_\mu^a ( \lambda_1 \beta_1) \, 
  A_\nu^b (\lambda_2 \beta_2) \Big]  \ket{0} \, + \, \ldots \nonumber \\
  && = \, 1 \, - g^2 \, \big( T_1^a \big)^{f_1 g_1} \, \big( T_2^b \big)^{f_2 g_2} \, 
  \beta_1^\mu \beta_2^\nu \int_0^\infty d \lambda_1 \, d \lambda_2 \,
  \int \frac{d^d k}{(2 \pi)^d} \, \bigg( \frac{- {\rm i} g_{\mu \nu} \, 
  \delta^{ab}}{k^2 + {\rm i} \eta}\bigg) \,
  {\rm e}^{{\rm i} k \cdot (\lambda_1 \beta_1 - \lambda_2 \beta_2)} \, + \, \ldots
  \nonumber \\
  && = \, 1 \, + \, {\rm i} g^2 \, T_1 \cdot  T_2 \, \beta_1 \cdot  \beta_2
  \int \frac{d^d k}{(2 \pi)^d} \, 
  \frac{1}{(k^2 + {\rm i} \eta) (k \cdot \beta_1 - {\rm i} \eta) 
  (k \cdot \beta_2 + {\rm i } \eta)} \, + \, \ldots \, . 
\eeq
We omitted self-energy corrections, which arise by expanding one of the two
Wilson lines to ${\cal O} (g^2)$, and vanish in a massless theory, where the Wilson
lines lie on the light cone. In the second line of \eq{twolines} we recognised the
coordinate-space gluon propagator, and in the last line we performed the parameter
integrals, and we defined the scalar product $T_1 \cdot T_2 \equiv T_1^a T_{2, \, a}$. 
We note that one could also perform the calculation by using the explicit 
expression of the coordinate-space gluon propagator: this possibility is exploited
and discussed in detail in \secn{MultiPart}. We recognise the result of \eq{eq:eikint},
but this  time with open colour indices, which can later be contracted with open
colour indices of the hard matching function.

The arguments above represent an important step forward on the way to a 
complete infrared factorisation for form factors, and later for fixed-angle scattering 
amplitudes: having identified a set of operator matrix elements (such as 
\eq{twolines}) that reproduce all the leading-power soft singularities of the 
form factor, we can promote the diagrammatic factorisation represented in 
Fig.~\ref{redgendiag}, where subgraphs are still linked by colour and spin indices, 
to the complete factorisation depicted in Fig.~\ref{fig:reduced_regions}(a), 
where we can now interpret the double lines as Wilson lines, and the soft
subgraph is promoted to a {\it soft function}, 
\beq
  \langle 0 |  T \Big[ \Phi_{\beta_1} (\infty, 0) \,  \Phi_{\beta_2} (\infty, 0) \Big] 
  | 0 \rangle \, ,
\label{hintSoftfofa}
\eeq
responsible for all divergent configuration originating from soft gluons. Hard 
and collinear integration regions will of course be misrepresented by the soft 
function, and will have to be included in other terms in the factorisation.

As far as collinear regions are concerned, the discussion leading to 
\eq{finnumcoll} provides a clear suggestion for how to proceed: collinear gluons
couple eikonally to anti-collinear lines, so we can build a {\it jet function}
by replacing anti-collinear lines by a Wilson line, while retaining spin and
colour information for the emitting collinear line. This results in the further
factorisation depicted in Fig.~\ref{fig:reduced_regions} (b). Tentatively, one 
could introduce an object of the form
\beq 
  \langle 0 \, | \, \Phi_{\beta_2} (0, \infty) \, \psi (0) \, | p_1, s_1 \rangle \, ,
\label{tentJqfofa}
\eeq
where a quark of momentum $p_1$ and spin polarisation $s_1$ in the initial
state is annihilated by the quark field at the origin, and a Wilson line replaces
the incoming antiquark. This function, however, has the drawback that it still
contains collinear divergences associated with the antiquark, since the Wilson
line is light-like in the massless case, and gluons attaching to it will have
anti-collinear enhancements, not fully matching those of the original amplitude.
The solution in this case is very simple: we can just replace the light-like
Wilson line along the $\beta_2$ direction with a `massive' Wilson line with
a direction $n^\mu$, with $n^2 \neq 0$. The resulting function will have the 
correct collinear singularities associated with  the quark direction, but only
finite contributions from the anti-collinear momentum region. As before, the 
resulting function will again misrepresent the soft and hard momentum 
integration regions, which are correctly approximated by other terms in 
the factorisation. In particular, we note that gluons that are both soft and 
either collinear or anti-collinear are correctly approximated both in the soft
and in the collinear factors. Including both factors will result in double-counting,
and will require a subtraction. In the next section, we will tackle these issues, 
and we will provide a complete picture of infrared factorisation for form factors.


\section{Factorisation, evolution and  resummation: form factors}
\label{FactEvo}

The tools developed in \secn{AllOrd} have motivated, if not proved in detail, a 
factorisation formula for form factors, where all soft and collinear divergences are 
accounted for in terms of universal function, which can be expressed as matrix 
elements of fields and Wilson lines. This factorisation formula will be discussed 
in greater detail in \secn{FactoForm}. For the moment, we wish to emphasise 
that, once the hard work of proving factorisation has been completed, there are 
low-hanging fruits to be reaped: every factorisation theorem, in fact, implies
a set of evolution equations, and the solution of these equation leads to a
partial summation of perturbation theory. Before we apply this to soft-collinear 
factorisation of scattering amplitudes, we wish to recall briefly how this happens
in standard applications, following~\cite{Contopanagos:1996nh}.

The simplest and best known example of this process is of course the renormalisation 
of UV divergences, which amounts to the factorisation of singular cutoff dependence 
into a finite set of universal renormalisation constants. Assuming that perturbative 
renormalisability has been proven for the theory at hand, consider the relation between 
bare and renormalised Green functions, 
\beq
  G_0^{(n)} \big( p_i, \Lambda, g_0 \big) \, = \, \prod_{i = 1}^n \, Z_i^{1/2} 
  \bigg( \frac{\Lambda}{\mu}, g(\mu) \bigg) \, G_R^{(n)} \big( p_i, \mu, g(\mu) \big) \, ,
\label{renormfact}
\eeq
where $Z_i$ are renormalisation constants for the fields appearing in the correlator, 
$g_0$ is a bare coupling, $\Lambda$ is an ultraviolet cutoff (for example the inverse 
of a lattice spacing), and $\mu$ is the renormalisation scale. Proving that
\eq{renormfact} holds, possibly with the added burden of showing that other useful
properties of the theory, such as unitarity, or gauge symmetry, are preserved, is in 
general difficult. When this has been achieved, however, one can mine \eq{renormfact}
for further information. In order to achieve the factorisation, it is strictly necessary
to introduce a additional energy scale, $\mu$, which, intuitively, we place somewhere 
between the laboratory energy scales, given by the Mandelstam invariants $p_i 
\cdot p_j$, and the cutoff scale. Renormalisability means that all singular cutoff 
dependence is confined to the field renormalisation constants $Z_i$. Once that 
is established, a (Callan-Symanzik) evolution equation immediately follows, by
simply noting that the  {\it l.h.s} of \eq{renormfact} does not depend on the 
renormalisation scale $\mu$, which has been introduced as a necessary artefact
of renormalisation. Scale dependence must therefore cancel between the factors
on the {\it r.h.s.}, so that
\beq
  \frac{d G_0^{(n)}}{d \mu} \, = \, 0 \, \qquad \longrightarrow \qquad 
  \frac{d \log G_R^{(n)}}{d \log \mu} \, = \, - \sum_{i = 1}^n \gamma_i \big( g(\mu) \big) \, ,
\label{rgeq}
\eeq
where the anomalous dimensions $\gamma_i$ are defined by
\beq
  \gamma_i \big( g(\mu) \big) \, \equiv \, \frac{1}{2} \, \frac{d \log Z_i}{d \log \mu} \, .
\label{gammai}
\eeq
The simple functional dependence of the anomalous dimensions is dictated by
separation of variables in \eq{rgeq}: $\gamma_i$ cannot depend on the cutoff 
$\Lambda$, because the renormalised Green function $G_R^{(n)}$ does not, 
and it cannot depend on the momenta $p_i$, because $Z_i$ does not. Anomalous
dimensions can only depend on arguments that are in common between the
{\it l.h.s.} and the {\it r.h.s.} of \eq{rgeq}, in this case just on the renormalised
coupling $g(\mu)$. Armed with an evolution equation, one can solve it, and,
in this case, `resum' logarithms of the renormalisation scale $\mu$. Since the
renormalised Green function $G_R^{(n)}$ depends on $\mu$ through the
coupling and through ratios of the form $p_i \cdot  p_j/\mu^2$, solving the 
equation gives precious information on the dependence of the correlator
on external energy scales.

A second standard example of the same phenomenon is collinear factorisation
in high-energy scattering, which we briefly introduced in \secn{CollDiv}. \eq{partmod}
is an all-order factorisation theorem, which can be proved in perturbation theory 
either by hard diagrammatic work, or by using more sophisticated tools such as 
the light-cone expansion. It states that collinearly enhanced contributions can be 
factored from the cross section in the form of a convolution, and can be organised 
into a set of parton distribution functions $f_{j/i}$. If desired, the convolution can be 
diagonalised by a Mellin transform, using the definition
\beq
  \widetilde{W} (N) \, \equiv \, \int_0^1 z^{N - 1} \, W(z) \, .
\label{MellTransf}
\eeq
Applying this definition to \eq{partmod}, for a single particle flavour, results in 
the simple factorisation
\beq
  \widetilde{W} \bigg( N, \frac{Q^2}{m^2}, \alpha_s (Q^2) \bigg) \, = \,
  \widetilde{H} \bigg( N, \frac{Q^2}{\mu_f^2}, \alpha_s (Q^2) \bigg) \,
  \widetilde{f} \bigg( N, \frac{\mu_f^2}{m^2}, \alpha_s (Q^2) \bigg) \, ,
\label{F2fact}
\eeq
where $Q$ is the hard scale, we used a particle mass $m$ as a collinear regulator, 
replacing dimensional regularisation, and we noted that the factorisation procedure 
requires the introduction of a factorisation scale $\mu_f$. In \eq{F2fact}, the singular 
dependence on the collinear cutoff $m$ is collected in the (Mellin-space) parton 
distribution $\widetilde{f}$, while the Wilson coefficient $\widetilde{H}$ is free of 
singularities as $m \to 0$. Also in this case, DGLAP evolution follows from the 
observation that the full partonic structure function does not depend on the factorisation 
scale. Then
\beq
  \frac{d \widetilde{W}}{d \mu_f} \, = \, 0 \, \qquad \longrightarrow \qquad 
  \frac{d \log \widetilde{f}}{d \log \mu_f} \, = \, - \, \gamma_N \big( \alpha_s (Q^2) 
  \big) \, ,
\label{rAPeq}
\eeq
where
\beq
  \gamma_N \big( \alpha_s (Q^2) \big) \, \equiv \, \frac{d}{d \log \mu_f} \,
  \widetilde{H} \bigg( N, \frac{Q^2}{\mu_f^2}, \alpha_s (Q^2) \bigg) \, .
\label{gammaN}
\eeq
Once again, the anomalous dimension $\gamma_N$ (the Mellin transform of the 
DGLAP splitting function) can depend only on variables that are common to the
two factors on the {\it r.h.s.} of \eq{F2fact}, in this case $N$  and $\alpha_s$. 
Solving the DGLAP equation leads to the resummation of collinear
logarithms, which effectively disappear from the calculation of the DIS cross 
section, as they are absorbed into the parton distribution evaluated at the proper 
scale.

Soft-collinear factorisation of scattering amplitudes generalises these familiar 
examples. Considering the form factor as a first example, we note that a double
factorisation has been performed, extracting a soft factor, and a collinear factor 
for each external leg. We expect therefore two evolution equations, matching the
double-logarithmic nature of the amplitude we are considering. This is indeed 
what we will find, after examining the  factorisation in more detail.


\subsection{Soft-collinear factorisation of a form factor}
\label{FactoForm}

Let us summarise the path that leads to the soft-collinear factorisation of the form 
factor, as outlined in \secn{AllOrd}. First, the Landau equations, as implemented
in the Coleman-Norton physical picture, identify soft and collinear gluons as the 
only potential source of divergences in the (renormalised) massless form factor. 
Next, power counting reveals that the connections between hard, soft and collinear 
subdiagrams are strongly constrained: at leading power in the normal variables, 
no lines can directly connect soft and hard factors; in a physical gauge, only one 
line can connect collinear and hard factors (while in a covariant gauge this line 
can be supplemented by any number of scalar-polarised gluons, which 
however decouple from the hard subgraph after summing over diagrams, by 
means of the Ward identity); soft gluons at wide angles couple to collinear lines 
only with the anti-collinear components of their momenta, and they are longitudinally 
polarised along the anti-collinear direction, so that they effectively couple to a single 
Wilson line along the jet direction. The last step in \secn{AllOrd} was the use of explicit 
diagrammatic arguments in order to identify operator matrix elements carrying the 
same singularities as the form factor in the soft and collinear sectors. This led to 
the identification of the {\it soft function} as a matrix element of light-like Wilson 
lines extending along the classical trajectories of the incoming particles. Taking as 
an example an incoming quark-antiquark pair, we define then
\beq
  {\cal S} \left( \beta_1 \cdot \beta_2, \alpha_s(\mu^2), \eps \right)  \, \equiv \,  
  \langle 0 |  T \Big[ \Phi_{\beta_1} (\infty, 0) \,  \Phi_{\beta_2} (\infty, 0) 
  \Big] | 0 \rangle \, .
\label{Softfofa}
\eeq
Similarly, for collinear singularities, after the decoupling from the anti-collinear line
and from wide-angle soft gluons, it becomes clear that collinear divergences are
uniquely  associated with each incoming hard particle, and they can be simulated
with a {\it jet function} of the form
\beq
 {\cal J} \bigg( \frac{(p \cdot n)^2}{n^2 \mu^2}, 
  \alpha_s (\mu^2), \eps \bigg)  u_s (p) \, \, \equiv \, 
  \langle 0 \, | \, T \Big[ \Phi_n (\infty, 0) \, \psi (0) \Big] \, | p,s \rangle \, ,
\label{Jqfofa}
\eeq
and a similar one for the antiquark. Contrary to the soft function in \eq{Softfofa}, the 
jet function in \eq{Jqfofa} does not contain colour correlations between the quark and 
the antiquark, but it is spin-dependent. In general, one will also need a definition for 
a gluon jet function: in order to provide it, one cannot simply replace the quark field 
by a gluon field in \eq{Jqfofa}, since the  result would not be gauge invariant. A 
possible definition bypassing this problem was proposed in~\cite{Becher:2010pd,
Magnea:2018ebr}, and is given by
\beq 
  g \, {\cal J}_{g}^{\mu \nu} \! \left( \frac{(k \cdot n)^2}{n^2 \mu^2}, 
  \alpha_s (\mu^2), \eps \right) \, \varepsilon^{(\lambda)}_\mu (k) 
  \, \equiv \, 
  \bra{0} T \Big[ \Phi_n (\infty, 0) \, {\rm i} D^\nu \, 
  \Phi_n (\infty, x) \Big] \ket{k, \lambda} \Big|_{x=0} \, ,
\label{Jgdef}
\eeq
where the Wilson line in the $n$ direction acts directly as the gluon source.

Finally, we turn to the issue of subtracting the potential double counting of the
momentum regions involving gluons that are both soft and collinear to the incoming 
partons. Such regions indeed appear both in the soft function in \eq{Softfofa} and 
in the jets, Eqns.~\ref{Jqfofa} and \ref{Jgdef}. It would appear to be non-trivial to 
subtract collinear configurations from the soft function, however the simple form 
of the jet functions leads to a straightforward proposal for subtracting their soft 
limits: one may simply divide each jet by its own soft approximation, which is 
naturally given by a very similar matrix element, with the field of the hard parton 
replaced by its own Wilson line. Thus we define the {\it eikonal jet function}
\beq
  {\cal J}_{\rm E} \bigg( \frac{(\beta \cdot n)^2}{n^2}, 
  \alpha_s (\mu^2), \eps \bigg)  & \equiv & \langle 0 | T \Big[ \, \Phi_\beta (\infty, 0) \, 
  \Phi_n (\infty, 0) \Big]\, | 0 \rangle \, .
\label{Jqeikfofa}
\eeq
As expected from the general properties of the soft approximation, eikonal jets
are independent of spin, and they depend on colour only through the representation
in which the Wilson lines are defined. The fact that the proper way to `subtract'
the soft-collinear region is to {\it divide} by the eikonal jets will be better justified
in what follows, where we will see that jet and eikonal jet functions {\it exponentiate},
and they are normalised to unity at lowest order: by taking a ratio, we are effectively 
subtracting their perturbative exponents.

Not suprisingly, closely related definitions of soft and jet functions arise in the SCET
approach to infrared factorisation (see, for example, Refs.~\cite{Becher:2009qa,
Beneke:2002ni,Feige:2014wja,Becher:2014oda,Bauer:2002nz}), and they have been 
extensively investigated both at amplitude and at cross-section level\footnote{We 
note in passing a significant difference between the soft functions we discuss 
here for fixed-angle scattering amplitudes and those featuring at cross-section 
level. For cross sections, soft functions are not universal, since the phase-space 
integration of real soft radiation is weighted with the selected observable. Thus, in
principle, every observable requires a new calculation. Representative examples
beyond NLO can be found in Refs.~\cite{Becher:2005pd,Jouttenus:2011wh,Kelley:2011ng,
Li:2011zp,Becher:2012za,Czakon:2013hxa,Bonvini:2014tea,Boughezal:2015eha,
Echevarria:2015byo,Campbell:2017hsw,Moult:2018jzp,Angeles-Martinez:2018mqh,
Bell:2018oqa,Liu:2020wmp}, but the literature is extensive. Cross-section-level jet functions, 
on the other hand, retain a degree of universality, and have been computed to high orders 
in SCET and in QCD, see for example Refs.~\cite{Becher:2006qw,Jain:2008gb,Bruser:2018rad,
Hoang:2019fze,Banerjee:2018ozf}. Finally, we note that, in the context of resummation,
it is useful to define non-universal initial state functions -- essentially modified parton 
distribution functions -- that reflect the phase-space constraint of the observable
under study. This technique was pioneered in Ref.~\cite{Sterman:1986aj}. In SCET,
the appropriate functions are called {\it beam functions}, and they have been extensively
studied in recent years~\cite{Stewart:2009yx,Stewart:2010qs,Gaunt:2014xga,Gaunt:2014cfa,
Gaunt:2014xxa,Melnikov:2018jxb,Melnikov:2019pdm,Behring:2019quf,Gaunt:2020xlc,
Ebert:2020unb}.}. In the context of SCET, soft and jet functions are defined in terms of 
soft and collinear fields, which directly enter the SCET lagrangian. Practical calculations 
at leading power in the two formalisms are closely related, but the double-counting 
problem for the soft-collinear region is dealt with in a different way, using the so-called 
{\it zero-bin} subtraction, discussed for example in Refs.~\cite{Manohar:2006nz,
Idilbi:2007ff,Idilbi:2007yi,Chiu:2009yx}.

\begin{figure}
\centering
  \includegraphics[height=3cm,width=12cm]{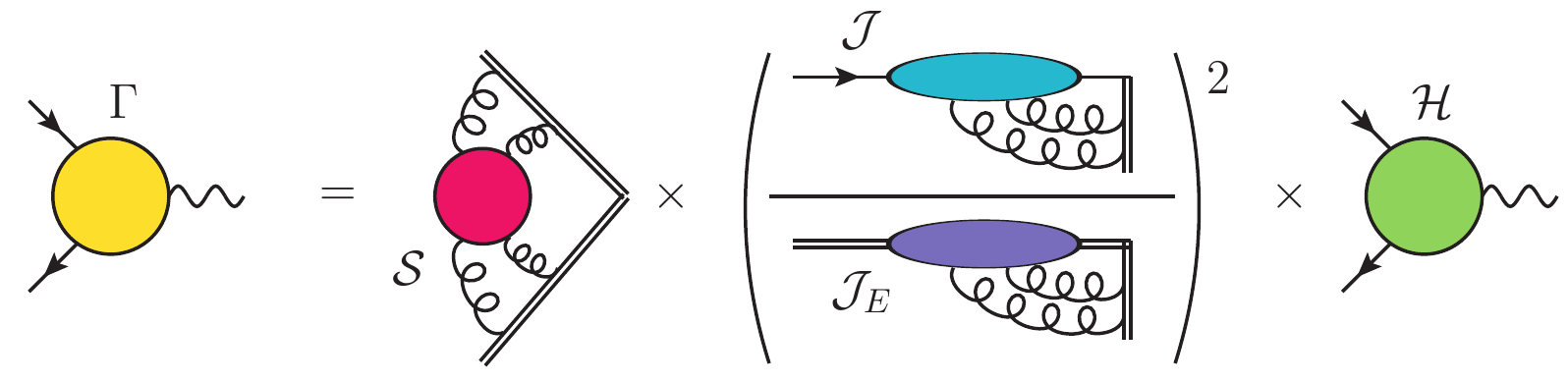}
  \caption{Pictorial representation of soft-collinear factorisation for the 
  quark form factor.}
\label{fofafacfig}
\end{figure}

Before making several comments about the physical meaning and implications of 
our definitions in \eq{Softfofa}, \eq{Jqfofa}, and \eq{Jqeikfofa}, we now pause to 
write down a precise form for the infrared factorisation of the form factor in terms 
of our soft and jet functions, which is illustrated pictorially in Fig.~\ref{fofafacfig}.
We write
\beq
\label{fofafac}
  \Gamma \left(\frac{Q^2}{\mu^2}, \alpha_s (\mu^2), \eps \right) & = & 
  \prod_{i=1}^2 \, \frac{{\cal J}_i \left( \frac{(p_i \cdot n_i)^2}{n_i^2 \mu^2},
  \alpha_s (\mu^2), \eps  \right)}{{\cal J}_{E, \, i} \,
  \left( \frac{(\beta_i \cdot n_i)^2}{n_i^2} , \alpha_s(\mu^2), \eps  \right)} \\
  && \hspace{2cm} \times \, 
  {\cal S} \Big( \beta_1 \cdot \beta_2, \alpha_s(\mu^2), \eps \Big) \,
  {\cal H} \left( \frac{Q^2}{\mu^2}, \frac{(p_i \cdot n_i)^2}{n_i^2 \mu^2}, 
  \alpha_s(\mu^2), \eps \right) \, .
  \nonumber
\eeq
Here the hard function ${\cal H}$, which is finite as $\eps \to 0$, plays the role of a 
matching coefficient: order by order in perturbation theory, soft and collinear factors
miss or misrepresent finite contributions from hard exchanges, and these can be 
reinstated by computing ${\cal H}$, which is defined by subtraction of the singular 
contributions from the full form factor. A detailed illustration of this factorisation at 
the one-loop order was given in Ref.~\cite{Dixon:2008gr}.
It is important to spend some time on \eq{fofafac}, since it forms the basis for further
generalisations to fixed-angle scattering amplitudes. We begin by noting that the
factorisation of collinear effects into jet functions has required the introduction of
a vector $n_i^\mu$ ($i = 1,2$) for each hard parton. The introduction of these
vectors can be understood in three complementary ways.
\begin{itemize}
\item First of all, such vectors are needed in order to ensure gauge invariance
of the jet definition, \eq{Jqfofa}. In the absence of the Wilson line, the jet function
would transform as the quark field under a gauge transformation. The Wilson line
takes this gauge variation at the origin, and transports it out to infinite distance, 
where it vanishes, since we are working perturbatively and considering only
globally trivial gauge fields\footnote{It should be noted that neglecting `large' 
gauge transformations, which do not vanish at infinity, can be considered as 
part of the problem that originates infrared divergences in the first place. This
point of view is developed in the {\it celestial} approach to infrared singularities,
see for example~\cite{Strominger:2017zoo}.}.
\item The one-loop collinear calculation leading to \eq{finnumcoll} illustrates
that the vectors $n_i$, {\it de facto}, replace the anti-collinear parton. Collinear 
gluons absorbed at late times (or emitted at early times) are effectively blind to the
detailed structure of non-collinear parts of the amplitude, so that those parts can 
be replaced, at leading-power accuracy, with a Wilson-line {\it absorber}. We recall 
again that it is important in principle to keep $n_i^2 \neq 0$ (a typical 
choice~\cite{Collins:1989bt} for the form factor being $n_1 = \beta_1 - \beta_2 
= - n_2$). This avoids the inclusion of spurious collinear divergences in the jet 
function, arising from gluon emission from the Wilson line, which are not present 
in the form factor and would need to be subtracted.
\item A third way to understand the $n_i$'s is to think of them as {\it factorisation
vectors}, generalising the idea of a factorisation scale. Since we are trying to 
distinguish collinear emissions from wide-angle ones, it is perhaps not surprising
that one needs to introduce a vector in order to draw a cone around the direction
of the emitter, specified by $\beta_i$: in this interpretation, one may decide to
assign a gluon to the collinear region if, say, $\beta_i \cdot k < n_i \cdot k$.
\end{itemize}

\noindent
The next set of observations on Eqs.~(\ref{Softfofa} - \ref{fofafac}) concerns their
functional dependences. The fact that jet functions can depend only on the variable
$(p \cdot n)^2/(n^2 \mu^2)$ (and similarly eikonal jets can only depend on $(\beta 
\cdot n)^2/n^2$) stems from their invariance under rescalings of the form $n \to 
\kappa n$, which is a property of semi-infinite straight Wilson lines stretching 
along the direction $n$, with $n^2 \neq 0$. Indeed, a correlator of two Wilson lines, 
in directions $n_i$, $i = 1,2$, with $n_i^2 \neq 0$, can only depend on the Minkowskian
cusp angle between the two directions, defined in a Lorentz-invariant way by 
$\gamma_{12} = (n_1 \cdot n_2)^2/(n_1^2 n_2^2)$. In the case of light-like Wilson 
lines, such as those appearing in the soft function and in the eikonal jet function, 
the cusp angles are ill-defined, so that one might expect correlators involving such 
lines to be just  numbers, free of kinematic dependence. As we have seen in 
\secn{eikint}, however, this is not the case, since the rescaling symmetry of 
Wilson lines correlators is broken by the presence of an extra collinear divergence. 
As a consequence, the soft function for massless partons acquires a dependence 
on $\beta_1 \cdot \beta_2$, as shown in \eq{Softfofa}, and the eikonal jets 
correspondingly acquire a dependence on $(\beta_i \cdot n_i)^2/n_i^2$. It is 
clear from the outset that this `anomalous' scale dependence must be comparatively 
simple, since it is entirely associated with the soft-collinear double pole of the form 
factor. This double singularity is governed by the light-like cusp anomalous dimension, 
as we will see in \secn{ResuForm}.

A consequence of this analysis, which will be crucial in applications to multi-parton 
amplitudes, is the following. If we can construct quantities where soft-collinear 
double poles cancel, then, for these quantities, the anomalous scale dependence 
must cancel as well. By inspection of \eq{fofafac}, we can readily identify two 
such quantities.
\begin{itemize}
\item The ratio of the partonic jet and the corresponding eikonal jet, 
${\cal J}_i/{\cal J}_{E, \, i}$, which is responsible for hard-collinear corrections, 
carrying a single collinear pole per loop. In this case, at least at one loop, it
is easy to envisage how the cancellation of the anomalous scale dependence 
takes place: both jets depend logarithmically on their arguments, and one 
can use
\beq
  \log \bigg( \frac{(p_i \cdot n_i)^2}{n_i^2 \mu^2} \bigg) - 
  \log \bigg( \frac{(\beta_i \cdot n_i)^2}{n_i^2} \bigg) \, \sim \, 
  \log \bigg( \frac{Q^2}{\mu^2} \bigg) \, ,
\label{ratjet}
\eeq
as the two logarithms must have the same coefficient, which stems from the 
soft-collinear double pole. Explicit expressions for the quark jet and eikonal jet,
at one loop, verifying \eq{ratjet}, are given in Ref.~\cite{Dixon:2008gr}.
\item Less trivially, one can define a {\it reduced soft function}, by taking the ratio
of the soft function with the product  of the two eikonal jets,
\beq
  \widehat{\cal S} \, \big( \rho_{12}, \alpha_s (\mu^2), \epsilon \big) \, \equiv \,
  \frac{{\cal S} \Big( \beta_1 \cdot \beta_2, \alpha_s(\mu^2), \eps \Big)}{{\cal J}_{E, 
  \, 1} \! \left( \frac{(\beta_1 \cdot n_1)^2}{n_1^2} , \alpha_s(\mu^2), \eps  \right) \,
  {\cal J}_{E, \, 2} \! \left( \frac{(\beta_2 \cdot n_2)^2}{n_2^2} , 
  \alpha_s(\mu^2), \eps  \right)} \, ,
\label{redsoft}
\eeq
which describes soft radiation at wide angles with respect to the hard partons.
Also this quantity provides a single (soft) pole per loop, and therefore must be
rescaling invariant under $\beta_i \to \kappa_i \beta_i$. This is achieved if 
$\widehat{\cal S}$ is a function of an invariant variable, which we take here
to be\footnote{It is possible, and often useful, to attach to scalar products in 
\eq{rho12} appropriate phases, dictating the rules for analytic continuation
from time-like to space-like kinematics, as done for example in~\cite{Gardi:2009qi}.
Since here we are only interested in  scaling properties, for simplicity we omit
the phase information.} 
\beq
  \rho_{12} \, \equiv \, \frac{ \big( \beta_1 \cdot \beta_2 \big)^2 \, n_1^2 \,  
  n_2^2}{\big( \beta_1 \cdot n_1 \big)^2 \big( \beta_2 \cdot n_2 \big)^2} \, .
\label{rho12}
\eeq
Once again, this happens naturally if the dependence of each function in the
ratio~(\ref{redsoft}) is simply logarithmic, and the logarithms have the same 
coefficients, tied to the soft-collinear double pole. Clearly, the cancellation implied
in \eq{redsoft} is more intricate than the jet case: imposing this constraint will
have highly non-trivial consequences in the case of multi-parton amplitudes.
\end{itemize}
\noindent
To conclude our analysis of \eq{fofafac}, we note the mechanism of cancellation
for the dependence on the factorisation vectors $n_i$: poles in $\eps$ carrying
$n_i$ dependence must cancel between the partonic and the eikonal jet, through
a mechanism akin to \eq{ratjet}. On the other hand the hard function ${\cal H}$,
which is finite as $\eps \to 0$, must depend on $p_i \cdot n_i$ (or equivalently
$\beta_i \cdot n_i$) in order to cancel the same dependence in the finite parts 
of the jet functions.


\subsection{Evolution and resummation of infrared poles}
\label{ResuForm}

Based on the discussion at the beginning of the present Section, we expect 
the soft-collinear factorisation displayed in \eq{fofafac} to lead to an evolution 
equation, and thus to a resummation, in this case of infrared poles in 
dimensional regularisation. The basic observation is that, in order to achieve
that double factorisation into hard, soft and collinear factors, it was necessary
to introduce a factorisation scale separating soft and hard momenta (which
we can take equal to the renormalisation scale), and also to introduce the 
`factorisation vectors' $n_i$, separating collinear from wide-angle radiation.
Needless to say, the full form factor does not depend on any of these
quantities, a statement which directly leads to two evolution equations,
which can then be fruitfully combined.

Beginning with the factorisation/renormalisation scale dependence, the fact that 
we are considering a matrix element of a conserved current leads directly to the
RG equation  
\beq
  \mu \, \frac{d }{d \mu} \, \Gamma \bigg( \frac{Q^2}{\mu^2}, 
  \alpha_s (\mu^2), \eps \bigg)  \, = \, 0 \, .
\label{RGfofa}
\eeq
If we now rewrite \eq{fofafac} as
\beq
  \Gamma \bigg( \frac{Q^2}{\mu^2} \bigg) \, = \, \widehat{\cal S} \big( \rho_{12} \big) \,
  {\cal J}_1 \bigg( \frac{( p_1 \cdot n_1)^2}{n_1^2 \mu^2} \bigg) \, 
  {\cal J}_2 \bigg( \frac{( p_2 \cdot n_2)^2}{n_2^2 \mu^2} \bigg) \,
  {\cal H} \bigg( \frac{Q^2}{\mu^2}, \frac{( p_i \cdot n_i )^2}{n_i^2 \mu^2} \bigg) \, ,
\label{fofafac2}
\eeq
where we omitted the dependence on the coupling and on $\eps$ for brevity, we 
can define three anomalous dimensions for ${\cal J}_i$, $\widehat{\cal S}$, and 
${\cal H}$, by taking logarithmic derivatives with respect to the scale, as
\beq
\label{andimdef}
  \mu \frac{d}{d \mu} \ln {\cal J} & = & - \, \gamma_{\cal J} (\alpha_s) \, , \nonumber \\
  \mu \frac{d}{d \mu} \ln \widehat{\cal S} & = & - \, \gamma_{\widehat{\cal S}} 
  \left( \rho_{12}, \alpha_s \right) \, ,\\
  \mu \frac{d}{d \mu} \ln{\cal H} & = & - \, \gamma_{\cal H} \left( \rho_{12}, 
  \alpha_s \right) \, .
  \nonumber
\eeq
Note that all anomalous dimensions are finite, since the functions we are differentiating
have a single pole per loop. Furthermore, note that the jet anomalous dimension 
$\gamma_{\cal J}$ depends only on the coupling: the partonic jet \eq{Jqfofa} is not
scale-less, therefore UV poles have no direct relation to IR poles, and the anomalous 
dimension is given by the coefficient of the pole in the UV counterterm; on the other
hand, the anomalous dimension for the reduced soft function has a residual kinematic
dependence, since, for scale-less quantities, UV singularities are strictly related
to IR ones: in this case, there is dependence on the rescaling-invariant variable
$\rho_{12}$, leftover after the cancellation of soft-collinear double poles. Finally, 
the renormalisation group equation~(\ref{RGfofa}) implies 
\beq
  \gamma_{\widehat{\cal S}} \left( \rho_{12}, \alpha_s \right) + 
  \gamma_{\cal H}  \left( \rho_{12}, \alpha_s \right) + 2 \gamma_{\cal J} (\alpha_s)
  \, = \, 0 \, ;
\label{cancandin}
\eeq
in particular, since $\gamma_{\cal J}$ is independent of $\rho_{12}$, the residual
kinematic dependence must cancel between $\gamma_{\cal H}$ and 
$\gamma_{\widehat{\cal S}}$.

Turning now to the dependence on the factorisation vectors $n_i$, first of all we note
that this dependence is always through the scalar variables $x_i = \left( \beta_i 
\cdot n_i \right)^2/n_i^2$ (recall that $p_i = \mu \beta_i$). Since the complete
form factor does not depend on $n_i$, we can write
\beq
  x_i \frac{\partial}{\partial x_i} \ln \Gamma \bigg( \frac{Q^2}{\mu^2}, 
  \alpha_s (\mu^2), \eps \bigg)  \, = \, 0 \, ,
\label{sudakev}
\eeq
where we let the derivative act on the logarithm of the form factor in order to turn
the factorisation in \eq{fofafac} into a sum of terms. Taking into account the fact
that the soft function does not depend upon $n_i$, we find that
\beq
  x_i \frac{\partial}{\partial x_i} \ln {\cal J}_i \, = \, - \frac{1}{2} \, x_i 
  \frac{\partial}{\partial x_i} \ln {\cal H} + x_i \frac{\partial}{\partial x_i} 
  \ln {\cal J}_{E, i} \, .
\label{upacklog}  
\eeq
\eq{upacklog} is very important, because it achieves the separation of a complicated
problem into two simpler ones. The first term on the {\it r.h.s.} of \eq{upacklog} can
be intricate at high orders, since ${\cal H}$ is a matching function, and thus inherits
some of the complexity of the full amplitude: on the other hand, ${\cal H}$ is finite
as $\epsilon \to 0$, so the same must be true for its derivative in \eq{upacklog}.
The second term on the {\it r.h.s.} of \eq{upacklog}, on the contrary, is divergent
as $\eps \to 0$ (indeed, it contains only poles in $\epsilon$, since ${\cal J}$ is a 
pure counterterm in dimensional regularisation); however, its functional form must 
be extremely simple: indeed, ${\cal J}$ would be just a number, with
no kinematic dependence whatsoever, were it not for the anomalous breaking of
rescaling invariance due to the collinear poles associated with the $\beta$ direction.
Order by order in perturbation theory, these soft-collinear poles are determined 
by the cusp anomalous dimension, and the ensuing kinematic dependence 
must be single-logarithmic\footnote{For a detailed treatment of the eikonal jet, 
see~\cite{Gardi:2009qi}; the first all-order perturbative analysis of this kind of matrix 
element was given in~\cite{Collins:1981uk}.}. It follows that the logarithmic derivative
of the eikonal jet ${\cal J}_i$ in \eq{upacklog} must be independent of the kinematic 
variable $x_i$. These considerations can be made explicit by defining two functions,
corresponding to the two terms on the {\it r.h.s.} of \eq{upacklog}. For the eikonal 
jet we define
\beq
  x_i \frac{\partial}{\partial x_i} \, \ln {\cal J}_{E, i} \, = \, \frac{1}{2} \, \ck 
  \Big( \alpha_s(\mu^2),\eps  \Big) \, ,
\label{Kdef}
\eeq
while for the hard matching function ${\cal H}$ we define 
\beq
  x_i \frac{\partial}{\partial x_i} \, \ln {\cal H} \, = \, - \, {\cal G}_i 
  \Big( x_i, \alpha_s(\mu^2), \eps  \Big) \, .
\eeq
\eq{upacklog} can then be written as
\beq
  x_i \frac{\partial}{\partial x_i} \, \ln {\cal J}_i \, = \, \frac{1}{2} \, \bigg[ \,
  \ck \Big( \alpha_s(\mu^2), \eps \Big) \, + \, 
  {\cal G}_i \Big( x_i, \alpha_s(\mu^2), \eps \Big)
  \bigg] \, ,
\label{KplusGforJ}
\eeq
\eq{KplusGforJ} is an example of a class of equations associated with the soft-collinear
evolution of cross-sections and amplitudes in QCD, which we will call `Collins-Soper
equations' as they were first derived in Ref.~\cite{Collins:1981uk}; they were later 
widely used for resummations of Sudakov logarithms and infrared singularities 
(see, for example, \cite{Sterman:1986aj,Collins:1989bt,Contopanagos:1996nh,
Ravindran:2005vv,Ravindran:2006cg,Dixon:2008gr}). In this form, \eq{KplusGforJ}
appears rather uninteresting, describing evolution with respect to an unphysical 
quantity. As is typically the case for RG equations, however, \eq{KplusGforJ}, 
when combined with \eq{RGfofa}, leads directly to a much more powerful and 
interesting equation for the evolution of the full form factor with respect to the 
physical scale $Q^2$. Indeed, we can write
\beq
 Q^2 \frac{\partial}{\partial Q^2} \, \ln \Gamma \bigg( \frac{Q^2}{\mu^2}, 
  \alpha_s (\mu^2), \eps \bigg)  & = & Q^2 \frac{\partial}{\partial Q^2} \, \ln {\cal H} \, +
  \, \sum_{i = 1}^2 Q^2 \frac{\partial}{\partial Q^2} \, \ln {\cal J}_i 
  \nonumber \\
  & = & - \, \mu^2 \frac{\partial}{\partial \mu^2} \, \ln {\cal H} \, +
  \, \sum_{i = 1}^2 x_i \frac{\partial}{\partial x_i} \, \ln {\cal J}_i \, ,
\label{Evofofa1}
\eeq
where in the first line we have used the fact that purely eikonal functions do not
depend on the scale $Q^2$, and in the second line we used dimensional analysis,
and the fact that jets depend on $Q^2$ only through $(p \cdot n)^2$. Now for the
$\mu$ dependence of the hard function we can use \eq{andimdef} and \eq{cancandin},
with the result
\beq
 - \, \mu^2 \frac{\partial}{\partial \mu^2} \, \ln {\cal H} & = & 
 - \frac{1}{2} \mu \frac{d}{d \mu} \, \ln {\cal H} \, + \, \frac{1}{2} \beta (\epsilon, \alpha_s)
  \frac{d}{d \alpha_s} \ln {\cal H} \nonumber \\
  & = & 
  \frac{1}{2} \beta (\epsilon, \alpha_s) \frac{d}{d \alpha_s} \ln {\cal H} - 
  \frac{1}{2}  \gamma_{\widehat{\cal S}} - \gamma_{\cal J} \, ,
\label{Evofofa2}
\eeq
while \eq{KplusGforJ} gives the $x_i$ dependence of the jets. The results can be 
summarised in a new, much more powerful Collins-Soper equation for the
full form factor
\beq
 Q^2 \frac{\partial}{\partial Q^2} \, \ln \Gamma \bigg( \frac{Q^2}{\mu^2}, 
  \alpha_s (\mu^2), \eps \bigg)  \, = \,   \frac{1}{2} \, \bigg[ \,
  K \Big( \alpha_s(\mu^2), \eps \Big) \, + \, 
  G \bigg( \frac{Q^2}{\mu^2}, \alpha_s (\mu^2), \eps \bigg)     
  \bigg] \, ,
\label{Evofofafin}
\eeq
where we defined
\beq
  K \Big( \alpha_s(\mu^2), \eps \Big) & \equiv & 
  2 {\cal K} \Big( \alpha_s(\mu^2), \eps \Big) \, , \nonumber \\
  G \bigg( \frac{Q^2}{\mu^2}, \alpha_s (\mu^2), \eps \bigg) & \equiv &
  \beta (\epsilon, \alpha_s) \frac{d}{d \alpha_s} \ln {\cal H}
  - \gamma_{\cal S} - 2 \gamma_{\cal J} + \frac{1}{2} \sum_{i = 1}^2 
  {\cal G}_i \Big( x_i, \alpha_s(\mu^2), \eps \Big) \, .
\label{defKG}
\eeq
Once again, $K$ is a pure counterterm, and independent of kinematics, while $G$
retains kinematic information, but is finite as $\epsilon \to 0$. There is a final nugget 
of information to be  extracted from \eq{RGfofa}: since the form factor is not affected
by an overall renormalisation, the functions $K$ and $G$ can renormalise additively,
but their renormalisations must cancel. Thus
\beq
  \mu \frac{d}{d \mu} \, G \bigg( \frac{Q^2}{\mu^2}, \alpha_s (\mu^2), \eps \bigg) & = &
  - \, \mu \frac{d}{d \mu} \, K \Big( \alpha_s(\mu^2), \eps \Big) \nonumber \\
  & = & 
  - \, \beta (\epsilon, \alpha_s) \, \frac{d}{d \alpha_s} K \Big( \alpha_s(\mu^2), \eps \Big)
  \, \equiv \, \gamma_K \big( \alpha_s (\mu^2) \big) \, ,
\label{RGadd}
\eeq
where in the second line we used the fact that $K$ has no explicit scale dependence.

For the purposes of our review, \eq{RGadd} can be taken as an operational definition
of the {\it light-like cusp anomalous dimension} $\gamma_K (\alpha_s)$. This is, however,
such an important object for infrared studies of perturbative gauge theories that it 
deserves a brief dedicated discussion, before we continue with our analysis of the
exponentiation of the form factor.

The function $\gamma_K (\alpha_s)$ appears in several cornerstone gauge-theory
calculations. As shown in Ref.~\cite{Korchemsky:1988si}, it gives the coefficient of the 
soft singularity of DGLAP splitting functions (discussed here in \secn{CollDiv} and later
in \secn{Subtra}) as the collinear momentum fraction $z \to 1$, to all orders in perturbation 
theory. As a consequence, it resums leading threshold logarithms in many crucial
hadronic cross sections (see, for example,~\cite{Sterman:1986aj,Catani:1989ne,
Korchemsky:1992xv,Korchemsky:1993uz}): in that context, it is often denoted by 
$A(\alpha_s)$. A closely related function appears in the resummation of leading 
transverse momentum logarithms in the hadronic production of colour-singlet final 
states~\cite{Collins:1984kg,Catani:2003zt}, however the two functions begin to differ 
at the three-loop order~\cite{Becher:2010tm}. Finally, $\gamma_K (\alpha_s)$ plays 
a crucial role for Regge behaviour in the high-energy limit~\cite{Korchemsky:1993hr,
Korchemskaya:1994qp,Korchemskaya:1996je,DelDuca:2011ae,DelDuca:2011wkl}, 
and of course, as we discuss below, for the all-order structure of infrared divergences 
in gauge-theory amplitudes.

The reason for the ubiquitous and crucial role played by the cusp anomalous dimension is 
not difficult to understand, and emerges from our discussion in \secn{WiliEik}. In the soft 
approximation, hard emitters can be replaced by Wilson lines directed along their classical 
trajectories; then, by the mechanism discussed here in \secn{eikint}, infrared poles of the 
original amplitude can be replaced by ultraviolet poles of the corresponding Wilson-line 
correlator~\cite{Korchemsky:1985xj}; such correlators (up to collinear divergences) are 
multiplicatively renormalisable~\cite{Polyakov:1980ca,Arefeva:1980zd,Brandt:1981kf,
Korchemskaya:1992je}; finally, one finds that the anomalous dimension controlling the 
ultraviolet divergences arising when two light-like Wilson lines meet at a cusp is given 
precisely by $\gamma_K(\alpha_s)$. More generally, as noted above in \secn{FactoForm}, 
when two straight Wilson lines directed along non-light-like directions $n_1^\mu$ and $n_2^\mu$ 
meet, forming a Minkowskian angle given by $\gamma_{12} = (n_1 \cdot n_2)^2/(n_1^2 n_2^2)$, 
the ultraviolet divergences of the corresponding correlator are controlled by a function 
$\Gamma_{\rm cusp} (\gamma_{12}, \alpha_s)$, sometimes called {\it angle-dependent} 
cusp anomalous dimension~\cite{Korchemsky:1985xu}. Not surprisingly, this function 
plays an essential role for the infrared behaviour of scattering amplitudes and form 
factors involving massive particles~\cite{Korchemsky:1991zp,Neubert:1993mb,
Manohar:2000dt,Grozin:2004yc}. To all orders, the function $\Gamma_{\rm cusp} 
(\gamma_{12}, \alpha_s)$ displays a collinear singularity as $n_i^2 \to 0$, or 
$\gamma_{12} \to \infty$, diverging as the logarithm of $\gamma_{12}$. The 
coefficient of this logarithm gives the light-like anomalous dimension $\gamma_K 
(\alpha_s)$.

Quite naturally, perturbative calculations of these vital anomalous dimensions
have been the focus of intense activity in the past decades. The light-like cusp
anomalous dimension can be extracted from calculations of collinear splitting 
kernels, so the two-loop result can be traced to Ref.~\cite{Curci:1980uw}, and,
in the context of resummation, to Ref.~\cite{Kodaira:1981nh}; similarly, the 
three-loop result emerges from Ref.~\cite{Moch:2004pa} (see also~\cite{Berger:2002sv}). 
At four loops, the result has been computed directly from correlation functions 
with lagrangian insertions in Ref.~\cite{Henn:2019swt}, and extracted from form 
factor calculations in Ref.~\cite{vonManteuffel:2020vjv}. The full angle-dependent 
cusp anomalous dimension was computed at two loops in Ref.~\cite{Korchemsky:1987wg} 
(see also~\cite{Kidonakis:2009ev}), and at three loops in Refs.~\cite{Grozin:2014hna,
Grozin:2015kna}. At four loops, efforts are ongoing, and a wealth of partial information 
is already available~\cite{Henn:2016wlm,Henn:2016men,Grozin:2017css,Bruser:2019auj,
Bruser:2020bsh}. Perhaps most remarkably, in the special case of ${\cal N} = 4$ 
Super-Yang-Mills theory, the integrability of the planar limit gives access to fully 
non-perturbative information: gluon amplitudes can be studied in the strong-coupling 
regime~\cite{Alday:2007hr}, the light-like cusp anomalous dimension can be 
defined~\cite{Alday:2007mf}, and it can be shown that it obeys a non-perturbative 
equation~\cite{Beisert:2006ez}, which can be analysed both at weak~\cite{Bern:2006ew,
Correa:2012nk,Henn:2012qz} and at strong coupling~\cite{Basso:2007wd,Correa:2012hh}. 
We will further discuss the remarkable role played by the cusp for the infrared limit 
of gauge theory amplitudes in \secn{MultiPart}, where we will also perform in detail 
the one-loop, angle-dependent calculation.

Returning now to the form factor problem, we note that \eq{RGadd} can be readily 
solved for $K$, with the result
\beq
  K \Big( \alpha_s(\mu^2), \eps \Big) \, = \, - \frac{1}{4} \int_0^{\mu^2} 
  \frac{d \lambda^2}{\lambda^2} \, \gamma_K \big( \overline{\alpha} (\lambda^2, \epsilon) 
  \big) \, .
\label{KfromgamK}
\eeq
Alternatively, it is possible to proceed order by order in perturbation theory, and 
recursively determine the perturbative coefficients of $K$ in terms of those of
the $\beta$ function and of the cusp anomalous dimension $\gamma_K$, as
was done in Ref.~\cite{Magnea:2000ss}. Up to three loops, the result takes 
the form
\beq
  K (\as, \e) & = & \frac{\alpha_s}{\pi} \, 
  \frac{\gamma_K^{(1)}}{4 \epsilon} \, + \left(\frac{\alpha_s}{\pi}\right)^2 \,
  \left( \frac{\gamma_K^{(2)}}{8 \epsilon} -
  \frac{b_0 \, \gamma_K^{(1)}}{32 \epsilon^2} \right) \nonumber \\
  && \, + \left(\frac{\alpha_s}{\pi}\right)^3 \left( \frac{\gamma_K^{(3)}}{12 \epsilon} -
  \frac{b_0 \, \gamma_K^{(2)} + b_1 \, \gamma_K^{(1)}}{48 \epsilon^2} 
  + \frac{b_0^2 \, \gamma_K^{(1)}}{192 \epsilon^3} \right) \, + \ord (\alpha_s^4) \, .
\label{KNNLO}
\eeq
The function $K$ has a long history in the context of perturbative QCD (see, 
for example, Ref.~\cite{Collins:1981uk}), and plays an important role also for 
multi-particle scattering amplitudes, and in the high-energy limit, as we will 
see in \secn{MultiPart}.

Solving \eq{Evofofafin} is a standard exercise, but in this case the consistent 
use of dimensional regularisation yields a very significant simplification: indeed, 
working with $\epsilon < 0$, one can use the fact that the $d$-dimensional
strong coupling vanishes in the infrared to impose the simple boundary 
condition
\beq
  \Gamma \Big( 0, \alpha_s (\mu^2), \epsilon \Big) \, = \, 
  \Gamma \Big( 1, \overline{\alpha} (0, \epsilon), \epsilon \Big) \, = \,  1 \, ,
\label{initcon}
\eeq
where on the {\it l.h.s} we have set $Q^2 = 0$ for fixed $\mu$, while in the 
second step we have first set $\mu^2 = Q^2$, and then taken $\mu \to 0$.
Thanks to \eq{initcon}, the form factor is a pure exponential with no prefactor,
and can be written as~\cite{Magnea:1990zb}
\beq
  \Gamma \bigg( \frac{Q^2}{\mu^2}, \alpha_s (\mu^2), \epsilon \bigg) \! \! & = & \! \!
  \exp \Bigg[ \frac{1}{2} \int_0^{- Q^2} \frac{d \xi^2}{\xi^2} 
  \Bigg( K \Big( \alpha_s (\mu^2), \epsilon \Big)  + 
  G \Big( \! - \!1, \overline{\alpha} ( \xi^2, \epsilon ), \epsilon \Big) \nonumber \\
  && \hspace{2cm}
  + \, \frac{1}{2} \int_{\xi^2}^{\mu^2} \frac{d \lambda^2}{\lambda^2} \,
  \gamma_K \big( \overline{\alpha} ( \lambda^2, \epsilon ) \big) \Bigg) \Bigg] \, ,
\label{fofasol1}
\eeq
where we have used \eq{RGadd} to evolve $G$ from the scale $\xi^2$ to 
the scale $\mu^2$, and we have chosen to integrate to the scale $(-Q^2)$
to emphasise that the form factor is real for negative $Q^2$: in this way,
$G(-1)$ is real, and all phases can be explicitly obtained by analytic continuation.
We note that in \eq{fofasol1}, in its present form, one needs a non-trivial cancellation 
of ill-defined contributions between the first term, which diverges at the lower 
limit of integration, since $K$ does not depend on $\xi$, and the contribution 
of the upper limit of integration of the $\lambda$ integral, which also does 
not depend on $\xi$. This cancellation can be performed analytically by 
using \eq{KfromgamK}, which easily leads to the final expression for the 
form factor,
\beq
  \hspace{-8mm}
  \Gamma \bigg( \frac{Q^2}{\mu^2}, \alpha_s (\mu^2), \epsilon \bigg) 
  \! \! \! \! & = & \! \! \! \!
  \exp \Bigg[ \frac{1}{2} \int_0^{- Q^2} \frac{d \xi^2}{\xi^2} 
  \Bigg(  \! G \Big( \! - \! 1, \overline{\alpha} ( \xi^2, \epsilon ), \epsilon \Big) - 
  \frac{1}{2} \, \gamma_K \big( \overline{\alpha} ( \xi^2, \epsilon ) \big) 
  \ln \! \bigg( \frac{-Q^2}{\xi^2} \bigg) \!\! \Bigg) \Bigg] \! .
\label{fofasol2}
\eeq
As promised, all the ingredients entering the exponent in \eq{fofasol2} are 
finite as $\epsilon \to 0$, and all infrared poles, to all orders in perturbation
theory, are generated by the scale integration.

It is worthwhile at this point to pause for a few considerations on \eq{fofasol2},
which is the simplest case of exponentiation of infrared poles for non-abelian
scattering amplitudes\footnote{Early references for the exponentiation of QCD
form factors include~\cite{Sterman:1977pz,Collins:1980ih,Sen:1981sd}.}, 
anticipating many features that will be generalised to multi-particle amplitudes 
in \secn{MultiPart}.
\begin{itemize}
\item Perhaps the first point to emphasise is the fact that \eq{fofasol2} is 
{\it predictive}: we are not simply shifting the problem from the calculation 
of $\Gamma$ to the calculation of its logarithm. To see this, note that the
direct computation of the form factor order by order yields two infrared poles per 
loop, so that one finds $\epsilon^{- 2 n}$ poles at ${\cal O} (\alpha_s^n)$.
The exponent in \eq{fofasol2}, on the other hand, has only one infrared 
pole per loop beyond one loop: the logarithm of the form factor has singularities
up to $\epsilon^{-n -1}$ at ${\cal O} (\alpha_s^n)$. Clearly, all poles of 
the form $\alpha_s^n \epsilon^{-p}$ with $n+1 < p \leq 2 n$ are generated 
at lower orders and can be obtained by exponentiation: for example, the
one-loop calculation of the form factor predicts the leading poles to all orders 
in perturbation theory. One can in fact look at \eq{fofasol1} a little closer, and 
note that a singularity is generated in the exponent by each integration, so that one
immediately detects an exponentiated double pole. The remaining poles
in the exponent, $\epsilon^{-p}$ with $2 < p \leq n+1$, are generated by the
running of the the $d$-dimensional coupling: for example, note that the 
expansion of \eq{oloruco} in powers of $\alpha_s (\mu_0^2)$ is finite order
by order as $\epsilon \to 0$, but the integration of each term over the scale 
$\mu$ generates poles proportional to powers of $b_0$. Higher-order terms 
in the $\beta$ function have a similar effect. This predictive structure has 
been both verified and exploited in finite-order calculations up to four 
loops~\cite{Moch:2005tm,Dixon:2017nat,Henn:2019swt,vonManteuffel:2020vjv,
Agarwal:2021zft}.
\item These considerations lead to a second significant observation: the form 
factor - perhaps not surprisingly - becomes extremely simple in the case of 
gauge theories that are conformal in $d = 4$, such as ${\cal N} = 4$ Super-Yang-Mills
theory, and several of its possible deformations. For these theories, the 
four-dimensional $\beta$ function vanishes to all orders, so that one has
\beq
  \beta \big( \epsilon, \alpha_s \big) \, = \,  - 2 \epsilon \alpha_s \, ,
\label{betaconf}
\eeq
and \eq{treeruco} holds exactly. As a consequence, the scale integrals in \eq{fofasol1},
or equivalently \eq{fofasol2}, can be performed explicitly. Expanding the cusp
anomalous dimension and the the function $G$ as
\beq
 \gamma_K (\alpha_s) \, = \, \sum_{n = 1}^\infty \bigg( \frac{\alpha_s}{\pi} \bigg)^n
 \gamma_K^{(n)} \, , \qquad 
 G \big( -1, \alpha_s, \epsilon \big) \, = \, \sum_{n = 1}^\infty \bigg( \frac{\alpha_s}{\pi} \bigg)^n
 G^{(n)} (\epsilon) \, ,
\label{expcuspG}
\eeq
one readily finds~\cite{Bern:2005iz}
\beq
  \ln \Bigg[ \Gamma \bigg( \frac{Q^2}{\mu^2}, \alpha_s (\mu^2), \epsilon \bigg) \Bigg]
  \!\!\! & = &\!\!\!  - \, \frac{1}{2} \sum_{n = 1}^\infty \bigg( \frac{\alpha_s (\mu^2)}{\pi} 
  \bigg)^{\! n}
  \bigg( \frac{\mu^2}{- Q^2} \bigg)^{\! n \epsilon} 
  \Bigg[ \frac{\gamma_K^{(n)}}{2 \, n^2 \epsilon^2} + \frac{G^{(n)} (\epsilon)}{n \epsilon}
  \Bigg] \, .
\label{symfofa}
\eeq
Using \eq{treeruco}, one easily sees that \eq{symfofa} displays exact RG invariance 
(it is independent of $\mu$), as expected. \eq{symfofa} has important consequences:
indeed, it is sufficient to determine the infrared singularity structure of all scattering 
amplitudes in ${\cal N} = 4$ SYM in the planar limit. This is easily understood by 
noting that, for planar gluon amplitudes, soft gluons can only connect two external 
states that are consecutive in the selected cyclic ordering, so that the infrared 
divergences of the full planar amplitude are just given by a product of (gluon) form 
factors. This fact was exploited in Ref.~\cite{Bern:2005iz} to constrain the {\it `BDS'}
ansatz for $n$-point planar amplitudes in ${\cal N} = 4$ SYM, which was later proved 
to give the exact all-order result for $n = 4$ and $n = 5$~\cite{Drummond:2007au,
Drummond:2008vq}.
\item A non-trivial practical feature of \eq{fofasol2} is the fact that it can be directly
compared to finite-order results obtained evaluating Feynman diagrams in dimensional 
regularisation. Using different infrared regulators, the exponential containing singular
terms is multiplied by an `initial condition' which depends on a factorisation scale, and
in general admits its own perturbative expansion. Using \eq{fofasol2}, on the other 
hand, a finite-order calculation can directly be compared to the corresponding expansion 
of the exponential, thus extracting the perturbative coefficients of $\gamma_K$ 
and  $G$. Non-singular contributions to the form factor are encoded in ${\cal O} 
(\epsilon^p)$ terms of $G$, with $p > 0$, and they also `exponentiate', yielding
a weak prediction for higher-order finite contributions~\cite{Parisi:1979xd,Sterman:1986aj,
Eynck:2003fn,Ahrens:2008qu}. 
\item A final feature of \eq{fofasol2} which is worth noting is the fact that it allows
for a straightforward extraction of the analytic continuation of the form factor from
negative values of $Q^2$ (where $\Gamma$ is real) to positive ones~\cite{Magnea:1990zb}. 
Indeed, the dependence on $Q^2$  is just in the upper limit of integration, so that the 
ratio of the time-like form factor to the space-like one is simply given by the exponential
of the same integral, stretching from $+ Q^2$ to $- Q^2$. Since at the origin one only
finds integrable singularities, one can deform the integration contour to a half-circle
of radius $Q^2$ in the complex $\xi^2$ plane, showing that this ratio is completely
dominated by perturbative contributions in asymptotically free theories. Furthermore,
one can show~\cite{Magnea:1990zb} that the poles of this ratio can be collected in 
an overall divergent phase, given by the counterterm function $K$,
\beq
  \frac{\Gamma \big( Q^2, \alpha_s \big)}{\Gamma \big( \! - \! Q^2, \alpha_s \big)} 
  \Bigg|_{\rm poles} \, = \, \exp \bigg[ {\rm i} \, \frac{\pi}{2} \, K\big( \alpha_s(Q^2), \epsilon 
  \big) \bigg] \, ,
\label{coulombphase}
\eeq
so that the modulus of the ratio is finite. This is relevant for phenomenological 
applications, since the modulus of the ratio (or related quantities) appears in 
cross-sections for vector boson or Higgs production~\cite{Parisi:1979xd,
Sterman:1986aj,Eynck:2003fn}, and `resums' large constants such as the 
factor of $\pi^2$ appearing in \eq{pisquare}. In the case of ${\cal N} = 4$ SYM, 
starting from \eq{symfofa}, one can also get a strikingly simple and elegant expression
for finite terms, which are given by
\beq
  \left| \frac{\Gamma \big( Q^2, \alpha_s \big)}{\Gamma \big( \! - \! Q^2, \alpha_s \big)} 
  \right|^2 \, = \, \exp \bigg[ \frac{\pi^2}{4} \gamma_K (\alpha_s) \bigg] \, .
\label{ratfofasym}
\eeq
Note that the {\it r.h.s.} is independent of $Q^2$, as expected for a four-dimensional 
quantity in a conformal theory. Since all quantities in \eq{ratfofasym} are well-defined 
at the non-perturbative level,  and since the {\it r.h.s.} provides a finite, unambiguous 
resummation of perturbation theory, \eq{ratfofasym} can be argued to be an 
exact non-perturbative result. In some cases, such non-perturbative results 
can be tested against strong-coupling calculations based on the AdS/CFT 
correspondence~\cite{Alday:2007hr,Alday:2007mf,Alday:2009zf}.
\end{itemize}
Having completed our survey of the exponentiation of form factors, we now move
on to discuss the general case of fixed-angle scattering amplitudes.


\section{Fixed-angle scattering amplitudes}
\label{MultiPart}

We consider now the case of multi-particle, fixed-angle scattering amplitudes in a
general massless gauge theory. We write such amplitudes as
\beq
  {\cal A}_{n}^{\, a_1 \ldots a_n} \left( \frac{p_i}{\mu} , \alpha_s(\mu^2), \epsilon \right) \, ,
\label{genamp}
\eeq
implying that the normalisation has been chosen in order to work with a dimensionless 
quantity. The spin structure is understood, but we are displaying the colour indices,
which in general can belong to different representations of the gauge group. The 
fixed-angle assumption amounts to the statement that all Mandelstam invariants
$s_{ij} = 2 p_i \cdot p_j$ are parametrically of the same size, say $\left| s_{ij} \right| 
\sim Q^2$, $\forall \, i,j$. This means that there are no strong scale hierarchies: the
cases in which some momentum $p_i$ becomes soft, or two momenta $p_i$ and 
$p_j$ become collinear, must be treated separately.

For amplitudes in this class, it is not too difficult to generalise the reasoning leading 
to the soft-collinear factorisation of the form factor, \eq{fofafac}. A diagrammatic 
analysis, using the Landau equations and the Coleman-Norton physical picture,
confirms that non-integrable singularities arise only from soft and collinear 
configurations; soft gluons, and separately collinear gluons, factorise from the 
hard part by the same reasoning that was applied to the form factor; finally,
so long as external particles are not collinear to each other, the factorisation of
soft gluons from each jet at leading power is supported by the same diagrammatic
arguments and Ward identities. From a physical point of view, at leading power,
jets moving in different directions can interact only at the hard scattering, and soft 
particles cannot resolve either the details of the hard interaction, or the fine structure
of streams of massless particles moving collinearly. The most delicate issue in
this generalisation is whether different loop momentum regions, other than soft 
or collinear, could give raise to singularities. Specifically, recall that the soft region
for loop momentum $k$ is defined by the scaling
\beq
  k^\mu \, = \, \{ k^+, k^-, {\bf k}_\perp \} \, \quad \to \quad \lambda k^\mu \, , 
  \qquad \lambda \to 0 \, .
\label{softscal}
\eeq
In the case of collinear scaling, we can separately consider each hard particle in the 
amplitude, and in each case define the collinear direction as one of the two directions
spanning the light-cone: for example, for a given loop momentum $k$, when considering 
the collinear region $k^\mu \parallel p_i^\mu$, we can choose a frame where $p_i^\mu = 
\{p_i^+, 0^-, {\bf 0}_\perp \}$; then the collinear scaling is
\beq
  k^\mu \, = \, \{ k^+, k^-, {\bf k}_\perp \} \, \quad \to \quad 
  \{ k^+, \lambda k^-, \sqrt{\lambda} {\bf k}_\perp \} \, , 
  \qquad \lambda \to 0 \, ,
\label{collscal2}
\eeq
which leads to $k^2 = 2 k^+ k^- - {\bf k}_\perp^2$ vanishing uniformly as $\lambda$.
One must wonder whether different momentum scalings could lead to singular 
contributions: a notable possibility\footnote{Finer distinctions are possible and often 
useful. One may for example consider soft gluons whose momentum components 
scale like the transverse momentum in \eq{collscal2}, rather then the anti-collinear
component $k_-$. In the context of SCET, such gluons are referred to as {\it soft}, 
while the scaling in \eq{softscal} is called {\it ultrasoft}. The version of SCET that
distinguishes the two scalings is called SCET$_I$, and it can mapped to the theory
with a single soft scaling (SCET$_{II}$) by a suitable matching procedure. One 
may also consider {\it Coulomb} gluons, whose (soft) spatial momentum components 
dominate their energy in a selected frame. These scalings will not be needed in what 
follows.} is given by soft wide-angle gluons, scaling as
\beq
  k^\mu \, = \, \{ k^+, k^-, {\bf k}_\perp \} \, , \quad \to \quad 
  \{ \lambda^p k^+, \lambda^p k^-, \lambda {\bf k}_\perp \} \, , \quad p \geq 2 \, ,
  \qquad \lambda \to 0 \, , 
\label{glaubscal}
\eeq
with respect to one of the hard parton directions. For fixed-angle amplitudes, such
gluons essentially have vanishing longitudinal components, and are usually called
{\it Glauber} gluons. Glauber gluons are known to give important contributions at 
cross-section level, and the cancellation of the corresponding divergences is a crucial 
step in the proof of collinear factorisation for collider processes~\cite{Collins:1988ig,
Aybat:2008ct}; furthermore, they can contribute to infrared poles of scattering amplitudes 
when the external particles are allowed to become collinear, while some of the Mandelstam 
invariants become space-like~\cite{Catani:2011st}. For fixed-angle scattering amplitudes, 
however, Glauber gluons do not contribute at leading power~\cite{Botts:1989kf}.

With these premises, it is easy to propose a generalisation of \eq{fofafac} for 
multi-particle amplitudes. First of all, we expect that collinear dynamics will be captured 
by jet functions, which are essentially single-particle quantities, and thus cannot 
change the colour content of the outgoing state; soft gluons, on the other hand, 
can connect any pair of hard particles, and they will change the colour of those 
particles even if they carry a vanishing energy. We expect then that the soft function 
in \eq{fofafac} will need to be promoted to a colour operator acting on the colour indices
of all external particles. The emergent form of soft-collinear factorisation for fixed-angle 
multi-particle scattering amplitudes in massless gauge theories is then
\beq
  {\cal A}_n \left( \frac{p_i}{\mu}, \as (\mu), \eps \right) & = & 
  \prod_{i = 1}^n \frac{{\cal J}_i \left( \frac{\left( p_i \cdot n_i \right)^2}{n_i^2 \mu^2},
  \as(\mu^2), \eps \right)}{{\cal J}_{E, i} \left( \frac{\left( \beta_i \cdot n_i \right)^2}{n_i^2},
  \as(\mu^2), \eps \right)} \nonumber \\
  && \times \, {\cal S}_n \big( \beta_i \cdot \beta_j, \as (\mu^2), \eps \big) \, 
  {\cal H}_n \! \left( \! \frac{p_i \cdot p_j}{\mu^2}, 
  \frac{\left( p_i \cdot n_i \right)^2}{n_i^2 \mu^2}, \as(\mu^2), \eps \right) \, ,
\label{ampnfact}
\eeq
where we introduced a jet function, and its eikonal counterpart, for each external
hard particle, with the definitions given in \eq{Jqfofa} and \eq{Jqeikfofa}, and we
defined the $n$-particle soft function as the natural generalisation of \eq{Softfofa},
\beq
  {\cal S} \left( \beta_i \cdot \beta_j, \alpha_s(\mu^2), \eps \right)  \, \equiv \,  
  \langle 0 | \, T \bigg[ \prod_{k = 1}^n \Phi_{\beta_k} (\infty, 0) \bigg] | 0 \rangle \, .
\label{Softnpart}
\eeq
The factorisation in \eq{ampnfact} is supported by the exhaustive diagrammatic 
analysis carried out in Ref.~\cite{Ma:2019hjq}, which leads to a BPHZ-like forest 
formula for soft and collinear singularities in fixed-angle scattering amplitudes.
This analysis makes use of the coordinate-space formalism developed in 
Ref.~\cite{Erdogan:2014gha}, and generalises the early results of~\cite{Collins:1981uk,
Sen:1982bt}. An independent derivation of an analogous factorisation in the 
context of SCET was given in~\cite{Feige:2014wja}.

In order to proceed to a more detailed analysis of the factorisation, we first need 
to be much more concrete concerning the treatment of colour flow in \eq{ampnfact}, 
clarifying the action of the soft colour operator ${\cal S}$ on the hard part of the 
amplitude, ${\cal H}$: this was trivial for form factors, where particles with opposite 
colour charges annihilate into a colour-singlet state, but can be very intricate for 
general scattering amplitudes. With this in mind, before proceeding to discuss the 
dynamical aspects of the soft-collinear factorisation in \eq{ampnfact}, we briefly 
digress to describe the two main methods commonly used to handle colour in this 
general case.


\subsection{Handling color structures}
\label{ColStru}

The colour structure of a generic multi-particle non-abelian scattering amplitude
is an interesting group theory problem, solved in principle with standard tools, 
but not easy to implement in practice when the particle number grows.
In general, each color index in \eq{genamp} can belong to a different representation 
of the gauge group $r_i$, $i = 1, \ldots, n$. The amplitude ${\cal A}$ is a tensor
in the vector space ${\cal V}_n \equiv r_1 \otimes \ldots \otimes r_n$, and colour 
conservation implies that it must be an invariant tensor. The task is then to decompose 
the reducible representation ${\cal V}_n$ into a sum of irreducible representations,
and impose the colour-conservation constraint on the result. A practical way to do 
this is to pick a {\it channel}, {\it i.e.} a set of particles taken as `incoming', and build 
the tensor product of the corresponding representations; one then repeats the 
construction for the remaining, `outgoing', particles, and selects the irreducible
representations that appear in both lists: those are the possible colour states that 
can be exchanged in that channel.  A thorough analysis of this construction was 
performed in Ref.~\cite{Beneke:2009rj} (see also~\cite{DelDuca:2011ae}): using 
Clebsch-Gordan coefficients, one can construct projection operators identifying
each possible colour flow in the selected channel. 

In the present context, our concern is to understand how the colour structure is 
implemented in the soft-collinear factorisation of the amplitude, \eq{ampnfact}.
In order to represent the action of the soft operator ${\cal S}_n$ on the hard matching 
coefficient ${\cal H}_n$, two formalisms are commonly adopted, one based upon a
choice of basis in the space of colour tensors available for the selected process, 
and one formulated in terms of basis-independent {\it colour insertion operators}. 
For completeness, we briefly present below the basics of both formalisms.


\subsubsection{Colour tensor bases}
\label{ColTensBas}

\begin{figure}
\centering
  \includegraphics[height=2cm,width=8cm]{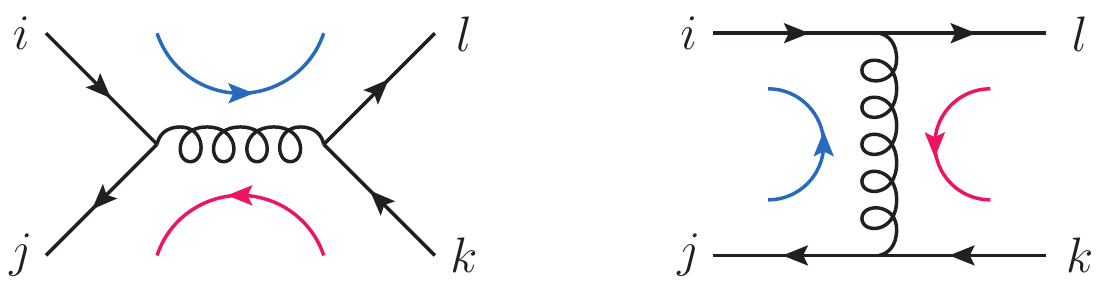}
  \caption{Feynman diagrams for tree-level quark-antiquark scattering, and their 
  colour flow at leading power in $N_c$.}
  \label{qqbartree}
\end{figure}

Using the techniques discussed in Refs.~\cite{Beneke:2009rj,DelDuca:2011ae},
it is possible to identify, in any selected channel, a basis of colour tensors spanning
the representations that can be exchanged in that channel. Denoting the basis
tensors by $c_L^{\,\, a_1 \ldots a_n}$, we can then write the amplitude as
\beq
  {\cal A}_n^{\, a_1 \ldots a_n} \left( \frac{p_i}{\mu} , \alpha_s(\mu^2), \epsilon \right)
  \, = \, \sum_L {\cal A}_n^L  \left( \frac{p_i}{\mu} , \alpha_s(\mu^2), \epsilon \right)
  \, c_L^{\,\, a_1 \ldots a_n} \, .
\label{colortens}
\eeq
The tensors $c_L^{\,\, a_1 \ldots a_n}$ are quadratic combinations of the 
Clebsch-Gordan coefficients appearing in the decomposition of ${\cal V}_n$ in
a direct sum of irreducible representations. They can be  chosen to be
orthonormal, in the sense that
\beq
  \sum_{\{a_i\}} c_L^{\,\, a_1 \ldots a_n} \, \left( c_M^{\,\, a_1 \ldots a_n} \right)^*
  \, = \, \delta_{LM} \, .
\label{ONtens}
\eeq
To illustrate this, consider the simple case of $q \bar{q}$ scattering: at tree level,
the contributing diagrams are depicted in Fig.~\ref{qqbartree}. Applying the Feynman 
rules, and the Fierz identity
\beq
  \big( T_a \big)_{ij} \, \big( T^a \big)_{kl} \, = \, \frac{1}{2} \bigg( 
  \delta_{il} \delta_{jk} - \frac{1}{N_c} \delta_{ij} \delta_{kl} \bigg) \, ,
\label{Fierz}
\eeq
one immediately identifies a possible set of basis color tensors as
\beq
  \hat{c}_1^{\,ijkl} \, = \, \delta^{il} \delta^{jk} \, , \qquad
  \hat{c}_2^{\,ijkl} \, = \, \delta^{ij} \delta^{kl} \, ,
\label{colbas1}
\eeq
corresponding to leading-color flow in the $s$ channel and in the $t$ channel, 
respectively. It is clear that this set forms a basis for higher-order corrections
as well, since the third possible combination of Kronecker $\delta$ functions
in the fundamental representation is forbidden, as it would require a quark to 
turn into an antiquark. More formally, one would start by noting that the 
representations entering the scattering (for example in the $s$ channel) give
$\underline{\bf 3} \otimes \underline{\bf 3}^* = \underline{\bf 1} \oplus 
\underline{\bf 8}$: one then expects a set of only two basis tensors. The
correspondence between the basis tensors and the representation content
is made more transparent by transforming the basis in \eq{colbas1} to an 
orthonormal one, according to \eq{ONtens}, by using again \eq{Fierz}. 
One can then pick
\beq
  c_1^{\,ijkl} \, = \, \frac{2}{\sqrt{N_c^2 - 1}} \big( T_a \big)_{ji} \, 
  \big( T^a \big)_{kl} \, , \qquad
  c_2^{\,ijkl} \, = \, \frac{1}{N_c} \, \delta^{ij} \delta^{kl} \, , 
\label{colbas2}
\eeq
corresponding to octet and singlet $s$-channel exchanges, respectively. Once a basis 
has been selected, the amplitude can be thought of as a vector in the vector space 
spanned by the $c_L$ tensors. In the chosen basis, the soft operator is represented 
by a matrix, acting on a vector of hard matching coefficients. To be precise, we note 
that the Wilson lines in \eq{Softnpart} have open colour indices at both ends, and 
both sets of indices can be projected on the basis tensors. We define then the 
matrix elements ${\cal S}^{\,\, L}_{n \,\, K}$ of the soft operator, in the chosen 
basis, by
\beq
  \sum_L c_L^{\,\, a_1 \ldots a_n} \, {\cal S}^{\,\, L}_{n \,\, K} \big( \beta_i \cdot \beta_j, 
  \alpha_s (\mu^2), \epsilon \big) \, = \, \sum_{b_1, \ldots, b_n} 
  \langle 0 |  \prod_{k = 1}^n \Big[ \Phi_{\beta_k} (\infty, 0) 
  \Big]^{a_k}_{\,\,\,\, b_k} | 0 \rangle \, 
  c_K^{\,\, b_1 \ldots b_n} \, ;
\label{softnproj}
\eeq
in \eq{ampnfact}, the soft matrix thus defined acts on a vector of finite matching 
coefficients ${\cal H}^K$.

Working with explicit colour bases is useful for practical applications, 
for example for the implementation of soft gluon resummations in hadron 
scattering~\cite{Kidonakis:1997gm,Kidonakis:1998bk,Kidonakis:1998nf,
Laenen:1998qw,Sjodahl:2009wx}. The formalism, however, does not easily 
lend itself to the general treatment of $n$-point amplitudes: the size of the 
matrices involved grows steeply with the number of particles, and colour 
conservation is implemented essentially on a process-by-process basis. 
For example, the treatment of $gg \to  gg$ scattering in $SU(N_c)$ requires 
9$\times$9 matrices (although some simplification are possible~\cite{Dokshitzer:2005ig,
Seymour:2008xr}), while already for amplitudes involving five gluons ($gg \to ggg$) 
the matrices involved have dimension $d = 44$~\cite{Sjodahl:2008fz}. We now 
proceed to describe a formalism which is better suited for generic particle multiplicities.


\subsubsection{Colour insertion operators}
\label{ColInsOp}

Considering soft-gluon corrections, the factorisation formula in \eq{ampnfact} 
formalises an intuitive understanding: soft divergences arise when soft gluons
(real or virtual) are emitted from an on-shell particle. Essentially, this means that
soft divergences are associated with emissions from external legs, so that one 
can think of the soft operator as acting `from the outside' on the hard factor, which
contains corrections associated with `inner', off-shell virtual exchanges. It then 
makes sense, at leading power, to think of the soft emission as a colour operator, 
inserting a soft gluon on the non-radiative amplitude, according to the scheme
\beq
  {\cal A}_{n + 1}^{\,\,a \,b_1 \ldots \, b_n} \bigg|_{\rm soft} \, \propto \, 
  \sum_{i = 1}^n  \Big[ {\bf T}_i^{\, a} \Big]^{b_i}_{\, \, \, c_i} 
  {\cal A}_n^{\,\,b_1 \ldots \, c_i \ldots \, b_n} \, .
\label{softradscheme}
\eeq
The colour operators ${\bf T}_i$ act on the tensor product space ${\cal V}_n$,
and the index $i$ identifies the specific factor in the product on which the operator 
acts non-trivially, corresponding to the emission of a gluon with (adjoint) index $a$
from particle $i$. In order to be more precise, and to illustrate how this operator 
formalism emerges~\cite{Bassetto:1984ik,Catani:1996vz}, consider the case of 
soft gluon emission from an outgoing quark, in a generic tree-level matrix element 
involving $n$ coloured particles, plus the emitted gluon. Let the $n$ `Born-level' 
particles have momenta $p_i$, colour indices $c_i$ and polarisation $\lambda_i$, 
and let the radiated gluon be characterised by momentum $k$, colour $c$ and 
polarisation $\lambda$. The only diagrams surviving at leading power in $k$ are 
those where the soft gluon is emitted by the on-shell external legs, so we can easily 
isolate the contribution of a specific outgoing quark, say the one carrying momentum 
$p_i$. The relevant contribution to the emission amplitude is given by
\beq
  g \mu^\epsilon \, \overline{u}_{s_i} (p_i) \, \gamma_\alpha \, 
  \frac{\slash{p}_i + \slash{k}}{2 p_i \cdot k} \,  \big( T^c \big)_{c_i d_i}
  \widehat{\cal A}^{\,\,\, c_1 \ldots \, d_i \ldots \, c_n}_{\, s_1 \dots s_n} 
  \big(\{ p_j\}, k \big) \,  
  \epsilon_\lambda^{* \, \alpha} (k) \, ,
\label{one_emiss_quark}
\eeq
where $\widehat{\cal A}$, while not quite a scattering amplitude itself, since the $i$-th 
leg is off-shell for generic $k$, represents all the remaining factors of the radiative 
amplitude. Now, taking the soft limit $k^\mu \to 0$, we can apply the eikonal 
approximation to the factors associated with the $i$-th particle, and, at leading 
power in $k$, we can neglect $k$ in the factor $\widehat{\cal A}$, which thus
becomes precisely the `Born' amplitude for the original $n$ on-shell particles.
Following the steps already outlined in Section 1 in the case of QED, this leads 
to the expression
\beq
  g \mu^\epsilon \, \frac{\beta_i \cdot  \epsilon_\lambda^*(k)}{\beta_i \cdot k}  
  \, \big( T^c \big)_{c_i d_i}
  \big({\cal A}_n \big)^{\, c_1 \ldots \, d_i \ldots \, c_n}_{s_1 \dots s_n} \big( \{ p_j\} \big) 
  \, \equiv \, g \mu^\epsilon \, \frac{\beta_i \cdot  \epsilon_\lambda^* (k)}{\beta_i \cdot k} 
  \, \, {\bf T}_i \, {\cal A}_n \big( \{ p_j\} \big) \, .
\label{defopoq}
\eeq
We have thus identified the precise expression for the colour operator ${\bf T}_i$, when the 
$i$-th particle is an outgoing quark: in that case, ${\bf T}_i$ is the colour generator 
$T^a_{\,\,c d}$, in the fundamental representation. Performing the same calculation for 
an outgoing antiquark, and an outgoing gluon, one finds the identifications
\beq
  {\bf T}_i \Big|_{\rm q, \, out} \, \to \, T^a_{\,\,c d} \, , \qquad
  {\bf T}_i \Big|_{\rm \bar{q}, \, out} \, \to \, - \, T^a_{\,\, d c} \, , \qquad
  {\bf T}_i \Big|_{\rm g, \, out} \, \to \, - \, {\rm i} f^a_{\,\,c d} \, ,
\label{colopid}
\eeq
with the convention that the colour index $d$ is the one to be contracted with the 
Born amplitude; the action on incoming particles is defined by crossing symmetry: 
it is unchanged for gluons, while the expressions for quarks and antiquarks are 
interchanged. Notice that the negative sign for antiquarks in \eq{colopid} emerges 
from the orientation of the fermion line of the Born amplitude, which is opposite 
to the (outgoing) momentum flow for an antiparticle.

In case the soft gluon is emitted into the final state, this discussion sketches the
derivation of the tree-level current for the emission of a soft-gluon with momentum $k$,
\beq
  {\bf J}^\mu (k) \, = \, g \mu^\epsilon \sum_{i =1}^n \frac{\beta_i^\mu}{\beta_i \cdot k} \,
  {\bf T}_i \, ,
\label{softcurrtree}
\eeq
which will be discussed in greater detail in \secn{Subtra}. If the soft gluon is virtual,
it will be reabsorbed by another hard particle, leading to the colour-dipole structure
${\bf T}_i \cdot {\bf T}_j$, where the product implies a sum over the soft gluon 
colour indices. Since the colour operators ${\bf T}_i$ act non-trivially only on the 
colour index of particle $i$, operators acting on different particles commute, and 
they satisfy the algebra
\beq
  {\bf T}_i \cdot {\bf T}_j \, = \, {\bf T}_j \cdot {\bf T}_i \, \qquad ( i \neq j ) \, ;  \qquad
  {\bf T}_i \cdot {\bf T}_i \, \equiv \, {\bf T}_i^2 \, = \, C^{(2)}_{r_i} \, ,
\label{colopalg}
\eeq
where $C^{(2)}_{r_i}$ is the quadratic Casimir eigenvalue for the representation $r_i$,
so that $C^{(2)}_A \equiv C_A = N_c$ and $C^{(2)}_F \equiv C_F = (N_c^2 - 1)/2 N_c$ 
for the adjoint and the fundamental representations, respectively. The statement that 
the scattering amplitude is an invariant tensor of the colour group is represented in 
this formalism by the fact that, when acting on the non-radiative amplitude, the 
colour operators satisfy
\beq
  \sum_{i = 1}^n {\bf T}_i \, = \, 0 \, ,
\label{colopgaugein}
\eeq
which, as we will see, poses a non-trivial constraint on the structure of the soft 
operator at high orders. This brief introduction suggests that the colour operator
notation, while somewhat more formal, and thus perhaps less suited to concrete
applications, is very powerful to construct general statements about scattering
amplitudes with a generic number of particles. In what follows, we will often 
leave the colour structure implicit, treating the soft matrix as a formal colour 
operator, as in \eq{ampnfact}: when needed, we will adapt our notation, either 
by choosing a basis or by explicitly introducing colour operators.


\subsection{The dipole formula and beyond}
\label{DipFor}

Having dealt with ways to handle the colour structure of the factorised amplitude,
we now turn back to explore the all-important dynamical consequences of the 
factorisation in \eq{ampnfact}. As we will see, most of the considerations that
were discussed in \secn{FactoForm} in the case of form factors still apply for
the general case of fixed-angle amplitudes, but they turn out to have much more 
powerful consequences.

The first point to notice is that the soft function defined in \eq{Softnpart}, like its
two-line counterpart in \eq{Softfofa}, obeys a renormalisation group equation,
which however now has a matrix form. Technically, multiplicative renormalisability
of (straight) Wilson-line correlators holds when the lines are not on the light cone,
or in the presence of a collinear regulator: formally, for the moment, we are going 
to use dimensional regularisation as a collinear regulator, while later, for concrete
calculations in \secn{CompMatr}, it will turn out to be more practical to tilt Wilson 
lines off the light cone and reserve dimensional continuation for the control of UV  
divergences. With this understanding, the renormalisation group equation for the
soft operator (in a selected colour basis) reads
\beq
  \mu \frac{d}{d \mu} \, {\cal S}_{L K} \left( \beta_i \cdot \beta_j, 
  \as (\mu^2), \eps \right) \, = \, - \, {\cal S}_L^{\,\,\,M} \left( \beta_i \cdot \beta_j, 
  \as (\mu^2), \eps \right) \Gamma^{({\cal S})}_{M K} \left( \beta_i \cdot \beta_j, 
  \as (\mu^2), \eps \right) \, ,
\label{softmatrRG}
\eeq
which defines the anomalous dimension matrix $\Gamma^{({\cal S})}$. At this stage,
$\Gamma^{({\cal S})}$ is not finite, but carries collinear poles in $\epsilon$. The
solution to \eq{softmatrRG} can readily be written as
\beq
  {\cal S} \left( \beta_i \cdot \beta_j, \as (\mu^2), \eps \right) \, = \, 
  {\cal P} \exp \left[ - \frac12 \int_0^{\mu^2} \frac{d \xi^2}{\xi^2} \,
  \Gamma^{({\cal S})} \big( \beta_i \cdot \beta_j, 
  \as (\xi^2, \eps), \eps \big) \right] \, ,
\label{softmatrsol}
\eeq
where we used the vanishing of the $d$-dimensional coupling in the infrared,
and we have shuffled the tree-level colour structure into the hard coefficient
function.

It would be nice, of course, to deal with a finite anomalous dimension matrix,
and this is where the full power of the soft-collinear factorisation of the amplitude 
can be brought to bear. Indeed, as was the case for form factors, we can organise
\eq{ampnfact} either by building the hard-collinear combinations ${\cal J}_i/{\cal 
J}_{E, i}$, which have no soft poles, or, much more interestingly, by defining
the reduced soft matrix
\beq
  \widehat{\cal S}_{L K} \big( \rho_{i j}, \as (\mu^2), \eps \big) \, \equiv \, 
  \frac{{\cal S}_{L K} \left( \beta_i \cdot \beta_j, \as (\mu^2), \eps \right)}{\prod_{i = 1}^n
  {\cal J}_{E, i} \left( \frac{\left( \beta_i \cdot n_i \right)^2}{n_i^2},
  \as(\mu^2), \eps \right)}  \, .
\label{softmatrred}
\eeq
which must be free of collinear poles, and is responsible for wide-angle soft 
singularities. We see, however, that in this case the cancellation is much more
surprising, since the numerator of \eq{softmatrred} is a matrix, while the denominator
is just a number. Clearly, this poses strong constraints on the soft operator: for
example, as the notation suggests, the anomalous dependence on the light-like
vectors $\beta_i$ must cancel along with  the collinear poles, which forces
the reduced soft function to depend on $n_i$ only through the rescaling-invariant 
ratios introduced in \eq{rho12},
\beq
  \rho_{ij} \, \equiv \, \frac{ \left( \beta_i \cdot \beta_j \right)^2 \, n_i^2 \, 
  n_j^2}{\left( \beta_i \cdot n_i \right)^2 \left( \beta_j \cdot n_j \right)^2} \, .
\label{rhoij}
\eeq
The full power of the constraints arising from \eq{softmatrred} is best displayed
at the level of the anomalous dimensions. One can write the renormalisation 
group equation for the reduced soft function as
\beq
  \mu \frac{d}{d \mu} \, \widehat{\cal S}_{L K} \left( \rho_{i j}, 
  \as (\mu^2), \eps \right) \, = \, - \, \widehat{\cal S}_L^{\,\,\,M} \left( \rho_{i j}, 
  \as (\mu^2), \eps \right) \Gamma^{(\widehat{\cal S})}_{M K} \left( \rho_{i j}, 
  \as (\mu^2) \right) \, ,
\label{softmatrredRG}
\eeq
and then note that the matrix $\Gamma^{(\widehat{\cal S})}$ can be written as
\beq 
  \Gamma^{(\widehat{\cal S})}_{K L} \left( \rho_{i j}, \as (\mu^2) \right) \, = \,
  \Gamma^{({\cal S})}_{K L} \left( \beta_i \cdot \beta_j, \as (\mu^2), \eps \right) - 
  \delta_{K L} \, \sum_{i = 1}^n \gamma_{{\cal J}_E} \! \left( \! \frac{\left( 
  \beta_i \cdot n_i \right)^2}{n_i^2}, \as(\mu^2), \eps \right) \, ,
\label{GammaGamma}
\eeq
where $\gamma_{{\cal J}_E}$ is the anomalous dimension for the eikonal jet 
function, which carries the anomalous dependence on the jet direction $\beta_i$,
arising from soft-collinear double poles. It is immediately clear that three
separate powerful constraints on the soft anomalous dimension matrix 
$\Gamma^{({\cal S})}$ arise from \eq{GammaGamma}.
\begin{itemize}
\item Collinear poles in $\Gamma^{({\cal S})}$ must be  confined to the diagonal 
entries of the matrix, and they must be determined solely by the light-like cusp 
anomalous dimension $\gamma_K$, which governs soft-collinear singularites, 
and thus fixes $\gamma_{{\cal J}_E}$~\cite{Gardi:2009qi}.
\item Finite diagonal terms in $\Gamma^{({\cal S})}$ must combine with
finite terms in $\gamma_{{\cal J}_E}$ to form the rescaling-invariant ratios
$\rho_{ij}$ in \eq{rhoij}, as was the case for the form factor.
\item Off-diagonal terms in $\Gamma^{({\cal S})}$ must be finite, and 
furthermore they must, by themselves, be rescaling-invariant functions 
of the four-velocities $\beta_i$. Thus, they can only depend on
{\it conformal-invariant cross ratios} of the form
\beq
  \rho_{ijkl} \, = \, \frac{\beta_i \cdot \beta_j \, \beta_k \cdot 
  \beta_l}{\beta_i \cdot \beta_k \, \beta_j \cdot \beta_l} \, .
\label{rhoijkl}
\eeq
\end{itemize}
In order to translate these qualitative statements into a precise mathematical
formulation, we can take a derivative of the matrix $\Gamma^{(\widehat{\cal S})}$
with respect the arguments $x_i = \left( \beta_i \cdot n_i \right)^2/n_i^2$,
which only occur in the eikonal jets. We find
\beq
  x_i \frac{\del}{\del x_i} \, \Gamma_{K L}^{(\widehat{\cal S})} \left(\rho_{i j}, \as \right)
  \, = \, - \, \delta_{K L} \,\,   x_i \frac{\del}{\del x_i} \, \gamma_{J_E} \, = \, 
  - \, \frac{1}{4} \, \gamma_K ( \as) \, \delta_{K L} \, ,
\label{masteq0}
\eeq
where the last equality expresses the fact that kinematic dependence in 
the eikonal jets can only arise through the superposition of soft and collinear 
singularities, and is thus determined by the light-like cusp anomalous 
dimension $\gamma_K ( \as)$. The precise form of the equality follows from 
a detailed analysis of the RG equation for the eikonal jet~\cite{Gardi:2009qi}, 
as well as from the general arguments leading to Collins-Soper equations, 
as first laid out in Ref.~\cite{Collins:1981uk}. We know, however, that 
$\Gamma^{(\widehat{\cal S})}$ can only depend on $x_i$ through the 
combinations $\rho_{ij}$ defined in \eq{rhoij}. Using the chain rule we then 
conclude that~\cite{Gardi:2009qi,Becher:2009qa}
\beq
  \sum_{i = 1}^n \sum_{j \neq i} \, \frac{\del}{\del \ln \rho_{i j}} \, 
  \Gamma_{K L}^{(\widehat{\cal S})} \left(\rho_{i j}, \as \right)
  \, = \, \frac{1}{4} \, \gamma_K ( \as) \, \delta_{K L} \, .
\label{masteq}
\eeq
\eq{masteq} governs the interplay of colour and kinematics in the soft anomalous
dimension matrix to all orders in perturbation theory, and is an exact result.

Our remaining task is now to solve \eq{masteq}. As we are dealing with a linear
inhomogeneous partial differential equation, the general solution can be expressed
as the sum of a particular solution of the inhomogeneous problem, plus a generic
solution of the corresponding homogeneous equation. In order to proceed, we
now shift from the basis-dependent notation employed thus far to the colour-operator
notation: upon making a single approximation, this will allow us to write a particular 
solution of \eq{masteq} in a transparent and elegant way. We begin by noting that 
the cusp anomalous dimension in colour representation $r$ can be computed
from the form factor for particles trasforming under $r$, and admits a {\it Casimir 
expansion}~\cite{Cvitanovic:1976am,deAzcarraga:1997ya,vanRitbergen:1998pn,
Cvitanovic:2008zz}
\beq
  \gamma_K^{(r)} (\as) \, = \,  
  \sum_{p, i}^\infty C^{\, (2 p, i)}_r \,\, \widehat{\gamma}^{\, (2 p, i)}_K (\as) \, ,
\label{gencusp}
\eeq
where $C^{\,(n, i)}_r$ are eigenvalues of $n$-th order Casimir operators of the
Lie algebra, in representation $r$, and the functions $\widehat{\gamma}^{\, (2 p, i)}_K$
are independent of $r$. The index $i$ takes into account possible dependence on 
the matter content of the particular gauge theory under consideration: in ordinary
QCD, for example, one finds loop-level contributions involving both the adjoint and the
fundamental representations. For $p > 1$, building a Casimir invariant 
of order $2 p$ requires $2 p$ gluon attachments to the Wilson lines, and a further 
$2 p$ vertices (for example forming a loop) in order to build a symmetric colour 
tensor saturating the open colour indices. Thus, for example, quartic Casimir
contributions (corresponding to $p = 2$) will start at four loops, sixth-order 
Casimir contributions ($p = 3$) at six loops, and so on; quadratic Casimirs are 
special because they are built with the Lie algebra metric tensor $\delta _{ab}$, 
so that the corresponding contribution starts at one loop. Finite-order calculations 
indeed confirm that the cusp anomalous dimension obeys {\it Casimir scaling} up 
to three loops ({\it i.e.} it is proportional to the quadratic Casimir eigenvalue 
$C^{(2)}_r$), while it develops quartic Casimir components at four loops, as 
expected~\cite{Moch:2017uml,Grozin:2017css,Boels:2017skl,Moch:2018wjh,
Henn:2019rmi,Henn:2019swt,Huber:2019fxe,vonManteuffel:2020vjv}.

Given this general structure, it is both phenomenologically and theoretically 
interesting to begin by focusing on the quadratic Casimir component of the 
cusp, which provides the full answer up to three loops. This component yields a 
simple solution to \eq{masteq}, since one can write (using \eq{colopalg})
\beq
  \gamma_K^{(r)} (\as) \, = \, C^{(2)}_r \, \widehat{\gamma}_K (\as) \, = \, 
  \widehat{\gamma}_K (\as) \, {\bf T}_i \cdot {\bf T}_i \, ,
\label{Casimirscal}
\eeq
where  for simplicity we dropped the Casimir-counting indices in \eq{gencusp}.
It is then easy to verify, using \eq{colopgaugein}, that the `dipole sum'
\beq
  \Gamma_{\rm dip}^{(\widehat{\cal S})} \left( \rho_{i j}, \as \right) \, = \,
  - \, \frac{1}{8} \, \widehat{\gamma}_K (\as) \sum_{i = 1}^n \sum_{j \neq i} 
  \ln \rho_{i j} \, {\bf T}_i \cdot {\bf T}_j \, + \, \delta^{\, (\widehat{\cal S})}
  (\as) \, \sum_{i = 1}^n {\bf T}_i \cdot {\bf T}_i \, ,
\label{firstdip}
\eeq
is a solution of \eq{masteq}, with the function $\delta^{\, (\widehat{\cal S})}$
playing the role of an integration constant. Before highlighting the important properties
of \eq{firstdip}, it is useful to embed it in the full factorisation of the amplitude, in
order to get rid of the unphysical dependence on the auxiliary vectors $n_i$, which 
still feature in \eq{firstdip} through the variables $\rho_{ij}$. Indeed, as already 
noted at the end of \secn{FactoForm}, $n_i$-dependent singular terms must 
cancel in the ratio of jets and eikonal jets, whereas finite $n_i$-dependent terms
must cancel between the jets and the hard matching function ${\cal H}$. Performing
these cancellations, order by order in perturbation theory, yields a simplified form
for the soft-collinear factorisation of fixed-angle amplitudes, which can be written 
as~\cite{Becher:2009qa,Gardi:2009zv}
\beq 
  {\cal A}_n \left( \frac{p_i}{\mu}, \as(\mu^2), \eps \right) \, = \, 
  {\cal Z}_n \left( \frac{p_i}{\mu}, \as(\mu^2), \eps  \right) 
  {\cal F}_n \left( \frac{p_i}{\mu}, \as(\mu^2), \eps  \right) \, ,
\label{IRfact}
\eeq
where ${\cal Z}_n$ is a colour operator generating all infrared singularities, and
acting on the finite vector ${\cal F}_n$. Since the non-trivial colour structure of
${\cal Z}_n$ arises entirely from the soft factor, it will obey a matrix renormalisation 
group equation of the same form as \eq{softmatrredRG}, and the solution will
again be an exponential, which we can write as
\beq
  {\cal Z}_n \left(\frac{p_i}{\mu}, \as (\mu^2), \eps \right) \, = \,  
  {\cal P} \exp \left[ \frac{1}{2} \int_0^{\mu^2} \frac{d \lambda^2}{\lambda^2} \, \,
  \Gamma_n \left(\frac{p_i}{\lambda}, \as \! \left( \lambda^2, \eps \right) \right) \right] \, ,
\label{RGsol}
\eeq
with the full infrared anomalous dimension matrix $\Gamma_n$ satisfying 
a constraint equation inherited from \eq{masteq}. The general solution to that 
equation will take the form
\beq
  \Gamma_n \left(\frac{p_i}{\mu}, \as(\mu^2) \right) \, = \, 
  \Gamma_n^{\rm dip} \left(\frac{s_{ij}}{\mu^2}, \alpha_s(\mu^2) \right) \, + \, 
  \Delta_n \left( \rho_{i j k l}, \as (\mu^2) \right) \, ,
\label{FullGamma}
\eeq
with $\Gamma_n^{\rm dip}$ a particular solution of the inhomogeneous equation,
akin to \eq{firstdip}, and $\Delta_n$ the general solution of the homogeneous 
problem. Considering first the dipole term, after the cancellation of of the dependence
on the auxiliary vectors, the logarithms of the scaling variables $\rho_{ij}$ must
be replaced with logarithms of the corresponding Mandelstam invariants. At this 
point we can also reinstate the analytic properties, which we have neglected so 
far, recalling from \secn{FactoForm} that time-like invariants carry imaginary 
parts. With these considerations, we can finally write the {\it dipole formula} 
as~\cite{Becher:2009cu,Gardi:2009qi,Becher:2009qa,Gardi:2009zv}
\beq
  \Gamma_n^{\rm dip} \bigg(\frac{s_{ij}}{\mu^2}, \as (\mu^2) \bigg) \, = \,  
  \frac{1}{2} \, \widehat{\gamma}_K \left( \as (\mu^2) \right) \sum_{i=1}^n
  \sum_{j = i+1}^n \log \left( \frac{s_{i j} \, 
  {\rm e}^{{\rm i} \pi \lambda_{ij}}}{\mu^2} \right) {\bf T}_i \cdot {\bf T}_j 
  + \sum_{i = 1}^n \gamma_i \left( \as(\mu^2) \right) \, ,
\label{GammaDip}
\eeq
where $\gamma_i$ are colour-diagonal anomalous dimensions arising from
jet functions, and the phases are given by $\lambda_{ij} = 1$, if particles $i$ 
and $j$ are either both in the final state or both in the initial state, while 
$\lambda_{ij} = 0$ otherwise.

Turning now to the homogeneous problem, we note that the operator
$\Delta_n$, in the language of \eq{masteq}, must satisfy
\beq
  \sum_{i = 1}^n \sum_{j \neq i} \, \frac{\del}{\del \ln \rho_{i j}} \, \Delta_n \, = \, 0 \, .
\label{eqDelta}
\eeq
As anticipated by the notation in \eq{FullGamma}, the general solution to this
equation expresses the fact that $\Delta_n$ must be scale invariant in each
momentum variable, which in turn implies that it can depend on Mandelstam
invariants only through the conformal cross ratios defined in \eq{rhoijkl}.
Since $\Delta_n$ involves correlations of at least four particles, we can 
conclude that \eq{GammaDip} gives the exact all-order result for the infrared
anomalous dimension for $n = 2,3$. Furthermore, for $n \geq 4$, we note
that correlations involving four hard particles can only arise starting at three 
loops. We conclude that \eq{GammaDip} provides the exact result, up to two
loops, for any number of hard particles: this is non-trivial, since at two loops
three-particle correlations can arise (and indeed they do arise when hard
particles are massive~\cite{Becher:2009kw}, as we briefly discuss in \secn{Twolosad}). 
This confirms, and further explains, the result of the direct calculation performed 
in~\cite{Aybat:2006mz,Aybat:2006wq}. Further constraints on $\Delta$ at three 
loops from various kinematic limits and symmetry arguments were studied in 
Refs.~\cite{Dixon:2009ur,Dixon:2009gx}, highlighting the limited span of the 
space of functions available for such corrections.

The lowest order contribution to the conformal correction $\Delta_n$, at the
three-loop order, was finally calculated for the first time in Ref.~\cite{Almelid:2015jia},
and verified by an explicit amplitude calculation in ${\cal N} = 4$ Super-Yang-Mills
theory in Ref.~\cite{Henn:2016jdu}. It takes the form of quadrupole colour 
correlations, with a kinematic dependence expressed in terms of a single 
combination of weight-five polylogarithms, of remarkable simplicity and elegance. 
We write
\beq
  \Delta_n \big( \rho_{ijkl}, \alpha_s \big) \, = \, \left( \frac{\alpha_s}{\pi} \right)^3 
  \Delta_n^{(3)} \big( \rho_{ijkl} \big) \, + \, {\cal O}(\alpha_s^4) \, ,
\label{Deltaexp}
\eeq
where~\cite{Almelid:2015jia}
\beq
\label{Delta3}
  \Delta_n^{(3)} (\rho_{ijkl}) & = & \frac{1}{4} \, f_{abe} f^e_{{\phantom e} c d} \,
  \Bigg\{ - C \, \sum_{i = 1}^n \, \sum_{\substack{{1 \leq j < k \leq n} \\ j,k \neq i}}
  \left\{ {\rm \bf T}_i^a,  {\rm \bf T}_i^d \right\} {\rm \bf T}_j^b {\rm \bf T}_k^c \\
  \nonumber && \hspace{-2cm} 
  + \, {\sum_{1 \leq i < j < k < l \leq n}} \bigg[
  {\rm \bf T}_i^a  {\rm \bf T}_j^b   {\rm \bf T}_k^c {\rm \bf T}_l^d  \,\, 
  {\cal F}(\rho_{ikjl},\rho_{iljk}) \, + \, 
  {\rm \bf T}_i^a  {\rm \bf T}_k^b {\rm \bf T}_j^c   {\rm \bf T}_l^d  \,\, 
  {\cal F}(\rho_{ijkl},\rho_{ilkj}) \\
  \nonumber && \qquad \qquad
  + \,\, {\rm \bf T}_i^a   {\rm \bf T}_l^b  {\rm \bf T}_j^c    {\rm \bf T}_k^d 
  \,\, {\cal F}(\rho_{ijlk},\rho_{iklj}) \bigg]       
  \Bigg\} \, .
\eeq
One readily observes that the form of $\Delta$ is heavily constrained by Bose 
symmetry. The first line in \eq{Delta3} is a constant (kinematic-independent)
colour tensor, with
\beq
  C \, = \, \zeta_5 + 2 \zeta_2 \zeta_3 \, .
\label{Cdef}
\eeq
The specific value of the constant $C$ is crucial to preserve the property of 
collinear factorisation, to which  we will return in \secn{Subtra}: in the limit in
which two (or more) hard particles become collinear, the scattering amplitude 
is expected to factorise, yielding a lower-multiplicity amplitude, multiplied
times a splitting kernel, and the splitting kernel is expected to depend only
on the quantum numbers of the collinear set. The value of the constant $C$
guarantees the necessary cancellations to enforce this result at three  
loops~\cite{Almelid:2015jia}. In order to display the kinematic dependence 
of $\Delta$, it is useful to introduce auxiliary variables $\{z_{ijkl},\bar{z}_{ijkl}\}$,
defined by the relations
\beq
  z_{ijkl} \, \bar{z}_{ijkl} \, = \, \rho_{ijkl} \, , \qquad
  \big( 1 - z_{ijkl} \big) \big( 1 - \bar{z}_{ijkl} \big) \, = \, \rho_{ilkj} \, .
\label{zdef} 
\eeq
In terms of these variables, we can express the kinematic function ${\cal F}$  as
\beq
  {\cal F} \big( \rho_{ijkl}, \rho_{ilkj} \big) \, = \, F \big( 1 - z_{ijkl} \big) - 
  F \big( z_{ijkl} \big) \, ,
\label{calFdef}
\eeq
where, finally, one finds
\beq
  F(z) \, = {\cal L}_{10101}(z) + 2 \zeta_2 \Big[ {\cal L}_{001}(z)
  + {\cal L}_{100}(z) \Big] \, .
\label{Fdef}
\eeq
The functions ${\cal L}_w (z)$ (where $w$ is a word composed of zeroes and 
ones) are {\it single-valued harmonic polylogarithms} (SVHPL)~\cite{Brown:2004}. 
These are special combinations of {\it harmonic polylogarithms}~\cite{Remiddi:1999ew}
with the property of being single-valued in the kinematic region where $\bar{z}$ is 
equal to the complex conjugate of $z$: this region is a subset of the Euclidean 
region, where all Mandelstam invariants are spacelike, $s_{ij} < 0$. Unitarity 
of massless scattering amplitudes dictates that they can only have singularities 
due to the vanishing of Mandelstam invariants, which do not occur in the Euclidean 
region of fixed angle scattering. Thus, the single-valuedness of the function $F$ 
directly reflects the analytic structure of the underlying amplitude. Interestingly,
the landmark result in \eq{Delta3} can be derived without resorting to the explicit 
calculation of the relevant Feynman diagrams, but rather by using bootstrap methods,
{\it i.e.} by identifying the space of functions which can contribute to $\Delta^{(3)}$,
and subsequently imposing kinematic constraints arising from known symmetries
and limiting behaviours of the scattering amplitude, such as high-energy and collinear
limits. This was done successfully at the three-loop order in Ref.~\cite{Almelid:2017qju}, 
reproducing Eqs.~(\ref{Delta3}-\ref{Fdef}).

We emphasise that, in this Section, we have chosen to arrive at the general expression
for the infrared anomalous dimension matrix in \eq{FullGamma} following the path
of factorisation. \eq{FullGamma}, with the definition of the dipole contribution in 
\eq{GammaDip}, are thus valid to all orders in perturbation theory. The historical 
path was quite different, starting with studies at two loops. The general structure of
infrared divergences in multi-particle scattering amplitudes for massless non-abelian 
gauge theories at two-loops was first displayed in the seminal paper by Catani,
Ref.~\cite{Catani:1998bh}, and subsequently derived from factorisation in  
Ref.~\cite{Sterman:2002qn}. At that time, the precise form of soft single-pole 
contributions at two loops could not be explicitly determined, and in principle
it was natural to expect a colour-tripole contribution at that level. Explicit  
calculations of four-point scattering amplitudes and splitting functions at two 
loops~\cite{Anastasiou:2000ue,Anastasiou:2000kg,Glover:2004si,Anastasiou:2001sv,
Glover:2001af,Bern:2000dn,Bern:2002tk,Bern:2003ck,Bern:2004cz} indeed found 
single-pole contributions with a colour-tripole structure. It turns out however that 
these contributions do not arise at the level of the soft anomalous dimension (in a 
minimal scheme), but can be generated when the infrared factorisation formula
is expanded to finite orders, and the one-loop hard part has been defined in
a non-minimal scheme, such as the one employed in Ref.~\cite{Catani:1998bh}.
The absence of tripole contributions in the soft anomalous dimension at two loops
was later explicitly verified by direct calculation in Refs.~\cite{Aybat:2006mz,
Aybat:2006wq}. Finally, Refs.~\cite{Becher:2009cu,Gardi:2009qi,Becher:2009qa,
Gardi:2009zv} provided the underlying symmetry argument for the cancellation,
which leads to the all-order structure in displayed in \eq{FullGamma}. Importantly,
it has been conjectured in \cite{Vladimirov:2016dll,Vladimirov:2017ksc} that the 
cancellation extends to all odd colour multipoles, which would for example forbid 
the presence of a penta-pole contribution to $\Delta_n$ at four loops.

The frontier of current research is the evaluation of the infrared anomalous dimension 
matrix at the four-loop level for massless gauge theories, and at the three-loop level
in the presence of massive coloured particles. At four loops, the occurrence of quartic 
Casimir operators in the cusp anomalous dimensions provides interesting new 
constraints, in particular arising from the requirements of collinear factorisation. 
The infrared anomalous dimension matrix for multi-particle amplitudes will also contain 
quartic Casimir contributions at four loops, with a more intricate colour structure 
as compared with the cusp. These  two sets of contributions must however be 
subtly correlated in order to preserve known all-order properties of the amplitudes
in collinear and high-energy limits. These constraints have been studied in
Refs.~\cite{Ahrens:2012qz,Becher:2019avh}, where a general ansatz for the 
infrared matrix was provided. It is possible that a bootstrap approach will succeed 
in determining all the unknown functions in the ansatz in~\cite{Becher:2019avh},
if a sufficient number of constraints can be imposed. Interestingly, the first direct
evidence for the existence of terms beyond the dipole formula came from the
analysis of the high-energy limit in Ref.~\cite{Caron-Huot:2013fea}, which 
uncovered a single-pole contribution at four loops, of the quartic Casimir type. 
More recently, further constraints on the infrared matrix, arising from the 
high-energy limit, were uncovered in Refs.~\cite{Caron-Huot:2017fxr,
Caron-Huot:2017zfo,Caron-Huot:2020grv,Falcioni:2020lvv,Maher:2021nlo,
Falcioni:2021buo}:  the high-energy limit of the infrared matrix is now known 
to all-orders at next-to-leading logarithmic (NLL) accuracy, while NNLL contributions 
constrain the quartic Casimir component at four loops.

Before moving on to illustrating techniques which are useful for explicit computation 
of the infrared matrix in \secn{CompMatr}, we conclude this Section by highlighting
some interesting consequences of the dipole approximation, \eq{GammaDip}, first
in the high-energy limit, and then in the language of the celestial approach to
the infrared limit.


\subsubsection{Taking the high-energy limit}
\label{HighEn}

Our discussion of scattering amplitudes so far has been limited by the {\it fixed-angle}
assumption, which in principle excludes very interesting and important configurations, 
such as the cases of soft and collinear radiation (to be discussed in \secn{Subtra}), and 
the high-energy limit, where the total energy $s$ of the process is much larger than
all $t$- and $u$-channel Mandelstam invariants involving both initial and final state 
particles. It is important to note, however, that once we have achieved a prediction
for infrared poles for a generic configuration of momenta, we are free to approach
(though not to reach) singular limits, where certain Mandelstam invariants are much
larger than others. Near these limits, the corresponding logarithms will become dominant, 
but the predictions concerning infrared poles will continue to apply: indeed, selecting dominant 
terms in an expansion in external invariants cannot generate new infrared singularities.
The failure of the infrared factorisation that we have described so far will, instead, 
manifest itself in the fact that we will be unable to predict the infrared-finite parts of 
the dominant logarithmic terms in the relevant limit. This simple consideration opens 
the way for cross-fertilisation between factorisations, and resummations, obtained 
in different limiting configurations. In what follows, we will briefly summarise 
an example of such a cross-fertilisation, examining how infrared factorisation 
constrains and complements the high-energy limit.

The high-energy, or {\it Regge} limit of gauge amplitudes has been the subject 
of a vast body of studies over a span of several decades (see, for example,
\cite{Eden:1966dnq,Collins:1977jy}, and, more recently, \cite{DelDuca:1995hf,
Forshaw:1997dc}). For the sake of illustration, here we will focus on the case of 
the four-gluon amplitude in a massless gauge theory, although the results generalise 
both to quark amplitudes and to multi-particle scattering. As with any renormalised 
massless four-point amplitude, the four-gluon amplitude is a function of the 
Mandelstam invariants $s$, $t$ and $u$, satisfying $s + t + u = 0$, and of the 
renormalisation scale $\mu$. The high-energy limit is defined, in the physical 
region, by taking $s \gg -t > 0$. In this limit, the amplitude is dominated by the 
exchange of a spin-1 gluon in the $t$ channel, and loop corrections generate 
large logarithms of the ratio $s/(-t)$. Leading and next-to-leading logarithms can 
be resummed to all orders thanks to the process of {\it Reggeisation}, which
roughly amounts to the replacement of the $t$-channel gluon propagator in the 
tree-level amplitude according to the rule~\cite{Kuraev:1977fs}
\beq
  \frac{1}{t} \,\, \to \,\, \frac{1}{t} \, \left( \frac{s}{-t} \right)^{\alpha(t)} \, ,
\label{reggeise}
\eeq
where $\alpha(t)$ is the {\it Regge trajectory}, which can be  computed in perturbation 
theory, and is characterised by the presence of infrared poles. At one-loop, for 
example, one can write
\beq
  \alpha(t) \, = \, \left( \frac{\mu^2}{-t} \right)^{\!  \epsilon} \, \frac{\alpha_s (\mu^2)}{\pi}
  \,\,  \alpha^{(1)} + \ord \left( \alpha_s^2 \right) \, ,
\label{oloregge}
\eeq
with $\alpha^{(1)} = C_A/(2 \epsilon)$. At the $n$-loop order, $\alpha^{(n)}$ has poles 
up to $1/\epsilon^n$, as well as finite parts. 

To be more precise, consider the scattering process $g(k_1) + g(k_2) \to g(k_3) + g(k_4)$. 
At leading-logarithmic (LL) accuracy~\cite{Balitsky:1979ap}, and next-to-leading 
logarithmic (NLL) accuracy for the real part of the amplitude~\cite{Fadin:2006bj}, one 
can write the factorised expression
\beq
  {\cal A}_{4g} \left(\frac{s}{\mu^2}, \frac{t}{\mu^2}, \as, \epsilon \right)
  & = & 4 \pi \as \,\, \frac{s}{t} \,\,
  \bigg[ \Big( {\bf  T}^a \Big)_{a_1 a_3}
  C_{\lambda_1 \lambda_3} (k_1, k_3) \bigg] \nonumber  \\
  && \hspace{-2cm} \times
  \left[ \left( \frac{s}{- t} \right)^{\alpha (t)}  + \left( \frac{- s}{- t} \right)^{\alpha(t)} \right]
  \bigg[ \Big( {\bf  T}_a \Big)_{a_2 a_4}
  C_{\lambda_2 \lambda_4} (k_2, k_4) \bigg] \, ,
\label{ReggeFact}
\eeq
where $s = (k_1 + k_2)^2$, $t = (k_1 - k_3)^2$, $a_i$ and $\lambda_i$ are respectively
the colour and polarisation labels of the gluons, and the colour operators ${\bf T}^a_{bc}
= - {\rm i} f^a_{bc}$ in the adjoint representation. The functions $C_{\lambda_i \lambda_j}
(k_i, k_j)$ depend on gluon helicities, but not on the center-of-mass energy
$\sqrt{s}$, and are called {\it impact factors}: they describe the separate dynamical 
evolution of the two colliding particles as they are scattered by a small angle through
the exchange of a reggeized gluon. The central factor in \eq{ReggeFact}, which is
responsible for the $s$ dependence, takes into account the symmetry of the amplitude
under $s \leftrightarrow u$, and the fact that $u = -s$ at leading power in the high-energy 
limit.

The fact that the Regge trajectory $\alpha(t)$ is infrared singular implies that the 
coefficients of the high-energy logarithms $\log(s/t)$ will contain infrared poles. 
These poles can be predicted by infrared factorisation: we expect therefore that 
the high-energy limit of \eq{RGsol} will reflect the structure of the factorisation 
in \eq{ReggeFact}. Focusing on the dipole contribution to the infrared anomalous 
dimension, \eq{GammaDip}, it is easy to verify that this is indeed the case. At
leading power in $t/s$, one finds that the infrared operator in \eq{RGsol}, in the
case of the four-gluon amplitude, factorises in the form~\cite{DelDuca:2011wkl,
DelDuca:2011ae}
\beq
  {\cal Z}_4 \left(\frac{s}{\mu^2}, \frac{t}{\mu^2}, \as \right) \, = \, 
  \exp \big[ \! - 2 \pi {\rm i} \, K (\as) \big] \,
  {\cal Z}_{4, \, C} \left(\frac{t}{\mu^2}, \as \right) \,   
  \widetilde{\cal Z} \left(\frac{s}{t}, \as \right) + \ord \left(\frac{t}{s} \right) \, ,
\label{Zfact1}
\eeq
where $K(\as)$ was defined in \eq{KfromgamK}. The energy-independent factor
${\cal Z}_{4, \, C}$  in \eq{Zfact1} is a product of four `jet' factors, associated with 
each one of the external particles, and is naturally interpreted as a divergent 
contribution to the impact factors in \eq{ReggeFact}. It can be written as
\beq
  {\cal Z}_{4, \, C} \left(\frac{t}{\mu^2}, \as \right) \, = \,
  \exp \Bigg\{ 2 \Bigg[ K (\as) \log \left( \frac{-t}{\mu^2} 
  \right) + D \left( \as \right) \Bigg]
  + 4 B_g \left( \as \right) \Bigg\} \, , 
\label{Z1}
\eeq
where we defined the functions
\beq
  D \left( \as \right) & = &
  - \frac{1}{4} \int_0^{\mu^2} \frac{d \lambda^2}{\lambda^2} \,
  \gamma_K \big( \alpha_s (\lambda^2) \big) 
  \log \left(\frac{\mu^2}{\lambda^2}\right), \nonumber\\
  B_g \left( \as \right) & = &
  - \frac{1}{2} \int_0^{\mu^2} \frac{d \lambda^2}{\lambda^2} \,
  \gamma_g \big( \alpha_s (\lambda^2) \big) \, ,
\label{intandim}
\eeq
and $\gamma_g$ is the gluon jet anomalous dimension. Recalling \eq{doupolefromruco},
one sees that the function $D(\alpha_s)$ encodes double poles of soft-collinear origin,
which do not display colour correlations, as expected from the general structure of 
infrared factorisation. Energy dependence and colour correlations in the high-energy 
limit are confined to the factor $\widetilde{\cal Z}$ in \eq{Zfact1}. In order to write
it  explicitly, it is useful to define colour charges associated with exchanges in the 
$s$, $t$ and $u$ channels, according to~\cite{Dokshitzer:2005ig}
\beq
  {\bf T}_s & \equiv &  {\bf T}_1 + {\bf T}_2 \, = \, - \big( {\bf T}_3 + {\bf T}_4 \big) \, ,
  \nonumber \\
  {\bf T}_t & \equiv &  {\bf T}_1 + {\bf T}_3 \, = \, - \big( {\bf T}_3 + {\bf T}_4 \big) \, ,
  \nonumber \\
  {\bf T}_u & \equiv &  {\bf T}_1 + {\bf T}_4 \, = \, - \big( {\bf T}_2 + {\bf T}_3 \big) \, .
\label{stucol}
\eeq
Due to colour conservation, these charges obey a sum rule formally similar to
the sum of Mandelstam invariants, with masses replaced by quadratic Casimir 
eigenvalues. One verifies that
\beq
  {\bf T}_s^2 + {\bf T}_t^2 + {\bf T}_u^2 \, = \, \sum_{i = 1}^4 C^{(2)}_i \, .
\label{colCas}
\eeq
In terms of these colour operators, one finds
\beq
  \widetilde{{\cal Z}} \left(\frac{s}{t}, \as \right)
  \, = \, \exp \left\{ \widehat{K} ( \as )
  \left[ \log \left( \frac{s}{-t } \right) {\bf T}_t^2 + i \pi {\bf T}_s^2\right] \right\} \, ,
\label{widetildeZ}
\eeq
where $\widehat{K} ( \as )$ is the function in \eq{KfromgamK} with the quadratic 
Casimir of the appropriate representation scaled out (in the present case, 
$\widehat{K} ( \as ) = K ( \as )/C_A$). Leading high-energy logarithms accompanied
by infrared poles are generated exclusively from the first term in the exponent, and 
a direct matching with \eq{ReggeFact} yields an expression for the divergent part
of the Regge trajectory $\alpha(t)$. Indeed, when the colour operator $\widetilde{\cal Z}$
acts on a tree-level exchange dominated at leading power by $t$-channel octet 
exchange (as is the case for gluon scattering), one can replace the colour operator
${\bf T}_t^2$ by its eigenvalue $C_A$. One then finds that
\beq
  \alpha(t) \, = \, K \big( \alpha_s (-t), \epsilon \big) \, ,
\label{ReggeTraj}
\eeq
where we naturally chose the renormalisation scale as $\mu^2 = |t|$. \eq{ReggeTraj}
readily reproduces \eq{oloregge} at one loop, and provides an all-order expression
for infrared poles at leading-logarithmic accuracy. Note that an expression analogous 
to \eq{ReggeTraj} was derived much earlier, in Refs.~\cite{Korchemsky:1993hr,
Korchemskaya:1994qp,Korchemskaya:1996je}, with a different approach to the 
high-energy limit. Recognising that high-energy scattering for $s \gg -t$ corresponds 
to small scattering angles, the colliding particles are approximated with {\it infinite} 
(as opposed to semi-infinite) Wilson lines, corresponding to the classical trajectories 
for forward scattering, and separated by a fixed impact parameter in the transverse 
plane. Arguments closer to the ones discussed here were also given in 
Ref.~\cite{Sotiropoulos:1993rd}.

In recent years, constraints linking the high-energy limit of scattering amplitudes 
with infrared factorisation have been successfully exploited to gain a number of 
important insights. For example, Refs.~\cite{DelDuca:2013ara,DelDuca:2014cya} 
showed how infrared information can be used to go beyond the (next-to-)leading 
logarithmic approximation of \eq{ReggeFact}, uncovering contributions that are 
not described by simple gluon reggeisation, and in particular explaining the 
breakdown of \eq{ReggeFact} for the real part of the amplitude at NLL accuracy, 
first observed in Ref.~\cite{DelDuca:2001gu}. The case of ${\cal N} = 4$ Super-Yang-Mills
theory was considered in Refs.~\cite{Naculich:2007ub,Naculich:2009cv}, focusing 
in particular on an interesting set of sub-leading colour contributions~\cite{Naculich:2013xa}. 
On the other hand, powerful tools are available to study the high-energy limit of scattering 
amplitudes with a much greater accuracy and generality than the approximation described 
by \eq{ReggeFact}. In particular, over the years, an effective theory for the high-energy limit 
in terms of multiple parallel Wilson lines interacting at finite impact parameters has been
developed, leading to the Balitsky-JIMWLK equation~\cite{Mueller:1994jq,Balitsky:1995ub,
Jalilian-Marian:1996mkd,Jalilian-Marian:1997jhx,Balitsky:1998ya,Balitsky:1998kc,
Kovchegov:1999yj}. This framework was recast in Ref.~\cite{Caron-Huot:2013fea} in a 
formulation allowing for a much more direct comparison with infrared factorisation: this 
made possible the first direct calculation of a contribution to the soft anomalous dimension
matrix going beyond the dipole formula, at the four-loop level. A systematic development
of the framework proposed in~\cite{Caron-Huot:2013fea} has led to the determination
of the soft anomalous dimension matrix for $2 \to 2$ scattering amplitudes at NLL
accuracy to all orders in perturbation theory~\cite{Caron-Huot:2017fxr,Caron-Huot:2017zfo,
Caron-Huot:2020grv}, and at NNLL accuracy up to four loops~\cite{Falcioni:2020lvv,
Maher:2021nlo,Falcioni:2021buo}. This result, in turn, establishes for the first time 
the presence of quartic Casimir contributions to the soft anomalous dimension matrix, 
beyond those induced by the cusp anomalous dimension.


\subsubsection{A celestial view}
\label{CeleVi}

We conclude this Section by providing an example of how the novel approach
to the infrared limit introduced in Refs.~\cite{Strominger:2013jfa,Strominger:2013lka}, 
and reviewed in Refs.~\cite{Strominger:2017zoo,Pasterski:2021rjz}, can provide 
fresh and surprising insights on the physics and mathematics of soft interactions. 
We will use the dipole formula, \eq{GammaDip}, as an example. Following 
Ref.~\cite{Magnea:2021fvy}, we will begin by rewriting the dipole formula in 
coordinates appropriate to the {\it celestial sphere}. We will see that this 
formulation leads to a remarkable simplification of the infrared operator 
${\cal Z}_n$, defined in \eq{RGsol}, in the dipole approximation. We will 
then observe that colour correlations, which originate exclusively from 
wide-angle soft radiation, can be computed in terms of a simple two-dimensional 
conformal field theory on the celestial sphere: a remarkable result, which might 
well lead to further insights beyond the dipole approximation, and new 
computational techniques.
 
As a first step, it is useful to disentangle the explicit dependence on the running 
renormalisation scale $\lambda$ in the integrand in \eq{RGsol} from colour correlations. 
To this end, note that the soft anomalous dimension matrix, given by \eq{GammaDip} 
in the dipole approximation, must be evaluated at the running scale $\lambda$ when 
substituted in \eq{RGsol}, where $\lambda$ is the integration variable. It is then useful 
to rewrite the arguments of the logarithms in \eq{RGsol} in terms of the fixed scale 
$\mu$, and then use colour conservation, expressed by \eq{colopgaugein}. This leads to
\beq
\label{Gammasimp}
  \Gamma_n^{\rm \, dip} \left( \frac{s_{ij}}{\lambda^2}, \alpha_s (\lambda, \e) \right) 
  & = & \frac{1}{2} \, \widehat{\gamma}_K \big(\alpha_s (\lambda, \e) \big)
  \sum_{i = 1}^n \sum_{j = i + 1}^n \ln \left( \frac{- s_{ij} + {\rm i} \eta}{\mu^2} \right)
  {\bf T}_i \cdot {\bf T}_j  \\
  & & - \, \sum_{i = 1}^n \gamma_i \big(\alpha_s (\lambda, 
  \e) \big) 
  \, - \, \frac{1}{4} \, \widehat{\gamma}_K \big(\alpha_s (\lambda, 
  \e) \big) \ln \left(\frac{\mu^2}{\lambda^2}\right)
  \sum_{i = 1}^n C_i^{(2)} \, . \nonumber
\eeq
\eq{Gammasimp} reflects the fact that colour correlations arise only from the exchange 
of soft, wide-angle gluons, and therefore they involve only {\it single} soft poles, generated
by the first line in \eq{Gammasimp}. Collinear effects must be free of colour correlations,
as dictated by the form of the factorisation in \eq{ampnfact}: this is reflected in the
second line in \eq{Gammasimp}, which is a sum of contributions from each external 
particle, including double-pole contributions arising from the last term.

The second step, which is natural in the approach of Refs.~\cite{Strominger:2013jfa,
Strominger:2013lka}, is to parametrise particle momenta in celestial coordinates,
using
\beq
  p_i^\mu \, = \, \omega_i \, \Big\{ 1 + z_i \zbar_i, \, z_i + \zbar_i, \, 
  - {\rm i} \big( z_i - \zbar_i \big), \, 1 - z_i \zbar_i \Big\} \, .
\label{spherepar}
\eeq
\eq{spherepar} describes a massless momentum $p_i^\mu$ in terms of a light-cone 
energy $\omega_i$ and a dimensionless complex coordinate $z_i$ identifying the 
direction of $p_i^\mu$. More precisely, note that the set of light rays emanating from 
the origin in four-dimensional Minkowsky space is in one-to-one correspondence with 
the points of a sphere ${\cal S}_2$. At asymptotically large light-cone times, the sphere 
can be mapped to the complex plane, and $z_i$ identifies the intersection of the light 
ray corresponding to $p_i^\mu$ with this asymptotic {\it celestial} sphere. Importantly,
the Lorentz group $SO(1,3)$ acts on the celestial coordinates $z_i$ as $SL(2, {\bf C})$,
providing a first suggestion that the theory describing the asymptotic dynamics could
be a two-dimensional conformal field theory. For our present  purposes, an important 
immediate consequence of the parametrisation in \eq{spherepar} is the expression
for the Mandelstam invariants, which are given by
\beq
  s_{ij} \, = \, 2 p_i \cdot p_j \, = \, 4 \omega_i \omega_j \big| z_i - z_j \big|^2 \, .
\label{sij}
\eeq
This implies that the logarithms appearing in the colour-correlated part of 
\eq{Gammasimp} will decompose as
\beq
  \log \left( \frac{- s_{ij} + {\rm i} \eta}{\mu^2} \right) \, = \, 
  \log \Big( \big| z_i - z_j \big|^2 \Big) + 
  \log \left( \frac{\omega_i}{\mu} \right) + \log \left( \frac{\omega_j}{\mu} \right) 
  +  2 \log 2 + {\rm i} \pi \, .
\label{energies}
\eeq
We can now neatly separate colour-correlated from colour-singlet contributions to 
the soft anomalous dimension matrix, writing
\beq
  \Gamma_n^{\rm \, dip} \left( \frac{s_{ij}}{\lambda^2}, \alpha_s (\lambda, \e) \right) 
  \, \equiv \,
  \Gamma_n^{\rm \, corr} \Big( z_{i j}, \alpha_s (\lambda, \e) \Big) 
  \, + \, \Gamma_n^{\rm \, singl} \bigg( \frac{\omega_i}{\lambda}, 
  \alpha_s (\lambda, \e) \bigg) \, ,
\label{alGammasimp}
\eeq
where we defined $z_{ij} \equiv z_i - z_j$.  The colour-singlet contribution is given by
\beq
  \Gamma_n^{\rm \, singl} \bigg( \frac{\omega_i}{\lambda}, 
  \alpha_s (\lambda, \e) \bigg) \, = \, 
  - \, \sum_{i = 1}^n \gamma_i \big(\alpha_s (\lambda, \e) \big) \,
  \, - \, \frac{1}{4} \, \widehat{\gamma}_K \big(\alpha_s (\lambda, 
  \e) \big) \sum_{i = 1}^n \,
  \ln \left( \frac{- 4 \hspace{1pt}  \omega_i^2 + {\rm i} \eta}{\lambda^2} \right)
  C_i^{(2)} \, ,
\label{FinSingl}
\eeq
while colour correlations are generated by
\beq
  \Gamma_n^{\rm \, corr} \Big( z_{ij}, \alpha_s (\lambda, \e) \Big) 
  \, = \, 
  \frac{1}{2} \, \widehat{\gamma}_K \big(\alpha_s (\lambda, \e) \big)
  \sum_{i = 1}^n \sum_{j = i + 1}^n \ln \Big( \big| z_{ij} |^2 \Big) \, 
  {\bf T}_i \cdot {\bf T}_j \, .
\label{FinCorr}
\eeq
\eq{FinCorr} is remarkable, showing that the dependence on the running scale 
$\lambda$, the coupling, and the infrared regulator is universal, and completely 
factorised from colour correlations. As a consequence, one can write the colour-correlated 
part of the infrared operator ${\cal Z}_n$ in a strikingly simple way, as
\beq
  {\cal Z}_n^{\rm \, corr} \Big( z_{ij}, \alpha_s(\mu), \e \Big) & \equiv & 
  \exp \left[ \int_0^{\mu} \frac{d \lambda}{\lambda} \,\, 
  \widehat{\Gamma}_n^{\rm \, corr} \Big( z_{ij}, \alpha_s (\lambda, \e) \Big) \right] 
  \nonumber \\
  & = & 
  \exp \bigg[ \! - \! \widehat{K} \big( \alpha_s (\mu), \e \big) \, \sum_{i = 1}^n 
  \sum_{j = i + 1}^n 
  \ln \Big( \big| z_{ij} |^2 \Big) \, {\bf T}_i \cdot {\bf T}_j \bigg] \, .
\label{ZCorr}
\eeq
The celestial parametrisation given in \eq{spherepar} thus shows that the 
factorisation of infrared singularities in the universal function $\widehat{K}$
is completely general, and not limited to the high-energy limit of four-point amplitudes
given in \eq{widetildeZ}. This universal role of the function $\widehat{K}$, which
can be seen as the scale average of the light-like cusp anomalous dimension
in the infrared range, provides support for a long-standing suggestion: that the
light-like cusp should indeed be taken as a proper definition of the gauge coupling
in the infrared regime~\cite{Catani:1990rr,Erdogan:2011yc,Grozin:2014hna,Grozin:2015kna,
Banfi:2018mcq,Catani:2019rvy}.

Taking now a more directly {\it celestial} viewpoint, we note that the expression in 
\eq{ZCorr}, regarded as a function on the punctured Riemann sphere, bears a striking 
similarity to a correlator of primary fields in a two-dimensional conformal field theory. 
This similarity was noticed, in the case of QED, in~\cite{Nande:2017dba}, and, in greater 
detail, in \cite{Kalyanapuram:2020epb}. Note however that in QED the infrared 
operator ${\cal Z}_n$ is one-loop exact, whereas \eq{ZCorr} organises highly 
non-trivial all-order corrections in the non-abelian theory. In order to explore this
remarkable connection, following Ref.~\cite{Magnea:2021fvy}, consider a theory
of free bosons, spanning the adjoint representation of the gauge algebra, with
the action
\beq
  S [ \phi ] \, = \,  \frac{1}{2 \pi} \int d^2 z \, \partial_z \phi^a (z, \zbar) \, 
  \barpartial \phi_a (z, \zbar) \, ,
\label{action}
\eeq
with $a = 1, \ldots, N_c^2 - 1$ for $SU(N)$.  As a conformal theory on the sphere,
the action in \eq{action} is of course well-known~\cite{Ginsparg:1988ui}, not least 
because it maps to the action for the free bosonic string, if the adjoint index $a$ is 
replaced by a space-time index $\mu$ (see, for example,~\cite{Polchinski:1998rq}). 
The theory is completely solvable: the energy-momentum tensor is traceless and 
conserved, and furthermore the manifest global translation invariance of \eq{action}
implies the existence of two conserved Noether currents, respectively holomorphic 
and anti-holomorphic, given by
\beq
  j^a (z) \, = \, \partial_z \phi^a (z, \zbar) \, , \qquad 
  \tilde{j}^a (\zbar) \, = \, \partial_{\zbar} \phi^a (z, \zbar) \, .
\label{maykac}
\eeq
By analogy with the bosonic string, and generalising the QED results of 
Refs.~\cite{Nande:2017dba,Kalyanapuram:2020epb} to the non-abelian theory, 
one can introduce matrix-valued vertex operators of the form
\beq
  V (z, \zbar) \, \equiv \, :  {\rm e}^{{\rm i} \kappa \, 
  {\bf T}_{\! z} \cdot \, {\bf \phi} (z, \zbar)}  : \,\, ,
\label{vertex}
\eeq
where the colon denotes normal ordering, $\kappa$ is a normalisation constant, 
and ${\bf T}_z$ is a colour operator as defined in \secn{ColInsOp}. Note that colour 
operators are associated with points on the sphere: operators acting at different 
points commute, as dictated by the structure of the gauge-theory factorisation that 
we are trying to reproduce.

It is far from obvious that \eq{vertex} defines a proper conformal primary field,
since $V(z, \zbar)$ is a matrix in colour space, however it can be readily verified
that the vertex operator has a well-defined conformal weight, given by $h = \kappa^2
C_r^{(2)}/4 > 0$, where $C_r^{(2)}$ is the quadratic Casimir eigenvalue of the
representation chosen for the operator ${\bf T}_z$. Similarly, one can verify that
the two-point function of two vertex operators has the appropriate power-law
behaviour 
\beq
  \big \langle V (z_1, \zbar_1) V (z_2, \zbar_2) \big \rangle \, \sim \, 
  | z_{12} |^{- 4 h} \, .
\label{twopf}
\eeq
Computing correlators of vertex operators of the form of \eq{vertex} in the theory
defined by \eq{action} is a straighforward exercise, and one finds
\beq
  {\cal C}_n \Big( \{ z_i \}, \kappa \Big) \, \equiv \, 
  \Big \langle \prod_{i = 1}^n V (z_i, \zbar_i) \Big \rangle \, = \, C(N_c) \,
  \exp \left[ \frac{\kappa^2}{2} \sum_{i = 1}^n \sum_{j = i + 1}^n 
  \ln \Big( \big| z_{ij} |^2 \Big) \, {\bf T}_i \cdot {\bf T}_j  \right] \, ,
\label{riscorr}
\eeq
where $C(N_c)$ is an overall normalisation depending only on the dimension of the
gauge algebra. Manifestly, the correlator ${\cal C}_n$ reproduces \eq{ZCorr}, with
the identification $\kappa^2 = - 2 \widehat{K} (\alpha_s, \e)$.

Further checks of the power of the celestial approach are possible. One may 
consider limits where one of the $n$ hard particles becomes soft, or where two 
particles become collinear. Strictly speaking, these limits lie outside the range
of validity of the fixed-angle approximation, as was the case for the high-energy 
limit discussed in \secn{HighEn}. One may however study how these limits are
approached, and the predictions for infrared poles will remain valid near the limits.
We will discuss the factorisation properties of real radiation in soft and collinear 
limits in greater detail in \secn{Subtra}, however, for completeness, we will 
describe here how some crucial aspects of that factorisation emerge from the
celestial viewpoint.

The first interesting case is the limit in which a single hard particle becomes 
soft. At tree level, this limit is described by a simple factorisation, discussed
already in~\cite{Weinberg:1965nx} for the abelian theory, and presented here
in \eq{eq:Real_tree_fact} of \secn{Subtra}. As expected from the general reasoning 
of Ref.~\cite{Strominger:2013lka}, this tree-level soft factorisation can be
derived in the celestial context as a Ward identity for the translation symmetry
of the action in \eq{action}. Indeed, computing a correlator of vertex operators
with the insertion of a holomorphic current, by means of the conformal OPE,
yields the result
\beq
   \Big \langle \partial_z \phi^a (z, \zbar) \, \prod_{i = 1}^n V (z_i, \zbar_i) 
   \Big \rangle \, \simeq \, - \frac{\rm i}{2} \sum_{i = 1}^{n} 
   \frac{{\bf T}_i^{\, a}}{z - z_i} \, {\cal C}_n \Big( \{ z_i \}, \kappa \Big) \, ,
\label{softglucur}
\eeq
which reproduces \eq{eq:Real_tree_fact}, upon projecting on one of the two
physical polarisations for the soft gluon, with the tree-level soft current given by
\eq{softcurrtree}. The poles as $z \to z_i$ are {\it collinear} poles, corresponding
to the singularities of the soft current as the soft momentum $k$ becomes
parallel to one of the hard momenta $p_i$. A Ward identity equivalent to 
\eq{softglucur} was derived in Ref.~\cite{He:2015zea} (see also Refs.~\cite{Cheung:2016iub,
Nande:2017dba,Fan:2019emx}) using a current built with asymptotic expressions 
for the gauge fields near null infinity.

Collinear limits for the soft anomalous dimension matrix in the dipole approximation 
were studied in Ref.~\cite{Becher:2009qa}, and are discussed here in \secn{Subtra}.
In such limits, it is useful to construct a {\it splitting} anomalous dimension, defined by
\beq
  \Gamma_{\rm Sp} \big( p_1, p_2 \big) \, \equiv \,
  \Gamma_n \big( p_1, p_2, \ldots, p_n \big) - 
  \Gamma_{n - 1} \big( p, p_3, \ldots, p_n \big) 
  \Big|_{{\bf T}_p \to {\bf T}_1 + {\bf T}_2} \, ,
\label{GamSplit}
\eeq
where $p_1$ and $p_2$ are the momenta becoming collinear, and $p$ is their 
common collinear limit. The statement of collinear factorisation at amplitude 
level\footnote{The statement is strictly valid when all coloured particles are
{\it outgoing}~\cite{Catani:2011st}.} is the fact that the splitting anomalous 
dimension $\Gamma_{\rm Sp}$ depends only on the quantum numbers of 
the two collinear particles. Using a precise definition of the collinear limit
on the celestial sphere~\cite{Magnea:2021fvy}, one may compute collinear 
limits of the celestial correlator ${\cal C}_n$ by means of the conformal OPE.
Upon reinstating the energy dependence that was factored in the singlet 
contribution, \eq{FinSingl}, one may verify that collinear factorisation is built 
in the conformal correlation function, and one recovers the known {\it all-order}
expression for the splitting anomalous dimension in the dipole approximation,
first derived in Ref.~\cite{Becher:2009qa},
\beq
  \Gamma_{\rm Sp.} \big( p_1, p_2 \big) & = & \frac{1}{2} \,
  \widehat{\gamma}_K ( \alpha_s ) \bigg[ \ln \bigg( 
  \frac{- s_{12} + {\rm i} \eta}{\mu^2} \bigg) {\bf T}_1 \cdot {\bf T}_2 
  - \ln x \,\, {\bf T}_1 \cdot \big( {\bf T}_1 + \, {\bf T}_2 \big)
  \nonumber \\ && \hspace{3cm}  
  - \, \ln (1- x) \,\, {\bf T}_2 \cdot \big( {\bf T}_1 + {\bf T}_2 \big)
  \bigg] \, ,
\label{GammaSplitFin}
\eeq
where $x$ and $1-x$ are the energy fractions carried by the two collinear
particles $p_1$ and $p_2$, respectively.

The emerging connection between the infrared properties of massless 
non-abelian gauge theories and two-dimensional conformal field theories
is certainly one of the most striking aspects of the approach pioneered in
Refs.~\cite{Strominger:2013jfa,Strominger:2013lka}. It can be more precisely 
formalised by taking a Mellin transform of momentum-space scattering 
amplitudes with respect to light-cone energies, which yields the so-called
{\it celestial amplitudes} (see, for example,~\cite{Pasterski:2016qvg,
Pasterski:2017kqt,Pasterski:2017ylz,Arkani-Hamed:2020gyp}), which 
have simple transformation properties under the two-dimensional conformal 
group. Within this approach, results similar, and in part complementary, to 
the ones discussed in this Section, were obtained in Ref.~\cite{Gonzalez:2021dxw},
and further extended in Ref.~\cite{Nastase:2021izh}, studying special colour 
configurations and the high-energy limit of four-point amplitudes\footnote{Furthermore,
an intense research activity continues on the general properties of celestial 
gauge amplitudes and their connections to the underlying conformal theory:
see, for example, Refs.~\cite{Schreiber:2017jsr,Pate:2019mfs,Nandan:2019jas,
Pate:2019lpp,Fan:2020xjj,Guevara:2021abz,Fan:2021isc,Crawley:2021ivb,Fan:2021pbp,
Kapec:2021eug,Adamo:2021zpw,Strominger:2021mtt}.}. In principle, existing 
gauge-theory data can be used to explore the non-linear generalisation of 
\eq{action}, which would be needed in order to extend this analysis beyond 
the dipole approximation. If this generalisation can be identified, conformal 
field theory techniques may well provide powerful new tools for the study of 
infrared factorisation, at high orders in perturbation theory, and possibly beyond.


\subsection{Computing the infrared anomalous dimension matrix}
\label{CompMatr}

It should be clear from the arguments of \secn{DipFor} that the infrared anomalous 
dimension matrix defined in \eq{FullGamma} is the cornerstone for the discussion
of soft and collinear divergences in gauge theory scattering amplitudes, as well as
a crucial ingredient for the resummation of large logarithms in many cross sections
of phenomenological interest; furthermore, it gives a perturbative window to explore
the connection between colour exchanges and kinematic configurations, providing
insights that have implications to all orders in perturbation theory. It is clear therefore
that it is very worthwhile to develop systematic tools to compute $\Gamma_n$ to
high orders in perturbation theory.

To this end, one may first note that the hard collinear components of the infrared 
matrix, essentially contained in the jet anomalous dimensions $\gamma_i$ in
\eq{GammaDip}, are colour-diagonal, and can therefore be extracted in a 
straightforward manner from form factor data. Non-trivial colour structures,
and the subtle correlations between colour and kinematics, arise from the 
soft operator defined in \eq{Softnpart}, and this Section will therefore focus on 
the properties of correlators of semi-infinite Wilson lines, and on techniques to 
evaluate them.

As discussed already in \secn{eikint} at the one-loop level, soft operators of the 
form of \eq{Softnpart} are highly singular, being affected by ultraviolet, soft, and, 
in case $\beta_i^2 = 0$, collinear divergences. As a consequence, special care is 
required to evaluate them~\cite{Gardi:2013saa,Mitov:2010xw,Henn:2013wfa,
Falcioni:2014pka}. In \secn{eikint}, it was possible to exploit the simple form
of the one-loop correction in order disentangle the different singularities, and
thus extract the relevant UV counterterm, which still contains a collinear pole,
as displayed in \eq{oneloopct}. At higher orders, this simple-minded treatment
becomes very difficult, as different singular regions in the relevant integrals
overlap in an intricate way. It is then necessary to introduce different regulators
in order to safely extract the anomalous dimension from the correlator.

As a first step, it is natural to regulate collinear divergences (for massless 
amplitudes) by setting $\beta_i^2 \neq 0$, {\it i.e.} tilting the Wilson lines
off the light cone. This clearly has an intrinsic interest, since infrared divergences
arising from soft gluon exchanges affect also amplitudes involving massive  
particles, and they are described precisely by this kind of soft function. Results
relevant for the massless case can then be extracted by carefully taking the
$\beta_i^2 \to 0$ limit~\cite{Almelid:2015jia}. An immediate consequence of
the introduction of the collinear regulator is that all issues related to collinear
divergences, and in particular the anomalous dependence of the soft function 
on the invariants $\beta_i \cdot \beta_j$, are eliminated. In the massive case, 
the soft function can only depend on rescaling-invariant variables, which can 
be expressed in terms of the Minkowskian angles
\beq
  \gamma_{ij} \, \equiv \, \frac{\beta_i \cdot \beta_j}{
  \sqrt{\beta_i^2 \beta_j^2}}
  \, = \, \frac{p_i \cdot p_j}{\sqrt{p_i^2 p_j^2}} \, .
\label{gamma_ij}
\eeq
Even for {\it massive} Wilson lines, it remains true that all loop corrections to the 
soft function in \eq{Softnpart}, before renormalisation, vanish in dimensional 
regularisation, since they are given by scale-less integrals. We must therefore 
concentrate on the calculation of UV counterterms, as was done in \secn{eikint}.
To this end, it is useful to introduce an infrared (soft) regulator, while retaining 
dimensional regularisation only for ultraviolet singularities. The bare, infrared-regulated
correlator will have  the same UV divergences as its unregulated counterpart: it 
can then be evaluated and renormalised, yielding the desired answer.

An elegant way to proceed was proposed in Ref.~\cite{Gardi:2011yz,Gardi:2013saa}, 
and exploited for explicit calculations up to 3 loops in Refs.~\cite{Almelid:2015jia,
Falcioni:2014pka}. In this approach, one introduces the infrared regulator at the
level of individual Wilson lines, forcing an exponential suppression of gluon emission 
at large distances by means of the definition
\beq
  \Phi_{\beta_i}^{(m)} \, = \, {\cal P} \exp \left[ \, {\rm i} g \mu^\eps 
  \int_0^\infty d \lambda \, \beta_i \cdot A \left( \lambda \beta_i \right) \, 
  {\rm e}^{- m \lambda \sqrt{\beta_i^2}} \,
  \right] \, ,
\label{Phidef}
\eeq
and then defining the regulated soft function by
\beq
  {\cal S}_n^{\,(m)} \left( \gamma_{ij}, \alpha_s(\mu), \eps, \frac{m}{\mu} \right) \, \equiv \, 
  \left\langle 0 \left| \, T \left[ \Phi_{\beta_1}^{(m)} \otimes \Phi_{\beta_2}^{(m)} \otimes
  \ldots \otimes \Phi_{\beta_L}^{(m)} \right] \right| 0 \right\rangle \, ,
\label{Sregdef}
\eeq
so that one recovers the unregulated Wilson line as $m \to 0$. The crucial advantage
of the regulator defined in \eq{Phidef} is that it preserves the rescaling invariance of
the correlator under $\beta_i \to \kappa_i \beta_i$, so that all the ensuing constraints
on its functional form also apply to the regulated definition. If one were to take seriously
the $m$ dependence, this kind of long-distance suppression would raise issues of
gauge invariance, however, in a minimal renormalisation scheme, all dependence 
on $m$ cancels in the renormalised correlator, so the proposed regularisation remains 
viable.

In order to proceed, we make use of the fact that Wilson line correlators are 
multiplicatively renormalisable~\cite{Korchemsky:1987wg,Polyakov:1980ca,
Arefeva:1980zd,Dotsenko:1979wb,Brandt:1981kf}, which, in the present case, 
means that we can define the renormalised correlator through the matrix equation
\beq
  {\cal S}_n^{\, (\rm ren.)} \left( \gamma_{ij}, \alpha_s(\mu^2), \eps, \frac{m}{\mu} 
  \right) \, = \, {\cal S}_n^{\, (m)} \left( \gamma_{ij}, \alpha_s(\mu^2), \eps, \frac{m}{\mu} 
  \right) \, {\cal Z}_n^{\, \cal S} \left( \gamma_{ij}, \alpha_s(\mu^2), \eps \right) \, .
\label{Srendef}
\eeq
We can now focus on the calculation of ${\cal Z}_n^{\cal S}$, which, as we will verify, is 
independent of $m$. Indeed, as stated above, the UV divergences of ${\cal S}_n^{\, (m)}$
coincide with those of the original soft function; furthermore, as we discussed, in the limit 
$m \to 0$ the bare correlator reduces to the unit matrix (since all loop corrections vanish 
in dimensional regularisation), so that, in that limit, ${\cal Z}_n^{\cal S}$ coincides with 
the renormalised soft function that we wish to evaluate.

The renormalization matrix ${\cal Z}_n^{\cal S}$ satisfies a renormalisation group 
equation of the same form as \eq{softmatrRG}, which we write as
\beq
  \mu \frac{d }{d \mu} \, {\cal Z}_n^{\, \cal S} \left( \gamma_{ij}, \alpha_s(\mu^2), \eps \right) 
  \, = \, - \, {\cal Z}_n^{\, \cal S} \left( \gamma_{ij}, \alpha_s(\mu^2), \eps \right) 
  \Gamma_n^{\, \cal S} \left(\gamma_{ij}, \alpha_s(\mu^2) \right) \, ,
\label{Zeq}
\eeq
which defines the soft anomalous dimension matrix $\Gamma_n^{\, \cal S}$: it 
is a finite matrix, independent of $\epsilon$, with a direct  physical meaning for 
the scattering of massive coloured particles, while in the massless case it provides 
a regularised version of the singular matrix introduced in \eq{softmatrRG}; it 
compactly encodes the ultraviolet singularities of ${\cal Z}_n^{\cal S}$, and 
thus of ${\cal S}_n$. We note that the ordering on the {\it r.h.s} of \eq{Zeq} 
is important, since both ${\cal Z}_N^{\cal S}$ and $\Gamma_n^{\, \cal S}$ are 
matrix valued, and therefore do not commute in general. This has important 
consequences on the explicit form of the solution of \eq{Zeq}, which are worth 
discussing before getting into the details of the evaluation.

The non-commutativity of the factors in \eq{Srendef} and in \eq{Zeq} influences
the calculation of the anomalous dimension matrix in two different ways. To 
illustrate them, let us for the moment assume, as suggested by \eq{softmatrsol} 
and \eq{RGsol}, and as will be discussed in detail in \secn{ExpoColo}, that all 
factors in \eq{Srendef} can be written in exponential form, as
\beq
  {\cal S}_n^{\, (\rm ren.)} \, = \, \exp \left[ W_n^{\, (\rm ren.)} \right] \, , \qquad
  {\cal S}_n^{\, (m)} \, = \, \exp \left[ W_n^{\, (m)} \right] \, , \qquad
  {\cal Z}_n^{\, \cal S} \, = \, \exp \big[ \zeta_n \big] \, .
\label{defexp}
\eeq
Now the first thing to notice is that the matrix nature of ${\cal Z}_n^{\, \cal S}$
complicates the relationship between the perturbative coefficients of $\Gamma_n^{\, 
\cal S}$ and those of $\zeta^{\, \cal S}_n$. To see this one may write~\cite{Gardi:2011yz}
\beq
  \Gamma_n^{\, \cal S} & = & - \, \left( {\cal Z}_n^{\, \cal S} \right)^{-1}
  \, \frac{d {\cal Z}_n^{\, \cal S}}{d \ln \mu} \, = \,
  \int_0^1 d \tau  \, {\rm e}^{- \tau \zeta_n} \, 
  \frac{d \zeta_n}{d \ln \mu} \, {\rm e}^{\tau \zeta_n} \nonumber \\ 
  & = & - \,  \frac{d \zeta_n}{d \ln \mu} + \frac{1}{2} \left[ \zeta_n ,  
  \frac{d \zeta_n}{d \ln \mu} \right] - \frac{1}{6} \left[ \zeta_n ,  \left[ \zeta_n,
  \frac{d \zeta_n}{d \ln \mu} \right] \right] + \ldots \, .
\label{calcGamma}
\eeq
Next one may use the fact that, in a minimal subtraction scheme, $\zeta_n$ 
can depend on the renormalisation scale only through the coupling; defining
the perturbative coefficients by 
\beq
  \Gamma_n^{\, \cal S} \, = \, \sum_{k = 1}^\infty \bigg( \frac{\alpha_s}{\pi} \bigg)^k
  \Gamma_{\!{\cal S}, n}^{\,(k)} \, , \qquad 
  \zeta_n  \, = \, \sum_{k = 1}^\infty \bigg( \frac{\alpha_s}{\pi} \bigg)^k
  \zeta_n^{\, (k)} \, ,
\label{pertexpGamze}
\eeq
and using \eq{beta} and \eq{calcGamma}, one easily finds at low orders
\beq
  \Gamma_{\!{\cal S}, n}^{\,(1)} \! & = & \! 2 \epsilon \zeta_n^{\,(1)} \, , \nonumber \\
  \Gamma_{\!{\cal S}, n}^{\,(2)} \! & = & \! 4 \epsilon \zeta_n^{\,(2)} + 
  \frac{b_0}{2} \zeta_n^{(1)} \, , \nonumber \\ 
  \Gamma_{\!{\cal S}, n}^{\,(3)} \! & = & \! 6 \epsilon \zeta_n^{\,(3)} 
  - \epsilon \left[ \zeta_n^{(1)}, \zeta_n^{(2)} \right] + b_0 \zeta_n^{(2)}
  + \frac{b_1}{2} \zeta_n^{(1)} \, ,
\label{Gamtozet}
\eeq
with further nested commutators appearing at higher orders. Inverting \eq{Gamtozet},
one finds, as customary in minimal schemes, that the $k$-loop anomalous dimension 
matrix is proportional to the single pole of $\zeta_n^{\,(k)}$; furthermore, $\zeta_n^{\,(k)}$ 
is found to have UV poles up to $\epsilon^{-k}$, with coefficients which are determined 
recursively from lower-order results~\cite{Gardi:2011yz}.

The second non-trivial consequence of the matrix nature of soft anomalous dimensions
is the relation between the logarithm of the renormalised correlator and that of its
regularised counterpart. In fact, substituting \eq{defexp} into \eq{Srendef}, one finds
that~\cite{Mitov:2010rp,Gardi:2011yz}
\beq
  W_n^{\, (\rm ren.)} \, = \, W_n^{\, (m)} + \zeta_n + \
  {\cal C} \Big( W_n^{\, (m)}, \zeta_n \Big) \, ,
\label{relbetexp}
\eeq
where ${\cal C}$ is a series of nested commutators, determined as usual by the 
Baker-Campbell-Haussdorf formula. As is often the case when employing dimensional
regularisation and minimal schemes, \eq{relbetexp} provides non-trivial constraints:
indeed, the renormalised matrix $W_n^{\, (\rm ren.)}$ must be finite as $\epsilon 
\to 0$, while the matrix $\zeta_n$ contains only poles in $\epsilon$. In the presence
of the commutator terms building up the function ${\cal C}$, this means, remarkably,
that terms proportional to {\it positive} powers of $\epsilon$ in the regularised matrix
$W_n^{\, (m)}$ can interfere with the poles in $\zeta_n$ to give finite contributions
to the renormalised correlator, and thus to the anomalous dimension matrix.
Concretely, one may expand the regularised matrix $W_n^{\, (m)}$ in powers
of $\alpha_s$ and $\epsilon$, as
\beq
  W_n^{\, (m)} \, = \, \sum_{k = 1}^{\infty} \bigg( \frac{\alpha_s}{\pi} \bigg)^{\! k}
  \sum_{p = - k}^\infty \eps^p \, W_n^{\, (k,p)} \, .
\label{expinaande}
\eeq
Using the finiteness of \eq{relbetexp}, as well as \eq{Gamtozet}, one can finally
express the perturbative coefficients of the anomalous dimension matrix directly
in terms of the regularized matrix $W_n^{\, (m)}$, which is the object one actually 
computes. At low orders, one finds~\cite{Mitov:2010rp,Gardi:2011yz}
\beq
  \Gamma_{\!{\cal S}, n}^{\,(1)} &=& - 2 W_n^{(1,-1)} \nonumber \\
  \Gamma_{\!{\cal S}, n}^{\,(2)} &=& - 4 W_n^{(2,-1)} - 2 \left[ W_n^{(1,-1)}, W_n^{(1,0)} \right] 
  \nonumber \\
  \Gamma_{\!{\cal S}, n}^{\,(3)} &=& - 6 W_n^{(3,-1)}  + \frac{3}{8} \, b_0 
  \left[ W_n^{(1,-1)}, W_n^{(1,1)} \right] + 3 \left[ W_n^{(1,0)}, W_n^{(2,-1)} \right]
  + 3 \left[ W_n^{(2,0)}, W_n^{(1,-1)} \right]  \nonumber  \\
  && + \left[ W_n^{(1,0)}, \left[ W_n^{(1,-1)}, W_n^{(1,0)} \right]  \right]
  - \left[ W_n^{(1,-1)}, \left[W_n^{(1,-1)}, W_n^{(1,1)} \right] \right] \, ,
\label{Gamfromreg}
\eeq
which can be straightforwardly (if tediously) extended to higher orders. Having
established this general framework, one can now proceed to evaluate the soft 
anomalous dimension matrix $\Gamma_n^{\, \cal S}$ from the regularised correlator,
with diagrammatic methods. The problem has two aspects: the determination and 
analysis of the possible colour structures appearing in $\Gamma_n^{\, \cal S}$, 
and the calculation of the corresponding kinematic factors. We will first study colour 
structures in \secn{ExpoColo} and in \secn{Cwebs}, and then we will give two simple 
examples of the calculation of kinematic factors in \secn{cusponeloop} and in 
\secn{Twolosad}.


\subsubsection{Diagrammatic exponentiation of Wilson-line correlators}
\label{ExpoColo}

So far in our analysis we have shown that Wilson-line correlators exponentiate
on the basis of factorisation and renormalisation group arguments. The resulting
exponentiation has proven to be non-trivial: the logarithm of the correlator 
is significantly less singular than the correlator itself, and it can be  expressed 
as a scale integral of a finite anomalous dimension colour matrix. We now
turn to the discussion of a second, independent method to prove exponentiation,
which is based on purely diagrammatic arguments, and leads to a systematic
way to compute directly the perturbative exponent.

We begin by stating the most general form of the result, which holds not only
for correlators of straight, semi-infinite Wilson lines, such as the ones we have 
been discussing so far, but actually extends to generic Wilson lines, following
curved paths $\gamma$ which can be open or closed. For the purposes of 
the present discussion we consider then correlators of the form
\beq
  {\cal S}_n \big( \! \left\{ \gamma_i \right\} \! \big) \, \equiv \, \bra{0} 
  \, T \bigg[ \prod_{k = 1}^n
  \Phi_{r_k} \left(  \gamma_k \right) \bigg] \ket{0} \, ,
\label{genWLC}
\eeq
where
\beq
  \Phi_r \left(  \gamma \right) \, \equiv \, P \exp \left[ {\rm i} g \!
  \int_\gamma d x \cdot {\bf A}_r (x) \right]
\label{genWL}
\eeq
is the Wilson line operator defined on the curve $\gamma$ and in the representation
$r$ of the gauge group. All such correlators can be written in exponential form as
\beq
  {\cal S}_n \big( \! \left\{ \gamma_i \right\} \! \big) \, = \, \exp \Big[ { W}_n 
  \big( \! \left\{ \gamma_i \right\} \! \big) \Big]  \, .
\label{diaxp}
\eeq
We already know, from renormalisation group arguments, that the exponentiation 
is non trivial. The new statement now is that the exponent $W_n$ can be directly 
computed in terms of a proper subset of the Feynman diagrams that form the
perturbative expansion of the original correlator ${\cal S}_n$; these diagrams
will however appear in the exponent with modified colour factors, which can be
determined from the original Feynman rules, as discussed below.

This general result has a long history. In the abelian theory, diagrammatic 
exponentiation is implicit in the original arguments on the cancellation of 
divergences~\cite{Bloch:1937pw,Yennie:1961ad,Grammer:1973db}, but 
can actually be derived in general on the basis of simple combinatoric 
arguments~\cite{Laenen:2010uz}. The result is that the diagrams contributing
to the exponent are those building up {\it connected photon correlators}, and 
they appear in the exponent with unchanged weights (since `colour' is trivial).
In the non-abelian theory, the case of two Wilson lines in a colour-singlet 
configuration was the first to be tackled~\cite{Gatheral,Frenkel-1984,Sterman-1981}.
In that case, diagrams contributing to the exponent were called {\it webs},
and they are characterised topologically as being two-eikonal-irreducible:
they are diagrams that are not partitioned into disjoint subdiagrams when
each Wilson line is cut exactly once. Web diagrams contribute to the exponent
with modified colour factors, which will emerge from our general treatment 
below. Exponentiation in the two-line case is reviewed in~\cite{Laenen:2010uz,
Berger:2003zh}. The general result that we are discussing, for an arbitrary 
number of generic Wilson lines, was derived with essentially combinatoric 
arguments in~\cite{Gardi:2010rn,Mitov:2010rp}, and further refined in~\cite{
Gardi:2011yz,Gardi:2011wa,Dukes:2013gea} (see also~\cite{Dukes:2016ger}), culminating
in the non-abelian exponentiation theorem proved in~\cite{Gardi:2013ita}, 
which we will summarise below: as we will see, this result involves a non-trivial
generalisation of the concept of {\it web}\footnote{For introductory reviews of the
web idea, see for example~\cite{White:2015wha,White:2018vfr}.}. In what follows, 
for ease of notation, we will drop the subscript $n$ that indicates the number of Wilson 
lines, and we will also leave the dependence on the contours $\gamma_i$ implicit.

Let $D$ be a Feynman diagram contributing to the correlator ${\cal  S}$. 
Each such diagram is a product of a kinematic factor ${\cal  K}(D)$ 
and a colour factor $C(D)$. For straight Wilson lines, the kinematic factor 
is a function of  the  four velocities $\beta_i$; more generally, it is a functional 
of the contours $\gamma_i$. Diagrammatic exponentiation means that 
the logarithm of the correlator, $W$, can be written as a linear combination 
of the same Feynman diagrams, however with modified colour factors. 
We write
\beq
  W \, = \, \sum_D {\cal K} (D)  \, \widetilde{C}(D) \, ,
\label{webeq}
\eeq
where $\widetilde{C}(D)$ is referred to as \textit{Exponentiated Colour 
Factor} (ECF) for diagram $D$. The crucial point in \eq{webeq} is of 
course that a large number of diagrams have vanishing ECFs, and therefore 
do not contribute to $W$: for example, for $n = 2$, Refs.~\cite{Gatheral,
Frenkel-1984,Sterman-1981} show that all two-eikonal-reducible diagrams 
have $\widetilde{C}(D) = 0$. In the general case involving $n$ Wilson 
lines, ECFs are linear combinations of the ordinary color factors of sets 
of diagrams that differ only by the order of their gluon attachments to the 
Wilson lines. This naturally groups the Feynman diagrams that appear 
at any given order into subsets, and each such subset\footnote{Note that 
in the two-line case the term {\it web} was used for individual diagrams, not 
for sets of diagrams. For an abelian theory, webs are (sets of) connected 
photon diagrams.} is called a {\it web}, which we denote by $w$. We then 
use the definition~\cite{Gardi:2010rn}
\begin{itemize}
\item[] {\bf Web}: A set of Feynman diagrams contributing to a Wilson-line 
correlator that can be obtained from any representative diagram by permuting 
the gluon attachments to the Wilson lines.
\end{itemize}

\begin{figure}[h]
\centering
\begin{subfigure}{0.3\textwidth}
	\centering
	{\includegraphics[height=4cm,width=4cm]{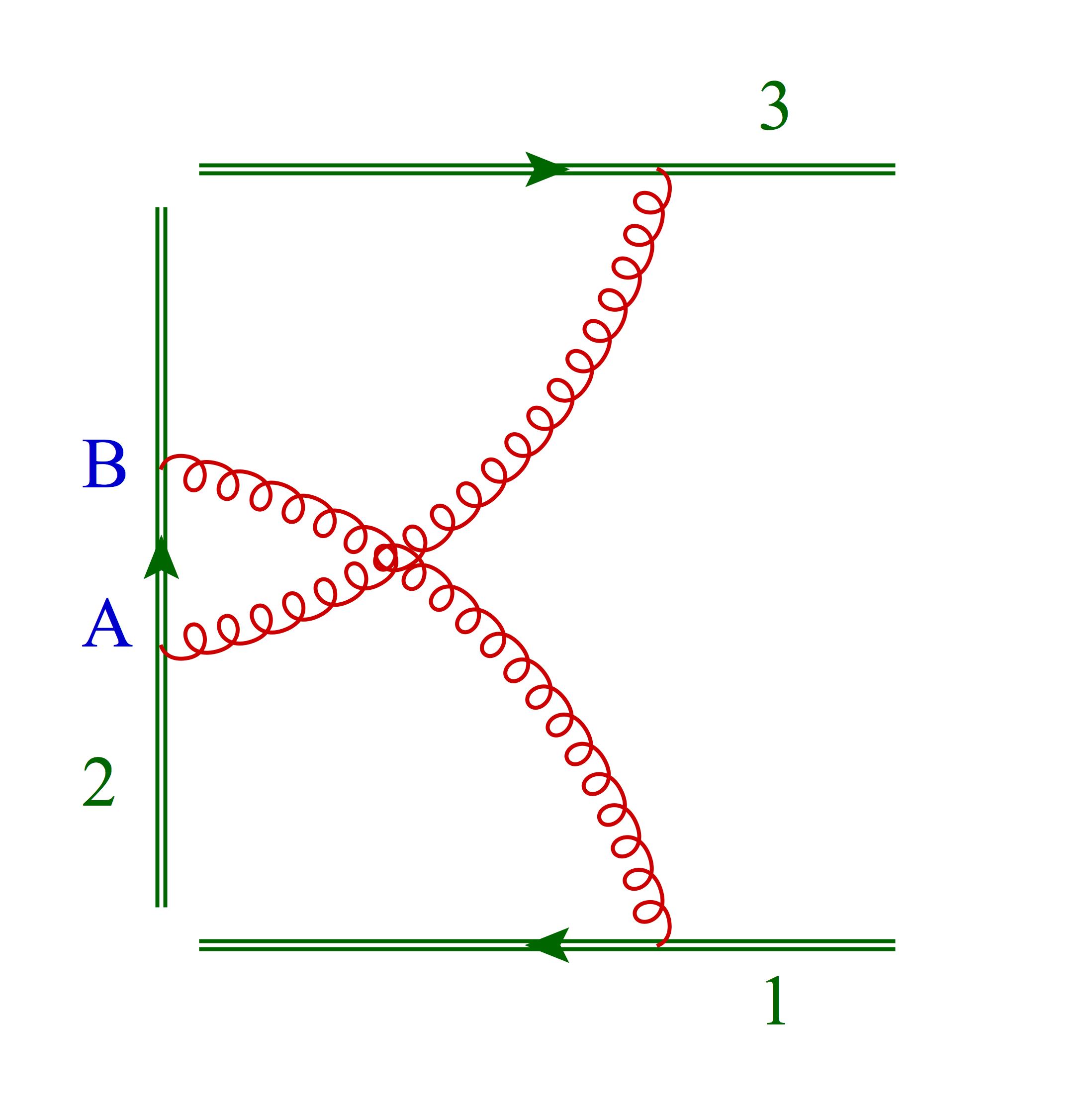} }
\caption{}
        \label{fig:Intro7}
        \end{subfigure}
\begin{subfigure}{0.3\textwidth}
        \centering               
	{\includegraphics[height=4cm,width=4cm]{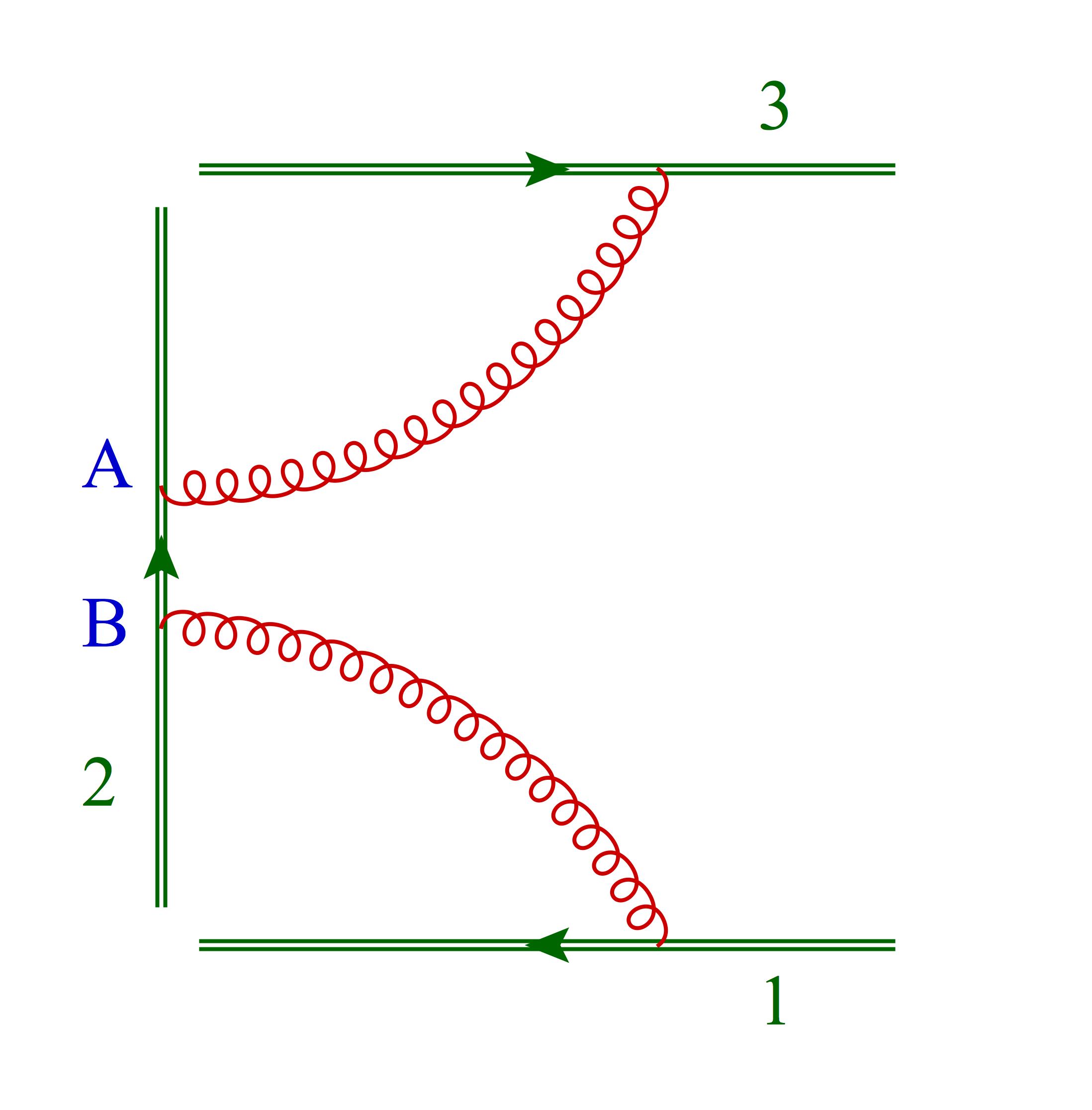} }
\caption{}
        \label{fig:Intro6}
        \end{subfigure}
	\caption{A simple two-loop, three-line web involving two Feynman diagrams. 
        Note that Wilson lines are oriented, and labelled by integers in green; multiple gluon 
	attachments to a given line (whose permutations generate the web) are labelled 
	by capital letters in blue.}
\label{threelegweb}
\end{figure}

\noindent Consequently, the sum in \eq{webeq} can be organized as a sum 
over webs, and, for each diagram $D$ of a given web $w$, the ECF can be
expressed as 
\beq
  \widetilde{C} (D)  \, = \, \sum_{D' \in w} R_w (D, D') \, C(D') \, ,
\label{eq:ecf}
\eeq
where the sum is over all diagrams belonging to web $w$, and $R_w (D, D')$ 
is called {\it web mixing matrix}. Thus, each web $w$ can be written as
\beq
  w \, = \, \sum_{D \in  w} {\cal  K} (D) \, \widetilde{C} (D) \, = \,
  \sum_{D,D' \in  w} {\cal K} (D) \, R_w (D, D') \, C (D')  \, .
\label{eq:webredef}
\eeq
In this language, $W = \sum w$, where the sum extends to all webs, order 
by order in perturbation theory, and now we can express the correlator in 
\eq{diaxp} as
\beq
  {\cal S} \, = \, \exp \left[ \sum_w
  \sum_{D,D' \in  w} {\cal K} (D) \, R_w (D, D') \, C (D')
  \right] \, .
\label{Snwebs}
\eeq
A simple example of a two-loop, three-line web involving two diagrams is presented 
in Fig.~\ref{threelegweb}. There are only 2 diagrams in this web, as two permutations 
are possible only on the second Wilson line.
Web mixing matrices are clearly crucial quantities for the purpose of computing
Wilson-line correlators, and therefore, in particular, the soft anomalous dimension matrix.  
Their properties were extensively studied in Refs.~\cite{Gardi:2011yz,Gardi:2010rn,Mitov:2010rp,
Gardi:2011wa,Dukes:2013gea,Dukes:2016ger,Gardi:2013ita,Dukes:2013wa},
while an interesting alternative approach to exponentiation was developed in 
Refs.~\cite{Vladimirov:2014wga,Vladimirov:2015fea}. We conclude this subsection 
by listing the properties of the mixing matrices that appear in \eq{eq:ecf}, while in 
\secn{Repli} we will describe an algorithm to compute them explicitly, after introducing
a more suitable diagrammatic organisation of the perturbative exponent in \secn{Cwebs}.
Properties 1 to 3 below were proved in Refs.~\cite{Gardi:2011yz,Gardi:2011wa,
Gardi:2013ita}, while Property 4 was conjectured in \cite{Gardi:2011yz}, and verified 
to hold up to four loops in Refs.~\cite{Agarwal:2020nyc,Agarwal:2021him}.

\begin{enumerate}

\item {\bf Idempotency.} Web mixing matrices are idempotent, {\it i.e.} 
\beq
  R_w^2 \, = \,  R_w \, .
\label{idempo} 
\eeq
This property implies that the eigenvalues can only be either 1 or 0: in other words, 
web mixing matrices are {\it projection operators}, selecting a subset of the available
colour factors. If we denote by $Y_w$ the matrix diagonalising $R_w$, we have then
\beq
  Y_w R_w Y_w^{-1} \, = \, {\rm diag}  \left( \lambda_1, ... , \lambda_{p_w} \right)  
  \, = \, {\bf 1}_{r_w} \oplus {\bf  0}_{p_w - r_w} \,  ,
\label{Yw}
\eeq
where $p_w$ is the number of diagrams for web $w$, and thus the dimension
of the matrix $R_w$, while $r_w < p_w$ is the rank  of $R_w$.  
We can now write \eq{eq:webredef}  as
\beq 
\label{eq:Web}
  w & = &  \big( {\cal K}_w^T \, Y_w^{-1} \big) \, Y_w R_w Y_w^{-1} \, 
  \big( Y_w \, C_w \big) \nonumber \\ 
  & = &  \sum_{h =  1}^{r_w} \big(  {\cal K}_w^T Y_w^{-1}  \big)_h 
  \big( Y_w C_w \big)_h  \, ,
\eeq
where ${\cal K}_w$ is the vector of kinematic factors and $C_w$ the vector of colour 
factors for web $w$. It is clear that only $r_w$ ECFs will contribute to web $w$.

\item {\bf Non-abelian exponentiation.} The non-abelian exponentiation 
theorem~\cite{Gardi:2013ita} identifies the colour factors which survive the
projection discussed above: to all orders in perturbation theory, they are the
colour factors which, by the Feynman rules, would be associated to connected 
gluon subdiagrams.

\item {\bf Row sum rule.} The elements of web mixing matrices obey the row sum 
rule 
\beq
  \sum_{D'} R_w (D, D') \, = \, 0 \, ,
\label{rowsum}
\eeq
for any web $w$ with $p_w > 1$, implying that at least one of the eigenvalues of  
any non-trivial web mixing matrix must vanish.

\item {\bf Column sum rule.} One may also envisage the web mixing matrix 
as acting on the vector of kinematic factors for the diagrams of web $w$. Taking 
this viewpoint, the projection effected by the matrix $R_w$ has another non-trivial 
effect: it selects combinations of kinematic factors that do not contain ultraviolet 
sub-divergences\footnote{We refer here to UV divergences arising from sub-diagrams 
involving the Wilson lines: interactions away from the Wilson lines will still involve
the usual gauge-theory UV divergences, which are dealt with by means of ordinary
renormalisation techniques.}. This is crucial for the consistency of the method we 
are outlining: thanks to this property, at each successive order in perturbation theory 
webs contribute a single extra UV pole, so that the corresponding counterterm can 
be extracted, in a minimal scheme, with no dependence on the infrared regulator.
The absence of sub-divergences was established for the two-line case in 
Refs.~\cite{Gatheral,Frenkel-1984,Sterman-1981}; for $n>2$ Wilson lines 
a complete all-order proof is still lacking, but the property holds for all webs
up to four loops~\cite{Gardi:2011yz,Agarwal:2020nyc,Agarwal:2021him}.  
A necessary ingredient for the proof is that the elements of web mixing matrices 
obey a column sum rule, which can be characterised as follows.

Given a diagram $D$, consider the set of its connected sub-diagrams (after  
removing the Wilson lines), $\{ D_{\rm c}^i \subset D  \}$; we say that a 
connected sub-diagram $D_{\rm c}^i$ can be shrunk to the common origin 
of the Wilson lines if all the vertices connecting the sub-diagram to the Wilson 
lines can be moved to the origin without encountering vertices associated
with other sub-diagrams. 

For any given diagram $D$, we then define the {\it column weight} of diagram 
$D$, $s(D)$, as the number of different ways in which the connected sub-diagrams 
$D_{\rm c}^i$ of $D$ can be {\it sequentially} shrunk, so that all gluon attachments 
to Wilson lines in $D$ are moved to their common origin. Thus, if all gluon attachments 
are entangled, so that no subdiagram can be shrunk without shrinking the whole 
diagram, then $s(D) = 0$. On the other hand, if, for example, a single subdiagram 
$D_{\rm c}^i$ can be shrunk without affecting the others, this provides a non-trivial 
sequence for the the shrinking of the whole diagram, so that $s(D) = 1$: this is 
the situation for the two diagrams portrayed in Fig.~\ref{threelegweb}. 

Armed with these definitions, we can state the (conjectured) column sum rule 
for web mixing matrices as
\beq
 \sum_{D \in w} \, s(D) \, R_w (D, D') \, = \, 0 \, .
\label{colsum}
\eeq
The connection of this rule to UV sub-divergences is clear in a coordinate-space
picture~\cite{Erdogan:2014gha}: in this picture, one understands UV divergences 
as arising from short distances between interaction vertices, and `shrinkable' 
sub-diagrams are naturally associated to UV sub-divergences. \eq{colsum} 
guarantees that these sub-diagrams are projected out of the webs.

\end{enumerate} 


\subsubsection{From gluon webs to correlator webs}
\label{Cwebs}

Even from the preliminary description given in \secn{ExpoColo}, it should be 
apparent that the diagrammatic structure of the perturbative exponent has two
independent components: on the one hand, connected gluon subdiagrams
with external (off-shell) gluons, and on the other hand the arrangement of
possible attachments of these gluons to the Wilson lines. Motivated by this 
observation, Ref.~\cite{Agarwal:2020nyc} proposed to organise the perturbative 
exponent not in terms of sets of diagrams, but in terms of sets of connected 
gluon correlators, which were called {\it Cwebs}. Conceptually, Cwebs are
the proper building blocks for the logarithm of the Wilson-line correlator:
indeed, their introduction simplifies considerably the counting and organisation 
of diagrammatic contributions, especially at high orders, where radiative corrections
to gluon sub-diagrams become important and proliferate; furthermore, using 
Cwebs does not affect the definition and structure of web mixing matrices, 
which are derived exclusively from the ordering of gluon attachments to the 
Wilson lines, and are not affected by gluon interactions away from the Wilson 
lines. In this Section, we will briefly review the properties of Cwebs, while in 
\secn{Repli} we will describe an algorithm to compute Cweb mixing matrices,
which is a simple adaptation of the algorithm given in Ref.~\cite{Gardi:2010rn},
implicitly showing that Cwebs are a natural and proper generalisation of webs.
\begin{figure}[h]
\begin{center}
\includegraphics[height=4cm,width=4cm]{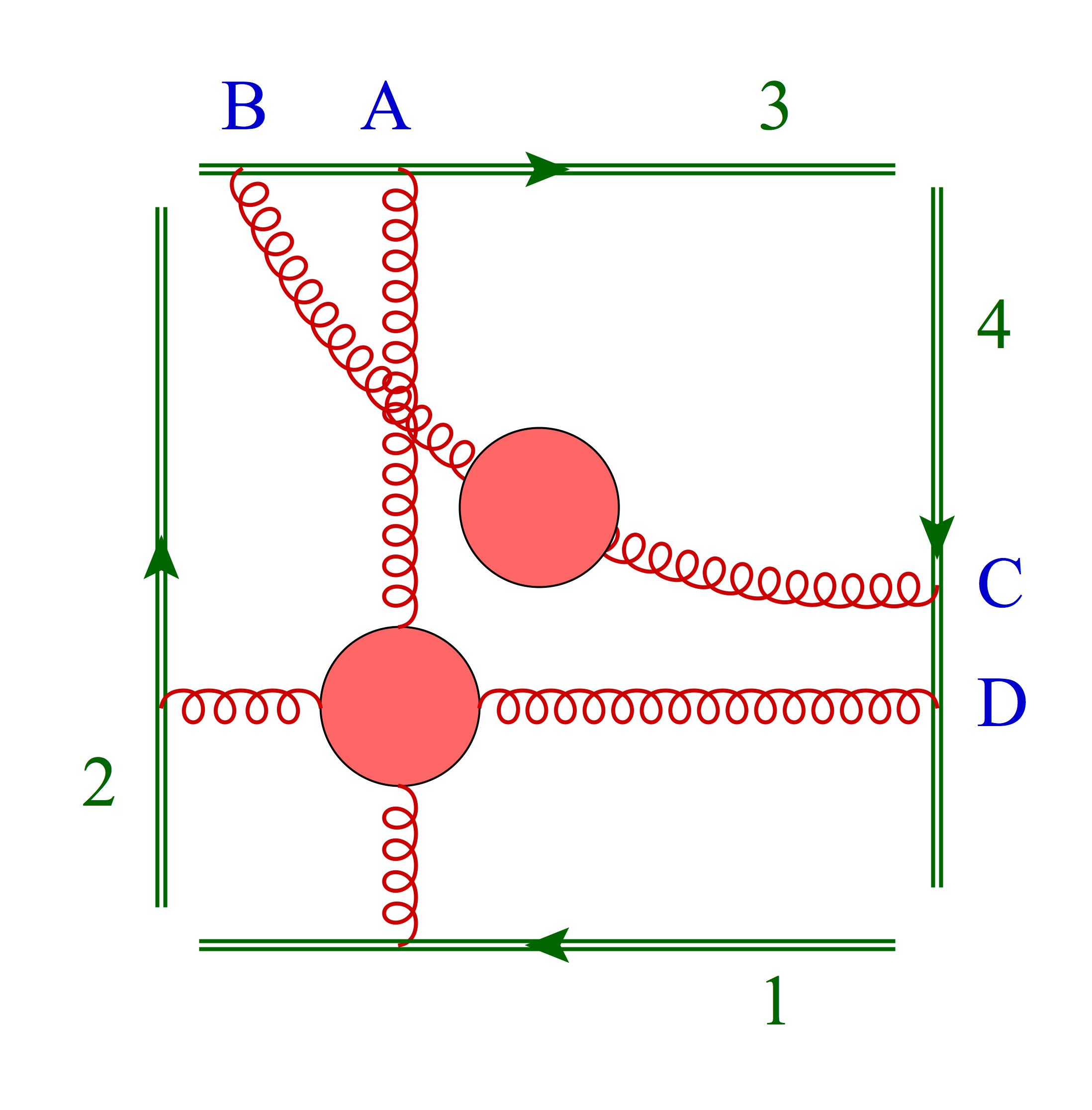}
\end{center}
\caption{A four-line Cweb containing one four-gluon correlator and one 
two-gluon correlator.}
\label{cwebex}
\end{figure}

We begin with the definition 

\begin{itemize}
\item[] {\bf Correlator Web (Cweb)}: A set of skeleton diagrams, built out of 
connected gluon correlators attached to Wilson lines, and closed under permutations 
of the gluon attachments to each Wilson line. 
\end{itemize}

\noindent As an example, a four-line Cweb built with two connected gluon correlators 
is shown in Fig.~\ref{cwebex}. Clearly, Cwebs are not fixed-order quantities, but 
admit their own perturbative expansion in powers of the gauge coupling $g$. 
Below, we will use the notation $W_n^{(c_2, \ldots , c_p)} (k_1, \ldots  , k_n)$ 
for a Cweb constructed out of $c_m$ $m$-point connected gluon correlators 
($m = 2, \ldots, p$), where $k_i$ is the number of gluon attachments to the $i$-th 
Wilson line\footnote{At sufficiently high-orders, this notation does not uniquely 
identify Cwebs, since different Cwebs can be constructed out of the same set 
of correlators, and with the same number of attachments to each Wilson line. 
The proposed notation will however suffice for our purposes here.}. Note that 
there is an obvious degeneracy in the counting of Cwebs, since Cwebs that 
differ only by a permutation of their Wilson lines are structurally identical, and 
it is trivial to include them in the calculation of the full correlator, simply by 
summing over Wilson-line labels. For the sake of classification, we can then 
focus on a specific Wilson-line ordering, choosing for example $k_1 \leq k_2 
\leq \ldots \leq k_n$. As an example, in this notation the Cweb in Fig.~\ref{cwebex} 
is denoted by $W_4^{(1,0,1)}(1,1,2,2)$. Taking into account the fact that the 
perturbative expansion for an $m$-point connected gluon correlator starts 
at ${\cal O} (g^{m - 2})$, while each attachment to a Wilson line carries a 
further power of $g$, the perturbative expansion for a Cweb can be written as
\beq
  W_n^{(c_2, \ldots , c_p)} (k_1, \ldots  , k_n)  \, = \, 
  g^{\, \sum_{i = 1}^n k_i \, + \,  \sum_{r = 2}^p c_r (r - 2)} \, \sum_{j = 0}^\infty \,
  W_{n, \, j}^{(c_2, \ldots , c_p)} (k_1, \ldots  , k_n) \, g^{2 j} \, ,
\label{pertCweb}
\eeq
which defines the perturbative coefficients $W_{n, \, j}^{(c_2, \ldots , c_p)} 
(k_1, \ldots  , k_n)$. With this in mind, it is natural to classify Cwebs based on 
the perturbative order where they receive their lowest-order contribution, which 
is given by the power of $g$ in the prefactor of \eq{pertCweb}; one may then 
easily design a recursive procedure for construction of Cwebs.

\begin{figure}[t]
     \begin{subfigure}{0.3\textwidth}
    \centering
        \includegraphics[width=0.8\textwidth]{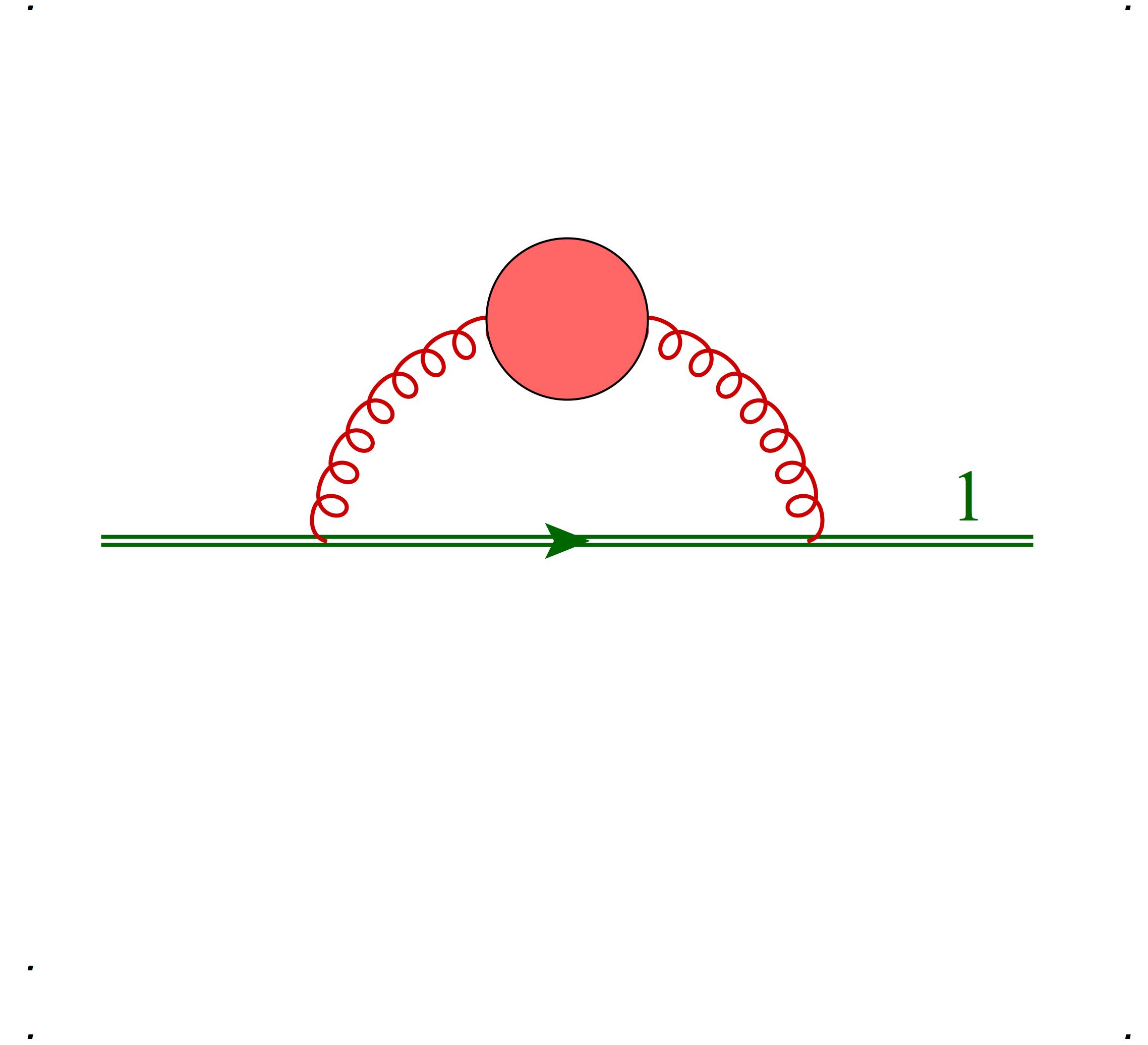}
        \caption{}
        \label{fig:Cwebg2-1}
    \end{subfigure}
\centering
    \begin{subfigure}{0.3\textwidth}
    \centering
        \includegraphics[width=0.8\textwidth]{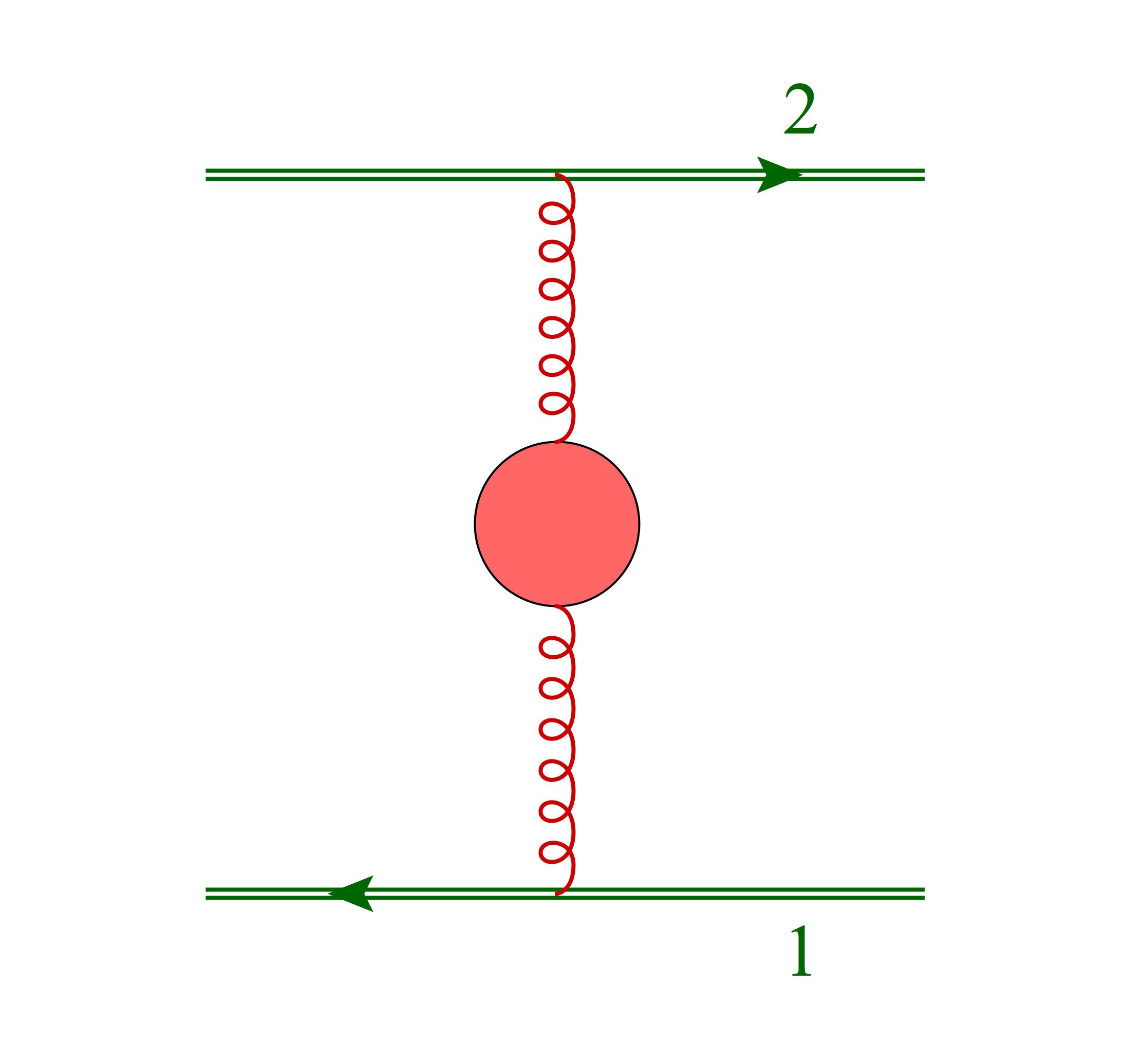}
        \caption{}
        \label{fig:Cwebg2-2}
    \end{subfigure}
      \caption{The only two Cwebs whose perturbative expansion starts at 
	${\cal O}(g^2)$.}
  \label{cwebg2}
\end{figure}

Starting at lowest order, we notice that only two Cwebs appear at order $g^2$: 
the self-energy insertion with a two-point connected gluon correlator attached 
to a single Wilson line, which we denote by  $W_1^{(1)} (2)$, and the configuration 
where the two-gluon correlator joins two Wilson lines, denoted by $W_2^{(1)} (1,1)$. 
These two Cwebs are depicted in Fig.~\ref{cwebg2}. 
Notice that, if the Wilson lines are light-like, the self-energy Cweb vanishes identically, 
since, by the eikonal Feynman rules, it is proportional to the square of the Wilson-line 
four-velocity vector, $\beta^2$: in that case, one is left with just one non-vanishing 
${\cal  O}(g^2)$ Cweb\footnote{Webs where at least one gluon begins and ends 
on the same (non-light-like) Wilson line have been recently studied in detail in 
Ref.~\cite{Gardi:2021gzz}.}. Starting from this  initial condition, one may recursively construct 
all Cwebs starting at higher perturbative orders: indeed, having constructed all Cwebs 
at ${\cal O} (g^{2 n})$, there are three ways of adding a power of $g^2$.
\begin{itemize}
\item Add a two-gluon connected correlator  connecting any two Wilson lines
(including `new' Wilson lines that had no attachments in lower-order Cwebs).
\item Connect an existing $m$-point correlator to any Wilson line (again, 
including Wilson lines with no attachments at lower orders), turning  it into
an  $(m+1)$-point correlator.
\item Connect an existing $m$-point correlator to an existing $n$-point 
correlator, resulting in an $(n+m)$-point correlator.
\end{itemize}
Executing all possible moves is clearly redundant, since the same Cweb is 
generated more than once through different sequences of moves: upon performing 
all moves, one must therefore remove Cwebs that have been counted more than 
once. The procedure can be considerably streamlined by taking into account 
known properties of webs, which naturally generalise to Cwebs. For example,
webs (and thus Cwebs) that are given by the product of two or more
disconnected lower-order webs (so that there are subsets of Wilson lines
not joined by any gluons) do not contribute to the logarithm of the correlator, 
${\cal W}_n$; furthermore, as mentioned above, in a massless theory all 
self-energy Cwebs, where all gluon lines attach to the same Wilson line, 
vanish as a consequence of the eikonal Feynman rules. Thus, any Cweb 
containing a connected gluon correlator attaching to a single Wilson line will 
vanish. Using these recursive rules, it is easy to enumerate inequivalent 
Cwebs at low orders. In the massless theory, for example, one finds four 
inequivalent Cwebs at ${\cal O}(g^4)$, fourteen new Cwebs at ${\cal O} (g^6)$, 
and a total of sixty new Cwebs at ${\cal  O}(g^8)$: they are discussed in detail 
in Ref.~\cite{Agarwal:2020nyc}.

At each order in $g$, Cwebs contain webs, all sharing the same mixing matrix,
since the colour algebra associated with permutations of attachments to the 
Wilson lines is unaffected by radiative corrections to the connected correlators
comprising the Cweb. We can thus construct Cweb mixing matrices by the same 
technique devised in~\cite{Gardi:2010rn} for webs, which we now discuss.


\subsubsection{A replica algorithm to generate web mixing matrices}
\label{Repli} 

As is  the case for other combinatorial problems involving exponentiation, such 
as the construction of effective actions, Wilson-line correlators can be studied  
by means of algorithms constructed with the replica method~\cite{MezaPariVira}.
In this section, we will summarise the application of the replica method to the 
construction of Cweb mixing matrices, following Refs.~\cite{Laenen:2008gt,
Gardi:2010rn}.

In order to introduce the method, consider the path integral expression
for the Wilson-line correlator in \eq{genWLC}
\beq
  {\cal S}_n \left( \gamma_i \right) \, = \, \exp \Big[ W_n (\gamma_i) \Big] 
  \, = \, \int {\cal D} A_\mu^a  \,\,
  {\rm e}^{{\rm i} S \left(  A_\mu^a \right)} \,  \prod_{k = 1}^n
  \Phi \left(  \gamma_k \right) \, ,
\label{genWNCpath}
\eeq
where $S$ is the classical action\footnote{We display only gluon fields as arguments 
of the action, since matter fields are irrelevant to the following arguments: they can
be integrated out, and their effects can be included in $S$, as only gluons couple 
to Wilson lines, and matter fields appear only in loops.}.  In order to compute 
$W_n (\gamma_i)$, one starts by building a {\it replicated theory}, replacing 
the single gluon field $A_\mu^a$ with $N_r$ identical copies $A_\mu^{a, \, i}$  
($i = 1, \ldots, N_r$), which do not interact with each other. Under this replacement, 
one has $S \left(  A_\mu^a \right)  \rightarrow \sum_{i = 1}^{N_r} S \left(  A_\mu^{a, \, i} 
\right)$; if, furthermore, we replace each Wilson line in \eq{genWNCpath} by the 
product of $N_r$ Wilson lines, each belonging to one replica of the theory, one 
readily sees that the replicated correlator is given by
\beq
  {\cal S}_n^{\, {\rm repl.}} \left( \gamma_i \right) \, = \,   \Big[ 
  {\cal S}_n \left( \gamma_i \right) \Big]^{N_r} \, = \, \exp \Big[ N_r \,
  W_n (\gamma_i) \Big] \, =  \, {\bf 1} + N_r \, W_n (\gamma_i) 
  + {\cal O} (N_r^2) \, .
\label{exprepl}
\eeq
\eq{exprepl} provides a straightforward method to compute $W_n (\gamma_i)$:
one may compute the replicated correlator (essentially a combinatorial problem, as we
will see), and then extract from the result the term of order $N_r$. 

An important point is that, while gluon fields in different replicas do not interact, they 
all belong to the same gauge group: therefore, the colour matrices associated with 
their attachments to a Wilson line do not commute, and their ordering is relevant.
On the other hand, in a Cweb, each connected gluon correlator can be assigned a 
unique replica number, since there are no interaction vertices connecting different
replicas.  One can then relate the contributions of different skeleton diagrams to 
the replicated correlator in \eq{exprepl} to those of the same diagrams in the 
un-replicated theory, by means of combinatorial factors, counting the multiplicities 
associated with the presence of different replicas.  The computation of $W_n 
(\gamma_i)$ in terms of the skeleton diagrams contributing to ${\cal S}_n (\gamma_i)$ 
is thus reduced to the computation of these combinatorial factors. The necessary 
steps, listed below, were identified for webs in Refs.~\cite{Laenen:2008gt,Gardi:2010rn}, 
and naturally adapt to the language of Cwebs.

\begin{itemize}

\item For any given Cweb, assign a replica number $i$ ($1 \leq i \leq N_r$) to each 
connected gluon correlator present in the Cweb. 

\item Define a {\it replica ordering operator} $R$, acting on color generators on 
each Wilson line as
\beq
  R \Big( {\bf T}_k^{(i)} {\bf T}_k^{(j)} \Big) \! & = & \! {\bf T}_k^{(j)} {\bf T}_k^{(i)} 
  \quad \textrm{if} \,\, i>j \, ,  \\
  R \Big( {\bf T}_k^{(i)} {\bf T}_k^{(j)} \Big) \! & = & \!{\bf T}_k^{(i)} {\bf T}_k^{(j)} 
  \quad \textrm{if} \,\,  i \leq j \, , \nonumber
\label{repordef}
\eeq
where  ${\bf T}_k^{(i)}$ is the group generator associated with the emission of a 
gluon in replica $i$ from the  $k$-th Wilson line. The operator $R$ effectively replaces 
the selected skeleton diagram with another one drawn from the same Cweb: as
a consequence, one may verify that the colour factor in the replicated theory is a 
linear combination of the colour factors of all skeleton diagrams in the Cweb, 
with multiplicities given by the number of possible replica orderings of the gluon
attachments on every Wilson line.

\item In order to proceed, one needs to list all possible hierarchies of the replica 
numbers associated with each correlator in the Cweb: these hierarchies in fact
define the action of the $R$ operator on each skeleton diagram of the Cweb.
For a Cweb built out of $m$ connected correlators, the number $h(m)$ of possible 
hierarchies between the $m$ replica numbers of the correlators grows rapidly 
with $m$, and is not entirely trivial to determine, since the case of equal replica 
numbers must be treated separately. The result is however well known~\cite{IntSeq}: 
$h(m)$ are the so-called Fubini numbers\footnote{The Fubini numbers admit a 
generating function, and they can be defined by
\beq
   \frac{1}{2 - \exp (x)} - 1 \, \equiv \, \sum_{m=1}^\infty h(m) \, \frac{x^m}{m!} \,\, .
\label{genfu}
\eeq}. 
In the first instances, for $m =  \{1,2,3,4,5,6\}$ one finds $h(m) = \{1,3,13,75,541,
4683\}$. For every fixed hierarchy $h$, the number relevant for the determination 
of the colour factor in the replicated theory is the multiplicity with which that 
hierarchy can occur in the presence of $N_r$ replicas, which we denote by 
$M_{N_r}(h)$. It is not difficult to see that it is given by
\beq
  M_{N_r}(h) \, = \, \frac{N_r!}{\big( N_r - n_r(h) \big)! \,\, n_r(h)!}  \, ,
\label{multhi}
\eeq
where $n_r(h)$ is the number of distinct replicas present for hierarchy $h$. 
To give a concrete example, for $m = 5$, labelling the replica numbers of 
the five correlators with $r_k$, $(k = 1, \ldots, 5)$, and picking  the hierarchy
$r_1 = r_2 <  r_3 = r_4 < r_5$, we have $n_r(h) = 3$, and thus $M_{N_r} (h)
= N_r (N_r - 1) (N_r - 2)/6$.

\item Finally, the exponentiated color factors for a skeleton diagram $D$ is 
given by
\beq
  C_{N_r}^{\, {\rm repl.}}  (D) \, = \, \sum_h M_{N_r} (h) \, R \big[ C(D) \big| h 
  \big]  \, ,
\label{expocolf}
\eeq
where $R \big[ C(D) \big|  h  \big]$ is the color factor of the skeleton diagram 
obtained from $D$ through  the action of the replica-ordering operator $R$, in
the case of hierarchy $h$. The Cweb mixing matrix is built by picking terms 
that are linear in $N_r$ out of the coefficients $M_{N_r}(h)$, which are polynomials 
in $N_r$. Note that in the presence of $m$-point correlators, with $m \geq 4$, 
each correlator can contribute different `internal' colour factors, for example 
different permutations of products of structure constants. Since, however, the 
information on the mixing matrix is encoded in the coefficients $M_{N_r}(h)$, 
the different colour factors arising from the internal structure of the correlators 
can be treated one by one, without affecting the mixing of the diagrams.

\end{itemize} 

\noindent
To conclude our discussion of exponentiated colour factors, we briefly comment
on the dimensionality of Cweb mixing matrices. One would na\"ively imagine
that it should be given by the product of the numbers of possible gluon permutations
on each Wilson line. This however neglects two important subtleties. First, if 
a set of, say, $k$ gluons emerging from a given correlator attaches on a single
Wilson line, their permutations are irrelevant, since the correlator is Bose symmetric.
One should not then count permutations on each line, but rather shuffles of
the subsets of gluons emerging from each correlator and attaching to a given line.
Furthermore, there is another important degeneracy, which already appears in 
the simple two-line Cweb $W^{(2)}_2 (2,2)$ at two loops: counting shuffles 
separately on each line would yield, in that case, the result $d = 4$ for the 
dimensionality of the mixing matrix, which is wrong, because the two shuffles 
on the second Wilson line can be obtained from the shuffles on the first line 
by exchanging the two gluons, which is manifestly a symmetry of the Cweb. 
In order to take into account this degeneracy, one must divide by the number 
of available permutations of subsets of $m$-point correlators that have the 
property of being attached to the same sets of Wilson lines.

The discussion in the last three sections has been general and rather formal,
but also incomplete, since of course much of the dynamical content of the
soft anomalous dimension matrix resides in the kinematic factors, which
emerge from loop integrals of rapidly increasing complexity at high orders.
To fill these gaps, we now give two simple examples of the evaluation of
these kinematic factors at one and two loops. At the two-loop level, this
will also serve to illustrate in a concrete case the interplay of colour and 
kinematics which has been discussed formally in the previous sections.


\subsubsection{The one-loop angle-dependent cusp anomalous dimension}
\label{cusponeloop}

We begin by describing the calculation of the only non-vanishing one-loop diagram
contributing to the soft anomalous dimension matrix in the massless case. 
The relevant diagram corresponds to the tree-level contribution to the Cweb in 
Fig.~\ref{fig:Cwebg2-2}, and of course it is a web on its own. Clearly, this 
diagram is very simple, and it can be computed in a number of ways - indeed, we 
performed an equivalent calculation in \secn{WiliEik} with light-like Wilson lines. 
We display the complete calculation here, with collinear divergences regulated
by using `massive' Wilson lines, and soft divergences regulated by the exponential
suppression defined in \eq{Phidef}, in order to illustrate techniques that can be  
profitably used at higher orders as well. Besides, the calculation yields the 
one-loop result for the {\it angle-dependent} cusp anomalous dimension, which is
a fundamental quantity for the scattering of massive particles and in many other 
contexts~\cite{Correa:2012nk}. Unlike \secn{WiliEik}, we will proceed by
recognising in the second line of \eq{twolines} the coordinate-space gluon propagator,
$D_{\mu \nu}(x)$, evaluated for $x = \lambda_i \beta_i - \lambda_j \beta_j$. The
expression for this propagator in $d = 4 - 2 \epsilon$ is well known, and is given by
\beq
  D_{\mu \nu}^{a b} (x) \, = \,  - \, \frac{\Gamma(1 - \eps)} {4 \, \pi^{2 - \eps}} \,
  g_{\mu \nu} \, \delta^{a b} \, \frac{1}{(- x^2 + i \eta)^{1 - \eps}} \, .
\label{gluprop}
\eeq
Using this expression leaves only two parameter integrals to be performed.
Arguably, this `coordinate space' approach to eikonal integrals simplifies their
calculations also at higher orders, and it was indeed adopted for the state-of-the-art
calculation of Ref.~\cite{Almelid:2015jia}.

In Cweb notation, the diagram in Fig.~\ref{cwebg2} (b), including radiative 
corrections to the gluon propagator, can be written as
\beq
  W_{2, \, ij}^{(1)} (1,1) \, = \,  {\bf T}_i \cdot {\bf T}_j \;
  {\cal K}^{(1)}_2 \bigg(\gamma_{ij}, \frac{\mu^2}{m^2} , \eps \bigg) \, ,
\label{easyweb}
\eeq
where ${\cal K}^{(1)}_2 (\gamma_{ij}, \mu^2/m^2 , \eps)$ gives the regularised kinematic 
factor. The color structure is straightforward and there is no mixing matrix involved,
since this is a single-diagram web. With the infrared regulator in place, and having
introduced the gluon propagator in \eq{gluprop}, the one-loop kinematic factor reads  
\beq
  \hspace{-2mm}
  {\cal K}^{(1)}_{2,1} \bigg( \gamma_{ij}, \frac{\mu^2}{m^2}, \eps \bigg) =  
  g_s^2 \mu^{2 \eps} \, \frac{\Gamma(1 - \eps)} {4 \, \pi^{2 - \eps}}  \, \beta_i \cdot \beta_j
  \int_0^{\infty} \! d \lambda_i \int_0^{\infty} \! d \lambda_j \,  
  \frac{ {\rm e}^{- m \lambda_i \sqrt{\beta_i^2} -  m \lambda_j \sqrt{\beta_j^2}} }{\Big( \!
  - \left( \lambda_i \beta_i  - \lambda_j \beta_j \right)^2 + i \eta \Big)^{1 - \eps}} \, .
 \label{kintwolines}
\eeq
To perform the integral, it is clearly useful to rescale the variables and define
\beq
  \sigma \, = \, \lambda_i \sqrt{\beta_i^2} \, , \qquad  
  \tau \, = \, \lambda_j \sqrt{\beta_j^2} \, .
\label{sigmatau}
\eeq
We then introduce coordinates $\lambda$ and $x$, which are respectively normal 
and intrinsic to the pinch surface (in this case a point),
\beq
  \lambda \, = \, \sigma + \tau \, , \qquad \quad x \, = \, \frac{\sigma}{\sigma + \tau} \, ;
\label{lambdax}
\eeq
the normal coordinate $\lambda$ gives the ``overall distance'' of the gluon from 
the cusp, and ranges from 0 to $\infty$. The intrinsic angular coordinate ranges 
between 0 and 1, and measures the degree of collinearity of the gluon to either
Wilson line. The integral over $\lambda$ can be readily performed and gives 
a factor of $\Gamma(2 \eps)$ containing the ultraviolet pole. The integration over 
the intrinsic variable $x$ remains, and we can write
\beq
  {\cal K}^{(1)}_{2,1} \bigg( \gamma_{ij}, \frac{\mu^2}{m^2} , \eps \bigg) \, =  \,
  - \bigg( \frac{\mu^2}{m^2} \bigg)^\eps g_s^2 \,\, \frac{\Gamma(1 - \eps)}{8 \pi^{2 - \eps}}
  \, \Gamma(2 \eps) \, \gamma_{ij} \int_0^1 d x \, P \big( x, \gamma_{ij} \big) \, ,
\label{Pint}
\eeq  
where
\beq
  P \big( x, \gamma_{ij} \big) \, = \, \Big[ x^2 + (1 - x)^2 - x (1 - x) \gamma_{ij} 
  - {\rm i} \eta \Big]^{- 1 + \eps} \, .
\label{Pdef}
\eeq
The quadratic polynomial  in $x$ in \eq{Pdef} can be factored with the change
of variables
\beq
  \gamma_{ij} \, \equiv \, - \, \alpha_{ij} - \frac{1}{\alpha_{ij}} \, ,
\label{alpha}
\eeq
where we can take $\alpha_{ij}$ to lie inside the unit circle in the complex plane, thanks
to the symmetry under $\alpha_{ij} \to 1/\alpha_{ij}$. The $x$ integral can then be performed
exactly, and gives a hypergeometric function,  
\beq
  {\cal K}^{(1)}_{2,1} \bigg( \gamma_{ij}, \frac{\mu^2}{m^2} , \eps \bigg)
  \, = \, - \left(\frac{\mu^2}{m^2} \right)^\eps \, g_s^2 \, \frac{\Gamma(1-\eps)}{8 \pi^{2 - \eps}}
  \,\, \Gamma(2\eps) \, \gamma_{ij} \, {}_2 F_1 \bigg( 1,1 - \eps, \frac{3}{2}, \frac{1}{2} +
  \frac{\gamma_{ij}}{4} \bigg) \, .
\label{K1exact}
\eeq
Expanding \eq{K1exact} around $\eps=0$ we can easily obtain the coefficient of the 
simple pole in \eq{Pint}. Reinstating then the colour factor, making use of
\eq{Gamfromreg} at one loop, and summing over available colour dipoles, we 
obtain the soft anomalous dimension matrix at one loop, for massive Wilson 
lines. In terms of the $\alpha_{ij}$ variables, it is given by
\beq
  \Gamma^{(1)}_{\cal S} \, = \, \sum_{i <j} \, {\bf T}_i \cdot {\bf T}_j \,  
  \frac{1+\alpha_{ij}^2}{1-\alpha_{ij}^2} \, \ln \alpha_{ij} \, .
\label{Gamma1alpha}
\eeq
In this representation, the limit of light-like Wilson lines is reached as $\alpha_{ij}
\to 0$, corresponding to $\gamma_{ij} \to \infty$. The logarithmic singularity in that 
limit corresponds to the collinear divergence in \eq{twolines}, and the coefficient
of the logarithm in the $\alpha_{ij} \to 0$ limit gives the light-like cusp anomalous 
dimensions discussed in earlier Sections. An alternative representation is often
given, in terms of the Minkowski-space cusp angles between Wilson lines $i$ and 
$j$, which we denote by $\xi_{ij}$, defined by
\beq
  \xi_{ij} \, = \, \cosh^{-1} \! \bigg( \! - \frac{\gamma_{ij}}{2} \bigg) \, .
\eeq 
Using this variable one finds
\beq
  \Gamma^{(1)}_{\cal S}  \, = \, - \, \sum_{i<j} {\bf T}_i \cdot {\bf T}_j \, \,
  \xi_{ij} \coth \big( \xi_{ij} \big) \, .  
\label{Gamma1xi}
\eeq
\begin{figure}[b]
\begin{center}
\includegraphics[height=4cm,width=4cm]{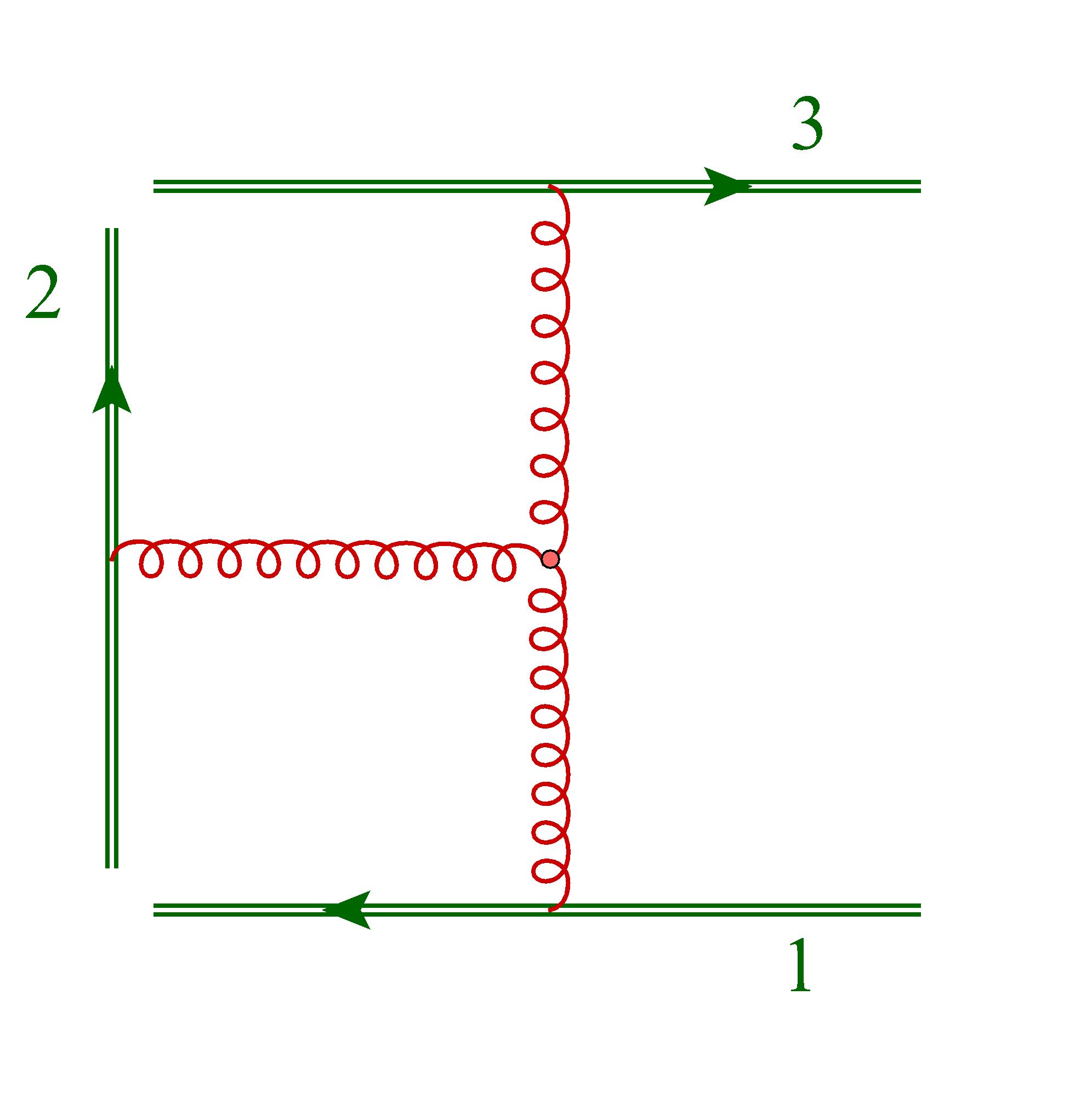}
\end{center}
\caption{A  two-loop, three-line web involving a 3-gluon vertex. }
\label{threelegweb2}
\end{figure}
As we will see below, the one-loop angle-dependent cusp anomalous dimension,
given here in \eq{Gamma1alpha} and in \eq{Gamma1xi} plays an important role also
for higher-order corrections to the soft anomalous dimension matrix for massive
particles.


\subsubsection{The two-loop soft anomalous dimension matrix}
\label{Twolosad}

\begin{table}
\begin{center}
\begin{tabular}{c|c|c}
$i$, $j$ & Replica-ordered colour factor & Multiplicity\\
\hline
$i=j$ & $C(a)$ & $N_r$\\
$i<j$ & $C(b)$ & $N_r (N_r - 1 )/2$\\
$i>j$ & $C(a)$ & $N_r (N_r - 1)/2$
\end{tabular}
\caption{Replica analysis of diagram $(a)$ of Fig.~\ref{threelegweb}, as
described in the text.}
\label{tab:rep2}
\end{center}
\end{table}

We now proceed to summarise the calculation of the two-loop soft anomalous 
dimension matrix, which will give us an opportunity to illustrate the application
of the replica method discussed in \secn{Repli} with the simplest non-trivial example.

At two-loops, a maximum of three Wilson lines can be connected, and we get a 
total of two distinct webs: one was shown in Fig.~\ref{threelegweb}, while
the other one, involving a single diagram with a three-gluon vertex, is shown 
in Fig.~\ref{threelegweb2}. Here we will discuss in some detail the web in
Fig.~\ref{threelegweb}, which is the lowest-order contribution to the Cweb $W_3^{(2)} 
(1,2,1)$, and is also an example of the class of webs called ``multiple gluon exchange
webs'' (MGEW), discussed in Ref.~\cite{Falcioni:2014pka}. The contribution of this 
two-loop web can  be written schematically as
\beq
  w_{121}^{(2)} \, = \, {\cal K} (a) \, \widetilde{C} (a) + {\cal K}(b) \, \widetilde{C} (b) \, ,
\label{repamp}
\eeq
where ${\cal K} (a)$ and ${\cal K} (b)$ are the kinematic contributions of the 
two diagrams of the web and $\widetilde{C}(a)$ and $\widetilde{C}(b)$ are the 
corresponding exponentiated color factors. Considering as an example diagram 
$(a)$ in Fig.~\ref{threelegweb}, and introducing $N_r$ replicas, we note that the 
replica-ordering operator will replace the colour factor of $(a)$ with that of $(b)$
in Fig.~\ref{threelegweb}, in the $N_r (N_r - 1)/2$ cases in which the replicas 
assigned to the two gluons are in the `wrong' order, as reported in Table~\ref{tab:rep2}. 
The colour factor of diagram $(a)$ in the replicated theory then reads
\beq
  C^{\, \rm repl.}_{N_r} (a) \, = \, N_r C (a) + \frac{N_r (N_r - 1)}{2} \big( C(a) +
  C(b) \big) \, = \, N_r \left(\frac{C(a) - C(b)}{2} \right) + {\cal O}(N_r^2) \, ,
\label{colfaca}
\eeq
and similarly for diagram $(b)$. The coefficients of  ${\cal O}(N_r)$ terms give us 
then the exponentiated color factors
\beq
  \widetilde{C} (a) \, = \, \left( \frac{C(a) - C(b)}{2} \right) \, = \, - \, \widetilde{C}(b) \, .
\label{colfacb}
\eeq
Thus
\beq
  w_{121}^{(2)} \, = \, \frac{1}{2} \, \big( C(a) - C(b) \big) \big( {\cal K}(a) - {\cal K}(b) \big) \, .
\label{expab}
\eeq
corresponding to the web mixing matrix
\begin{align}
\begin{split}
                &
                R_{121} \, = \, \frac{1}{2} \, \displaystyle{\left(\!\!
                        \begin{array}{cc}
                        1 & - 1 \\
                        - 1 & 1 \\
                        \end{array}
                        \!\!\right) } \, .
\label{R121}
\end{split}
\end{align}
We note in passing that, as expected, $R_{121}$ satisfies the row sum rule 
in \eq{rowsum}. In order to check the column sum rule in \eq{colsum}, we must 
construct  the vector $s(D)$ for the two diagrams in Fig.~\ref{threelegweb}. Taking
into account the orientation of the Wilson lines, one sees that for both diagrams
there is only one way in which the two gluons can be sequentially shrunk to 
the origin, so that we get
\beq
  s(a)  \, = \,  s(b) \, = \, 1 \, ,
\label{webvectex}
\eeq
and, also as expected, the column sum rule is verified. Concerning the exponentiated 
colour factors, one easily sees that the non-abelian exponentiation theorem is 
verified for this simple web. Indeed, the color factors of the diagrams (a) and (b) 
are
\beq
  C(a) \, = \, T_1^a  T_2^a T_2^b T_3^b \, ,  \qquad 
  C(b) \, = \, T_1^a T_2^b T_2^a T_3^b \, .
\label{colfactab}
\eeq
Recalling that color generators pertaining to the same Wilson line do not commute,
and using the colour algebra, we can write the exponentiated color factor as 
\beq
  \widetilde{C}(a) \, = \, - \frac{1}{2} \, {\rm i} f_{abc} T_1^aT_2^bT_3^c \, = \, - \, 
  \widetilde{C}(b) \, ,
\label{ecf}
\eeq
a result which is proportional to the colour factor of the fully connected diagram 
in Fig.~\ref{threelegweb2}.

We now turn to the computation of the kinematic factor of this web, following
Ref.~\cite{Gardi:2013saa}. This will allow us to illustrate the dynamical effects 
of the column sum rule on leading infrared poles. We follow the same strategy 
as in \secn{cusponeloop}, associating directly to each gluon its coordinate-space 
propagator, and thus leaving only parameter integrals to be performed. We 
parametrise the positions of gluon attachments on Wilson lines $\ii,\jj$ and $\kk$ 
by $s \beta_\ii$, $t_1 \beta_\jj$, $t_2 \beta_\jj$ and $u \beta_\kk$, respectively, 
and introduce exponential suppressions for each Wilson line. In diagram (a), the 
gluon that attaches at $t_1$ is further away from the origin than the one at $t_2$, 
which is accounted for by including step function $\theta(t_1 - t_2)$. We get then, 
for diagram (a),
\beq
  {\cal K}(a) \! & = & \! g_s^4  \mu^{4\eps} \, \frac{ \Gamma^2 (1-\eps)  }{ 16 \, \pi^{4 - 2 \eps}}   
  \, \beta_\ii \cdot \beta_\jj \, \beta_\jj \cdot \beta_\kk \, 
  \int_0^\infty ds \, dt_1 \, dt_2 \, du \, \, \theta( t_1 - t_2) \nonumber \\
  & & \times \, \frac{ \exp \Big[ {- m \Big( s \sqrt{\beta_\ii^2} + u \sqrt{\beta_\kk^2} +  
  \big( t_1 + t_2 \big) \sqrt{\beta_\jj^2} \Big)} \Big]} {\Big( - (s \beta_\ii  - t_1 \beta_\jj)^2 
  + {\rm i} \eta   \Big)^{1 - \eps}  
  \Big( - (u \beta_\kk  - t_2 \beta_\jj)^2 + i \eta \Big)^{1 - \eps}} \, , 
\eeq
while the contribution of diagram (b) is the same as the above, except that one 
needs to replace the step function by its complement, $\theta(t_2 - t_1)$. As 
before, we can exploit the rescaling invariance of the integrand to scale the 
variables according to
\beq
  \sigma \, = \, s \sqrt{\beta_\ii^2} \, , \qquad \tau_{1,2} \, = \, t_{1,2} \sqrt{\beta_\jj^2} \, , 
  \qquad \upsilon \, = \, u \sqrt{\beta_\kk^2} \, ,
\label{rescale3}  
\eeq
and we can define variables $\lambda_1$ and $\lambda_2$, corresponding to 
the average distance of the two gluons from the origin, as well as `angular'
variables $x_1$ and $x_2$, using
\beq
  \lambda_1 \, = \, \sigma + \tau_1 \, , \qquad \lambda_2  \, = \, \upsilon + \tau_2 \, , \\
  x_1 \, = \, \frac{\tau_1}{\sigma + \tau_1} \, , \qquad x_2 \, = \, 
  \frac{\tau_2}{\upsilon + \tau_2} \, .
\label{partsum}
\eeq
The overall UV divergence can then be extracted considering the overall distance of 
the two-gluon system from the origin, using
\beq
  \lambda \, = \, \lambda_1 + \lambda_2 \, ,  \qquad  
  \omega \, = \, \frac{\lambda_1}{\lambda_1 + \lambda_2} \, ,
\label{overalllam}
\eeq
Integration over $\lambda$ is now immediate, and gives the overall divergence 
in the form of a factor $\Gamma(4\eps)$. In this case, however, the integral over 
$\omega$ gives a sub-divergence for each graph: for graph $(a)$, one finds
a factor of
\beq
  \int_0^1 d \omega \frac{1}{\big[ \omega (1 - \omega) \big]^{1 - 2 \eps} }
  \, \, \theta \bigg( \frac{\omega}{1 - \omega} > \frac{x_2}{x_1} \bigg)
  \, = \, \frac{1}{2 \eps} + \phi (a) \, 
\label{omegafacta}
\eeq
where
\beq
  \phi (a) \, = \,  - \ln \left( \frac{x_2}{x_1} \right) + \left[ 4 \Li_2 \left( - \frac{x_1}{x_2} 
  \right) + \ln^2 \left( \frac{x_2}{x_1} \right) \right] \eps + \co (\eps^2) \, ,
\label{phia}
\eeq
whereas for diagram (b) one finds
\beq
  \int_0^1 d \omega \frac{1}{\big[ \omega (1 - \omega) \big]^{1 - 2 \eps} }
  \, \, \theta \bigg( \frac{\omega}{1 - \omega} < \frac{x_2}{x_1} \bigg)
  \, = \, \frac{1}{2 \eps} + \phi (b) \, 
\label{omegafactb}
\eeq
where
\beq
  \phi (b) \, = \, - \phi(a) - 4 \zeta_2 \eps + \co (\eps^2) \, .
\label{phib}
\eeq
The presence of subdivergences is related to the fact that, for each diagram, 
one can shrink the inner gluon line to the origin, while leaving the other gluon 
untouched. As a consequence, the kinematic contributions of each diagram 
have a double pole, which, however, cancels as expected when we combine 
the two diagrams to compute the kinematic factor of the whole web. We get
\beq
  {\cal K} (a) - {\cal K} (b) \, = \, 2 \kappa^2 \, \Gamma(4 \eps) \, \gamma_{\ii \jj} 
  \gamma_{\jj \kk} \int_0^1 dx_1 dx_2 \, P(x_1,\gamma_{\ii \jj}) \, 
  P(x_2,\gamma_{\jj \kk}) \, \phi(x_1,x_2; \eps) \, ,
\label{finkin}
\eeq
where 
\beq
  \kappa \! & \equiv & \! - \left( \frac{\mu^2}{m^2} \right)^\eps g_s^2 \, 
  \frac{\Gamma(1 - \eps)}{8 \pi^{2 - \eps}} \, , \nonumber \\
  \phi (x_1,x_2; \eps) \! & = & \! \ln \left( \frac{x_1}{x_2} \right)  + 
  \Big[ 4 \Li_2 \left( - \frac{x_1}{x_2} \right) + \ln^{2} \left( \frac{x_1}{x_2} \right)
  + 2 \zeta_2 \Big] \eps + \co (\eps^2) \, .
\label{defkin}
\eeq
This can be substituted into \eq{expab}, together with \eq{ecf},  to obtain 
the complete expression for the web $w_{121}^{(2)}$. In order to recover the
corresponding contribution to the soft anomalous dimension, one step still 
needs to be performed: we need to use \eq{Gamfromreg}, combining
the result for $w_{121}^{(2)}$ with the appropriate commutator involving the
one-loop result, evaluated to ${\cal O}(\eps^0)$. We need the combination
\beq
  \Gamma^{(2)}_{{\cal S}, 121} \, = \, - 4 w_{121}^{(2,-1)} - 2 \Big[ w^{(1,-1)}, 
  w^{(1,0)} \Big] \, ,
\label{combigam}
\eeq
where $w^{(1,k)}$ denotes the contribution of order $\eps^k$ to the one-loop web 
discussed in \secn{cusponeloop}. The remaining parameter integrals can be performed
with standard techniques~\cite{Gardi:2013saa,Falcioni:2014pka}, and the result
has a relatively simple form, which can be written as
\beq
  \Gamma^{(2)}_{{\cal S}, 121} \, = \, \frac{{\rm i}}{4} \,  f^{abc} T_\ii^a T_\jj^b T_\kk^c \,
  \, r ( \alpha_{\ii \jj} ) \, r ( \alpha_{\jj \kk} ) \Big[ 
  \ln (\alpha_{\ii \jj}) U (\alpha_{\jj \kk}) - \ln (\alpha_{\jj \kk}) U (\alpha_{\ii \jj})  
  \Big] \, ,
\label{finredweb}
\eeq
where $\alpha_{ij}$ is defined in \eq{alpha}, and the functions $r(\alpha)$ and 
$U(\alpha)$ are given by
\beq
  r (\alpha) & = & \frac{1 + \alpha^2}{1 - \alpha^2} \, , \nonumber \\
  U (\alpha) & = & 2 \Li_2 (\alpha^2) + 4 \ln \alpha \ln (1 - \alpha^2) - 2 \ln^2 \alpha 
  - 2 \zeta_2 \, .
\label{functinredweb}
\eeq
In addition to the contribution in \eq{combigam}, which we get from diagrams with 
two gluon attachments on leg number two, we also need to add the contributions 
arising from diagrams where legs number one and three have two gluon attachments.
The result of this sum can be written as
\beq
  \Gamma^{(2)}_{{\cal S}, \, {\rm MGEW}} \, = \, \frac{{\rm i}}{4} \,  
  f^{abc} T_\ii^a T_\jj^b T_\kk^c \,
  \sum_{i, j, k} \eps_{ijk} \,
  \, r ( \alpha_{ij} ) \, r ( \alpha_{jk} ) 
  \ln (\alpha_{ij}) \, U (\alpha_{jk}) \, ,
\label{finredweb2}
\eeq
where the subscript MGEW denotes the contribution of {\it multiple-gluon-exchange 
webs}, in the sense of Ref.~\cite{Falcioni:2014pka}. If $n$ Wilson lines are present, 
a further sum over all possible triples of Wilson lines yields
\beq
  \Gamma^{(2)}_{{\cal S}, \, {\rm MGEW}} \, = \, \frac{{\rm i}}{4} \, 
  \sum_{i>j>k=1}^n
   f^{abc} T_i^a T_j^b T_k^c \,
  \sum_{I, J, K \in \{ i, j, k \}} \!\!\! \eps_{IJK} \,
  \, r ( \alpha_{IJ} ) \, r ( \alpha_{JK} ) 
  \ln (\alpha_{IJ}) \, U (\alpha_{JK}) \, .
\label{finredweb3}
\eeq
The full two-loop soft anomalous dimension for three particles receives a further 
contribution, in addition to MGEWs, from the single-diagram web with a three-gluon 
vertex depicted in Fig.~\ref{threelegweb2}. As a consequence of the dipole formula, 
this web vanishes in light-like limit, as first discovered in Refs.~\cite{Aybat:2006mz,
Aybat:2006wq}. The case of Wilson lines off the light cone, of course, has great 
interest of its own, since it governs infrared divergences for amplitudes involving 
massive particles. At the one-loop level, the general form of these divergences 
was given in Ref.~\cite{Catani:2000ef}; at two-loops, the methods discussed here 
become relevant, and the calculation is far from trivial, requiring state-of-the-art 
integration techniques. The result, however, is remarkably simple~\cite{Ferroglia:2009ep,
Ferroglia:2009ii,Kidonakis:2009ev,Mitov:2009sv,Chien:2011wz}. For completeness, 
we give here the complete outcome, which can be written as
\beq
  \Gamma^{(2)}_{\cal S} \, = \, \Gamma^{(2)}_{{\cal S}, \, 3g} + 
  \Gamma^{(2)}_{{\cal S}, \, {\rm MGEW}} \, ,
\label{fullgamma2}
\eeq
with $\Gamma^{(2)}_{{\cal S}, {\rm MGEW}}$ given in \eq{finredweb}, while
\beq
  \Gamma^{(2)}_{{\cal S}, \, 3g} \, = \, \frac{{\rm i}}{2} \, 
    \sum_{i>j>k = 1}^n 
  f_{abc} T_i^a T_j^b T_k^c 
  \, \sum_{I, J, K \in \{ i, j, k \}} \!\!\! \eps_{IJK} \, \ln^2 (\alpha_{JK}) 
  \ln(\alpha_{IJ}) \, r(\alpha_{IJ}) \, ,  
\label{final}
\eeq
a strikingly compact and elegant expression. Comparing with \eq{Gamma1alpha}, we see that 
the result is a weighted sum of the three available angle-dependent cusp anomalous 
dimensions, strongly constrained by the overall Bose symmetry: an intriguing result, 
which is not yet fully understood.


\section{Real radiation: factorisation and subtraction}
\label{Subtra}

The main focus of our report is the infrared singularity structure of virtual corrections
to scattering amplitudes, which were discussed in considerable detail in the previous 
Sections. We learnt that infrared singularities arise from specific momentum configurations, 
corresponding to the exchange of soft or collinear virtual massless particles. Ultimately,
we saw that the long-distance sensitivity of a massless gauge amplitude is ruled by 
a small set of universal quantities, that are in turn determined by their soft and collinear 
anomalous dimensions. As a consequence, the IR content of virtual corrections is 
universal, {\it i.e.} it is independent of the details of the underlying hard process. 

In this Section, for completeness, we will discuss the factorisation properties
of soft and collinear real radiation, with a two-fold motivation. On the one hand,
factorisation for real radiation is a necessary complement of the factorisation of
virtual corrections, in light of the KLN theorem: in fact, the theorem states the 
cancellation of IR singularities among real and virtual corrections, regardless 
of the details of the resolved subprocess. Given the factorisation and universality 
properties of virtual corrections, it is natural to expect to be able to also identify 
and extract singular contributions to real radiation matrix elements in terms of 
universal kernels, multiplying lower-multiplicity matrix elements, which should be 
the only process-dependent component. This requires a delicate interplay between 
virtual and real corrections, with virtual singularities providing `sum-rule' constraints
on real radiation kernels, whose phase-space integrals must effect the predicted 
cancellation.

A second motivation is the fact that the factorisation of soft and collinear real 
radiation is a fundamental tool for phenomenological applications to collider 
observables. As we will discuss is some detail in \secn{SubtraNLO}, constructing
a precise theoretical prediction for an infrared-safe QCD observable requires
being able to implement the cancellation of infrared poles in an automatic and
efficient way. This is the goal pursued by {\it subtraction} methods, which have 
been the focus of a great deal of theoretical activity in the past decades.

In what follows, we will begin in \secn{Real} by sketching the factorisation properties 
of gauge amplitudes in soft and collinear limits, giving explicit examples at tree 
level and referring to the literature for known higher-order corrections. The simple 
and well-known results at lowest order will suggest the possibility of computing 
the factorisation kernels by means of matrix elements of fields and Wilson lines, 
closely related to the matrix elements discussed earlier in the context of virtual 
corrections. We will then proceed, in \secn{SubtraNLO}, to outline the basic ideas 
underlying the subtraction methods, which propose to combine real and virtual 
corrections in a universal way, delivering directly finite predictions for IR-safe 
observables, that can be evaluated numerically. In this context, in \secn{counterterm}, 
we will discuss a strategy to define and compute the real radiation kernels (and thus 
the local universal infrared counterterms required in the context of subtraction) 
in a systematic way, in principle at arbitrary order, so that the cancellation of 
real and virtual singularities can be made explicit from the outset. We will illustrate 
the results of this approach in \secn{StruSub}, up to NNLO in perturbation theory,
paying particular attention to strongly ordered multiple soft and collinear limits,
that play an important and intricate role for subtraction beyond NLO.


\subsection{Factorisation for soft and collinear real radiation}
\label{Real}

In the past decades, many important studies have focus on 
extracting the singular contributions to gauge amplitudes and cross sections 
in the limits where a number of massless particles become soft and/or collinear.
This vast research endeavour would deserve a review on its own, since the 
results were achieved with a variety of innovative techniques, and in many 
cases they are of great phenomenological relevance.

As expected from the KLN theorem, and from general properties of $S$-matrix 
elements, under infrared singular limits scattering amplitudes {\it quasi-completely} 
factorise into universal kernels multiplying lower-point amplitudes. To be precise, 
when a subset of particles becomes soft or collinear, certain Mandelstam invariants
will vanish: factorisation is a feature of leading-power contributions in those 
invariants; we describe it as {\it quasi-complete} in the sense that soft and
collinear kernels are still linked to lower-point amplitudes by either colour
correlations (in the soft case) or spin correlations (in the collinear case).
At cross-section level, the basic lowest-order factorisation properties have of 
course been well-known since the early days of perturbative QCD~\cite{Altarelli:1977zs,
Gribov:1972ri,Dokshitzer:1977sg,Mueller:1981sg,Dokshitzer:1991wu,Bassetto:1984ik,
Altarelli:1981ax,Ellis:1991qj}, and they are directly relevant for the cancellation of infrared 
divergences at NLO. Collinear splitting kernels for hadronic cross sections are 
also necessary ingredients for all applications of collinear factorisation in QCD, and 
they have by now been determined to three loops~\cite{Curci:1980uw,Furmanski:1980cm,
Moch:2004pa,Vogt:2004mw,Moch:2014sna}, with remarkable partial results at four and five 
loops~\cite{Davies:2016jie,Moch:2017uml,Moch:2018wjh,Herzog:2018kwj,Moch:2021qrk}.
Detailed studies at amplitude level are somewhat more recent: the general form 
of the factorisation of tree-level gauge amplitudes in soft and collinear limits was 
established in the late eighties, using recursion relations~\cite{Berends:1987me,
Berends:1988zn} or general expression for the amplitudes derived from string 
theory~\cite{Mangano:1987kp,Mangano:1990by}; a combined factorisation 
including both soft and collinear limits was later proposed in~\cite{Kosower:1997zr,
Kosower:2003bh}. Amplitude-level factorisation was extended to single-unresolved 
radiation at one-loop in~\cite{Bern:1993qk,Bern:1994zx,Bern:1995ix,Bern:1998sc,
Bern:1999ry,Catani:2000pi,Kosower:1999rx}, and to double-unresolved radiation 
at tree level in~\cite{Campbell:1997hg,Catani:1998nv,Catani:1999ss}: these results 
are necessary ingredients to implement the cancellation of infrared divergences 
at NNLO. At yet higher orders, a growing body of results is available, including 
an all-order proof of collinear factorisation in the planar limit~\cite{Kosower:1999xi}, 
and explicit results for multiple collinear limits at tree level~\cite{DelDuca:1999iql,
DelDuca:2019ggv,DelDuca:2020vst} and one loop~\cite{Catani:2003vu,
Sborlini:2013jba,Sborlini:2014mpa,Badger:2015cxa}, as well as for multiple 
soft limits at tree level~\cite{Anastasiou:2013srw,Catani:2019nqv} and one 
loop~\cite{Zhu:2020ftr,Catani:2021kcy}, single-soft limits at two loops~\cite{Duhr:2013msa,
Li:2013lsa}, and single-collinear limits at two loops~\cite{Badger:2004uk,
Bern:2004cz}. The organisation of colour degrees of freedom in multiple 
soft emissions has been further explored in Refs.~\cite{AngelesMartinez:2016iow,
Forshaw:2019ver}.

Of special interest for both the theory and the applications of 
factorisation are the results of Ref.~\cite{Catani:2011st}: they show that for
scattering amplitudes involving space-like as well as time-like collinear splittings,
the na\"ive form of collinear factorisation is violated; specifically, one finds
that the splitting kernel in such cases does not depend exclusively on the
quantum numbers of the two collinear particles, but there are leftover colour
correlations with the other particles participating in the scattering process.
This effect can be traced back to the existence of non-vanishing phases
in the divergent parts of the amplitude in the space-like case (see for example 
\eq{GammaDip}) and to the non-commutativity of the collinear limit with the 
$\epsilon$ expansion in dimensional regularisation (see for example the 
discussion in Ref.~\cite{Kosower:1999xi}). This breakdown of simple collinear
factorisation does not affect the discussion of fixed-angle scattering amplitudes 
developed in \secn{MultiPart}, but it has important phenomenological consequences,
since it is related to the appearance of unconventional logarithmic enhancements
in QCD cross sections with phase-space constraints, such as {\it non-global}
logarithms~\cite{Dasgupta:2001sh}, or {\it super-leading} logarithms~\cite{Forshaw:2006fk,
Forshaw:2008cq,Forshaw:2012bi}. The finiteness of sufficiently inclusive hadronic
cross sections relies upon the cancellation of these effects, and is proved to all orders
for processes with electroweak final states~\cite{Collins:1988ig,Collins:1989gx},
with techniques that extend to inclusive jet cross sections~\cite{Aybat:2008ct}.
The precise boundaries of applicability of these techniques to more intricate
or less inclusive observables have not however been determined in detail.

In what follows, we will not try to review in further detail this great body of results. 
Rather, we will focus on some simple examples at tree-level, and later at one loop, 
and we will infer a general proposal for the definition of soft and collinear real-radiation
kernels, bearing a close correspondence to the virtual soft and jet functions defined
in the previous section. As we will see, this will lead to a natural procedure for
the construction of local subtraction counterterms~\cite{Magnea:2018ebr}.

Following~\cite{Catani:1996vz}, we start by introducing a ket notation for scattering
amplitudes, which is particularly efficient for the discussion of real infrared emission. Consider
a generic scattering process involving massless final-state QCD partons with momenta 
$p_1, p_2, \dots $. Non-QCD particles carrying a total momentum $Q$ are always 
understood. To express colour, spin and flavour degrees of freedom we introduce 
different sets of indices: $\{c_1, c_2, \dots \}$ for colour, $\{s_1, s_2, \dots \}$ 
for spin, and $\{f_1, f_2, \dots\}$ for flavour. Given a set of basis vectors in 
colour and spin spaces $\{\ket{c_1, c_2, \dots} \otimes \ket{s_1, s_2, \dots}\}$, the 
scattering amplitude and the corresponding transition probability, with spin and 
colour indices summed, can be written as
\beq
\label{introket}
  \mathcal{A}_{\, f_1 f_2 \dots}^{\, c_1 c_2 \dots ; \, s_1  s_2 \dots} \, (p_1, p_2, \dots ) 
  & \equiv &
  \Big(\bra{c_1 c_2 \dots} \otimes \bra{s_1 s_2 \dots} \Big) 
  \ket{\mathcal{A}_{\, f_1 f_2 \dots}(p_1, p_2, \dots)} \, , 
  \nonumber \\
  \big| \mathcal{A}_{\, f_1 f_2 \dots}(p_1, p_2, \dots) \big|^2 & = & 
  \langle \mathcal{A}_{\, f_1 f_2 \dots}(p_1, p_2, \dots) | 
  \mathcal{A}_{\, f_1 f_2 \dots}(p_1, p_2, \dots)\rangle \, .
\eeq
The effects of radiation on the colour content of the amplitude will be treated by 
introducing colour-insertion operators, as discussed in \secn{ColStru}. If, for 
example, we consider the tree-level matrix element $\mathcal{A}_{\, g, \, a_1 \dots a_n}
(k, p_1, \dots, p_n)$, where the outgoing gluon $g$ carries momentum $k$, 
colour $c$ and spin polarisation $\lambda$, we know from our earlier discussions
that the leading singular contribution to this matrix element in the limit when the 
gluon becomes soft is given by
\beq
  \langle c | \otimes \langle \lambda \!
  \ket{ \,\mathcal{A}_{\, g, f_1 \dots  f_n}(k, p_1, \dots, p_n)}_{\rm soft} \, = \, 
  \epsilon_\lambda (k) \cdot J^c (k) \ket{ \mathcal{A}_{\, f_1 \dots f_n}
  (p_1, \dots, p_n)} \, ,
\label{eq:Real_tree_fact}
\eeq
where $J^c$ is the appropriate colour component of the eikonal current defined 
in \eq{softcurrtree},
\beq
  {\bf J}^\mu (k) \, = \, g \mu^\epsilon \, \sum_{i = 1}^n {\bf T}_i \; 
  \frac{p_i^\mu}{p_i \cdot k} \, ,
\label{curragain}
\eeq
which is transverse by colour conservation
\beq
  k \cdot {\bf J} (k) \, = \, g \mu^\epsilon  \sum_{i = 1}^n {\bf T}_i \, = \, 0 \, .
\label{currcons}
\eeq
Squaring the soft matrix element in \eq{eq:Real_tree_fact}, and exploiting colour 
conservation, one recovers the soft contribution to the cross section, as in
\eq{radmelsoft2}, which we rewrite here as
\beq
  \Big| \mathcal{A}_{\, g, \, f_1 \dots f_n} (k, p_1, \dots, p_n) \Big|_{\rm soft}^2
  \, = \, -  8 \pi \alpha_s \, \mu^{2 \eps} \sum_{i \neq j} \,
  \mathcal{I}^{(k)}_{ij}\;  \Big| \mathcal{A}^{\, ij}_{\, f_1 \dots f_n} (p_1, \dots, p_n) \Big|^2 \, ,
\quad 
\label{eq:real_soft_fact}
\eeq
where the eikonal prefactor is given by 
\beq
  \mathcal{I}^{(k)}_{ij} \, \equiv \, \frac{s_{ij}}{s_{ki} \, s_{kj}} \, ,
\label{eq:eikonal_current_tree}
\eeq
with $s_{k i} = 2 k \cdot p_i$, and where the squared amplitude on the {\it r.h.s.} 
of \eq{eq:real_soft_fact} is the colour-connected Born amplitude (without radiation), 
which in the present notation reads
\beq
  \big| \mathcal{A}^{\, ij}_{\, f_1 \dots f_n} \big|^2 & \equiv &
  \bra{ \mathcal{A}_{\, f_1 \dots f_n}} {\bf T}^i \cdot {\bf T}^j
  \ket{ \mathcal{A}_{\, f_1 \dots f_n}} \; .
\label{eq:colour_connected}
\eeq
Based on our experience with virtual corrections in \secn{universal_fun} and in 
\secn{MultiPart}, we would expect this result to emerge from a matrix element
of Wilson lines, replacing the hard particles appearing in the Born amplitude: the
expectation is borne out by the observation that the current ${\bf J}^\mu$ is 
a sum of eikonal vertices, and thus can be easily modelled in terms of Wilson 
lines. A simple explicit computation indeed shows that the matrix element 
\beq
  {\cal S}_{\, n, \, 1} \big(k ; \{\beta_i\} \big) \, \equiv \, \bra{k, \lambda} \, T \bigg[ 
  \prod_{i = 1}^n \Phi_{\beta_i} (0, \infty) \bigg] \ket{0} \, ,
\label{eq:NLO_oper}
\eeq
computed at tree level, precisely reproduces the soft factor in \eq{eq:Real_tree_fact}, 
and thus, upon squaring, \eq{eq:real_soft_fact}.

\begin{figure}
\begin{center}
  \includegraphics[height=3cm,width=12cm]{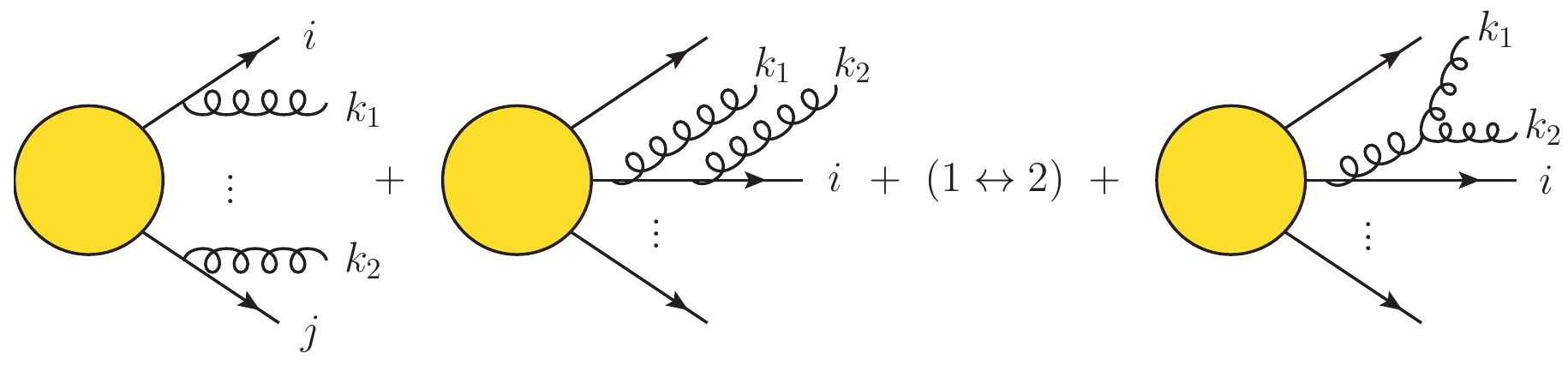}
  \caption{Double soft gluon emission from a generic hard amplitude.}
\label{doublesoft}
\end{center}
\end{figure}  
The analysis performed for virtual corrections, leading to the conclusion that soft
emissions can be described by means Wilson lines, makes us confident that the
simple argument leading to \eq{eq:Real_tree_fact}, and to the identification of
the single soft emission operator in \eq{eq:NLO_oper}, will generalise to the 
emission of multiple soft gluons at leading power. Indeed, the limit of QCD 
tree-amplitudes when two gluons become simultaneously soft was studied 
by Berends and Giele \cite{Berends:1988zn} and by Catani \cite{Proceedings:1992fla},
showing precisely a factorisation of this kind. Following Ref.~\cite{Catani:1999ss},
we introduce the double-soft emission current by writing
\beq
  && \langle a_1 a_2 | \otimes \langle \lambda_1 \lambda_2  
  \ket{ \mathcal{A}_{\, g g, f_1 \ldots f_n} (k_1, k_2, p_1, \dots, p_n)}_{\rm soft}
  \, = \, \epsilon^{\mu_1}_{\lambda_1} (k_1) \epsilon^{\mu_2}_{\lambda_2} (k_2)
  \nonumber \\ && \hspace{5cm} \times \, 
  J^{a_1 a_2}_{\mu_1 \mu_2} (k_1,k_2) 
  \ket{ \mathcal{A}_{\, f_1,\dots, f_n}(p_1, \dots, p_n)} \, .
\label{eq:Real_2glu_tree_fact}
\eeq
The double-soft current receives contribution from three different diagram topologies,
depicted in Fig.~\ref{doublesoft}. The explicit expression for the double-soft current 
$J_{a_1 a_2}^{\mu_1 \mu_2}$ is given by \cite{Catani:1999ss}
\beq
\label{eq:double_soft_J}
  J^{a_1 a_2}_{\mu_1 \mu_2} (k_1, k_2) & = & 4 \pi \as \mu^{2\eps}
  \Bigg\{ 
  \sum_{\ell \neq m} T_m^{a_1} \, T_\ell^{a_2} \, 
  \frac{p_m^{\mu_1}}{p_m \cdot k_1} \, 
  \frac{p_\ell^{\mu_2}}{p_\ell \cdot k_2} \\
  && \hspace{1cm} + \, \sum_m  \, 
  \bigg[ \, \frac{p_{m}^{\mu_1} \, p_m^{\mu_2}}{p_m \cdot 
  \left( k_1 + k_2 \right)}
   \left(
  \frac{T_m^{a_2} T_m^{a_1}}{p_m \cdot k_2} 
  \, + \, \left( 1 \leftrightarrow 2 \right) \right)
  \nonumber \\
  && \hspace{1cm} + \, \,
  {\rm i} f^{\, \, \, a_1 a_2}_a \, T_m^a \,\,
  \frac{p_m \cdot \left( k_2 - k_1 \right) 
  g^{\mu_1 \mu_2} 
  + 2 \, p_m^{\mu_1} \, k_1^{\mu_2} 
  - 2 \, p_m^{\mu_2} \, k_2^{\mu_1}}{2 k_1 \cdot k_2 \, 
  p_m \cdot \left( k_1 + k_2 \right)} \bigg] \Bigg\} \, . \nonumber 
\eeq
The correspondence between the expression in \eq{eq:double_soft_J} and the 
diagrams in Fig.~\ref{doublesoft} is straightforward: the first term stems from diagrams 
of type {\it (a)}, the second contribution reflects diagrams of type {\it(b)}, and 
finally the last term, proportional to $f_a^{\; \; a_1, a_2}$, clearly emerges 
from diagrams of type {\it (c)}. At squared-amplitude level, the factorisation 
formula for the configuration at hand can be written as
\beq
\label{doublesoftsq}
  &&
  \Big| \mathcal{A}_{\, g g, f_1 \ldots f_n} (k_1, k_2, p_1, l\dots, p_n) \Big|^2_{\rm soft}
  \, = \,  2 \left( 4\pi \as \mu^{2 \eps} \right)^2 \\ \nonumber
  && \qquad  \times \,
  \Bigg[ \sum_{i \neq j} \sum_{k \neq l} \, \mathcal{I}_{ij}^{(1)} \, \mathcal{I}_{kl}^{(2)} \, 
  \Big| \mathcal{A}_{\, f_1 \ldots f_n}^{ijkl} (p_1, \dots, p_n) \Big|^2
  + \sum_{i \neq j}  \, \mathcal{I}_{ij}^{(12)} \Big| 
  \mathcal{A}_{\,  f_1 \ldots f_n}^{ij} (p_1, \dots, p_n) \Big|^2 \Bigg] \, ,
\eeq
where the first term in square brackets is the factorised contribution, proportional 
to the double-colour-connected matrix element
\beq
  \big| \mathcal{A}_{\, f_1 \ldots f_n}^{ij kl} \Big|^2
  \, \equiv \, 
  \bra{\mathcal{A}_{\, f_1 \ldots f_n}} \Big\{ {\bf T}^i \cdot {\bf T}^j ,  
  {\bf T}^k \cdot {\bf T}^l \Big\} \ket{\mathcal{A}_{\, f_1 \ldots f_n}}  \, ,
\label{eq:double_colour_connected}
\eeq
while the double emission from a single colour dipole si expressed by the kernel
$\mathcal{I}_{ij}^{(12)}$, which can be read off from Eq.(109) in \cite{Catani:1999ss}, 
after appropriate relabelling of the momenta. 

As before, the double-soft current can be extracted from a matrix element of 
Wilson lines: quite naturally, we propose
\beq
  {\cal S}_{\, n, \, 2} \big(k_1, k_2 ; \{\beta_i\} \big) \, \equiv \, 
  \bra{k_1, \lambda_1; k_2, \lambda_2} \, T \bigg[ \prod_{i = 1}^n 
  \Phi_{\beta_i} (0, \infty) \bigg] \ket{0} \, .
  \label{eq:NNLO_oper}
\eeq
It is straightforward to show that \eq{eq:NNLO_oper}, evaluated at tree level, 
indeed reproduces the current $J^{a_1 a_2}_{\mu_1 \mu_2}$, contracted 
with gluon polarisation vectors as in \eq{eq:Real_2glu_tree_fact}. Indeed, 
expanding the product of Wilson lines to order $g^2$, there are three different 
configurations that return non-vanishing contributions: one may expand to 
first order in $g$ two of the $n$ Wilson lines, obtaining the first term in 
\eq{eq:double_soft_J}, or one may expand only one Wilson line to order $g$, 
and then include a Lagrangian interaction, to get the non-abelian component; 
finally, one can expand a single line up to order $g^2$: in this case, the path 
ordering that defines $\Phi(0, \infty)$ precisely returns the remaining contributions 
in the first line of \eq{eq:double_soft_J}.

It is important to stress that the singular soft configurations we have just described 
have been extracted by taking the two soft momenta to vanish at the same rate,
formally replacing $k_1 \to \lambda k_1$ and $k_2 \to \lambda k_2$, taking
$\lambda \to 0$ and retaining the leading power in $\lambda$. This we describe
as a {\it democratic} IR limit. In the context of subtraction, as we will see, it is
also important to identify {\it hierarchical}, or {\it strongly-ordered} limits, corresponding 
to configurations where one gluon is much softer than the other, {\it i.e.}  one of
the two momenta vanishes at a faster rate with respect to the other. The expression 
of the corresponding current can be easily deduced by taking the leading term 
in the $k_1$ ($k_2$) expansion of $J_{a_1 a_2}^{\mu_1 \mu_2}$ at amplitude 
level, and of $\mathcal{ I}_{cd}^{(12)}$ at cross-section level. One gets
\beq
\label{2sCGso}\nonumber
  \big(J^{\, \rm s.o.}\big)^{a_1 a_2}_{\mu_1 \mu_2} (k_1, k_2) \! & = & \!
  g_s \mu^\eps
  \Bigg(
  \sum_{m = 1}^n T^{a_2}_m \, \frac{p_{m, \, \mu_2}}{p_m \cdot k_2} \, 
  \delta^{a_1 a} + {\rm i} \, 
  f^{a_1 a_2 a} \, \frac{k_{1 , \, \mu_2}}{k_1 \cdot k_2} \Bigg) 
  \sum_{\ell = 1}^n \, T_{\ell, \, a} \, 
  \frac{p_{\ell, \, \mu_1}}{p_\ell \cdot k_1} \, , \nonumber \\
  \mathcal{I}_{ij}^{(12) \, {\rm s.o.}} \! & = & \! 
  - 2 C_A \,  \mathcal{I}_{ij}^{(2)} \; 
  \Big[ \mathcal{I}_{i 2}^{(1)}  + \mathcal{I}_{j 2}^{(1)}  
  - \mathcal{I}_{ij}^{(1)} \Big] \, .
\eeq
For the purposes of constructing subtraction counterterms in a systematic way, 
it would be interesting to be able to derive strongly ordered limits in terms of
matrix elements of Wilson lines, as done above for the democratic case. This
approach will be briefly discussed in \secn{StruSub}. To conclude our discussion
of soft factorisations for real radiation at tree level, we recall that the eikonal 
approximation is spin-independent, so the formulas that we have described
can be applied to soft gluon radiation from hard gluons, simply by placing
the colour generators in the appropriate (adjoint) representation; furthermore,
the case of massless quark radiation (through an intermediate gluon which splits 
into a quark-antiquark pair) can be treated similarly and does not present new
difficulties. We remark also that the soft factorisation is {\it quasi complete}, 
as noted at the beginning of this Section, in the sense that eikonal kernels 
and Born-like matrix elements are not entirely independent of each other, but
are still connected by colour correlations.

To complete the picture of tree-level infrared factorisation for real radiation,
we must consider collinear splittings as well. In this case, we will also find
a {\it quasi complete} factorisation, but with colour correlations replaced by 
spin correlations. The trivial colour structure of tree-level collinear factorisation
is best displayed by picking the most suitable colour basis: in this case,
a natural choice is the colour trace basis of Ref.~\cite{Mangano:1990by},
later extensively employed at loop level~\cite{Bern:1990ux,Bern:1994zx}.
For multi-gluon amplitudes, this organisation is intuitively motivated by
string theory, where tree-level gluon amplitudes are computed as expectation 
values of conformal vertex operators inserted on the boundaries of the 
tree-level string world sheet, which can be conformally mapped to a circle. 
Colour degrees of freedom are carried by {\it Chan-Paton factors}, which are
generators for the fundamental representation of the gauge group associated
with each external state. The full tree-level amplitude is then constructed
as a sum of sub-amplitudes, each corresponding to a cyclic ordering of 
the vertex operators on the boundary of the circle, and each accompanied
by the corresponding trace of Chan-Paton factors. For $n$ gluons, one may 
write
\beq
  {\cal A}_n^{a_1 \ldots a_n} (p_1, \ldots, p_n) \, = \, 
  \sum_{\sigma \in S_n/Z_n} {\rm Tr} \Big[ T^{a_{\sigma (1)}} \ldots T^{a_{\sigma (n)}}
  \Big] \, A_n \left( p_{\sigma(1)}, \ldots, p_{\sigma(n)} \right) \, ,
\label{tracebasis}
\eeq
where the sum is over $(n-1)!$ non-cyclic permutations of the $n$ external 
gluons, and we omitted polarisation indices for simplicity. This string-inspired 
organisation of gauge amplitudes can be generalised to loop amplitudes,
where at $\ell$ loops one must allow for products of up to $\ell + 1$ trace 
factors, and to amplitudes with massless fermions. 

The sub-amplitudes $A_n$ are gauge invariant, and they enjoy a number of 
properties that make them very useful for practical calculations~\cite{Mangano:1990by}.
In the present context, the most notable aspect is the fact that, for each
sub-amplitude, collinear poles arise only from configurations where the two
particles that become collinear are adjacent in the selected cyclic ordering.
Thus, for example, $A_5(p_1, p_2, p_3, p_4, p_5)$ will have a collinear pole
when $p_2$ becomes collinear to $p_3$, but not when $p_2$ becomes collinear 
to $p_5$: indeed, within the set of diagrams contributing to this ordering, there
is no diagram where $p_2$ and $p_5$ emerge from the same internal vertex,
so that no propagator becomes singular as $p_2$ becomes parallel to $p_5$.
Picking then a sub-amplitude where two momenta $p_i$ and $p_j$ are adjacent,
and the corresponding gluons carry helicities $s_i$ and $s_j$, one finds that,
in the collinear limit, the sub-amplitude factorises as
\beq
  A_n^{\, \ldots s_i s_j \ldots} ( \ldots, p_i, p_j, \ldots) \Big|_{p_i || \, p_j} \, = \, 
  \sum_{s = \pm} {\rm Split}_{-s}^{(0)} (p_i, s_i; p_j, s_j) \, A_{n - 1}^{\, \ldots s \ldots} 
  ( \ldots, p_i + p_j, \ldots)
\label{splitkoso}
\eeq
where the parent gluon carrying momentum $p_i + p_j$ is taken to carry outgoing
helicity $(+s)$, so that it enters the tree-level splitting amplitude Split$_{-s}^{(0)}$ 
with incoming helicity $(-s)$. The factorisation in \eq{splitkoso} encodes the 
singular behavior of $A_n$ as $s_{ij} = 2 p_i \cdot p_j \to 0$, which turns out to 
be proportional to $s_{ij}^{-1/2}$ (giving the expected $1/s_{ij}$ singularity at 
cross-section level). Tree-level splitting amplitudes in this basis were derived
in~\cite{Berends:1987me} and in ~\cite{Mangano:1990by}, and are listed in
those references for all flavour and spin pairings consistent with QCD couplings.
As an example, one can write the splitting amplitude for a gluon branching into 
two gluons as~\cite{Kosower:1999xi,Catani:2003vu}
\beq
\label{splitglue}
  {\rm Split}_{-s}^{(0)} (p_i, s_i; p_j, s_j) \! & = & \! g \mu^\epsilon \, \frac{2}{s_{ij}} \,
  \Big[ \epsilon_{s_i} (p_i) \cdot \epsilon_{s_j} (p_j) \,\, p_i \cdot
  \epsilon_{s} (p_i + p_j)  \\ 
  && \hspace{-2cm} + \, \epsilon_{s_i} (p_i) \cdot \epsilon_{s} (p_i + p_j) 
  \,\, p_i \cdot \epsilon_{s_j} (p_j) \, - \, \epsilon_{s_j} (p_j) \cdot 
  \epsilon_{s} (p_i + p_j) \, \,  p_j \cdot \epsilon_{s_i} (p_i) \Big] \, , \nonumber
\eeq
where the parent gluon is taken on shell ({\it i.e.} in the exact collinear limit),
and the $1/s_{ij}$ singularity is softened to a square root by the products of
polarisation vectors in square brackets.

If one does not wish to be constrained by the choice of a specific colour basis,
it is possible to rephrase the collinear factorisation of massless multi-particle
amplitudes in terms of colour operators (which can then be evaluated in 
any basis), as was done for soft emissions. In this case, one can write the
tree-level factorisation directly in colour space as
\beq
  \ket{ \,\mathcal{A}^{\ldots s_i s_j \ldots}_n (\ldots, p_i, p_j, \ldots )}_{p_i || p_j} \, = \, 
  {\bf Sp}^{(0)}_{-s} (p_i, s_i; p_j, s_j) 
  \ket{ \mathcal{A}^{\ldots s \ldots}_{n-1} (\ldots, p_i + p_j, \ldots)} \, ,
\label{splitcata}
\eeq
where the splitting matrix ${\bf Sp}^{(0)}_{-s}$ is now a colour operator acting on 
the right on the colour index of the parent particle, and on the left on the colour 
indices of the collinear pair. At tree level, the relation between the operator
${\bf Sp}^{(0)}_{-s}$ and the colourless splitting amplitude Split$_{-s}^{(0)}$
is trivial, since each possible flavour splitting can be proportional to only a
single colour tensor. Thus, for example, for a gluon with colour index $a$ 
splitting into gluons with colour indices $b$ and $c$, the splitting operator is 
obtained from \eq{splitglue} by simply inserting a factor of $( - {\rm i} f_{a b c} )$. 
More complicated colour structures will, of course, arise for multiple collinear
splittings at tree and loop level.

\begin{figure}[t]
\centering
\includegraphics[width=0.7\textwidth]{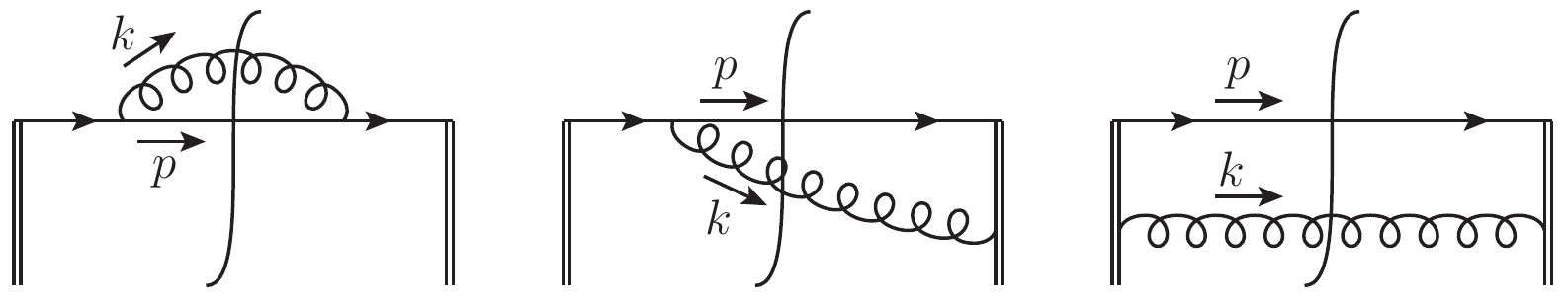}
\caption{Feynman graphs contributing to the radiative quark jet function, defined 
in \eq{qradjetsq1}, at lowest order.}
\label{fig:cut_one_loop_jet}
\end{figure}

To discuss collinear factorisation at cross-section level, taking into account the
non-trivial spin structure, it is useful to define a squared amplitude that is summed 
over all the spin indices, except for those of a select particle of flavour $f_1$, which 
will be interpreted as the parent particle in the subsequent collinear splitting. 
Following Ref.~\cite{Catani:1999ss}, we define
\beq
  \mathcal{T}_{\, f_1 \dots f_n}^{\, s_1 s'_1} (p_1, \ldots, p_n) \, = \,
  \sum_{s_2, \ldots, s_n} \sum_{c_i} \mathcal{A}_{\, f_1 \ldots f_n}^{\, c_1 \ldots c_n ;
  \, s_1 \ldots s_n} (p_1, \dots, p_n) 
  \Big[ \mathcal{A}_{\, f_1 \ldots f_n}^{\, c_1 \dots c_n ; \, s'_1 \ldots s_n} 
  (p_1, \dots, p_n) \Big]^\dag \, .
\label{eq:T}
\eeq
Next, we need to be more precise in parametrising the collinear limit.  For two 
particles of flavour $f_i$ and $f_j$, carrying momenta $p_i$ and $p_j$, we 
identify a light-like momentum $p^\mu$ as the collinear direction, and we pick 
an auxiliary light-like vector $n^\mu$. How the collinear direction is approached 
is specified by the transverse momentum vector $k^\mu_\bot$, by definition 
orthogonal to both $p^\mu$ and $n^\mu$. Each collinear parton carries a 
collinear momentum fraction which can be defined by $z_a = s_{an}/(s_{in}
+ s_{jn})$, with $a = i, j$ and $s_{an} = 2 p_a \cdot n$, so that $z_i + z_j = 1$; 
hence, one may use $z \equiv z_i$ and $z_j = 1 - z$. This results in the
Sudakov parametrisation
\beq
  p_i \, = \, z \, p^\mu + k_\bot^\mu - \frac{k_\bot^2}{z} \, 
  \frac{n^\mu}{2 p \cdot n} \, , \quad \!
  p_j \, = \, (1 - z) \, p^\mu - k_\bot - \frac{k_\bot^2}{1 - z} \, 
  \frac{n^\mu}{2 p\cdot n} \, , \quad \!
  s_{ij} \, = \, - \frac{k_\bot^2}{z (1 - z)} \, .
\label{eq:sudakov_NLO}
\eeq
The collinear limit is approached as $k_\bot \rightarrow 0$, and the leading-power
behaviour of the squared matrix element is encoded by~\cite{Catani:1999ss}
\beq
  \Big| \mathcal{A}_{\, f_1 \ldots f_n} (p_1, \dots, p_n) \Big|^2_{p_i || p_j} \, = \,  
  \frac{8 \pi \as \mu^{2 \eps}}{s_{ij}} \, \,
  \mathcal{ T}^{s s'}_{f, f_1 \ldots f_n} (p, p_1, \ldots, p_n) \, 
  \widehat{P}_{f_i f_j}^{\, s s'} \big(z, k_\bot; \eps \big) \, ,
\label{eq:coll_facto}
\eeq
where on the {\it r.h.s.} it is understood that particles $f_i$ and $f_j$, carrying 
momenta $p_i$ and $p_j$, are omitted from the list, and they have been replaced
by a single particle $f$ carrying momentum $p$. The possible flavour combinations
are determined by the Feynman rules of the theory under consideration. For QCD,
the kernels $ \widehat{P}_{a_i a_j}^{s s'}$ are the $d$-dimensional DGLAP splitting 
functions, represented as spin operators acting on the spin indices $s,s'$ of the 
spin tensor $\mathcal{T}$. The splitting functions are given by squares of the 
splitting amplitudes ${\bf Sp}^{(0)}_{-s}$, summed over colours, after extracting 
the coupling and the factor of $1/\sqrt{s_{ij}}$. Their explicit expressions are of 
course well known, and in the present normalisation they are given by
\beq
\label{eq:spin-dep_AP}
  \widehat{P}_{qg}^{ss'} (z, k_\bot; \eps) \! & = & \! \delta_{ss'} \, C_F 
  \bigg[\frac{1+z^2}{1-z}-\eps (1-z) \bigg] \, , \nonumber \\
  \widehat{P}_{gq}^{ss'} (z, k_\bot; \eps) \! & = & \! \delta_{ss'} \, C_F  
  \bigg[\frac{1+(1-z)^2}{z}-\eps z \bigg] \, , \\
  \widehat{P}_{q \bar q}^{\mu \nu} (z, k_\bot; \eps) \! & = & \! T_R 
  \bigg[ - g^{\mu \nu} + 4 z(1 - z) \frac{k_\bot^\mu k_\bot^\nu }{k_\bot^2} \bigg] \, ,
  \nonumber \\
  \widehat{P}_{gg}^{\mu \nu} (z, k_\bot; \eps) \! & = & \! 2 C_A 
  \bigg[ - g^{\mu \nu} \Big( \frac{z}{1 - z} + \frac{1 - z} z \Big) - 
  2 (1 - \eps) z (1 - z) \frac{k_\bot^\mu k_\bot^\nu }{k_\bot^2} \bigg] \, . \nonumber
\eeq
Notice that all the kernels are symmetric under the exchange of a quark with 
an anti-quark, {\it i.e.} $\widehat{P}_{Xq} = \widehat{P}_{X \bar q}$. From 
\eq{eq:spin-dep_AP} it is evident that the spin sensitivity of the kernels, and, 
consequently, the one of the spin tensor, is trivial in the case of a parent fermion, 
while gluon splittings retain a non-trivial azimuthal dependence. 
Also in the collinear sector, we are interested in expressing the singular kernels 
in terms of universal, gauge invariant operators. Let us begin by by considering 
gluon radiation from an outgoing quark: in this case, we may take inspiration 
from the virtual jet function defined in \eq{Jqfofa}, and implement collinear radiation, 
as was done for the soft current, by including an extra gluon in the final 
state. At amplitude level, this amounts to introducing the matrix element
\beq
\label{qradjetsq_amp} 
  \overline{u}_s (p) {\cal J}_{q, \, 1}^{\lambda} \big(k; p, n \big) \, \equiv \, 
  \bra{p, s; k, \lambda} \, T \Big[ \Phi_n (\infty, 0) \, \psi(0) \Big] \ket{0} \, .
\eeq
The function ${\cal J}_{q, \, 1}^{\lambda}$ is a spin matrix, and at tree level it
must reproduce the collinear singularity of the splitting amplitudes, stripped
of colour information. The dependence on the vector $n^\mu$ defining the Wilson 
line mimicks the gauge dependence within the families of axial gauges, and
choosing $n^2 = 0$ one can reproduce the Sudakov parametrisation of the 
collinear region. To test our ansatz, we can square \eq{qradjetsq_amp}, take
a Fourier transform, and introduce the (unpolarised) cross-section-level radiative 
jet function
\beq
\label{qradjetsq1}
  J_{q, \, 1} \big(k; l, p, n \big) \equiv \int d^d x \, {\rm e}^{{\rm i} l \cdot x} \, 
  \sum_{\lambda, s} \bra{0} \, T \, \Big[ \Phi_n (\infty, x) \, \psi(x) \Big] \ket{p, s; k, \lambda}
  \nonumber \\ 
  \hspace{3cm} \times \, \bra{p, s; k, \lambda} \, \overline{T} \, \Big[ \overline{\psi} (0) \, 
  \Phi_n (0, \infty) \Big] \ket{0} \, ,
\eeq
where the complex-conjugate matrix element is anti-time-ordered by the action 
of the operator $\overline{T}$, and the vector $l^\mu$ plays the role of the total 
momentum flowing into the final state. We can now perform a simple test of the 
correctness of our assumptions by computing, in Feynman gauge, the lowest 
perturbative order for \eq{qradjetsq1}. It receives contributions from three different 
Feynman diagrams, displayed in Fig.~\ref{fig:cut_one_loop_jet}. 
Taking, for simplicity, $n^2 = 0$, we find
\beq
  J_{q, \, 1}^{(0)} \big(k; l, p, n \big) \, = \, \frac{4 \pi \alpha_s}{(l^2)^2} \,  C_F \, 
  (2 \pi)^d \delta^d \left(l - p - k \right) 
  \left[ - \slash{l} \gamma_\mu \slash{p} \gamma^\mu \slash{l} 
  + \frac{l^2}{k \cdot  n}  
  \big( \,  \slash{l} \slash{n} \slash{p} 
  + \slash{p} \slash{n} \slash{l} \,  \big) \right] \, . 
\label{onecollglu}
\eeq
It is easy to verify the correspondence between the contributions of the three 
diagrams in Fig.~\ref{fig:cut_one_loop_jet} and the axial-gauge calculation of 
tree-level splitting kernel in Ref.~\cite{Catani:1999ss}. Notice however that 
in \eq{onecollglu} the collinear limit for $k$ and $p$, corresponding to $l^2 \to 0$, 
still has to be performed. This can be done exploiting the Sudakov parametrisation
in \eq{eq:sudakov_NLO}, and truncating the expressions for the two collinear 
momenta at leading order in the transverse momentum, as
\beq
  p^\mu \, = \, z l^\mu + {\cal O} \left( l_\perp \right) \, , \qquad
  k^\mu  \, = \, (1 - z) l^\mu + {\cal O} \left( l_\perp \right) \, , \qquad
  n^2 \, = \, 0 \, .
\label{Sudakov}
\eeq
At leading power in $l_\bot$ we easily find
\beq
  J_{q, \, 1}^{(0)} \big(k; l, p, n \big) \, = \, \frac{8 \pi \alpha_s}{l^2} \,  C_F \,
  (2 \pi)^d \, \delta^d \left(l - p - k \right)\, \slash{l} \,
  \left[ \frac{1 + z^2}{1 - z} - \epsilon \left( 1 - z \right) \right] \, ,
\label{AP0}
\eeq
up to corrections of relative order $l_\bot^2$. In the square bracket, as expected, 
we recognise the leading order unpolarised DGLAP splitting function $P_{qg}$, 
as reported in \eq{eq:spin-dep_AP}.

Clearly, the number and complexity of splitting functions at amplitude and cross-section 
level rapidly increases at higher orders, both because loop corrections must be included,
and because multiple real radiation comes into play; furthermore, when multiple collinear
particles are radiated, the identification of strongly-ordered limits becomes relevant.
The operator matrix element expressions that we have identified in these simple 
examples, however, point the way to a useful systematic approach, linking together
virtual corrections and real radiation. We will now proceed by first introducing the
problem of infrared subtraction is some detail, and then discussing a set of possible
definitions for real-radiation soft and collinear counterterms to all orders in perturbation 
theory.


\subsection{Introducing the problem of infrared subtraction at NLO}
\label{SubtraNLO}

The discussion in \secn{Real} lays the groundwork for constructing a practical and 
general implementation of the cancellation of IR divergences in IR-safe observables,
which follows from the KLN theorem discussed in \secn{KLN}. After our discussion 
of the general cancellation theorems, and our tour of many technical aspects of IR 
factorisation, the reader may well wonder why this cancellation should still be described
as a `problem'. Indeed, so long as one is computing total cross sections, or highly
inclusive observables, the cancellation between divergences coming from virtual 
corrections and those stemming from the phase space integration of unresolved 
radiation can be performed analytically, even to rather high orders in perturbation 
theory. What changes the game, and raises the stakes, is the nature of typical 
collider observables. Even the definition of IR-safe observables, and verifying their
IR-safety to all orders, become non-trivial questions. Experimental observables
are very often defined in terms of particle jets, and jets in turn are typically 
defined in terms of clustering algorithms; furthermore, experimental constraints
imply complex phase space cuts on kinematic variables, often arising at the level
of triggers for data collection: these must be implemented in efficient and versatile 
numerical codes. In practice, this means that phase-space integrals of real-radiation 
matrix elements must be performed numerically, and this in turn requires subtracting 
their singular contributions before performing the integration in $d=4$. Another way 
to state the problem is that, in a collider environment, versatility means being able to 
provide fully differential distribution for interesting observables, in order to be able 
to cope with different sets of experimental cuts. As a consequence, it becomes
necessary to perform the extraction of singular contributions in a universal way,
for arbitrary configurations of the resolved particles. The factorisation properties of
real radiation matrix elements, sketched in \secn{Real}, suggest that completing
this program is possible in principle, however a general and efficient practical 
implementation at high orders is far from easy to achieve.

In this section, we will present the basic ideas that have been applied to the cancellation 
problem, in the relatively simple context of NLO calculations, where, arguably, the 
problem is satisfactorily solved. In the past decades, two main strategies have been 
developed to solve the problem at NLO: the {\it slicing}~\cite{Giele:1991vf,Giele:1993dj,
Giele:1994xd} approach and the {\it subtraction}~\cite{Frixione:1995ms,
Catani:1996jh,Catani:1996vz,Catani:2002hc,Nagy:2003qn,Frederix:2009yq} approach. 
In what follows, we will mainly develop the subtraction viewpoint, but, in order to 
present the basic idea behind these methods, we illustrate them in the context of 
a toy model. Start by considering the integral
\beq
  I (F) \, = \, \lim_{\eps \rightarrow 0} \Bigg[\int_0^1 \frac{dx}{x} \, x^\eps \, F(x)
  - \frac1\eps \, F(0) \Bigg] \, ,
\label{eq:example}
\eeq
where $F(x)$ is a generic function admitting a Taylor expansion around $x=0$, 
and the first term mimicks the phase-space integration of the real radiation,
while the second term plays the role of the virtual correction. The goal of both 
{\it slicing} and {\it subtraction} schemes is computing $I(F)$ without relying 
on the analytic evaluation of the integral over $x$, which involves the arbitrarily
complicated function $F$. The {\it slicing} approach tackles the problem by 
splitting the integration domain by means of a small parameter $\delta \ll 1$, 
and working with a simple approximate expression for $F$ (for example $F(0)$)
in the region containing the singularity. One writes then
\beq
  \hspace{-2mm}
  I (F) \simeq \lim_{\eps \rightarrow 0} \Bigg[ \, F(0)
  \int_0^\delta \frac{dx}{x} \, x^\eps + \int_\delta^1 \frac{dx}{x} \, x^\eps \, F(x)
  - \frac1\eps \, F(0) \Bigg]  \, = \, F(0) \, \log \delta  
  + \int_\delta^1 \frac{dx}{x} \, x^\eps \, F(x) \, .
\label{eq:example2} 
\eeq
The {\it r.h.s.} of \eq{eq:example2} is manifestly finite as $\epsilon \to 0$, 
and therefore is suitable for numerical evaluation. The slicing procedure is 
simple and intuitive, but of course, depending on the required accuracy, one 
needs to worry about the residual dependence on the slicing parameter 
$\delta$, which needs to be small, but not too small. The problem is not
insurmountable, and can be significantly ameliorated by improving the 
estimate for $F(x)$ to be employed in the singular region (see, for 
example,~\cite{Boughezal:2016zws,Moult:2016fqy}).

To avoid problems associated with the sensitivity to the slicing parameter, 
the subtraction approach proceeds by subtracting from $F(x)$ its value in $x=0$,
and adding it back in integrated form, therefore leaving $I(F)$ unchanged. In formulae
\beq
  I (F) \, = \, \lim_{\eps \rightarrow 0} \Bigg[ \int_0^1 \frac{dx}{x} \, x^\eps 
  \Big( F(x) - F(0) \Big) + \int_0^1 \frac{dx}{x} \, x^\eps \, F(0)
  - \frac1\eps \, F(0) \Bigg]  \, .
\label{eq:subtra}
\eeq
Once again, the first term is finite in $\epsilon$ by construction, while the 
proposal is to evaluate the second one analytically to cancel the explicit 
pole of the last term. This is of course trivial in this toy example, but requires
considerable care in realistic calculations, as we will see. Let us stress that 
no approximations have been made to write \eq{eq:subtra}. At NLO, the 
subtraction schemes proposed by Catani and Seymour in~\cite{Catani:1996vz}, 
and by Frixione, Kunszt and Signer in~\cite{Frixione:1995ms} have been 
fully developed for general collider processes, and both are implemented 
in efficient event generators~\cite{Campbell:1999ah,Gleisberg:2007md, 
Frederix:2010cj,Frederix:2008hu,Czakon:2009ss,Frederix:2009yq}. One 
can argue that the IR subtraction problem can be considered solved in full 
generality at this accuracy.

Before moving on to higher orders, let us introduce some notation, in order
to describe the subtraction approach more precisely. Let us consider, for 
simplicity a production process involving $n$ massless final-state coloured 
particles at Born level, and let $\mathcal{A}_n (p_i)$, $i = 1, \ldots, n$, be 
the relevant scattering amplitude. We expand the amplitude in perturbation 
theory as 
\beq
  {\cal A}_n (p_i) \, = \, {\cal A}_n^{(0)} (p_i) \, + \, {\cal A}_n^{(1)} (p_i) \, + \,
  {\cal A}_n^{(2)} (p_i) \, + \, \ldots \, , 
\label{pertexpA}
\eeq
where $\mathcal{A}_n^{(k)}$ is the $k$-loop correction, including the appropriate 
power of the strong coupling constant. Moreover, consider a generic infrared-safe 
observable $X$, whose differential distribution is given by 
\beq
  \frac{d \sigma}{d X} \, = \, \frac{d \sigma_\LO}{d X} \, + \, 
  \frac{d \sigma_\NLO}{d X} \, + \, 
  \frac{d \sigma_\NNLO}{d X} \, + \, \ldots \, .
\label{pertexpsig}
\eeq
The leading-order contribution is proportional to the Born transition probability, 
$B_n \, \equiv \, \big| {\cal A}_n^{(0)} \big|^2$, and is IR finite. Starting with NLO, the 
infrared content of the distribution becomes relevant, and $d\sigma/dX$ must be 
computed by combining unresolved radiative contributions with loop corrections. 
Writing
\beq
  R_\npo \, \equiv \, \left| {\cal A}_\npo^{(0)} \right|^2 \, , \qquad
  V_n \, = \, 2 {\bf Re} \left[ {\cal A}_n^{(0) \dag} \, {\cal A}_n^{(1)} 
  \right] \, ,
\label{pertA2}
\eeq  
we know that infrared poles in $V_n$ will be cancelled by the integration of
unresolved radiation in $R_\npo$. At this stage, therefore, both contributions 
must be regulated, and we write the NLO distribution as 
\beq
  \frac{d \sigma_\NLO}{d X} & = & \lim_{d \to 4} 
  \Bigg\{ \! \int d \Phi_n \, V_n \, \delta_n (X) + 
  \int d \Phi_\npo \, R_\npo \, \delta_\npo (X) \Bigg\} \, , 
\label{pertO}  
\eeq
where $\delta_m (X) \equiv \delta (X - X_m)$ fixes $X_m$, the expression 
for the observable appropriate for an $m$-particle configuration, to the 
prescribed value $X$, and $d \Phi_m$ denotes the Lorentz-invariant phase 
space measure for $m$ massless final state particles.

Following the idea described in \eq{eq:subtra}, we introduce a {\it local 
counterterm} $K_\npo^{\one}$, which is required to mimic the singular IR 
behaviour of the real-radiation matrix element {\it locally} in phase 
space, and, at the same time, is expected to be simple enough to be 
analytically integrated in the radiative phase space $d \Phi_{\rm rad, 1} \equiv 
d \Phi_{\npo}/d \Phi_n$. If this is possible, one can compute the {\it integrated 
counterterm}
\beq
  I_n^{\, ({\bf 1})} \, = \, \int d \Phi_{\rm rad, 1} \, K_\npo^{\one}  \, ,
\label{intcountNLO}
\eeq
exposing explicit poles corresponding to the phase space singularities of 
the integrand, which, in turn, are the same as those of the real-radiation matrix 
element. It is now possible to construct a version of \eq{pertO} where 
virtual corrections and real contributions are separately finite, and therefore
phase space integrals can be performed numerically when needed:
\beq
  \frac{d \sigma_\NLO}{d X} \, = \, 
  \int d \pn \Big[ V_n + I_n^{\, ({\bf 1})} \Big] \, 
  \delta_n (X)  
  + \int d \Phi_\npo \, 
  \Big[  R_\npo \, \delta_\npo (X) - K_\npo^{\one} \,
  \delta_n (X) \Big] \, .
\label{subtNLO}
\eeq
The interpretation of Eq.\eqref{subtNLO} is the following: the combination 
$R_\npo - K_\npo^{\one}$ is free of phase-space singularities by construction, 
while $I_n$ cancels the explicit poles of the virtual correction, as a direct 
consequence of the KLN theorem. The sum $V_n + I_n$ is thus finite as 
$\epsilon \rightarrow 0$, and both combinations in square brackets are 
suitable for a numerical implementation. Note that IR safety of the observable
$X$ is necessary for the cancellation, which requires $\delta_\npo (X)$
to turn smoothly into $\delta_n (X)$ in all unresolved limits.

Clearly, the actual definition of an appropriate local counterterm is not unique, 
and characterises the subtraction scheme. A precise definition must include
a prescription to map the phase space for $\npo$ massless on-shell particles
in the radiative phase space to a configuration of $n$ massless on-shell particles
in the Born-level phase space, satisfying the same momentum conservation
condition, in order to work at all stages with well-defined matrix elements (see,
for example,~\cite{DelDuca:2019ctm}). Furthermore, care must be taken to 
include in the counterterm all singular limits of the matrix element, subtracting
all possible double countings of soft and collinear regions. To illustrate and 
clarify the procedure that we have just rather formally described, we now briefly 
summarise a possible sequence of steps to build a suitable NLO counterterm. 
More details on the procedure, and on its extension to NNLO, can be found 
in~\cite{Magnea:2018hab}. We note that the procedure described below is just an 
example, and a number of alternative strategies exist~\cite{TorresBobadilla:2020ekr}.

In order to isolate and enumerate the singular regions in the radiative phase 
space, following the logic of~\cite{Frixione:1995ms}, we begin by introducing a set 
of {\it sector functions} ${\cal W}_{ij}$, which constitute a partition of unity, and 
subdivide the phase space in cells where the only singularities are due to a 
selected particle $i$ becoming soft, or becoming collinear to a second particle 
$j$. As an example, following~\cite{Magnea:2018hab}, define
\beq
 e_i \, \equiv \, \frac{s_{qi}}{s} \, , \qquad w_{ij} \, \equiv \, \frac{s s_{ij}}{s_{qi} s_{qj}} \, ,
 \qquad \sigma_{ij} \, \equiv \, \frac{1}{e_i w_{ij}} \, ,
\label{sectprelim}
\eeq 
where, as we are considering final-state coloured particles only, we can 
work with a fixed total incoming momentum $q^\mu$, and we defined $s = q^2$
and $s_{q\ell} = 2 q \cdot p_\ell$. In terms of these variables, we can define
the sector functions as
\beq
  {\cal W}_{ij} \, \equiv \, \frac{\sigma_{ij}}{\sum_{k \neq l} \sigma_{kl}} \, ,
\label{sectfunc}
\eeq
obviously satisfying $\sum_{i \neq j} {\cal W}_{ij} = 1$. Next, we introduce two 
operators, ${\bf S}_i$ and ${\bf C}_{ij}$, which are defined to extract from the
functions on which they act the leading power in the Laurent expansion in the
normal variables defining the soft and collinear singularities, respectively. Thus,
for example, the action of ${\bf S}_i$ on the real radiation matrix element $R_\npo$
yields \eq{eq:real_soft_fact}, upon identifying particle $i$ with the gluon of momentum
$k$; similarly, the action of ${\bf C}_{ij}$ on $R_\npo$ yields \eq{eq:coll_facto}.
Acting on the sector functions ${\cal W}_{ij}$, the operators satisfy
\beq
  {\bf S}_i  \sum_{k \neq i} {\cal W}_{ik} \, = \, 1 \, , \qquad 
  {\bf C}_{ij} \Big[ {\cal W}_{ij} + {\cal W}_{ji} \Big] \, = \, 1 \, .
\label{SCsumrules}
\eeq
One easily verifies that the function $R_\npo {\cal W}_{ij}$ is only singular when 
particle $i$ becomes soft and when the pair $ij$ becomes collinear; \eq{SCsumrules}
then guarantees that, upon suitably summing over sectors, the full soft and collinear
singularities are recovered, and sector functions will not explicitly appear in the
soft and collinear counterterms that will need to be integrated.

At this stage, we can construct a {\it candidate counterterm}, by taking the 
combination
\beq
  K^{({\bf 1})}_\npo \, = \, \sum_i  \sum_{j \neq i} \Big( {\bf S}_i + {\bf C}_{ij}
  - {\bf S}_i {\bf C}_{ij} \Big) R_\npo {\cal W}_{ij} \, ,
\label{candcount}
\eeq
where the last term in parentheses subtracts the soft-collinear regions, which would
otherwise be doubly counted. Note that this requires that the operators ${\bf S}_i$ 
and ${\bf C}_{ij}$ be defined to commute. We describe \eq{candcount} as a {\it 
candidate counterterm} because it is not quite ready to be used directly in 
\eq{subtNLO}. The reason can be traced to the definition of the integrated 
counterterm $I_n^{({\bf 1})}$ in \eq{intcountNLO}, with its phase-space measure
$d \Phi_{{\rm rad}, \, 1}$. Operationally, in order to perform that integration 
in a universal, process-independent way, $K^{({\bf 1})}_\npo$ must factorise
a Born-level squared matrix element involving $n$ on-shell particles, whose
momenta must still satisfy the original momentum-conservation condition
in the full phase space $d \Phi_\npo$: only in this way the definition of
$d \Phi_{{\rm rad}, \, 1}$ makes sense, and the remaining integration in
$d \Phi_n$ can be performed at a later stage for both terms in \eq{subtNLO}
without ambiguities. To solve this problem, one needs to provide a mapping
of the $(\npo)$-particle phase space onto the $n$-particle phase space,
which must not affect soft and collinear limits at leading power. An example
of such a mapping is given by~\cite{Catani:1996vz}
\beq
  \overline{p}_i^{(abc)} \, = \, p_i \,\,\,\, {\rm if} \, i \neq a,b,c \, ; \quad 
  \overline{p}_b^{(abc)} \, = \, p_a + p_b - \frac{s_{ab}}{s_{ac} + s_{bc}} p_c \, ; \quad
  \overline{p}_c^{(abc)} \, = \, \frac{s_{abc}}{s_{ac} + s_{bc}} p_c \, ;
\label{mapNLO}
\eeq
one easily verifies that the $n$ momenta $\overline{p}_\ell^{(abc)}$, with 
$\ell = 1, \ldots, n$, are massless, their sum equals the sum of the original of
$\npo$ momenta $p_j$, with $j = 1, \ldots, \npo$, and finally the two sets coincide
when $p_a$ becomes soft and when $p_a$ becomes collinear to $p_b$. We 
can now turn our candidate counterterm in \eq{candcount} into a full-fledged 
local counterterm, suitable for direct use in \eq{subtNLO}, by taking the factorised 
expressions for soft and collinear limits (\eq{eq:real_soft_fact} and \eq{eq:coll_facto}, 
respectively) and evaluating the Born-level squared matrix elements on mapped 
momentum configurations of the form of \eq{mapNLO}, sector by sector in
the radiative phase space. Notice that there is ample freedom in the choice of 
phase space mappings (see, for example, Ref.~\cite{DelDuca:2019ctm}), but they
are not arbitrary: in particular, taking the limits ${\bf S}_i$ and ${\bf C}_{ij}$, or
their combinations, on the mapped counterterm, must yield the same result
as taking the limits on the candidate counterterm.

Having illustrated in some detail the structure of IR subtraction at NLO, we 
now discuss in general terms how this structure generalises to higher orders.
In \secn{counterterm}, we will see how the factorisation of virtual corrections
to scattering amplitudes provides the ingredients for defining soft and collinear
candidate counterterms to all orders in perturbation theory. Finally, in \secn{StruSub}
we will discuss the structure of counterterms needed at NNLO.


\subsection{Soft and collinear local counterterms: a general strategy}
\label{counterterm}

The results of \secn{Real} and \secn{SubtraNLO} strongly suggest that local 
infrared counterterms for real radiation should be computable in terms of
matrix elements of fields and Wilson lines, which could be easily matched
to those expressing infrared virtual corrections, allowing for a transparent 
cancellation of singularities.

In what follows, we introduce a theoretical framework based on this 
idea~\cite{Magnea:2018ebr}. Using as a guiding principle the factorisation 
structure of virtual gauge amplitudes, we propose a general definition of 
local soft and collinear candidate counterterms, valid to all orders in 
perturbation theory and for any number of real radiated particles. These 
candidate counterterms can be straightforwardly combined with their virtual 
counterparts, building up manifestly finite quantities. Of course, candidate 
counterterms will still need the appropriate phase space mappings to become 
part of a full-fledged subtraction algorithm.

Inspired by the NLO results in \secn{Real}, and in particular by the definitions
in \eq{eq:NLO_oper}, \eq{eq:NNLO_oper}, and \eq{qradjetsq_amp}, our strategy 
to define the soft and collinear counterterms can be schematically summarised 
as follows.
\begin{itemize}
\item We introduce radiative soft, collinear and soft-collinear functions at 
amplitude level, using matrix elements with the same structure as for virtual 
corrections, but including on-shell real radiation in the final state.
\item We square these matrix elements to build cross-section-level radiative 
soft and jet functions, local in the radiative phase space: these will be our 
candidate counterterms.
\item We show that our candidate counterterms, when summed over the
number of radiated particles, and integrated over the corresponding phase 
spaces, including the case of no radiation, build up quantities that are IR
finite to all orders in perturbation theory. This proves that our candidate 
counterterms, upon integration, will indeed cancel virtual IR poles order 
by order.
\end{itemize}
Beginning with soft radiation, consider the case of $n$ hard particles, 
represented by Wilson lines in the soft approximation, radiating $m$ soft 
gluons, and introduce the {\it eikonal form factors}
\beq
  {\cal S}_{n, \, m} \Big(\{k_m\}, \{\beta_i\} \Big) & = & 
  \bra{k_1, \lambda_1; \ldots ; k_m, \lambda_m} \, T \bigg[ 
  \prod_{i = 1}^n \Phi_{\beta_i} (\infty, 0) \bigg] \ket{0} 
  \nonumber \\
  & = & \sum_{p = 0}^\infty
  {\cal S}_{n, \, m}^{(p)} \Big(\{k_m\}, \{\beta_i\} \Big) \, ,
\label{softrad}
\eeq
where the second line defines the perturbative expansion of the matrix 
element in powers of the coupling $g$. Note that the definition in \eq{softrad} 
includes loop corrections to all orders, and can easily be modified to include 
soft final state fermions, by simply adding them to the list of final state 
particles. For $m=0$, we recover the definition of the virtual soft function 
for $n$ particles, given in in \eq{Softnpart}, while for $m=1,2$ we get 
\eq{eq:NLO_oper} and \eq{eq:NNLO_oper}. The working assumption behind
\eq{softrad} is that the exact amplitude for the emission of $m$ soft gluons 
from $n$ hard coloured particles obeys, to all orders, the factorisation
\beq
  {\cal A}_{n, \, m} \left(k_1, \ldots, k_m; p_i \right)
  \, \simeq  \, {\cal S}_{n, \, m} \left(k_1, \ldots, k_m; 
  \beta_i \right) \, {\cal H}_n (p_i) \, ,
\label{ampfact}
\eeq
with corrections that are finite in four dimensions, and integrable in the soft 
particle phase space. \eq{ampfact} is proven to all orders for $m = 0$, and 
it is consistent with all known perturbative results, in particular with the 
arguments of~\cite{Catani:2000pi,Campbell:1997hg,Catani:1999ss}. To 
check the consistency of this approach beyond tree level, one can focus
on the case of single soft radiation at one loop, where \eq{ampfact} reads
\beq
  {\cal A}_{n, \, 1}^{(1)} \left(k ; p_i \right) \, = \, {\cal S}_{n, \, 1}^{(0)} 
  \left(k; \beta_i \right) \, {\cal H}^{(1)}_n (p_i) \, + \,  
  {\cal S}_{n, \, 1}^{(1)} \left(k; \beta_i \right) \, {\cal H}^{(0)}_n (p_i) \, .
\label{expone1}
\eeq
The first term describes tree-level soft-gluon emission, multiplying the finite 
part of the one-loop correction to the Born process; the second term has 
both explicit soft poles and singular factors from single soft real radiation. 
\eq{expone1} should be compared with the soft factorisation proposed 
in~\cite{Catani:2000pi}, 
\beq
  {\cal A}_{n, \, 1} \left(k ; p_i \right) \, = \, 
  \epsilon^{* \, (\lambda)} (k) \cdot 
  J_{\mbox{\tiny{CG}}} \left( k, \beta_i \right) \, {\cal A}_n (p_i) \, ,
\label{twofact}
\eeq
where the Catani-Grazzini soft current $J_{\mbox{\tiny{CG}}} \! \left( k, 
\beta_i \right)$ multiplies the full $n$-particle amplitude, including loop 
corrections containing infrared poles. In order to map the two calculations, 
one may express the one-loop hard part ${\cal H}_n^{(1)} (p_i)$ using 
\eq{ampfact} for $m = 0$, with the result
\beq
  {\cal H}^{(1)}_n (p_i) \, = \, {\cal A}^{(1)}_n \left( p_i \right) -
  {\cal S}^{(1)}_n \left( \beta_i \right) \, {\cal A}^{(0)}_n (p_i)  \, ,
\label{exptwo}
\eeq
where we normalised ${\cal S}^{(0)}_n$ to the identity operator in colour space.
This leads to an expression for the Catani-Grazzini one-loop soft-gluon
current in terms of eikonal form factors, as
\beq
\label{CGolsc}
  \epsilon^{* \, (\lambda)} (k)  \cdot J^{(1)}_{\mbox{\tiny{CG}}} (k, \beta_i) \, 
  \,\, = \,\, {\cal S}_{n, \, 1}^{(1)} \left(k; \beta_i \right) \, - \,
  {\cal S}_{n, \, 1}^{(0)} \left(k; \beta_i \right) \, {\cal S}^{(1)}_n 
  \left( \beta_i \right) \, .
\eeq
Comparing \eq{CGolsc} with the calculation in~\cite{Catani:2000pi}, one 
easily recognises that the same combination of Feynman diagrams is 
involved, and one recovers the known result
\beq
  \left[ J^{(1)}_{CG} \right]^\mu_a (k, \beta_i) & = & - \frac{\as}{4 \pi} \, 
  \frac{1}{\eps^2} \, \frac{\Gamma^3 (1 - \eps) 
  \Gamma^2 (1 + \eps)}{\Gamma(1 - 2 \eps)} \nonumber \\
  && 
   \times \, {\rm i} f_a^{\,\, b c} 
  \sum_{i = 1}^n \sum_{j \neq i} T_i^b T_j^c
  \left( \frac{\beta_i^\mu}{\beta_i \cdot k} - 
  \frac{\beta_j^\mu}{\beta_j \cdot k} \right) 
  \left[ \frac{2 \pi \mu^2 \left( - \beta_i \cdot \beta_j \right)}{\beta_i \cdot k 
  \, \beta_j \cdot k} \right]^\eps \, .
\label{olsc}
\eeq
For the purposes of subtraction, we now need to build a cross-section-level 
quantity: the eikonal form factor has to be multiplied times its complex conjugate, 
and one may perform the (trivial) polarisation sum. The result is the {\it eikonal 
transition probability}
\beq
  && \hspace{-5mm}
  S_{n, \, m} \Big( \{k_m\}, \{\beta_i\} \Big) 
  \,  \equiv \, 
  \sum_{p = 0}^\infty \, S_{n, \, m}^{(p)} \Big( \{k_m\}, \{\beta_i\} \Big) 
  \label{softsigma} \\
  && 
  \equiv \, 
  \sum_{\{\lambda_i\}} 
  \bra{0} \overline{T} \left[  \prod_{i = 1}^n 
  \Phi_{\beta_i} (0, \infty) \right] \ket{k_1, \lambda_1; \ldots; k_m, \lambda_m}
  \bra{k_1, \lambda_1; \ldots; k_m, \lambda_m} T \left[ \prod_{i = 1}^n 
  \Phi_{\beta_i} (\infty, 0) \right] \ket{0} \, , \nonumber
\eeq
where we used again the operator $\overline{T}$, as in \eq{qradjetsq1}, and we 
exploited the Wilson line property that $\Phi^\dagger_n(a, b) = \Phi_n(b,a)$.  
One could now make use of the completeness of the Fock basis in the soft 
gluon Hilbert space
\beq
  {\bf 1} \, = \, \ket{0} \bra{0} + \sum_{\ell = 1}^\infty
  \bigg(\int \prod_{i=1}^\ell d \Phi_1 (k_i) \bigg)
  \ket{k_1, \dots, k_\ell}
  \bra{k_1, \dots, k_\ell} \, , 
\label{eq:1}
\eeq
where $d\Phi_1 (k) = d^{d-1} k/(2 k^0 (2 \pi)^{d-1})$, to perform the final-state 
sum, with the result
\beq
  \sum_{m = 0}^\infty \int d \Phi_m \, S_{n, \,m} \Big(\{k_m\};
  \{ \beta_i \} \Big) \, = \, 
  \bra{0} \,  \overline{T} \left[ \prod_{i = 1}^n 
  \Phi_{\beta_i} (0, \infty) \right] T \left[ \prod_{i = 1}^n 
  \Phi_{\beta_i} (\infty, 0) \right] \ket{0} \, .
\label{complete}
\eeq
with $d\Phi_m=\prod_{i=1}^m d \Phi_{1}(k_i)$. In \eq{complete} one recognises 
an eikonal total cross section, where all the coloured particles belong to the final 
state. Thanks to the standard cancellation theorems (which can be verified with 
power-counting arguments), such cross sections are IR finite to all orders in perturbation 
theory: this implies that the transition probability in \eq{softsigma}, which is fully
local in the radiation phase space, indeed provides a set of candidate counterterms
for soft divergences. Note that, as written, the phase-space sum in \eq{eq:1} will 
lead to ultraviolet singularities: these can be dealt with by means of a suitable 
regulator, without affecting infrared poles. Indeed, eikonal cross sections with 
the general structure of \eq{complete} have been extensively used in the past, 
in the context of the resummation program for threshold and transverse momentum 
logarithms in QCD (see, for example,~\cite{Sterman:1986aj,Laenen:2000ij}) and 
in SCET~\cite{Becher:2014oda}: in those cases, one incorporates phase-space constraints
by shifting the origin of the Wilson line in the complex-conjugate amplitude, and
taking a suitable Fourier transform. In the process, the UV region of the phase-space 
integral is automatically regulated. These manipulations do not affect the infrared
behaviour of the cross section.

We conclude that \eq{complete} provides a prescription to cancel virtual poles 
of soft origin against phase-space integrals of the appropriate set of radiative
soft functions, which approximates the relevant squared matrix element at 
leading power in the soft momenta. For example, we can expand \eq{complete}
at NLO, obtaining 
\beq
  S_{n, \, 0}^{(1)} + \int d\Phi_{1} (k) \, S_{n, \, 1}^{(0)}(k) \, = \, \text{finite} \, .
\label{eq:soft_compl_NLO}
\eeq
The first term coincides with the squared virtual function in \eq{Softfofa}, and 
therefore contains the same soft poles as the virtual matrix element. The 
second contribution provides a natural candidate for the integrated countertem.

The collinear component can be handled in analogy to the soft case. Starting 
from the definition in \eq{Jqfofa}, and generalising along the lines of \eq{qradjetsq_amp},
we introduce amplitude-level {\it radiative jet functions}~\cite{Bonocore:2015esa,
Bonocore:2016awd,Beneke:2019oqx,Liu:2020ydl,Liu:2021mac} for quark-induced 
processes, by including in the final state an arbitrary number of outgoing 
gluons (again, other particles can be included by simply modifying the 
final-state quantum numbers). We define 
\beq
  \overline{u}_s (p) \, {\cal J}_{q, \, m}^{\{\lambda_i\}} \big(\{k_m\}; p, n \big)
  & \equiv & \bra{p, s; k_1, \lambda_1; \ldots; k_m, \lambda_m}  \overline{\psi} (0) \, 
  \Phi_{n} (0, \infty) \ket{0} \nonumber
  \nonumber \\ 
  & \equiv & \overline{u}_s (p) \sum_{p = 0}^\infty 
  {\cal J}_{q, \, m}^{(p)} \big(\{k_m\}; p, n \big) \, ,
\label{qradjet}
\eeq
where, again, the second line defines the perturbative expansion of the 
jet function. At cross-section level, one needs to account for the conservation 
of the momentum flowing in the final state. To do so, in analogy with what is 
done for parton distributions, we shift the field position in the complex-conjugate 
amplitude, and we take a Fourier transform, defining the {\it collinear transition 
probability} as 
\beq
\label{qradjetsq}
  && 
  J_{q, \, m} \big(\{k_m\}; l, p, n \big) \, \equiv \,
  \sum_{p = 0}^\infty J_{q, \, m}^{(p)} \big(\{k_m\}; l, p, n \big) \\
  &&   \hspace{-5mm} \equiv \,
  \int d^d x \, {\rm e}^{{\rm i} l \cdot x} \sum_{\{ \lambda_m \}} \bra{0} 
  \overline{T} \Big[ \Phi_n (\infty, x) \, \psi(x) \Big] \ket{p, s; \{k_m, \lambda_m\}}
  \bra{p, s; \{k_m, \lambda_m\}} T \Big[ \overline{\psi} (0) \, 
  \Phi_n (0, \infty) \Big] \ket{0} \, , \nonumber
\eeq
where the momentum $l^\mu$ appearing in the exponent, after performing the
Fourier transform, is given by $l=p+\sum_i k_i$, and, as before, we displayed the
time-ordering operators. Note that we have performed the polarisation sum for 
simplicity, but this is not necessary, and one can consider each polarisation 
separately. As done for the soft transition probability, we can construct a finite
quantity from \eq{qradjetsq} by summing over the number of final-state particles 
(with the final-state hard quark always included), and integrating over the final-state
phase space. The result of the sum is dictated by unitarity, and it is given by the
imaginary part of a generalised two point function (see also~\cite{Falcioni:2019nxk}),
\beq
  && \sum_{m = 0}^\infty \int d \Phi_{m+1} \, 
  J_{q, \, m} \big(\{k_m\}; l, p, n \big) \, = \\ 
  && \hspace{5mm} = \, 
  {\rm Disc} \bigg[ \int d^d x \, {\rm e}^{{\rm i} l \cdot x} \,
  \bra{0} T \Big[ \Phi_n (\infty, x) \psi(x) \overline{\psi} (0) 
  \Phi_n (0, \infty) \Big] \ket{0}
  \!  \bigg] \, . \nonumber
\label{compcoll}
\eeq
The expression in \eq{compcoll} is finite by power counting, as a consequence 
of the fact that it is fully inclusive in the final state. Expanding to NLO, we find 
\beq
  \int d \Phi_{1} \, 
  J_{q, \, 0}^{(1)} \left(l,p, n \right)+
  \int d \Phi_{2} \, 
  J_{q, \, 1}^{(0)} \, \left(k_1; l, p, n \right)
  \, = \, \text{finite} \, ,
\label{eq:coll_compl_NLO}
\eeq
which establishes \eq{qradjetsq}, for $m = 1$ at this order, as a local collinear
counterterm. This construction can be directly generalised to collinear radiation 
from gluons, starting from \eq{Jgdef} for single radiation at amplitude level, and
then constructing gluon-induced collinear transition probabilities as was done 
above in the quark-induced case. Similarly, one can define radiative eikonal jets, 
mimicking the definition used for virtual corrections, \eq{Jqeikfofa}, and suitably 
adding radiated particles to the final state.

Let us conclude this Section by illustrating how these formal definitions of local
radiative soft and jet functions can be used to derive the structure of subtraction
counterterms, directly from the factorisation of virtual corrections to scattering 
amplitudes. The first step is to expand the infrared factorisation formula for
multi-particle scattering amplitudes, \eq{ampnfact}, at NLO, and then interfere 
it with the Born amplitude. In order to do that, we note that jets and eikonal jets 
can be normalised to unity at tree level; furthermore, one can easily verify, at
this order, that amplitude-level soft and jet functions readily combine to form 
their cross-section-level counterparts when the the amplitude is squared. In 
the language of \eq{pertA2}, and using the perturbative expansions defined in 
\eq{softsigma} and in \eq{qradjetsq}, one gets
\beq
  V_n \, = \, 2 {\bf Re} \left[ {\cal A}_n^{(0) \dag} \, {\cal A}_n^{(1)} 
  \right] 
  \, = \, {\cal H}^{(0) \, \dagger}_n S_{n, \, 0}^{(1)} {\cal H}^{(0)}_n
  \, + \, \sum_{i = 1}^n {\cal H}^{(0) \, \dagger}_n\left( J^{(1)}_{i, \, 0}  -
  J_{i, \, \E, \, 0}^{(1)} \right) S_{n, \, 0}^{(0)} {\cal H}^{(0)}_n
   \, + \, {\rm finite} \, ,
\label{virtolo}
\eeq
where the notation for every function is that $F^{(p)}_m$ denotes the contribution
at $p$ loops to the function involving $m$ radiations. One readily identifies the first 
term as encoding the soft divergences, including soft-collinear double poles, while
the second term is responsible for hard-collinear singularities.

Having identified explicit poles of $V_n$, we now simply need to pick the needed
candidate counterterms from our list: in this case, we know that the poles will be 
cancelled by the phase space integral dictated by \eq{eq:soft_compl_NLO} and 
by \eq{eq:coll_compl_NLO}, together with an analogous relation for the eikonal 
jet. It is then straightforward to define 
\beq
  I_n^{\one} \, \equiv \, \int d \Phi_{1} (k) \, K_\npo^{\one}(k) \, ,
\label{eq:I1_NLO}
\eeq
with
\beq
\label{Knpo1}
  K_\npo^{\one}(k) \, \equiv \,
  {\cal H}^{(0) \, \dagger}_n \, S_{n, \, 1}^{(0)}(k) \, {\cal H}^{(0) }_n +
  \sum_{i=1}^n {\cal H}^{(0) \, \dagger}_n \, 
  \Big( J_{i, \, 1}^{(0)}(k) - J_{i, \, \E, \, 1}^{(0)}(k)  \Big) \, 
  {\cal H}^{(0) }_n \, ,
\eeq
so that the combination $V_n \, + \, I^{\one}_n$ is finite in $d=4$ by construction, 
and the difference $R_{\npo} - K_{\npo}^{\one}$ is free of phase space 
divergences as required by the KLN theorem.

Clearly, at this stage \eq{Knpo1} defines only a {\it candidate} counterterm, in 
the language of \secn{SubtraNLO}, since the kinematics inside the matrix elements 
involved does not properly factorise on-shell, momentum conserving Born 
configurations. \eq{Knpo1} is however a good starting point to implement 
appropriate phase-space mappings: one can easily recognise the structure 
described at the end of \secn{SubtraNLO}, in \eq{candcount}, and furthermore, with 
an appropriate parametrisation, the phase-space measure $d \Phi_{1} (k)$ can 
be turned into $d \Phi_{\rm rad, 1}$, as needed for an exact implementation of 
subtraction. In essence, the factorised structure of virtual corrections has dictated 
the form of the real-radiation counterterms, with a transparent organisation of soft, 
collinear and soft-collinear contributions. As we will briefly see in the next section, 
this construction directly generalises to higher orders.


\subsection{The structure of infrared subtraction at NNLO}
\label{StruSub}

The complexity of the subtraction problem grows very steeply when moving from
NLO to NNLO and beyond, due to the increase in the number of overlapping and 
nested singular configuration, to the growth in complexity of soft and collinear 
splitting kernels, and to the need to perform intricate phase space and loop integrations, 
either analytically or numerically. These difficulties have been tackled with a variety 
of methods, and the construction of efficient subtraction algorithms at NNLO and 
beyond has been a very active field of research for nearly two decades. The 
literature is too vast to be comprehensively cited, but a sample of the techniques
that have been developed, adopting both the slicing and the subtraction philosophies,
together with some impressive phenomenological results, can be found in 
Refs.~\cite{Anastasiou:2003gr,Frixione:2004is,Somogyi:2005xz,GehrmannDeRidder:2005cm,
Catani:2007vq,Gehrmann-DeRidder:2007vsv,Gehrmann-DeRidder:2008qsl,
Catani:2009sm,Weinzierl:2009ms,Weinzierl:2009nz,Czakon:2010td,Boughezal:2011jf,
Czakon:2013goa,Czakon:2014oma,DelDuca:2015zqa,Boughezal:2015dva,Boughezal:2015dra,
Gaunt:2015pea,Cacciari:2015jma,Gehrmann-DeRidder:2015wbt,Bonciani:2015sha,
Czakon:2015owf,DelDuca:2016csb,DelDuca:2016ily,Sborlini:2016hat,Caola:2017dug,
Currie:2017eqf,Grazzini:2017mhc,Magnea:2018hab,Magnea:2018ebr,
Herzog:2018ily,Anastasiou:2018rib,Caola:2019nzf,Gehrmann-DeRidder:2019ibf,
Catani:2019hip,Caola:2019pfz,Czakon:2019tmo,Capatti:2019edf,DelDuca:2019ctm,
Chawdhry:2019bji,Capatti:2020ytd,Capatti:2020xjc,Anastasiou:2020sdt,Kallweit:2020gcp,
Magnea:2020trj,Chawdhry:2021hkp,Czakon:2021mjy}.  A more detailed discussion 
of recent developments, and a more complete list of references, are provided by 
the review~\cite{TorresBobadilla:2020ekr}.

Notwithstanding this remarkable variety of techniques, the NNLO IR subtraction 
problem cannot be considered solved, on a par with the NLO case. In our view,
a complete solution should satisfy a set of four requirements: complete {\it generality}
across all IR-safe observables with an arbitrary number of coloured
particles, exact {\it locality} of soft and collinear counterterms, {\it independence} of the 
results on external `slicing' parameters, and of course {\it computational efficiency};
to these, we would add the {\it analytic control} of all integrated counterterms, which
is bound to play an important role both for the speed of the procedure and for our
theoretical understanding of the problem. In light if these criteria, much work 
remains to be done at NNLO, even as the first developments and applications
at N$^3$LO begin to appear~\cite{Anastasiou:2015vya,Dreyer:2016oyx,Dulat:2017prg,
Mistlberger:2018etf,Currie:2018fgr,Cieri:2018oms,Dulat:2018bfe,Dreyer:2018qbw,
Gehrmann:2018odt,Duhr:2020seh,Duhr:2020sdp,Chen:2021isd,Chen:2021vtu}.

In this Section, we will address three issues concerning NNLO subtraction. First, 
we describe the structure of the problem, generalising the NLO discussion of 
\secn{SubtraNLO}; next, we will show how NNLO candidate counterterms can 
be identified by means of factorisation, following the prescription of \secn{counterterm};
finally, we will briefly consider the problem of characterising strongly-ordered
counterterms, which first arise at NNLO, by means of operator matrix elements.
Once again, we emphasise that our focus here is the connection between 
factorisation and subtraction, and not the analysis of specific algorithms.

At NNLO, the differential distribution for an IR-safe observable $X$ receives 
three distinct contribution, generalising \eq{pertO}: one must include the double-virtual 
corrections to the Born-level $n$-point matrix element, which we denote by $VV_{n}$, 
the one-loop corrections to the single-radiative matrix element, denoted by $RV_{\npo}$, 
and finally the double real-radiation contribution $RR_{\npt}$. All three contributions
present singularities, so they must be regularised and combined according to
\beq
  \frac{d \sigma_\NNLO}{d X} & = & \lim_{d \to 4} 
  \Bigg\{ \! \int d \Phi_n \, VV_n \, \delta_n (X) 
  + \! \int d \Phi_{n+1} \, 
  RV_\npo \, \delta_\npo (X) \nonumber \\ 
  && \hspace{15mm} + \int d \Phi_{n+2} \, RR_\npt \, \delta_\npt (X) 
  \Bigg\} \, .
\label{NNLOX}
\eeq 
All the elements appearing in \eq{NNLOX} can be expressed in terms of scattering 
amplitudes with the appropriate final state multiplicity as 
\beq
  RR_\npt \, = \, \left| {\cal A}_\npt^{(0)} \right|^2 \! , \quad
  VV_n  \, = \, \left| {\cal A}_n^{(1)} \right|^2 \! + 
  2 {\bf Re}  \left[ {\cal A}_n^{(0) \dagger} {\cal A}_n^{(2)} \right] ,\quad
  RV_\npo \, = \, 2 {\bf Re} \left[ {\cal A}_\npo^{(0) \dagger} \, 
  {\cal A}_\npo^{(1)} \right] . \quad
\label{defNNLO}
\eeq
The double-virtual matrix element features up to a quadruple pole in $\eps$, while 
the real-virtual correction contains up to a double pole in $\eps$, and up to two 
phase-space singularities. Finally, the double-real matrix element is finite in $d=4$, 
but is affected by up to four phase-space singularities. The explicit singularities of 
virtual origin have to be subtracted by means of local counterterms that can be 
defined by generalising the strategy introduced at NLO. 

On general grounds, one quickly understands that the structure of the subtraction 
procedure is much more intricate than was the case at NLO. In fact, the expression
for the fully subtracted distribution, which was given at NLO by \eq{subtNLO}, must
be generalised as
\beq
\label{subtNNLO} 
  \frac{d \sigma_\NNLO}{dX} \!\!\! & = &
  \!\!\!\! \int \! d \Phi_n
  \Big[ VV_n + I^{\two}_n + I^{\RV}_n \Big] \, \delta_n (X) \\
  & + &
  \!\!\!\! \int \! d \Phi_\npo \, \bigg[ 
  \Big( RV_{\npo} + I^{\one}_\npo \Big) \, \delta_\npo (X)
  \, - \, 
  \Big( K^{\RV}_\npo + I^{\otwo}_\npo \Big) \delta_n (X) \bigg] \nonumber \\
  & + &
  \!\!\!\! \int \!  d\Phi_\npt \, \bigg[ 
  RR_\npt \, \delta_\npt (X) \, - \,
  K^{\one}_\npt \, \delta_\npo (X) \, - \,
  \Big( K^{\two}_\npt - K^{\otwo}_\npt \Big) \, \delta_n (X) \bigg] \, , \nonumber
\eeq
involving four different local counterterm functions, and their integrals. To 
understand the structure of \eq{subtNNLO}, one can start by examining 
the last line. The double-real-radiation term $RR_\npt$ has a set of phase-space 
singularities when a {\it single} particle becomes unresolved: those are subtracted 
by the counterterm labelled $K^{\one}_\npt$. A second set of singularities 
arises when {\it two} particles become unresolved: those are subtracted by the 
counterterm $K^{\two}_\npt$. It is clear, however, that these two sets of 
singularities overlap: $K^{\one}_\npt$ will involve configurations where 
a second particle may become unresolved, and those are shared with 
$K^{\two}_\npt$. The counterterm $K^{\otwo}_\npt$ takes care of subtracting 
this double counting: one may intuitively understand this counterterm as the 
set of {\it strongly-ordered} soft or collinear configurations present in 
$K^{\two}_\npt$.

Turning to the second line of \eq{subtNNLO}, we note that $RV_\npo$ has phase-space
singularities when the single emitted particle becomes unresolved: these require a 
fourth local counterterm, which we label $K^{\RV}_\npo$. We must however still 
deal with the fact that $RV_\npo$ has explicit IR poles, which in turn implies that
such poles must appear also in the counterterm $K^{\RV}_\npo$. Explicit poles
in $RV_\npo$ will be cancelled by the phase-space integral of the single-unresolved 
counterterm $K^{\one}_\npt$, in a manner similar to what happens at NLO. This 
integrated counterterm, $I^{\one}_\npo$ in \eq{subtNNLO}, will however have its 
own set of phase-space singularities, arising when a second particle becomes 
unresolved. This problem can be solved by integrating the strongly-ordered
counterterm $K^{\otwo}_\npt$ over the phase space of the `most unresolved' 
particle, which yields the integrated counterterm $I^{\otwo}_\npo$. This integrated
counterterm must have poles matching those of $K^{\RV}_\npo$, and still retains 
the singular phase-space dependence on the second unresolved particle, matching 
that of $I^{\one}_\npo$. One sees that the second line in \eq{subtNNLO} involves 
a subtle balance of singularities:  both parentheses are free of IR poles, however 
phase-space singularities cancel between the first parenthesis and the second 
one. Furthermore, the pole cancellation between $RV_\npo$ and $I^{\one}_\npo$ 
is guaranteed by general theorems, but the definition of the local counterterms 
$K^{\RV}_\npo$ and $K^{\otwo}_\npt$ involves considerable freedom, and the 
choice must be fine-tuned in order not to spoil this delicate cancellation pattern.

To complete the subtraction procedure, one must still add to the terms enumerated 
so far the integrals of the double-unresolved counterterm $K^{\two}_\npt$ and of
the real-virtual counterterm $K^{\RV}_\npo$: we denote them by $I^{\two}_n$ 
and $I^{\RV}_n$, respectively. The general finiteness theorems for inclusive
cross sections guarantee that they will cancel the poles of the double-virtual
correction $VV_n$ in the first line of \eq{subtNNLO}.

Having established the general structure of the subtraction procedure at 
NNLO, one then needs to provide a precise definition of all the required 
counterterms. In what follows, we will summarise how a suitable set of definition
arises from factorisation, as was the case at NLO. Given the complexity of 
the problem, and the variety of singular limits involved, we will just give some 
selected examples, at the level of {\it candidate counterterms}, without addressing
the issue of phase-space parametrisations.

We first focus on the real-virtual correction. To organise its singularities, we 
can adapt the result obtained for the virtual matrix element at NLO, (see 
\eq{virtolo}) and implement it to include an extra hard radiation, substituting 
$n \rightarrow n+1$. This yields
\beq
  RV_\npo  \, = \,  
  {\cal H}^{(0) \, \dagger}_\npo \, S_{n+1, \, 0}^{(1)} \, {\cal H}^{(0)}_\npo
  \, + \, 
  \sum_{i = 1}^n 
  {\cal H}^{(0) \, \dagger}_\npo \, 
  \left( J^{(1)}_{i, \, 0}  - J_{i, \, \E, \, 0}^{(1)} \right) S_{n+1, \, 0}^{(0)} \,
  {\cal H}^{(0)}_\npo
  \, + \,  {\rm finite} \, .
\label{RVfact}
\eeq
In analogy with \eq{eq:I1_NLO} and \eq{Knpo1}, the finiteness of inclusive 
soft and collinear cross sections, embodied by \eq{eq:soft_compl_NLO} and 
\eq{eq:coll_compl_NLO}, suggests the definition of a candidate counterterm
for single-unresolved radiation as
\beq
\label{twoloct1}
  K_{n+2}^{\bf (1)}  \, = \, 
  {\cal H}^{(0) \,  \dagger}_\npo \, S_{\npo, \, 1}^{(0)} \, {\cal H}^{(0)}_\npo
  \, + \,
  \sum_{i = 1}^n 
  {\cal H}^{(0) \, \dagger}_\npo \Big( J^{(0)}_{i, \, 1} - J_{i, \, \E, \, 1}^{(0)} \Big)
  S_{\npo, \, 0}^{(0)} {\cal H}^{(0)}_\npo \, .
\eeq
With this definition, \eq{eq:soft_compl_NLO} and \eq{eq:coll_compl_NLO}
ensure that the integrated counterterm
\beq
  I^{\one}_\npo \, = \, \int d \Phi_1 (k) \, K^{\one}_\npt (k) \, ,
\eeq
cancels the IR poles of $RV_\npo$, so that the combination $RV_\npo + I^{\one}_\npo$
is finite in $d=4$. 

The construction of counterterms for double-virtual singularities is somewhat 
more involved: we summarise it below because it illustrates the simple and transparent
organisation of singularities which can be achieved by applying the factorisation 
approach. Expanding \eq{ampnfact} at NNLO, a number of contributions are 
generated, but they can easily be organised according to number and nature 
(soft or collinear) of the singular configurations involved in the process. The 
singular part of the double-virtual correction can be written as
\beq
\label{virtwolo}
  VV_{n} \bigg|_{\rm poles} = \, 
  VV_{n}^{(2 {\rm s})} 
  +  
  VV_{n}^{(1 {\rm s})} 
  +  
  \sum_{i = 1}^n 
  \bigg[
  VV_{n, \, i}^{(2 {\rm hc})} 
  +
  \! \sum_{j= i+1}^n \!
  VV_{n, \, ij}^{(2 {\rm hc})} 
  +  
  VV_{n, \, i}^{(1 {\rm hc},\, 1 {\rm s})} 
  +
  VV_{n, \, i}^{(1 {\rm hc})}  
  \bigg] \, ,
\eeq
where the superscripts characterise the nature of the singularity and the number
of virtual particles involved in it. Explicitly
\beq
\label{virtwolopieces}
  VV_{n}^{(2 {\rm s})} & = &  
  {\cal H}^{(0) \,  \dagger}_n \, S_{n, \, 0}^{(2)} \, {\cal H}^{(0)}_n \, , 
  \nonumber \\
  VV_{n}^{(1 {\rm s})}  & = & 
  {\cal H}^{(0) \,  \dagger}_n S_{n, \, 0}^{(1)} {\cal H}^{(1)}_n 
  \, + \,
  {\cal H}^{(1) \,  \dagger}_n S_{n, \, 0}^{(1)} \, {\cal H}^{(0)}_n  \, , 
  \nonumber \\
  VV_{n, \, i}^{(2 {\rm hc})} & = & 
  {\cal H}^{(0) \, \dagger}_n \bigg[ \!
  J^{(2)}_{i, \, 0} - J_{i, \, \E, \, 0}^{(2)} 
  \, - \, 
  J_{i, \, \E, \, 0}^{(1)} \left( J^{(1)}_{i, \, 0} - J_{i, \, \E, \, 0}^{(1)} \right) \!
  \bigg]
  S_{n, \, 0}^{(0)} \, {\cal H}^{(0)}_n \, , 
  \nonumber \\
  VV_{n, \, ij}^{(2 {\rm hc})} & = & 
  {\cal H}^{(0) \, \dagger}_n 
  \left( J^{(1)}_{i, \, 0} - J_{i, \, \E, \, 0}^{(1)} \right) \!
  \left( J^{(1)}_{j, \, 0} - J_{j, \, \E, \, 0}^{(1)} \right) 
  S_{n, \, 0}^{(0)} \, 
  {\cal H}^{(0)}_n \, , 
  \\
  VV_{n, \, i}^{({1\rm hc} ,\,  {1\rm s})} & = & {\cal H}^{(0) \,  \dagger}_n 
  \left( J^{(1)}_{i, \, 0} - J_{i, \, \E, \, 0}^{(1)} \right) S_{n, \, 0}^{(1)} \, 
  {\cal H}^{(0)}_n \, , 
  \nonumber \\
  VV_{n, \, i}^{(1{\rm hc})} & = &  
  {\cal H}^{(0) \,  \dagger}_n 
  \left( J^{(1)}_{i, \, 0} - J_{i, \, \E, \, 0}^{(1)} \right) S_{n, \, 0}^{(0)} \, 
  {\cal H}^{(1)}_n 
  \, + \,
  {\cal H}^{(1) \,  \dagger}_n 
  \left( J^{(1)}_{i, \, 0} - J_{i, \, \E, \, 0}^{(1)} \right) S_{n, \, 0}^{(0)} \, 
  {\cal H}^{(0)}_n \, . \nonumber
\eeq
Most of the contributions to \eq{virtwolopieces} have a transparent interpretation.
The first line collects poles due to two-loop soft exchanges, including soft-collinear 
ones, yielding singularities up to $1/\epsilon^4$; the second line describes one-loop
soft exchanges interfering with one-loop hard remainders, yielding poles up to
$1/\epsilon^2$. The last three lines collect hard collinear singularities: on the 
fourth line, a pair of independent one-loop hard collinear exchanges on lines $i$ 
and $j$, giving poles up to $1/\epsilon^2$; on the fifth line, a one-loop hard collinear 
exchange on line $i$, accompanied by a one-loop soft singularity, for a maximum
total singularity given by $1/\epsilon^3$; the last line gives single hard-collinear 
singularities interfering with finite one-loop remainders, with only $1/\epsilon$ 
poles.

\begin{figure}
  \centering
  {\includegraphics[scale=0.5]{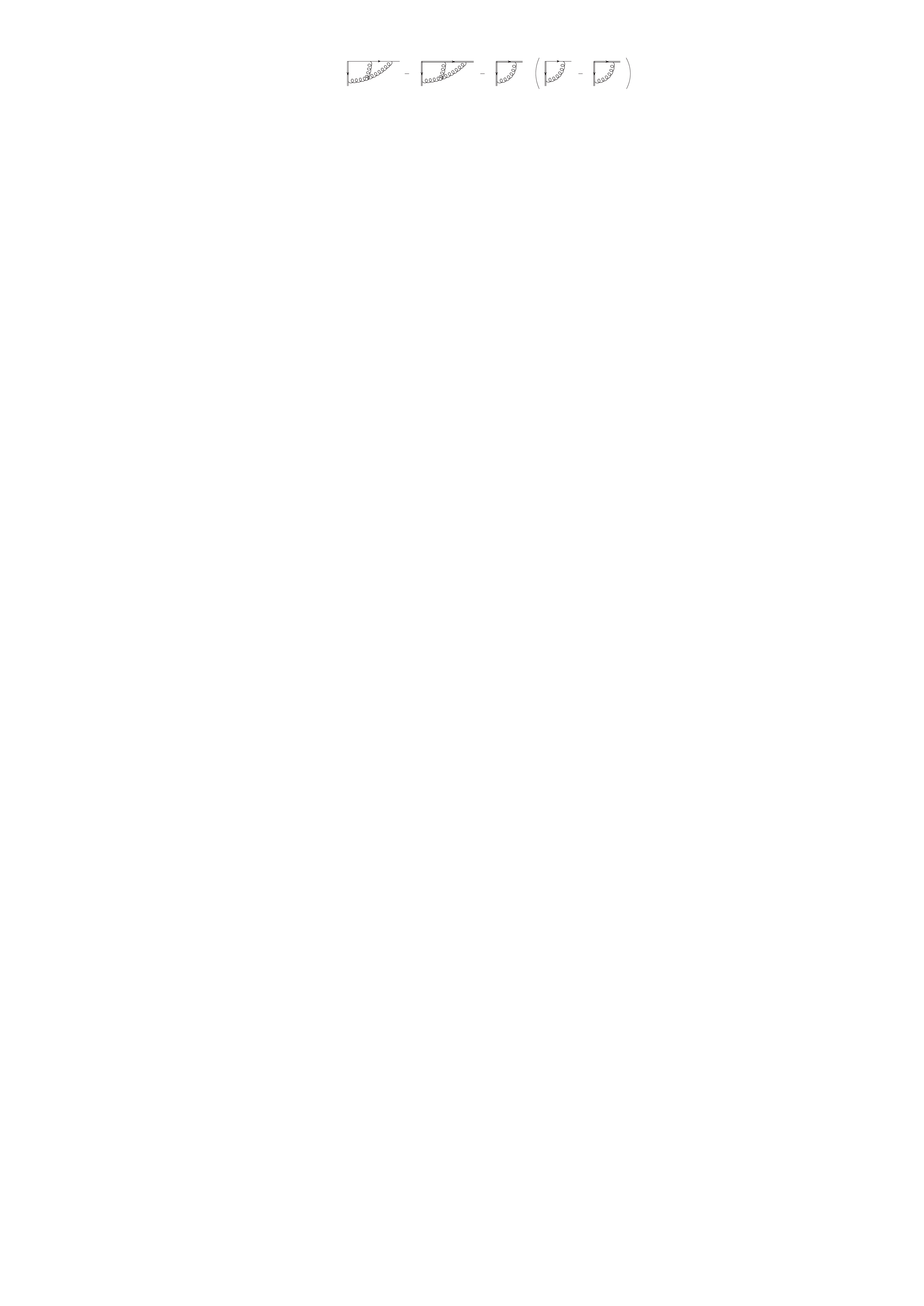}}
  \caption{Representative diagrams illustrating double-hard-collinear contributions 
  to the virtual two-loop amplitude along a single hard line.}
  \label{hard-coll}
\end{figure}

Perhaps the most interesting part of \eq{virtwolopieces} is the third line, describing
double hard collinear exchanges along line $i$, giving poles up to $1/\epsilon^2$:
the organisation of these singularities is non-trivial, and reflects the power of
factorisation. The terms contributing to $VV_{n, \, i}^{(2 {\rm hc})}$ are illustrated,
at amplitude level, in Fig.~\ref{hard-coll}: the first term is a two-loop contribution
to the full jet function, giving both collinear and soft poles; the second term subtracts 
soft-collinear contributions where both virtual partons are soft, as well as collinear.
This does not guarantee the complete cancellation of soft-collinear singularities, 
since they can also arise from configurations where only one virtual line is soft, 
while the other one is hard and collinear. These contributions are precisely reproduced 
by the third term: crucially, when one gluon is much softer that the other one, upon
summing over diagrams it will re-factorise from the two-loop jet, giving the product
of a one-loop eikonal jet times a one-loop hard-collinear combination.
Having at our disposal a full characterisation of double-virtual poles, we can 
now browse our list of local counterterms, given in \eq{softsigma} and \eq{qradjetsq},
and extract the combinations we need: we are guided by the finiteness conditions
\beq
  S_{n, 0}^{(2)} 
  \, + \, 
  \int d \Phi_1 (k) \, S_{n, \, 1}^{(1)}(k) 
  \, + \, 
  \int d \Phi_2 (k, k') \, S_{n, \, 2}^{(0)}(k, k') 
  & = & 
  {\rm finite} \, , \nonumber \\
    \int d \Phi_1 \, J^{(2)}_{i, \, 0} 
  \, + \, 
  \int d \Phi_2 (k) \, J_{i, \, 1}^{(1)}(k)
  \, + \, 
  \int d \Phi_3 (k, k') \, J_{i, \, 2}^{(0)}(k,k')
  & = & 
  {\rm finite} \, ,
\label{completeness2}
\eeq
(with a similar one for eikonal jets) which stem from the finiteness of \eq{complete} 
and \eq{compcoll}. Quite naturally, one-loop functions with single radiation are
assigned to the real-virtual counterterm $K_\npo^{\RV}$, while tree-level functions 
with double radiation are assigned to the double-unresolved counterterm 
$K_\npt^{({\bf 2})}$.

Using these criteria on \eq{virtwolopieces}, one finds that the double-unresolved 
counterterm in the $(\npt)$-particle phase space, $K^{\two}_\npt$, can be organised 
as
\beq
\label{twoloct2}
  K_{n+2}^{\bf (2)} \, = \, 
  K_{n+2}^{({\bf 2}, \, {\rm 2 s})} 
  \, + \, 
  \sum_{i = 1}^n
  \bigg[
  K_{n+2, \, i}^{({\bf 2}, \, \rm{2hc})} 
  \, + \, 
  \sum_{j = i+1}^n  K_{n+2, \, i j}^{({\bf 2}, \, \rm{2hc})} 
  \, + \,
  K_{\npt, \, i}^{({\bf 2}, \, \rm{1 hc, \, 1s})} \bigg] \, ,
\eeq
where the notation mimicks \eq{virtwolopieces}. Explicit expressions for all the 
different contributions in \eq{twoloct2} are too lengthy to list here, but they can be 
found in Ref.~\cite{Magnea:2018ebr}. To illustrate their structure, we just report 
here the double-soft contribution, and the double-hard-collinear contribution
arising from a single line. They are given by
\beq
  K^{({\bf 2}, \, {\rm 2s})}_\npt 
  & = & 
  {\cal H}^{(0) \,  \dagger}_n \, S_{n, \, 2}^{(0)} \, {\cal H}^{(0)}_n 
  \, , 
  \nonumber \\
  K^{({\bf 2}, \, {\rm 2hc})}_{\npt, \, i} 
  & = & 
  {\cal H}^{(0) \,  \dagger}_n
  \Big[ 
  J^{(0)}_{i, \, 2} - J_{i, \, \E, \, 2}^{(0)} 
  -   
  J_{i, \, \E, \, 1}^{(0)} \left( J^{(0)}_{i, \, 1} - J_{i, \, \E, \, 1}^{(0)} \right) \!
  \Big]
  S_{n, \, 0}^{(0)} \, {\cal H}^{(0)}_n \, .
\eeq
In the last line, we recognise the hard-collinear structure discussed above,
and illustrated in Fig.~\ref{hard-coll}, but this time applied to tree-level,
double-radiative contributions.

In a similar vein, the real-virtual counterterm $K_\npo^{\RV} $ has the structure
\beq
\label{twoloctRV}
  K_{n+1}^{\RV} \, = \,
  K_{n+1}^{({\bf  RV}, \, {\rm s})} 
  \, + \, 
  \sum_{i = 1}^n \bigg[
  K_{n+1, \, i}^{({\bf RV}, \, \rm{2hc})} 
  \, + \, 
  \! \sum_{j = i+1}^n\! 
  K_{n+1, \, i j}^{({\bf RV}, \, \rm{2hc})} 
  \, + \,
  K_{n+1, \, i}^{({\bf RV}, \, \rm{1hc , \, 1s})} 
  \, + \,
  K_{n+1, \, i}^{({\bf RV}, \, \rm{1hc})}  \bigg] \, ,
\eeq
where, this time, the notation `2hc' refers to a configuration with a hard collinear
real gluon paired with the hard collinear poles arising from the loop integration.
Once again we display only the soft component and the single-line hard-collinear 
component. They are
\beq
  K^{({\bf RV}, \, {\rm s})}_\npo 
  & = & 
  {\cal H}^{(0) \,  \dagger}_n \, S_{n, \, 1}^{(0)} \, {\cal H}^{(1)}_n 
  \, + \, 
  {\cal H}^{(1) \,  \dagger}_n \, S_{n, \, 1}^{(0)} \, {\cal H}^{(0)}_n 
  \, + \,
  {\cal H}^{(0) \,  \dagger}_n \, S_{n, \, 1}^{(1)} \, {\cal H}^{(0)}_n  \, ,  \\
  K^{({\bf RV}, \, {\rm 2hc})}_{\npo, \, i} 
  & = & 
  {\cal H}^{(0) \,  \dagger}_n
  \bigg[ J^{(1)}_{i, \, 1} - J_{i, \E, \, 1}^{(1)} -  
  \Big( J_{i, \, 0}^{(1)} -J^{(1)}_{i, \E, \, 0}\Big) J^{(0)}_{i, \E, \, 1} - 
  J_{i, \E, \, 0}^{(1)} \Big( J^{(0)}_{i, \, 1} -  J^{(0)}_{i, \E, \, 1} \Big) \bigg] 
  S_{n, \, 0}^{(0)} \, {\cal H}^{(0)}_n \, , \nonumber
\eeq
In the first line, we see that single soft radiation can be paired with the finite part
of a one-loop contribution, while the last term involves the one-loop correction
to the single-radiative soft function. In the second line, we again recognise a 
factorised structure similar to the one discussed in connection with \eq{virtwolopieces}:
the first term in square brackets combines a collinear loop with a collinear radiation,
including all soft regions; the second term subtracts the contribution where both the
virtual and the real gluon are soft; the third term subtracts the region where the
radiated gluon is soft, while the virtual one is hard and collinear, and finally the 
last term deals with the complementary region where the virtual gluon is soft,
but the radiated one is hard and collinear.

These examples show how a factorisation approach provides a set of transparent
expressions for candidate counterterms at NNLO, in a manner that can be
directly generalised to higher orders. To complete the picture, we must however
deal with the last remaining local counterterm, the $K^{\otwo}_\npt$ function,
which describes the overlap between single-unresolved and double-unresolved
configurations in the $(\npt)$-particle phase space. It is natural to define the 
strongly-ordered counterterm as the collection of the single-unresolved limits 
of $K^{\two}_\npt$: in this sense, one might do without an explicit definition
of $K^{\otwo}_\npt$ in terms  of operator matrix elements, since the relevant 
soft and collinear limits can be easily computed once $K^{\two}_\npt$ is known.
We have seen however, in \secn{counterterm}, that having a formal definition
of the counterterm as a matrix element allows to prove the cancellation of
divergences, to any order, in a straightforward way. It is therefore worthwhile 
to search for a general, formal definition of strongly-ordered local counterterms.

\begin{figure}
  \centering
  {\includegraphics[scale=0.6]{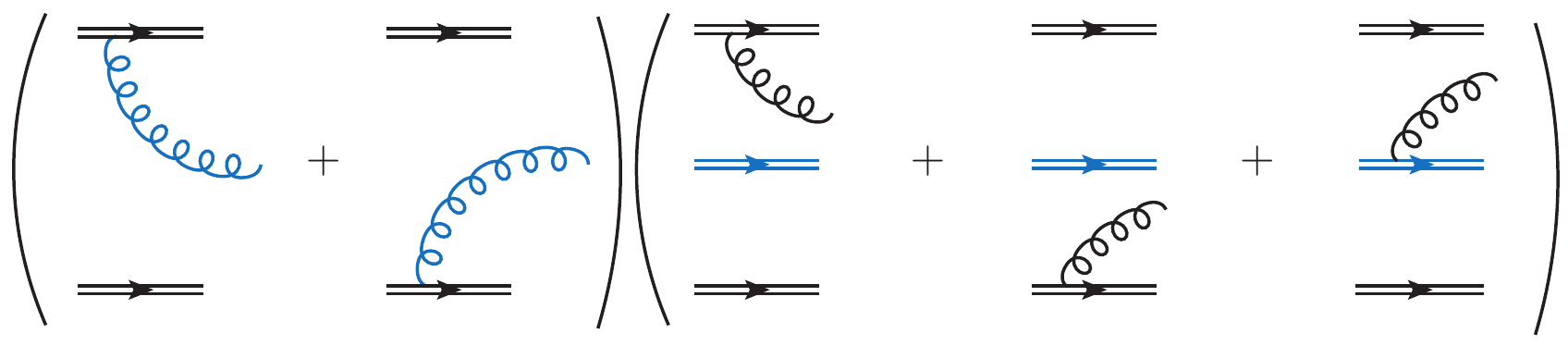}}
  \caption{A simple example of the re-factorisation of an eikonal form factor in
  a strongly-ordered limit, with $n=2$ Born-level particles.}
  \label{wilsonisation}
\end{figure}

Here, we will simply illustrate how such a definition can be constructed in the 
case of tree-level emissions of soft gluons. It is not difficult to envisage the 
structure of the result in this case: if two soft gluons are emitted from a system 
of $n$ hard particles, and one of the two gluons is much softer than the other, 
the softer gluon, at leading power, will effectively arise from a configuration
of $\npo$ hard particles, which will be represented by $\npo$ Wilson lines
in the soft approximation. This soft factorisation will leave behind a hard
matching function; however, since the exact result for the double soft emission
when both soft gluons scale in the same way is known, and is expressed
as a matrix element of $n$ Wilson lines, it is natural to guess that the hard
matching function for the softest emission will be given by the matrix element
for the emission of the harder gluon from the original $n$ Wilson lines. This
translates into the following tree-level ansatz for the strongly-ordered soft
function for double gluon emission from $n$ hard particles:
\beq
\label{softrads.o.}
  \Big[ {\cal S}^{(0)}_{n; \, 1, \, 1} \Big]^{a_i a_j}_{\{d_m e_m\}}
  \big(k_i, k_j; \{\beta_{m}\} \big) & \equiv &  
  \bra{k_j, a_j} \, 
  \Big[ \Phi_{\beta_i} (0,\infty) \Big]^{a_i b}
  \prod_{m = 1}^{n} \Big[ \Phi_{\beta_m} (0,\infty) \Big]_{d_m}^{\,\,\,\, c_m } 
  \ket{0} \\ \nonumber
  & & \hspace{1cm}
  \times \,
  \bra{k_i, b} \,  \prod_{m=1}^{n}
  \Big[ \Phi_{\beta_m} (0,\infty)  \Big]_{c_m e_m} \ket{0} \Big|_{\rm tree} 
   \\ \nonumber
  & = & \left[ {\cal S}^{(0)}_{n+1,1} \right]^{a_j, \, a_i b}_{\{d_m c_m\}}
  \big(k_j;  \beta_i, \{\beta_{m}\} \big) \, 
  \left[ {\cal S}^{(0)}_{n,1} \right]_{b, \, \{ c_m e_m \}} \! \big(k_i; \{\beta_{m}\} \big)\, ,
\eeq
where we wrote explicitly all colour indices, including the colour indices of the
two ordered soft gluons in the final state bra vectors. The physical interpretation 
of \eq{softrads.o.} is the following: in the limit $k_j \ll k_i$, the hierarchy between 
the two soft gluon momenta states that $k_i$ is soft with respect to the $n$ 
Born-level hard partons, but it is hard if compared with the softest radiation $k_j$. 
This means that gluon $j$ cannot resolve the different energy scales of the 
other $\npo$ partons, since their momenta are all much greater than $k_j$: 
thus, from the point of view of gluon $j$, gluon $i$ can be treated as another 
classical source of soft radiation, represented by its own Wilson line. According 
to this picture, in the first line of \eq{softrads.o.} the emission of gluon $j$ is 
modelled at leading power by a Wilson-line correlator, where one of the $n+1$ 
lines carries the quantum numbers of gluon $i$, including colour, and is directed 
along momentum $k_i$. Next, in the second line of \eq{softrads.o.}, the emission 
of gluon $i$ is described by a single-radiative soft function, involving the $n$ 
Born-level hard particles. This configuration is schematically represented in 
Fig.~\ref{wilsonisation}, in the simplest case with $n=2$. The blue curly line 
in the first bracket corresponds to gluon $i$, which behaves as a soft gluon 
when emitted from the Born particles, but acts as a hard particle, represented 
by the blue Wilson line in the second bracket, when gluon $j$ is emitted. 
Note that the colour structure of \eq{softrads.o.} is crucial: the first eikonal 
form factor acts as a colour operator on the second form factor, with the 
adjoint index $b$ of the Wilson line associated with gluon $i$ contracting 
with the index of the gluon state in the second line.
It is easy to verify that the definition in Eq.\eqref{softrads.o.} returns precisely
\beq
  \Big[{\cal S}^{(0)}_{n; \, 1, \, 1} \Big]^{a_i a_j}
  \big(k_i, k_j; \{\beta_{m}\} \big)=
  \epsilon^{* \, \mu_i} (k_i) \;
  \epsilon^{* \,  \mu_j} (k_j) \,
  \left[ J^{(0), \, \rm s.o.} \right]_{\mu_i \mu_j}^{a_i a_j} 
  \big(k_i, k_j; \{\beta_m\}\big) \, ,
\label{checksoftradso}
\eeq
where the strongly-ordered current on the {\it r.h.s.} was introduced in \eq{2sCGso}. 
The strongly-ordered double-soft counterterm $K^{\otwo , \, \rm s}_\npt$ is then
obtained by squaring \eq{softrads.o.}, in analogy to \eq{softsigma}. Clearly, the
re-factorisation in \eq{softrads.o.} is only the simplest example of an infinite tower
of similar expressions, when multiple soft gluons are emitted with different
hierarchical patterns. For example, it can be verified that formulas with the same 
structure as \eq{softrads.o.} correctly generate the strongly ordered limits for
triple and quadruple tree-level soft emissions, which can be checked with the 
complete limits computed in Ref.~\cite{Catani:2019nqv}. Furthermore, \eq{softrads.o.}
naturally lends itself to an all-order interpretation, where loop corrections can be
distributed between the two radiative soft functions on the {\it r.h.s}. Strongly-ordered 
collinear limits are somewhat more intricate, due to the need to perform spin sums,
and to the fact that the jet functions that we are employing contain collinear-finite
contributions: a similar tree-level pattern, however, emerges. We believe that
this kind of re-factorisation can shed light on the all-order structure of multiple
soft and collinear emissions, with significant applications for subtraction at high
orders, which are under study.


\section{Perspectives}
\label{Persp}

Our journey through the infrared comes to an end. The journey has been long --
it had to be, given the long history of the subject. Yet we regret that, limited by 
space and by the knowledge span of the authors, it has barely touched many 
interesting issues, and it has mentioned only in passing several important lines 
of enquiry. 

The first part of the journey has been an overview of history. We started in \secn{cataQED} 
by presenting the Bloch-Nordsieck theorem, which displays for the first time the 
mechanism for the cancellation of infrared divergences, and gives the first 
example of factorisation and exponentiation of real and virtual radiative corrections, 
in the context of QED. The physical origin of the problem -- the {\it mistake} that 
leads to the failure of the standard perturbative approach in the presence of 
massless particles -- was already correctly diagnosed at that time, as was the 
path to a general solution.

As we have seen, however, the infrared problem has been {\it solved many times},
from different conceptual angles, with increasing degrees of sophistication, 
and uncovering deeper layers of structure. In this respect, the history of the 
infrared problem parallels the history of the ultraviolet one, which also fruitfully
evolved through several levels of interpretation and many technical developments. 

The Bloch-Nordsieck approach reaches its maximal range with the KLN theorem, 
discussed in \secn{KLN}, which proves that the mechanism of real-virtual 
cancellation is completely general, and applies to any quantum-mechanical theory
containing states that are degenerate in energy. While this provides a solid foundation 
for all subsequent developments, it cannot be considered completely satisfactory,
since it applies only at the level of sufficiently inclusive cross sections, bypassing
the $S$-matrix, which historically has been -- and to this day remains -- a 
crucial quantity for most developments in quantum field theory. The first 
answer to this omission is the construction of a Hilbert space of coherent 
states, which leads to the formal definition of an infrared-finite $S$-matrix
both in abelian and in non-abelian theories, as reviewed in \secn{Coherent}.

The advent of non-abelian gauge theories posed novel problems at two different 
levels, and in turn led to a whole series of new developments that are far from 
exhausted. On the one hand, there are conceptual problems, arising directly at 
the non-perturbative level, at least in the case of confining gauge theories like 
QCD. Indeed, QCD has an added layer of mystery to it: quarks and gluons -- the 
perturbative degrees of freedom -- cannot be observed as asymptotic states, unlike 
their electroweak counterparts; they are confined. This drastic discrepancy between 
perturbative and non-perturbative asymptotic states leads to great difficulties both 
for a KLN approach -- since the sum over degenerate initial states containing 
short-distance degrees of freedom does not correspond directly to any physical 
observable -- and for the coherent state construction, which so far has been 
achieved only at the level of perturbation theory. On the other hand, there are 
technical and practical problems, increasingly related to the spectacular progress 
of accelerator experiments, now reaching extraordinary levels of precision even 
for intricate collider observables. A key technical problem is the unavoidable
presence of collinear divergences, directly related to the masslessness
of gauge bosons; a general practical problem is the development
of perturbative techniques to high enough orders, so as to match 
the precision reached by collider experiments.

Our current best answer to both the conceptual and the practical 
challenges posed by non-abelian gauge theories is the factorisation
program, both at the level of inclusive cross sections and at the
level of scattering amplitudes. Thus, most of our review has been 
devoted to the development of perturbative tools for factorisation, and 
to their concrete application to fixed-angle scattering amplitudes and to
real radiation of soft and collinear particles.

The physical ideas underpinning factorisation are as powerful as they are simple 
and intuitive. Processes taking place at very different length (or energy) scales 
must be decoupled to a large extent: thus for example we don't expect 
the size of the laboratory to play a role in elementary particle collisions, nor do 
we expect subatomic events to influence how rivers flow. Quantum mechanics 
challenges this simple picture, forcing us -- in principle -- to take into account 
all physical scales simultaneously. Recovering the factorised, semiclassical picture
with conventional perturbative tools is possible, but hard, since the contributions
of different physical scales are scattered across a plethora of Feynman diagrams, 
additively building up the observable at each perturbative order.

In the infrared domain, we have sketched the rather winding path that finally 
organises perturbative contributions to quantum amplitudes into a product of
factors, each responsible for phenomena occurring at a specific physical scale. 
After introducing some tools by means of one-loop examples in \secn{FinOrd}, 
we proceeded in \secn{AllOrd} by outlining an all-order diagrammatic 
analysis. Remarkably, the infinite taxonomy of Feynman diagrams can be tamed, 
and singular contributions located and organised. Once the physical scales 
relevant for a particular observable have been identified, one can proceed with 
the tools of effective field theories, which naturally lead to factorised expressions
for the observable. On the other hand, the same diagrammatic analysis that
succeeded in locating singular contributions motivates the construction of 
universal functions responsible for infrared enhancements, taking into account
the constraints of gauge invariance, as discussed in \secn{universal_fun}.

Once factorisation is achieved, be it in the infrared or in the ultraviolet, it provides
a precise mathematical expression for the idea of universality: that only limited
information can be transferred between physical scales separated by wide gaps.
In the ultraviolet, in the case of renormalizable theories, one finds that very 
high-energy degrees of freedom can influence low-energy scattering only by 
tuning coupling constants; in the infrared, one similarly finds that long-distance 
effects can be captured by a small set of functions, introduced here in \secn{FactoForm}, 
which can be defined independently of the particular scattering process under 
consideration.

Even more importantly, factorisation inevitably leads to the existence of evolution 
equations. Physical observables cannot depend on the artificial parameters that
theorists must introduce to define the relevant scale domains, whether they take 
the form of factorisation scales, separating ultraviolet from infrared phenomena, 
or factorisation vectors, separating collinear from wide-angle dynamics. This 
independence translates into evolution equations, discussed here in \secn{ResuForm}. 
Solving these equations, in turn, leads to a resummation of perturbation theory in 
the relevant kinematic domain. Universal quantities are then expressed in terms 
of anomalous dimensions, which naturally become the cornerstone of perturbative 
calculations. Already at the level of massless form factors, a prominent role is played 
in particular by the light-like cusp anomalous dimension $\gamma_K (\alpha_s)$, 
which governs the soft limit of virtual exchanges.

Moving on from form factors to multi-particle fixed-angle scattering amplitudes,
discussed in \secn{MultiPart}, one encounters a significant step, both conceptual 
and technical: soft anomalous dimensions become colour matrices, and their
high-order colour structure, which is highly non-trivial, becomes an important focus
of investigation. In \secn{DipFor} and in \secn{CompMatr} we studied the all-order 
structure of infrared anomalous dimensions matrices, and we introduced different 
methods for computing them, uncovering in the process remarkable links to
the Regge limit and to conformal field theories on the celestial sphere.

Factorisation and universality apply to real radiation too, as must be the case in 
order for the KLN theorem to work with the expected generality. The factorisation
of scattering amplitudes in soft and collinear limits, which is by itself a vast
field of research, has been briefly reviewed here in \secn{Real}. Even when 
the soft and collinear factorisation kernels are known to a given order, efficiently 
organising the KLN cancellation at high orders remains a challenge. In the 
remaining part of \secn{Subtra}, we have discussed modern approaches to 
tackle this {\it subtraction} problem. Our aim there was not to review existing 
methods, which were recently discussed in Ref.~\cite{TorresBobadilla:2020ekr};
rather, we tried to display the link between the factorisation of virtual corrections
and that of real emission, presenting a general strategy to define soft and
collinear local counterterms, which are universal, and constructed to manifestly
cancel virtual poles to any order. We also pointed out the significant technical
difficulties that must still be tackled, on the way to an efficient implementation,
even when the form of the local counterterms is known.

One may wonder at this point what the future holds, as the second century of
infrared studies lies just below the future horizon. There are certainly many open
lines of developments, as well as a few dreams.

A first obvious direction of further studies is dictated by the quest for precision
which characterises current high-energy research. Clearly, fully general and efficient
subtraction algorithms, working beyond NLO, will be needed in the upcoming phases 
of the LHC program, as well as for other present and future colliders. Given the
pace of present developments, and the resources that are devoted to this effort 
by many groups, it is to be fully expected that more than one such algorithm will
be made available at NNLO within the next few years. NNLO algorithms will then 
have to be fully integrated with event generators and parton showers at that accuracy, 
which in itself will be a non-trivial task. One may be somewhat bolder: since most
real-radiation kernels required for N$^3$LO calculations are already known, as is
the full set of three-loop infrared anomalous dimensions, it is to be expected that
a similar effort towards generality and efficiency will be applied at N$^3$LO. 
Along the way, many technical question will hopefully find answers: for example, 
concerning the all-order structure of soft and collinear local counterterms, including 
those for strongly-ordered configurations, and concerning the optimisation of 
the phase-space mappings required to match radiative and Born-level momentum
configurations. 

On the side of virtual corrections, the goal of having a full set of soft and collinear 
anomalous dimensions for massless scattering at the four-loop order appears 
reachable within a similar time scale. Achieving this result would have practical 
implications, allowing for soft gluon resummations (in the threshold, $p_T$ and 
high-energy domains) with impressive logarithmic accuracies, and permitting 
highly non-trivial studies of the convergence properties of resummed perturbation 
theory for complex collider processes. Furthermore, there are also interesting 
theoretical issues that start being accessible at such high orders. An important 
example is the role played by higher-order Casimir invariants, which appear for 
the first time at four loops, and subsequently build up their own perturbative 
expansion. The role of these complex colour structures in the factorisation 
program, and more generally in the dynamics of high-energy scattering, is
largely unexplored to this day.

Performing such high-order calculations for infrared divergences may require
new technical developments, on a par with those that have facilitated the 
calculation of multi-loop Feynman integrals in recent years. Empirically, current
techniques for the evaluation of Wilson-line correlators, which are based on 
Feynman diagrams, seem to suffer from the same drawbacks that have been
many times lamented for the case scattering amplitudes. Both colour and kinematic 
dependences of Wilson-line correlators display large-scale complexity in the  
intermediate stages of calculations, while the final results are remarkably simple
and elegant. This is certainly the case for the massless soft anomalous dimension
matrix at three loops, and for the massive correction at two loops. One feels that
the simplicity of the final result is at least partly unexplained, and obscured in the
various stages of the calculation by many unphysical dependences, such as the
gauge redundance and the reliance on different regulators to tame the intricate
singularity structure.

In this regard, Wilson line correlators that are relevant for the infrared limit differ
in two important ways from ordinary amplitudes. First, their integrands involve 
denominators linear in momenta: this is an important difference in view of applying
standard methods for the calculation of multi-loop Feynman integrals, such as
integration by parts and differential equations~\cite{Smirnov:2012gma}. While 
existing methods and codes can be generalised to include linear denominators, 
it is clear that they are not optimised for this purpose. A second aspect is the 
strong dependence of eikonal integrals on auxiliary soft and collinear regulators, 
such as the exponential regulator introduced in \eq{Phidef}: while the final results 
for anomalous dimensions are independent of the chosen regulators, intermediate 
steps in the calculation are heavily affected. Regularisation procedures that work 
well for the purposes of analytic calculations performed with pencil and paper at 
low orders are not necessarily optimal for automatic evaluation of high-order 
corrections using existing algorithms. Both these problems can probably be 
attacked by generalising to Wilson-line correlators some of the techniques 
that have been developed in recent years for the calculation of scattering 
amplitudes and form factors. In particular, generalised unitarity surely applies 
in the presence of Wilson lines~\cite{Laenen:2014jga,Laenen:2015jia},
albeit in a somewhat unconventional manner; similarly, general tools such as 
intersection theory~\cite{Mastrolia:2018uzb} are likely to be applicable also in this 
context; finally, a bootstrap approach, based on the identification of the function 
space comprising the correlators, followed by the implementation of symmetry, 
kinematic and dynamical constraints, has already been shown to be very powerful.

Moving on from technical aspects to more general and long-term prospects,
we believe that it is fair to say that infrared factorisation for real soft and collinear
radiation is less developed than is the case for virtual corrections to fixed-angle
scattering amplitudes. While general theorems exist, the detailed structure
of soft and collinear kernels at high orders becomes increasingly complex,
and only limited handles are available to organise and disentangle this complexity.
Clearly, to some extent this is an inevitable consequence of the fact that 
real radiation kernels are {\it differential} quantities, allowing for an intricate 
dependence on the momenta and quantum numbers of soft and collinear 
particles. On the other hand, the comparative simplicity of the singularity 
structure of virtual corrections should provide organising principles for real
radiation as well, given the general theorems on the cancellation of divergences.
At a minimum, as an effect of the KLN theorem, virtual corrections must provide
{\it sum rules} for real radiation kernels, whose phase-space integrals must 
cancel virtual poles: this viewpoint is at the basis of the organisation of 
subtraction counterterms discussed in \secn{counterterm}. One may, however, 
hope for a deeper understanding: for example, the exponentiation of virtual 
poles, which implies subtle correlations between results at different perturbative
orders, should be reflected in a similar organisation for real radiation corrections.
An interesting first step in this direction was taken in Ref.~\cite{Catani:2019nqv},
where a form of exponentiation of soft real radiation in terms of a generating 
functional was suggested. 

More generally, an important and largely unexplored issue is the connection 
between fixed-angle scattering and the limits in which strong hierarchies develop
between Mandelstam invariants, ultimately leading to singular behaviours. Soft 
and collinear configurations belong to this category, but also Regge and 
multi-Regge limits. Large logarithms that appear in these limits, endangering
the reliability of perturbation theory, are tightly connected with infrared 
enhancements: while this connection has been understood and exploited 
for a long time, it is likely that much more structure remains to be uncovered.

Finally, we turn to ultimate targets, beyond the range of predictable developments 
and into the sphere of speculations and perhaps dreams. While almost everything
we have written has been strictly perturbative, it must be noted that in the infrared 
domain one can typically derive {\it resummed} results which contain information
to all orders in perturbation theory. It is inevitable then to muse on the possibility
that, through an infrared portal, one might access truly non-perturbative information.
In a limited sense, this is certainly true and well known: indeed, as briefly discussed 
in the Introduction, summations of perturbation theory inevitably display the 
divergence of the perturbative expansion, and a detailed analysis of the divergent 
behaviour yields parametric information on non-perturbative corrections (see, for 
example,~\cite{Dasgupta:2001zv,Magnea:2002xt,Sterman:2004en}. On the other 
hand, it is perhaps reasonable to hope for much more detailed information. First 
of all, we note that infrared factorisation localises long-distance information in a 
set of universal functions which are given by matrix elements of local and non-local 
operators, and these matrix elements are in principle well defined in the full gauge 
theory, beyond the perturbative domain. These functions can then be studied with 
non-perturbative methods, at times with direct phenomenological applications, as 
was done for example in Ref.~\cite{Schlemmer:2021aij}. Furthermore, even when 
resumming infrared poles, which has been the focus of our review, one can uncover 
relations between matrix elements and anomalous dimensions which are expected 
to be valid beyond perturbation theory, as was the case in Ref.~\cite{Dixon:2008gr}, 
with the refinements later introduced in Ref.~\cite{Falcioni:2019nxk}. For special, 
highly symmetric gauge theories these relations can sometimes be tested 
non-perturbatively, as was done for example in Refs.~\cite{Alday:2007mf,
Alday:2009zf}. Note that the relations we are discussing connect
finite quantities, a prime example being \eq{ratfofasym}. 

For confining gauge theories, non-perturbative effects must ultimately entangle
colour and kinematic degrees of freedom at large distances, forcing the flow of 
soft radiation towards the formation of colour-singlet states. The first occurence
of such a colour-kinematic entanglement to all orders in perturbation theory takes 
place at the level of soft anomalous dimensions, as exemplified by \eq{GammaDip}
and \eq{Delta3}. Clearly, one sees no trace of confinement  at this level (indeed, 
the expressions we are discussing are common to confining gauge theories like
pure Yang-Mills, and to conformal gauge theories like ${\cal N} = 4$ super-Yang-Mills 
theory). One can however hope that the language of soft anomalous dimensions 
could be used to access and study non-perturbative aspects as well. The novel 
viewpoint introduced in Refs.~\cite{Strominger:2013lka} is perhaps the most 
promising in this regard, provided it can be consistently extended and understood
to include quantum corrections and collinear dynamics. As suggested by the 
discussion in \secn{CeleVi}, based on Ref.~\cite{Magnea:2021fvy}, this may 
indeed be possible. In the dipole approximation, the picture painted by the celestial 
approach is that of a Coulomb gas of two-dimensional coloured particles with
pairwise interactions on the celestial sphere; furthermore, we already know that 
this picture will receive corrections involving multi-particle interactions on the 
sphere, which could lead to clustering in appropriate circumstances. Most 
importantly, the two-dimensional theory on the sphere, once properly identified,
could be studied with non-perturbative tools: if it turned out to be locally conformal
invariant, which seems likely, it could well be solvable. This would provide a
completely new and remarkable window on the long-distance behaviour of 
non-abelian gauge theories.


\section*{Acknowledgements}
\label{Ackno}

L.M. thanks the Ministry of Human Resource Development (MHRD), Government of India
and IIT Hyderabad for kind hospitality and the opportunity to present the lectures
which formed the basis of the present review. AT and NA would like thank Sourav 
Pal for collaboration on projects that have been included in this review, for useful 
discussions, and help with one of the figures. AT would also like to thank the University 
of Turin and INFN Turin for warm hospitality during the course of this work. CSS would like to thank Calum Milloy for valuable comments on  \secn{Subtra}. AT and LM 
would like to thank MHRD Government of India, for the GIAN grant (171008M01), 
{\it The Infrared Structure of Perturbative Gauge Theories}, and for the SPARC grant  
SPARC/2018-2019/P578/SL, {\it Perturbative QCD for Precision Physics at the LHC}, 
which were crucial to the completion of the review. This work was partially supported 
by the Italian Ministry of University and Research (MIUR), grant PRIN 20172LNEEZ.
We also thank Eric Laenen, Paolo Torrielli and Leonardo Vernazza for useful suggestions 
to improve the manuscript.



\bibliographystyle{utphys.bst}

\bibliography{Review_Biblio}

\end{document}